\documentclass[a4paper,11pt]{article}
\usepackage{jcappub}
\pdfoutput=1
\usepackage[utf8]{inputenc}
\usepackage[T1]{fontenc}
\usepackage{braket}
\usepackage{graphicx}
\usepackage{epsfig}
\usepackage{amsmath}
\usepackage{slashed}
\usepackage{amsfonts} 
\usepackage{float} 
\usepackage{amssymb}
\usepackage{xcolor}
\usepackage{pbox}
\usepackage{subfigure}
\usepackage{enumitem}
\usepackage{array, multirow,makecell}
\usepackage{hyperref}
\usepackage{tabularx}
\usepackage{titlesec}
\usepackage{mathrsfs,mathtools}
\usepackage[normalem]{ulem}
\usepackage{natbib}
\usepackage{array,multirow,makecell}
\usepackage{lipsum}
\usepackage{comment}
\usepackage[section]{placeins}
\usepackage{cancel}
\usepackage[normalem]{ulem}
\usepackage{bm}
\usepackage{booktabs}
\usepackage{indentfirst}
\usepackage[usestackEOL]{stackengine}
\strutlongstacks{T}
\def\l{\lambda}

\def\th{\theta}

\newcommand{\ba}{\begin{array}}
	\newcommand{\ea}{\end{array}}
\newcommand{\ovr}{\overline}
\def\lsim{\raise0.3ex\hbox{$\;<$\kern-0.75em\raise-1.1ex\hbox{$\sim\;$}}}
\def\gsim{\raise0.3ex\hbox{$\;>$\kern-0.75em\raise-1.1ex\hbox{$\sim\;$}}}
\def\be{\begin{equation}}
	\def\ee{\end{equation}}
\def\bea{\begin{eqnarray}}
	\def\eea{\end{eqnarray}}
\def\nn{\nonumber}
\newcommand{\beq}{\begin{equation}}
	\newcommand{\eeq}{\end{equation}}
\newcommand{\beqn}{\begin{eqnarray}}
	\newcommand{\eeqn}{\end{eqnarray}}

\def\lsim{\raise0.3ex\hbox{$\;<$\kern-0.75em\raise-1.1ex\hbox{$\sim\;$}}}
\def\gsim{\raise0.3ex\hbox{$\;>$\kern-0.75em\raise-1.1ex\hbox{$\sim\;$}}}
\def\be{\begin{equation}}
	\def\ee{\end{equation}}
\def\bea{\begin{eqnarray}}
	\def\eea{\end{eqnarray}}
\def\nn{\nonumber}
\title{%\boldmath 
Prospecting bipartite Dark Matter through Gravitational Waves}

\author[a]{Pankaj Borah,}
\author[a]{Pradipta Ghosh,}
\author[b]{and Abhijit Kumar Saha}
\affiliation[a]{Department of Physics, Indian Institute of Technology Delhi, Hauz Khas, New Delhi 110 016, India}
\affiliation[b]{Institute of Physics, Sachivalaya Marg, Bhubaneswar, 751005, India}
\emailAdd{Pankaj.Borah@physics.iitd.ac.in}
\emailAdd{tphyspg@physics.iitd.ac.in}
\emailAdd{abhijit.saha@iopb.res.in}

\abstract{We explore the gravitational wave probes of a two-component dark matter framework, consisting of an $SU(2)_L$ triplet scalar and a Standard Model singlet fermion. The triplet scalar dark matter typically remains underabundant in the region below $\sim 1.9$ TeV, due to the strong $SU(2)_L$ gauge mediated interactions. We introduce a second dark matter component, an $SU(2)_L$ singlet vector-like Dirac fermion, to address this deficit in the dark matter relic abundance within a sub-TeV range. A key aspect of the proposed setup is the potential dark matter inter-conversion between the two components, which impacts the dark matter freeze-out dynamics and relic density of individual dark matter components. In such a scenario, we examine the properties of electroweak phase transition and identify the regions of parameter space that exhibit strong first-order phase transition. We estimate the resulting gravitational wave spectrum and its detectability, which could be probed through the conventional power-law-integrated sensitivity limits and the recently proposed peak-integrated sensitivity curves. Our analysis reveals that a novel region of the model's parameter space, compatible with dark matter observables, can generate a detectable gravitational wave spectrum, observable by upcoming space-based gravitational wave detectors such as LISA, BBO, DECIGO, and DECIGOcorr, while also offering complementary detection prospects in the dark matter and collider experiments. 
\\
}

\begin{document} 
\maketitle	
%%%%%%%%%%%%%%%%%%%%%%%%%%%%%%%%%%%%%%%%%%
%%%%%%%%%%%%%%%%%%%%%%%%%%%%%%%%%%%%%%%%%%%%%%%%%%%%%%%%%%%%%%%%%%%%%%%%%%%%%%%%%%%
\section{Introduction}\label{sec:intro}
%%%%%%%%%%%%%%%%%%%%%%%%%%%%%%%%%%%%%%%%%%%%%%%%%%%%%%%%%%%%%%%%%%%%%%%%%%%%%%%%%%%
The existence of dark matter (DM) in the Universe is well supported by observations from cosmological experiments such as the WMAP \cite{WMAP:2012fli,WMAP:2012nax} and Planck \cite{Planck:2018vyg}. Extending the Standard Model (SM) framework is essential to incorporate an experimentally viable DM candidate. A plausible but minimal particle physics DM model involves a Weakly Interacting Massive Particle (WIMP)~\cite{Steigman:1984ac}, which interacts with the SM particles through the weak force, see Refs. \cite{Arcadi:2017kky,Arcadi:2024ukq} for extensive reviews. The simplest DM model features an SM singlet scalar DM interacting with the SM via Higgs portal \cite{Silveira:1985rk,McDonald:1993ex,Burgess:2000yq,Barger:2008jx,Gonderinger:2009jp,Guo:2010hq}. However, this simplest beyond the SM (BSM) scenario faces limitations from collider searches, as it relies primarily on mono-$X$ signatures with missing energy, where $X$ represents a $jet$ \cite{CMS:2021far,ATLAS:2021kxv}, $W^\pm$ \cite{ATLAS:2020fgc,CMS:2021mjq}, $Z$ \cite{ATLAS:2018nda,CMS:2020ulv,ATLAS:2021gcn}, $\gamma$ \cite{CMS:2018ffd,ATLAS:2020uiq} or Higgs boson \cite{CMS:2019ykj,ATLAS:2021jbf,ATLAS:2021shl}. These signals, resulting from initial state radiation, are often overwhelmed by substantial background noise from the SM. Introducing a higher multiplet in the dark sector, including charged components, provides more promising opportunities for the future collider experiments, although it also faces stricter constraints from the DM detection experiments. A well-known example is the inert $SU(2)_L$ doublet model (IDM) \cite{Deshpande:1977rw,Ma:2006km,Barbieri:2006dq,LopezHonorez:2006gr,Hambye:2009pw} or the inert $SU(2)_L$ triplet model (ITM) \cite{Ross:1975fq,Cheng:1980qt,Gunion:1989ci,Cirelli:2005uq,Cirelli:2007xd,FileviezPerez:2008bj,Araki:2011hm}. However, due to strong $SU(2)_L$ gauge interactions, single-component IDMs or ITMs are highly constrained, leading to a ``desert region'' \cite{Cirelli:2005uq,Hambye:2009pw} where DM masses between $\approx\, 80-550$ GeV (for IDM) \cite{Cirelli:2005uq,LopezHonorez:2010tb} and up to $\sim 1.9$ TeV \cite{Cirelli:2005uq, Fischer:2011zz,Araki:2011hm, Khan:2016sxm} (for ITM) are excluded due to the DM underabundance.

Thus, a framework incorporating a sub-TeV $SU(2)_L$ multiplet DM typically requires an additional DM candidate to compensate for the deficit in total relic abundance, ensuring agreement with the observational data. Multipartite DM models \cite{Berezhiani:1989fp,Berezhiani:1990sy,Boehm:2003ha,Cirelli:2005uq,Cao:2007fy,Hambye:2009pw,Profumo:2009tb,Aoki:2012ub,Biswas:2013nn,Gu:2013iy,Aoki:2013gzs,
Kajiyama:2013rla,Bian:2013wna,Bhattacharya:2013hva,Geng:2013nda,Esch:2014jpa,Dienes:2014via,
Geng:2014dea,Bian:2014cja,Bhattacharya:2016ysw,DiFranzo:2016uzc,Aoki:2016glu,DuttaBanik:2016jzv,Borah:2017xgm,Herrero-Garcia:2017vrl,Pandey:2017quk,Ahmed:2017dbb,DuttaBanik:2018emv,Aoki:2018gjf,Chakraborti:2018lso,Bernal:2018aon,
Poulin:2018kap,YaserAyazi:2018lrv,Herrero-Garcia:2018qnz,Bhattacharya:2018cgx,Elahi:2019jeo,Borah:2019epq,Bhattacharya:2019fgs,Biswas:2019ygr,
Nanda:2019nqy,Yaguna:2019cvp,Becker:2019xzq,Berman:2020kko,Belanger:2020hyh,VanDong:2020bkg,Khalil:2020syr,DuttaBanik:2020jrj,Hernandez-Sanchez:2020aop,
Chakrabarty:2021kmr,DiazSaez:2021pmg,Yaguna:2021vhb,Yaguna:2021rds,Ho:2022erb,Das:2022oyx,Bhattacharya:2022wtr,PandaX:2022djq,BasiBeneito:2022qxd,Costa:2022lpy,Bhattacharya:2022qck,Hosseini:2023qwu,WileyDeal:2023trg,Alguero:2023zol,Ayazi:2024mdb,Qi:2024uiz,Boto:2024tzp,DuttaBanik:2024vro,Mescia:2024rki, Khan:2024biq,Kumar:2024xfb,Nagao:2024hit,Qi:2024zkr,Agudelo:2024luc} \footnote{Interested readers can look at Ref.~\cite{Ahmed:2017dbb} where different plausible bipartite configurations, e.g., multi-scalar DM, multi-fermion DM, multi-vector DM, scalar-fermion DM, scalar-vector DM, vector-fermion DM, etc., are given with references.}, in general, are advantageous in evading the strong direct detection (DD), e.g., XENON1T \cite{XENON:2019gfn}, XENOXnT \cite{XENON:2023cxc} LUX-ZEPLIN (LZ)-2022 \cite{LZ:2022lsv} updated as LZ-2024 \footnote{Presents the combined WS2022 and WS2024 analysis for $280$ live days.} \cite{LZCollaboration:2024lux}, DARWIN \cite{DARWIN:2016hyl}, etc., and indirect detection (ID), for example, the Earth-based set-ups like H.E.S.S. \cite{Abdalla:2016olq,Abdallah:2016ygi,HESS:2015cda,Abramowski:2014tra,Abramowski:2011hc}, CTA \cite{CTA:2015yxo,CTAConsortium:2010umy} or satellite-based experiments like FERMI-LAT \cite{Fermi-LAT:2016uux,Fermi-LAT:2015kyq,Fermi-LAT:2015att,Ackermann:2015tah,Ackermann:2013yva,Ackermann:2013uma,
Abramowski:2012au,Ackermann:2012nb,Fermi-LAT:2011vow,Abdo:2010ex}, AMS \cite{AMS:2021nhj,AMS:2002yni}, etc., bounds, as the detection cross-sections are normalized by the relative abundances of each component \cite{Cao:2007fy,Aoki:2012ub}. 
In this work, we propose a bipartite, i.e., two-component DM framework where the dark sector consists of an $SU(2)_L$ triplet hyperchrageless scalar $(\bm{T})$ and a SM gauge-singlet Dirac fermion $(\psi)$. Additionally, another real SM singlet scalar $(S)$ is introduced, serving as a portal between the singlet fermion DM and the SM particles and facilitating inter-conversion \cite{Liu:2011aa,Belanger:2011ww} between the triplet and singlet DM components, essential to evade the relic density bound.

Furthermore, extending the SM with a singlet scalar can aid in realizing a strong first-order phase transition (SFOPT), in the electroweak (EW)\footnote{In literature, a strong first-order electroweak phase transition (EWPT) is often also abbreviated as SFOEWPT.} sector \cite{Anderson:1991zb,Dine:1990fj,Dine:1991ck,Espinosa:1993bs,Ahriche:2007jp,Profumo:2007wc,Espinosa:2011ax,Profumo:2014opa,Chen:2017qcz}, (see also Ref.~\cite{Mazumdar:2018dfl,Athron:2023xlk} for a review of the plausible PT) along the directions of both the SM Higgs and the singlet fields. An SFOPT along the $SU(2)_L$ field direction is a key prerequisite for the electroweak baryogenesis (EWBG). Interested readers may look at Refs. \cite{Cohen:1993nk,Trodden:1998ym,Riotto:1998bt,Riotto:1999yt,Quiros:1999jp,Dine:2003ax,Cline:2006ts,Morrissey:2012db} for review of the EWBG. Additionally, during the phase transition (PT) \cite{Witten:1984rs}\footnote{See Ref. \cite{Weir:2017wfa} for a brief review.}, plausible emission of gravitational waves (GWs) \cite{Hogan:1986qda,Caprini:2009fx,Caprini:2015zlo}\footnote{The appearance of the GWs depends on bubble collision \cite{Turner:1990rc,Kosowsky:1991ua,Kosowsky:1992rz}, turbulence in primordial plasma \cite{Kamionkowski:1993fg,Kosowsky:2001xp,Dolgov:2002ra,Gogoberidze:2007an,Caprini:2009yp}, and sonic waves \cite{Hindmarsh:2013xza,Hindmarsh:2015qta,Hindmarsh:2016lnk}.}, could be detected by space-based GW interferometers, such as, SKA \cite{Janssen:2014dka}, $\mu$-Ares \cite{Sesana:2019vho}, LISA \cite{eLISA:2013xep,LISA:2017pwj}, BBO \cite{Crowder:2005nr,Corbin:2005ny,Harry:2006fi}, DECIGO, U-DECIGO, and U-DECIGO-corr \cite{Kudoh:2005as, Yagi:2011wg,Kawamura:2020pcg}, AEDGE \cite{AEDGE:2019nxb}, AION \cite{Badurina:2019hst}, CE \cite{LIGOScientific:2016wof}, ET \cite{Hild:2008ng}, future upgrades of LVK \cite{KAGRA:2021kbb, Jiang:2022uxp, LIGOScientific:2022sts}, etc. In the chosen framework, the singlet scalar plays a dual role: (i) it is crucial for the DM phenomenology, as mentioned earlier, and (ii) it helps to achieve an SFOPT, potentially generating observable GWs, thereby improving the discovery prospects of the DM model alongside the conventional spin-independent (SI) DD and collider experiments. This two-fold role of the scalar singlet makes the present study more involved by establishing a distinct correlation between the dark sector and the physics of EWPT.

%%%%%%
Extensions beyond the SM with a singlet scalar (or pseudoscalar) and a singlet fermion DM have been investigated in the literature, see, for example, Refs. \cite{Kim:2008pp,Baek:2011aa,Daikoku:2011mq,Baek:2012uj,Aoki:2012ub,Bhattacharya:2013hva,Bhattacharya:2013asa,Bae:2013hma,Aoki:2014lha,Esch:2014jpa,DuttaBanik:2016jzv,Khan:2017ygl,YaserAyazi:2018lrv}, mainly in the context of the DM phenomenology, although a connection with the non-zero neutrino masses and mixing has also been explored in some of the works. These bipartite SM-singlet scalar-fermion DM setups rely on scalar portal interactions, similar to the chosen framework. Besides, some of these can also yield detectable GW signals, following an SFOPT \cite{Fairbairn:2013uta,Li:2014wia,Beniwal:2018hyi,Ayazi:2024mdb}. A similar analysis with an $SU(2)_L$ triplet scalar would, however, appear rather challenging to accommodate simultaneously the correct DM physics and a detectable GW signature aided by an SFOPT, due to the $SU(2)_L$ interaction, which is very constraining for the DM sector. A few analyses \cite{Bandyopadhyay:2021ipw,Cao:2022ocg,Jangid:2023jya}, nevertheless, exist where an $SU(2)_L$ triplet scalar of various hypercharge assignations, together with another SM singlet scalar, can accommodate the correct DM phenomenology together with an SFOPT-assisted GWs. A concurrent analysis of an $SU(2)_L$ triplet scalar DM, together with an SM-singlet fermion DM, was not addressed before which we plan to explore in this work. Here, one can easily house the correct DM physics, including the desert region, along with SFOPT-driven detectable GWs at the cost of a few additional inputs.
Certain input parameters, as detailed later, play a key role in the DM phenomenology while exerting minimal influence on PT dynamics. This distinct feature alleviates the challenges of reconciling the DM phenomenology with a successful SFOPT. Likewise, some crucial coupling parameters driving PT dynamics have a limited impact on the DM relic abundance, DD, and ID prospects. However, as will be discussed later, we also identify certain model parameters and their correlations that have a non-trivial influence on both the DM and PT dynamics, shaping their interplay. For other plausible variants of extended SM frameworks, that can simultaneously accommodate DM physics, SFOPT and yield detectable GW signatures, see Refs.~\cite{Cline:1996mga,Schwaller:2015tja,Soni:2016yes,Chao:2017vrq,Beniwal:2017eik,Huang:2017kzu,Baker:2017zwx,Flauger:2017ged,Hashino:2018zsi,Croon:2018erz,
Okada:2018xdh,Baldes:2018nel,Fujikura:2018duw,Baldes:2018emh,Ellis:2018mja,Madge:2018gfl,Bian:2018bxr,Shajiee:2018jdq,Breitbach:2018ddu,Kannike:2019wsn,YaserAyazi:2019caf,Fairbairn:2019xog,Hasegawa:2019amx,Paul:2019pgt,
Athron:2019teq,Helmboldt:2019pan,Mohamadnejad:2019vzg,Dunsky:2019upk,Alves:2019igs,Addazi:2019dqt,Alanne:2019bsm,Baker:2019ndr,Kannike:2019mzk,Barman:2019oda,Bian:2019szo,Ghosh:2020ipy,Han:2020ekm,Chao:2020adk,Pandey:2020hoq,Alanne:2020jwx,Wang:2020wrk,Ertas:2021xeh,Asadi:2021pwo,
Ahmadvand:2021vxs,Mohamadnejad:2021tke,Chatterjee:2022pxf,Costa:2022oaa,Dent:2022bcd,
Goncalves:2022wbp,Biondini:2022ggt,Wang:2022akn,Morgante:2022zvc,Ghosh:2022fzp,Khoze:2022nyt,Arcadi:2022lpp,Abe:2023zja,Han:2023olf,Goncalves:2023svb,Arcadi:2023lwc,Hosseini:2023qwu,Kanemura:2023jiw,Gao:2023djs,Pasechnik:2023hwv,Bringmann:2023iuz,Gu:2023fqm,Benincasa:2023vyp,Dichtl:2023xqd,Bhattacharya:2023kws,Xu:2023lkf,Abe:2023yte,Ghoshal:2024gai,Ayazi:2024mdb,Ghosh:2024ing}.
%%%%%%

In this work, we proceed with a minimalistic framework consisting of a hypercharge-neutral $SU(2)_L$ triplet, a vector $SU(2)_L$ singlet fermion, and a real singlet scalar, and explore the DM phenomenology in detail. The interaction of the triplet DM with the SM is primarily driven by the $SU(2)_L$ gauge coupling, while the singlet fermion DM interacts with the SM through the Higgs portal, enabled by the mixing between the singlet scalar and the SM Higgs doublet. Notably, the scalar singlet also induces inter-conversion between the two DM components, significantly influencing the relic density of the heavier DM candidate. We then thoroughly investigate the singlet scalar’s impact on the EWPT along the SM Higgs direction. The other fields, including the DM candidates, contribute to the EWPT at the one-loop level in the total scalar potential, relevant at temperature, $T=0$, and also for $T>0$. The effective dimension of the relevant field space for the EWPT analysis in our case is two since the triplet DM does not acquire a non-zero vacuum expectation value (VEV) at $T=0$ and we do not consider any transition along the triplet DM at $T > 0$, however, it can modify the total effective potential through loop corrections, as already mentioned. We highlight the regions of parameter space that allow an SFOPT along the SM Higgs direction favouring EWBG while maintaining consistency with the DM relic abundance, SI DD limits from experiments like the XENON1T, LZ-2022, LZ-2024 and DARWIN, collider searches, besides the other relevant theoretical and experimental constraints. Finally, we discuss the prospects of detecting signals within the aforesaid experimentally feasible parameter space in the upcoming GW interferometers such as LISA, BBO, DECIGO, and DECIGOcorr, highlighting the complementarity of GW observations with the DM and collider experiments.

The paper is structured as follows: after introducing the chosen bipartite DM model in Sec. \ref{sec:Model}, a detailed depiction of the various possible theoretical and experimental bounds are given in Sec. \ref{sec:constraints}. The findings of our numerical analyses of the dark sector are detailed in Sec. \ref{sec:DM-pheno}, following a schematic as well as a dedicated parameter scan. The distinctive features of the EWPT and GWs, together with the plausible correlations with the DM sector, are investigated in Sec. \ref{sec:EWPT-GW}. This section also contains an in-depth scrutiny of the GW detection prospects, in light of the conventional as well as a recently proposed improved technique.  Sec. \ref{sec:conclu} houses the summary and conclusion of our analyses. The relevant complex details are relegated to the appendices.
%%%%%%%%%%%%%%%%%%%%%%%%%%%%%%%%%%%%%%%%%%%%%%%%%%%%%%%%%%%%%%%%%%%%%%%%%%%%%%%%%%%
\section{The Model}\label{sec:Model}
%%%%%%%%%%%%%%%%%%%%%%%%%%%%%%%%%%%%%%%%%%%%%%%%%%%%%%%%%%%%%%%%%%%%%%%%%%%%%%%%%%%
  
In this analysis, the SM framework is extended with an $SU(2)_L$ triplet scalar $\bm{T}$ having zero hypercharge \cite{Ross:1975fq,Gunion:1989ci}, an SM singlet real scalar $S$, and another SM singlet vector-like Dirac Fermion $\psi$. A discrete symmetry $\mathbb{Z}_2\times \mathbb{Z}'_2$ is also introduced for these BSM states to ensure the stability of the plausible DM candidates. Together with the SM states, the charge assignments for these BSM states are depicted in Table \ref{tab:discrete-charge}.
%%%%%%%%%%%%%%%%%%%%%%%%%%%%%%%%%%%%%%%%%
\begin{table}[h!]
	\begin{center}
		%\centering
		\begin{tabular}{cccccc}
			%\hline
			\hline
			\vspace{0.1cm}
			Field & $SU(3)_C$& $SU(2)_L$ & $U(1)_Y$ & $\mathbb{Z}_2$ & $\mathbb{Z}'_2$\\
            \hline
            $S$ & $1$ &$1$ & $0$ & $+$ & $+$ \\
            \hline
            $\bm{T}$ &$1$ & $3$ & $0$ & $-$ & $+$ \\
			\hline
			$\psi$ & $1$ & $1$ & $0$ & $+$ & $-$ \\
			\hline
		\end{tabular}
		\caption{Charge assignments for the BSM states under the chosen $\mathbb{Z}_2\times \mathbb{Z}'_2$ symmetry together with the SM charges from $SU(3)_C \times SU(2)_L\times U(1)_Y$. All the SM states are trivially charged under the $\mathbb{Z}_2\times \mathbb{Z}'_2$ symmetry, just like the $S$.}
		\label{tab:discrete-charge}
	\end{center}
\end{table}\\
%%%%%%%%%%%%%%%%%%%%%%%%%%%%%%%%%%%%%%%%%
The Lagrangian density for the scalar sector of the SM is now modified as
%%%%%%%%%%%%%%%%%%%%%%%%%%%%%%%%%%%%%%%%%
\beq\label{eq:Lag:total-scalar}
\mathscr{L}_{(H,S,{\bm{T}})} = |D^\mu H|^2 + (D^\mu S)^2 + {\rm{Tr}}|D^\mu {\bm{T}}|^2  -
V(H,S,{\bm{T}}),
\eeq
%%%%%%%%%%%%%%%%%%%%%%%%%%%%%%%%%%%%%%%%%
here,  $H$ represents the SM Higgs doublet and the various $D^\mu$s are written using the charge assignments shown in Table \ref{tab:discrete-charge}. The renormalizable scalar potential $V(H,S,{\bm{T}})$, invariant under the imposed $\mathbb{Z}_2\times \mathbb{Z}'_2$ symmetry is written as,
%%%%%%%%%%%%%%%%%%%%%%%%%%%%%%%%%%%%%%%%%
\beq\label{eq:pot:total-scalar}
V(H,S,{\bm{T}}) = V_H + V_S + V_{\bm{T}} + V_{\rm int},
\eeq
%%%%%%%%%%%%%%%%%%%%%%%%%%%%%%%%%%%%%%%%%
where different parts of the $V(H,S,{\bm{T}})$ are given by,
%%%%%%%%%%%%%%%%%%%%%%%%%%%%%%%%%%%%%%%%
\bea\label{eq:pot:different-part}
	V_H &&= -\mu_H^2 H^{\dagger} H + \l_H (H^{\dagger} H)^2,\nn\\
	V_S &&= -\frac{\mu_S^2}{2} S^2 - \frac{1}{3} \mu_3 S^3 + \frac{\l_S}{4} S^4,\nn\\
	V_{\bm{T}} &&= -\frac{\mu_{\bm{T}}^2}{2} {\rm Tr}[{\bm{T}}^{\dagger} {\bm{T}}] + \frac{\l_{\bm{T}}}{4} \Big({\rm Tr}[{\bm{T}}^{\dagger} {\bm{T}}]\Big)^2,
\eea
%%%%%%%%%%%%%%%%%%%%%%%%%%%%%%%%%%%%%%%%
and, the interaction potential $V_{\rm {int}}$ is expressed as,
%%%%%%%%%%%%%%%%%%%%%%%%%%%%%%%%%%%%%%%%%
\bea\label{eq:pot:interaction-potential}
V_{\rm int} &=& \mu_{HS} S (H^{\dagger} H) + \frac{\l_{SH}}{2} S^2 (H^{\dagger} H)  + \frac{\mu_{ST}}{2} S\, {\rm Tr}[\bm{T}^{\dagger} \bm{T}] + \frac{\l_{ST}}{4} S^2\, {\rm Tr}[\bm{T}^{\dagger} \bm{T}]\nn\\
&&+ \frac{\l_{HT}}{2} (H^{\dagger} H) {\rm Tr}[\bm{T}^{\dagger} \bm{T}].
\eea
%%%%%%%%%%%%%%%%%%%%%%%%%%%%%%%%%%%%%%%%
The fermionic Lagrangian density is written as follows:
\begin{align}{\label{eq:Lag:Fermion}}
	\mathscr{L}_{\psi} = \ovr{\psi} (i \slashed{\partial} - \mu_{\psi})\psi - g_S \ovr{\psi} \psi S.
\end{align}
%%%%%%%%%%%%%%%%%%%%%%%%%%%%%%%%%%%%%%%%%%%%%%%%%%%%%%%%%%%%%%%%%%%%%%%%%%%%%%%%%
It is evident from Eq. (\ref{eq:pot:interaction-potential}) and Eq. (\ref{eq:Lag:Fermion}) that the $S$ field acts like the bridge between the two DM sectors
comprising of ${\bm{T}}$ and $\psi$. The presence of a $S$ field facilitates conversion between the two DM species without including non-renormalizable interactions (suitably scaled to four dimensions) like $\ovr{\psi}\psi {\rm{Tr}}[{\bm{T}}^\dagger{\bm{T}}]$. Non-trivial $SU(2)_L$ charge for ${\bm{T}}$ (see Table \ref{tab:discrete-charge}) allows ${\bm{T}}$ to interact with the SM states through Higgs
and gauge bosons whereas $S$, being an $SU(2)_L$ singlet, can interact with the SM states only via Higgs boson, as depicted in Eq.~(\ref{eq:pot:interaction-potential}).

One should note that both $\mathscr{L}_\psi, \, \mathscr{L}_{(H,S,{\bm{T}})}$
are invariant under the $SU(3)_C\times SU(2)_L\times U(1)_Y \times \mathbb{Z}_2\times \mathbb{Z}'_2$. Thus, interactions like $H^{\dagger} {\bm{T}} H,\, S\, (H^{\dagger} {\bm{T}} H)$, etc., (see, for example, Ref.~\cite{FileviezPerez:2008bj}) that are invariant under the SM gauge groups but not for $\mathbb{Z}_2\times \mathbb{Z}'_2$ are forbidden. A term linear in $S$ is admissible in 
$V_S$ by all symmetries. Nevertheless, this can be absorbed by performing 
a constant shift in $S$ by redefining $\mu_H^2,\,\mu_S^2,\,\mu_3,\mu_{HS},\mu_{ST},\,g_S$. We have assumed that these parameters are defined after a constant shift.

The field components of $V(H,\, S,\,{\bm{T}})$ (see Eq. (\ref{eq:pot:total-scalar})), i.e., $H,\, S,\, {\bm{T}}$, once the Electro-Weak Symmetry Breaking (EWSB) occurs and the neutral CP-even components of fields acquire VEVs, will mix amongst to produce mass eigenstates, depending on their charge-parity (CP) arguments. We will continue this discussion in the next subsection with further details.

%%%%%%%%%%%%%%%%%%%%%%%%%%%%%%%%%%%%%%%%%%%%%%%%%%%%%%%%%%%%%%%%%%%%%%%%%%%%%%%%%
\subsection{ Masses and mixing in the scalar sector}\label{subsec:scalarmassmixing}
%%%%%%%%%%%%%%%%%%%%%%%%%%%%%%%%%%%%%%%%%%%%%%%%%%%%%%%%%%%%%%%%%%%%%%%%%%%%%%%%%
   The different scalar fields $H,\, S,\,{\bm{T}}$, in terms of the constituents, after the EWSB, are written as,
%%%%%%%%%%%%%%%%%%%%%%%%%%%%%%%%%%%%%%%
\beq\label{eq:scalar-field-basis}
H = \frac{1}{\sqrt{2}}
\begin{pmatrix}
	~G^{+}~ \\
	v+ ~h + i G^0~
\end{pmatrix},\quad
{\bm{T}} =
\frac{1}{\sqrt{2}}\begin{pmatrix}
	~T^0~ & ~ -\sqrt{2}T^+ ~ \\
	~-\sqrt{2}T^- ~ & ~ -T^0~
\end{pmatrix},\quad
S = v_s + s,
\eeq
%%%%%%%%%%%%%%%%%%%%%%%%%%%%%%%%%%%%%%%%
where $h,\, s,\, T^0$ are the CP-even neutral scalar fields, $G^+, G^0$ are the charged and neutral Goldstone bosons, and $T^\pm$ denote the charged scalar fields. The quantities $v=246~{\rm GeV},\,v_s$  represent VEVs for fields $H,\,S$, respectively. The $\mathbb{Z}_2\times \mathbb{Z}'_2$ charges for the BSM states, as shown in Table \ref{tab:discrete-charge}, guide us to identify $T^0,\,\psi$, the lightest electrically neutral states odd under $\mathbb{Z}_2,\,\mathbb{Z}'_2$, respectively, as the feasible DM candidates. Thus, after the EWSB, a vanishing VEV is assigned for $T^0$ \cite{Cirelli:2005uq,Cirelli:2007xd} which, besides $\l_{\bm {T}}>0$ to remain bounded from below, also demands $\mu^2_{\bm{T}} <0$ (see Eq. (\ref{eq:pot:different-part})).
%%%%%%%%%%%%%%%%%%%%%%%%%%%%%%%%%%%%%%%
%\bea\label{eq:scalar-field-unitary-gauge}
%H &= \frac{1}{\sqrt{2}}
%\begin{pmatrix}
%	~0~ \\
%	~h + v~
%\end{pmatrix},\quad
%S = s + v_s,
%\eea
%%%%%%%%%%%%%%%%%%%%%%%%%%%%%%%%%%%%%%%%
The fact $v_s\neq 0$ triggers mixing between states which, in the ${h,\,s}$ basis, yields a squared mass matrix
%%%%%%%%%%%%%%%%%%%%%%%%%%%%%%%%%%%%%%%%
\bea%\label{eq:scalar-mass-HS}
\label{eq:scalar-mass-HS-new}
\mathcal{M}^2 &=
%\begin{pmatrix}
%	~-\mu_H^2 + 3 \l_H v^2 +\mu_{HS} v_s + \frac{1}{2} \l_{SH} v_s^2~ & ~\mu_{HS} v+ \l_{SH} v v_s ~ \\
%	~\mu_{HS} v + \l_{SH} v v_s ~ & ~ -\mu_S^2 -2 \mu_3 v_s + 3 \l_S v_s^2 + \frac{1}{2} \l_{SH} v^2~
%\end{pmatrix},
\begin{pmatrix}
~\mathcal{M}_{hh}^2 ~ & ~\mathcal{M}_{hs}^2 ~ \\
	~\mathcal{M}_{sh}^2 ~ & ~\mathcal{M}_{ss}^2
\end{pmatrix},
\eea
%%%%%%%%%%%%%%%%%%%%%%%%%%%%%%%%%%%%%%%%
with
%%%%%%%%%%%%%%%%%%%%%%%%%%%%%%%%%%%%%%%%
\bea
\label{eq:mass-elements}
\mathcal{M}_{hh}^2 &=& ~2 \l_H v^2,\,\quad \mathcal{M}_{ss}^2=-\mu_3 v_s + 2 \l_S v_s^2 - \frac{\mu_{HS}}{2 v_s} v^2,\,\nn\\
\mathcal{M}_{hs}^2 &=& \mathcal{M}_{sh}^2 = \mu_{HS} v+\l_{SH} v v_s.
\eea
%%%%%%%%%%%%%%%%%%%%%%%%%%%%%%%%%%%%%%%%%%
These elements are derived after using the tadpole equation for $H~(S)$
to replace $\mu^2_H~(\mu^2_S)$ with $\l_H,\, \mu_{HS},\, \l_{SH}, v,\, v_s~(\mu_3,\, \l_S,\, \mu_{HS},\, \l_{SH}, v,\, v_s)$.
%%%%%%%%%%%%%%%%%%%%%%%%%%%%%%%%%%%%%%%
%%%%%%%%%%%%%%%%%%%%%%%%%%%%%%%%%%%%%%%%
The squared mass elements for the triplet scalar are,
%%%%%%%%%%%%%%%%%%%%%%%%%%%%%%%%%%%%%%%
\bea\label{eq:mass-triplet}
m_{T^0, T^{\pm}}^2 &= - \mu_{\bm{T}}^2 + \mu_{ST} v_s + \frac{1}{2} \left(\l_{HT} v^2 + \l_{ST} v_s^2 \right).
\eea
%%%%%%%%%%%%%%%%%%%%%%%%%%%%%%%%%%%%%%%
This tree-level degeneracy between $m_{T^0},\,m_{T^\pm}$ is lifted at the one-loop level \cite{Cirelli:2005uq, Cirelli:2009uv}, leading to a mass splitting $\Delta m = m_{T^\pm}-m_{T^0} \approx 166\, {\rm MeV}$ \cite{Cirelli:2005uq} when $m_{T^0} \gg m_W,\,m_Z$, i.e., the SM $W^\pm, \, Z^0$ masses. This small mass gap\footnote{The quoted number, i.e., $166$ MeV would change by a few MeVs once second-order corrections are also implemented \cite{Ibe:2012sx}. This information would appear resourceful when we address the collider constraints in subsection \ref{subsec:constraints:disapperaing-charged-track}.} assures that $T^0$ remains the lightest triplet state and justifies, as already stated, why it is considered as one of the two DM candidates.

The squared mass matrix as shown in Eq. (\ref{eq:scalar-mass-HS-new}) needs to be diagonalized to procure mass eigenstates $h_1,\, h_2$ with masses $m_{h_1},\, m_{h_2}$, respectively. This is mathematically represented as:
%%%%%%%%%%%%%%%%%%%%%%%%%%%%%%%%%%%%%%%%
\bea\label{eq:hsmixtheta}
\begin{pmatrix}
	~ h_1 ~ \\
	~ h_2 ~
\end{pmatrix}
&=
\begin{pmatrix}
\cos\theta~ & \sin\theta~ \\
	-\sin\theta ~ & ~\cos\theta
\end{pmatrix}
\begin{pmatrix}
	~ h ~ \\
	~ s ~
\end{pmatrix},
\eea
%%%%%%%%%%%%%%%%%%%%%%%%%%%%%%%%%%%%%%%%
where the mixing angle $\theta$ and the physical masses $m_{h_1},\, m_{h_2}$ are connected to the entries of $\mathcal{M}^2$ (see Eq. (\ref{eq:scalar-mass-HS-new})) as:
%%%%%%%%%%%%%%%%%%%%%%%%%%%%%%%%%%%%%%%%
\bea\label{eq:mass-eigen-HS}
\tan2\th &=& \frac{2 \mathcal{M}_{hs}^2}{\mathcal{M}_{ss}^2 - \mathcal{M}_{hh}^2},\nn\\
m_{h_1}^2 &=& \mathcal{M}_{hh}^2 \cos^2\th + \mathcal{M}_{ss}^2 \sin^2\th -\mathcal{M}_{hs}^2 \sin2\th, \nn\\
m_{h_2}^2 &=& \mathcal{M}_{hh}^2 \sin^2\th + \mathcal{M}_{ss}^2 \cos^2\th + \mathcal{M}_{hs}^2 \sin2\th,
\eea
%%%%%%%%%%%%%%%%%%%%%%%%%%%%%%%%%%%%%%%%%
%%%%%%%%%%%%%%%%%%%%%%%%%%%%%%%%%%%%%%%
where we have considered $m^2_{h_1}$ to be the lighter eigenvalue. Now one can use Eq. (\ref{eq:mass-elements}) to get the following relations
%%%%%%%%%%%%%%%%%%%%%%%%%%%%%%%%%%%%%%%%
\bea\label{eq:model-param-relation}
\l_H &&= \frac{\mathcal{M}_{hh}^2}{2 v^2} = \frac{m_{h_1}^2 \cos^2\th + m_{h_2}^2 \sin^2\th}{2 v^2},\\
\label{eq:muHS}
\mu_{HS} &&= -\frac{2 v_s}{v^2} \left( \mathcal{M}_{ss}^2 + \mu_3 v_s - 2 \l_S v_s^2 \right)\nn\\
&&= -\frac{2 v_s}{v^2} \left( m_{h_1}^2 \sin^2\th + m_{h_2}^2 \cos^2\th + \mu_3 v_s - 2 \l_S v_s^2 \right),\\
\label{eq:lambdaSH}
\l_{SH} &&= \frac{\mathcal{M}_{hs}^2}{v v_s} - \frac{\mu_{HS}}{v_s} \nn \\
&& = \frac{\big( m_{h_2}^2 - m_{h_1}^2 \big)\sin2\theta}{2 v v_s} -\frac{2}{v^2} \left( m_{h_1}^2 \sin^2\th + m_{h_2}^2 \cos^2\th + \mu_3 v_s - 2 \l_S v_s^2 \right).
\eea
%%%%%%%%%%%%%%%%%%%%%%%%%%%%%%%%%%%%%%%%
Finally, after the EWSB, the fermionic DM Lagrangian density in Eq.~(\ref{eq:Lag:Fermion}) becomes
%%%%%%%%%%%%%%%%%%%%%%%%%%%%%%%%%%%%%%%%
\begin{equation}\label{eq:Lag:final-fermion}
	\mathscr{L}_{\psi} = \ovr{\psi} (i\slashed{\partial} - m_\psi) \psi - g_S \ovr{\psi} \psi s,
\end{equation}
%%%%%%%%%%%%%%%%%%%%%%%%%%%%%%%%%%%%%%%%
where 
%%%%%%%%%%%%%%%%%%%%%%%%%%%%%%%%%%%%%%%%
\begin{equation}\label{eq:fermionDm-mass}
	m_\psi = \mu_\psi + g_S v_s
\end{equation}
%%%%%%%%%%%%%%%%%%%%%%%%%%%%%%%%%%%%%%%%
is the physical mass of the fermionic DM. 
%%%%%%%%%%%%%%%%%%%%%%%%%%%%%%%%%%%%%%%%

At this juncture, it is crucial to identify the {\it{independent input}} parameters that are relevant to the planned phenomenological analyses. First, we consider the scalar sector as shown in Eq. (\ref{eq:pot:total-scalar}). The potential $V(H,S,{\bm{T}})$ accommodates twelve parameters, namely $\mu_H,\, \l_H,\, \mu_S$, $\mu_3,\, \l_S$, $\mu_{\bm{T}}, \l_{\bm{T}},\, \mu_{HS},\, \l_{SH}$,   $\mu_{ST},\, \l_{ST}$, and $\l_{HT}$, as depicted by Eq. (\ref{eq:pot:different-part}) and Eq. (\ref{eq:pot:interaction-potential}). Simultaneous use of the tadpole equations, Eq. (\ref{eq:mass-triplet}), and Eqs. (\ref{eq:model-param-relation}) - (\ref{eq:lambdaSH}) help us the recast those twelve inputs as 
$m_{h_1},\, \sin\theta, m_{h_2}$, $v,\, v_s,\, \mu_3, \l_S$, $m_{T^0}, \l_{\bm{T}}, \mu_{ST},\, \l_{ST}$, and $\l_{HT}$. Now in the SM, $v=246$ GeV and for this study, we identify $h_1$ as the SM-like Higgs which gives $m_{h_1} = 125.20 \pm 0.11$ GeV 
\cite{ParticleDataGroup:2024cfk}, following the LHC measurements \cite{ATLAS:2023oaq,CMS:2020xrn}. Hence, effectively one ended up with ten free inputs for the scalar sector. A similar analysis, using Eq. (\ref{eq:Lag:Fermion}) and Eq. (\ref{eq:fermionDm-mass}), helps us to identify $m_\psi, \, g_S$ as the free inputs for the fermionic sector. Combining the above two pieces, one gets twelve free inputs for phenomenological analyses as will be addressed further in the next section.
%%%%%%%%%%%%%%%%%%%%%%%%%%%%%%%%%%%%%%%%
\beq\label{eq:model-parameter}
\big\{m_{h_2},\, m_{T^0},\,m_{\psi},\, v_s,\, \mu_3,\,\mu_{ST},\, \sin\theta,\,\l_S,\, g_S,\, \lambda_{HT},\, \lambda_{ST},\, \lambda_{\bm{T}}\big\}.
\eeq
%%%%%%%%%%%%%%%%%%%%%%%%%%%%%%%%%%%%%%%%%
%%%%%%%%%%%%%%%%%%%%%%%%%%%%%%%%%%%%%%%%%%%%%%%%%%%%%%%%%%%%%%%%%%%%%%%%%%%%%%%%%%
\section{Theoretical and experimental constraints}\label{sec:constraints}
%%%%%%%%%%%%%%%%%%%%%%%%%%%%%%%%%%%%%%%%%%%%%%%%%%%%%%%%%%%%%%%%%%%%%%%%%%%%%%%%%%
  
We have identified independent inputs in Eq. (\ref{eq:model-parameter}) which contains physical masses of the three BSM states, namely $m_{h_2},\, m_{T^0},\, m_\psi$, VEV for the singlet state $v_s$, mixing angle $(\sin\theta)$ between the two scalars $h_1,\, h_2$, Yukawa coupling $(g_S)$ for the fermionic DM $\psi$, and a few  other parameters, namely,
$\mu_3,\, \mu_{ST},\,\l_S,\, \l_{\bm{T}},\, \l_{HT},\, \l_{ST}$. We will use these inputs 
to scan the model parameter space, as will be addressed in subsection \ref{subsec:scan}.
However, to fix scan ranges we need to consider first the relevant theoretical and experimental constraints on these inputs which we plan to address subsequently.
%%%%%%%%%%%%%%%%%%%%%%%%%%%%%%%%%%%%%%%%
%\beq\label{eq:model-parameter}
%\big\{m_{h_2},\, m_{T^0},\,m_{\psi},\, v_s,\, \mu_3,\,\mu_{ST},\, \sin\theta,\,\l_S,\, g_S,\, \lambda_{HT},\, \lambda_{ST},\, \lambda_T\big\}
%\eeq
%%%%%%%%%%%%%%%%%%%%%%%%%%%%%%%%%%%%%%%%%

%%%%%%%%%%%%%%%%%%%%%%%%%%%%%%%%%%%%%%%%%%%%%
\subsection{Vacuum stability and perturbativity constraints:}
\label{subsec:constraints:vacuum-constraints}
%%%%%%%%%%%%%%%%%%%%%%%%%%%%%%%%%%%%%%%%%%%%%
  
On the theoretical frontier, first, we need to address the constraints arising from the need for vacuum stability and perturbative unitarity. The former demands stability of $V(H,\,S,\,{\bm{T}})$ (see Eq.~(\ref{eq:pot:total-scalar})). Following Ref. \cite{Kannike:2012pe} for the co-positivity conditions of vacuum stability, we can write:
%%%%%%%%%%%%%%%%%%%%%%%%%%%%%%%%%%%%%%%%%%%%
\bea
\label{eq:constraints:stability}
\l_{H}, \l_S, \l_{\bm{T}} \geq 0, \quad \l_{SH} \geq -2 \sqrt{\l_H \l_S}, \quad \l_{HT} \geq -2 \sqrt{\l_H \l_{\bm{T}}}, \quad \l_{ST} \geq -2 \sqrt{\l_S \l_{\bm{T}}}\,.
\eea
%%%%%%%%%%%%%%%%%%%%%%%%%%%%%%%%%%%%%%%%%%%%

A perturbative theory expects certain upper bounds for a few model parameters
such that higher-order effects appear sub-dominant compared to the tree/lower-order analyses. The same can be verified by employing the renormalization group equations (RGE)
for these parameters till the Planck scale (${\rm M_{Pl}}\sim 10^{19}$ GeV). The resultant 
bounds on the relevant parameters can be written as
%%%%%%%%%%%%%%%%%%%%%%%%%%%%%%%%%%%%%%%%%%%
\bea
\label{eq:constraints:perturbativity}
|\l_H|,\,|\l_S|,\, |\l_{\bm{T}}|, \, |\l_{SH}|,\, |\l_{HT}|,\, |\l_{ST}| \leq 4\pi, \quad |g_i|, |y_{ij}| \leq \sqrt{4\pi}, 
\eea
%%%%%%%%%%%%%%%%%%%%%%%%%%%%%%%%%%%%%%%%%%%
where $g_i$ represents the SM gauge couplings and $y_{ij}$ denotes the SM Yukawa couplings.
%%%%%%%%%%%%%%%%%%%%%%%%%%%%%%%%%%%%%%%%%%%%%%%%%%%%%%%%%%%%%%%%%%%%%%%%%%%%%%%%%
\subsection{Tree-level unitarity constraints}
\label{subsec:unitarity}
%%%%%%%%%%%%%%%%%%%%%%%%%%%%%%%%%%%%%%%%%%%%%%%%%%%%%%%%%%%%%%%%%%%%%%%%%%%%%%%%%
  
Besides Eq.\,(\ref{eq:constraints:perturbativity}), it is also necessary to consider constraints from the perturbative unitarity, which are associated with the S-matrix for scattering processes involving all possible two-particle initial and final states\footnote{See Eqs. (\ref{eq:appx:unitarity:neutral-basis}), (\ref{eq:appx:unitarity:charged-basis}) for possible neutral and charged states.}. At energy scales, very high compared to masses involved in the scattering process, the dominant contributions to the $2\to 2$ scalar scattering amplitudes are those mediated by the quartic scalar couplings \cite{Arhrib:2000is} and gauge bosons can be replaced by their corresponding Goldstone bosons. The conditions of the perturbative unitarity are satisfied provided the eigenvalues of the scattering amplitude matrix are less than $8\pi$ \cite{Cornwall:1974km,Lee:1977eg,Lee:1977yc,Kanemura:1993hm}.
In our model, the constraints imposed on the quartic couplings by the requirement of tree-level unitarity are given below
%%%%%%%%%%%%%%%%%%%%%%%%%%%%%%%%%%%%%%%%%%%%%
\beq\label{eq:constraints:unitarity}
|\l_{H}|, |\l_{\bm{T}}| < 4\pi, \quad  |\l_{SH}|, |\l_{HT}|, |\l_{ST}| < 8\pi, \quad
{\rm and} \quad |x_{1,2,3}| < 16 \pi,
\eeq
%%%%%%%%%%%%%%%%%%%%%%%%%%%%%%%%%%%%%%%%%%%%%
where, $|x_{1,2,3}|$ are the roots of the following cubic equation:
%%%%%%%%%%%%%%%%%%%%%%%%%%%%%%%%%%%%%%%%%%%%%
\bea\label{eq:unitary:x-values}
	x^3 &&+ x^2 (-12 \l_{H} -6 \l_S -10 \l_{\bm{T}}) \nn \\
	&&+ x (-12 \l_{HT}^2 +72 \l_H \l_S -4 \l_{SH}^2 -3 \l_{ST}^2 + 120 \l_H \l_{\bm{T}} + 60 \l_S \l_{\bm{T}}) \nn \\
	&&+ (72 \l_{HT}^2 \l_S - 24 \l_{HT} \l_{SH} \l_{ST} + 36 \l_H \l_{ST}^2 - 720 \l_H \l_S \l_{\bm{T}} + 40 \l_{SH}^2 \l_{\bm{T}}) = 0.
\eea
%%%%%%%%%%%%%%%%%%%%%%%%%%%%%%%%%%%%%%%%%%%%%

One also needs to focus on the process $\ovr{\psi}\psi \to \ovr{\psi}\psi$ to complete 
the discussion on the perturbative unitarity for an $s$-channel process with $s\gg m_s$.
Following Ref. \cite{Gopalakrishna:2017zku}, based on helicity amplitude calculation \cite{Chanowitz:1978uj,Chanowitz:1978mv} at the $s\gg m_s$ limit for the $\ovr{\psi}\psi \to \ovr{\psi}\psi$ process, in the light of the partial wave analysis, one finally gets
%%%%%%%%%%%%%%%%%%%%%%%%%%%%%%%%%%%%%%%%%%%%%
\beq\label{eq:constraints:unitarity2}
|g_S|  \leq \sqrt{16\pi},
\eeq
%%%%%%%%%%%%%%%%%%%%%%%%%%%%%%%%%%%%%%%%%%%%%
where $g_S$ denotes the fermionic DM Yukawa coupling as depicted in Eq. (\ref{eq:Lag:Fermion}). Further details on the perturbative unitarity are mentioned in Appendix.~\ref{appx:unitarity}.
%%%%%%%%%%%%%%%%%%%%%%%%%%%%%%%%%%%%%%%%%%%%%%%%%%%%%%%%%%%%%%%%%%%%%%%%%%%%%%%%%
\subsection{Electroweak precision observables}
\label{subsec:EWPO}
%%%%%%%%%%%%%%%%%%%%%%%%%%%%%%%%%%%%%%%%%%%%%%%%%%%%%%%%%%%%%%%%%%%%%%%%%%%%%%%%%
  It is important to consider the Electroweak Precision Observables (EWPOs) data while investigating a BSM scenario. These EWPOs are generally expressed in terms of three measurable quantities that parametrize the contributions from the BSM states to the SM EW radiative corrections, known as Peskin-Takeuchi or oblique parameters $S$, $T$, and $U$ \cite{Peskin:1991sw}. The presence of additional scalars in our setup can contribute to these parameters. One can write such contributions as,
%%%%%%%%%%%%%%%%%%%%%%%%%%%%%%%%%%%%%%%%%%%%
\bea
\Delta \mathcal{X} = \Delta \mathcal{X}_{\rm IT} + \Delta \mathcal{X}_{\rm rS},
\eea
%%%%%%%%%%%%%%%%%%%%%%%%%%%%%%%%%%%%%%%%%%%%
where $\Delta \mathcal{X} \equiv \mathcal{X} - \mathcal{X}^{\rm{SM}}$ for $\mathcal{X} = \{S, T, U\}$, and the subscripts IT and rS denotes the contribution from the ``inert triplet'' and the ``real scalar'' singlet, respectively. In other words, $\Delta \mathcal{X}$ represents the difference between the values of any oblique parameter evaluated for the BSM framework and the SM. The explicit forms of these contributions are mentioned in Appendix \ref{appx:EWPO-constraints}. Based on an analysis of EW precision data, including the recent CDF-II $W$-mass result \cite{CDF:2022hxs}, Ref.~\cite{Lu:2022bgw} recently provided the values for the $\Delta S, \Delta T$ and $\Delta U$ parameters. As mentioned in Ref.~\cite{Lu:2022bgw}, the global EW fit of the CDF $W$-boson mass strongly indicates the need for the non-degenerate multiplets beyond the SM, however, this feature is absent in our present framework. Since we do not attempt to address the CDF $W$-mass anomaly in this work, therefore, we stick to the global EW fit presented using the PDG data \cite{ParticleDataGroup:2020ssz,ParticleDataGroup:2024cfk}, and the measurements for $\Delta S, \Delta T$ and $\Delta U$ parameters are given as \cite{Lu:2022bgw},
%%%%%%%%%%%%%%%%%%%%%%%%%%%%%%%%%%%%%%%%%%%%
\beq
\label{eq:constrants:STU-param}
\Delta S = 0.06 \pm 0.10,\quad \Delta T =  0.11 \pm 0.12, \quad \Delta U = -0.02 \pm 0.09.
\eeq
%%%%%%%%%%%%%%%%%%%%%%%%%%%%%%%%%%%%%%%%%%%%
The correlation among different oblique parameters are $\rho_{ST} = 0.90,\, \rho_{SU} = -0.57$ and $\rho_{TU} = -0.82$ \cite{Lu:2022bgw}. With $\Delta U$ fixed to be zero, the central values of $\Delta S$ and $\Delta T$ from Ref.~\cite{Lu:2022bgw} are given as,
%%%%%%%%%%%%%%%%%%%%%%%%%%%%%%%%%%%%%%%%%%%
\bea
\label{eq:constraints:ST-param}
\Delta S = 0.05 \pm 0.08, \quad \Delta T = 0.09 \pm 0.07,\quad {\rm with}~~ \rho_{ST} = 0.92.
\eea
%%%%%%%%%%%%%%%%%%%%%%%%%%%%%%%%%%%%%%%%%%%

We consider a $\chi^2$-test with the two remaining {\it d.o.f.} on our parameter space which constraints the $\Delta S, \Delta T$ parameters as follows \cite{Batra:2022arl}\footnote{The missing factor in Eq. (43) of Ref. \cite{Batra:2022arl} has been rectified in Eq.~(\ref{eq:constraints:STU-param-chi2-test}).},
%%%%%%%%%%%%%%%%%%%%%%%%%%%%%%%%%%%%%%%%%%%
\bea
\label{eq:constraints:STU-param-chi2-test}
\chi^2 \geq \frac{1}{(1-\rho_{ST}^2)} \left[ \frac{(\Delta S - S_0)^2}{\sigma_S^2} + \frac{(\Delta T - T_0)^2}{\sigma_T^2} - 2 \rho_{ST} \frac{(\Delta S - S_0) (\Delta T - T_0)}{\sigma_S \sigma_T} \right],
\eea
%%%%%%%%%%%%%%%%%%%%%%%%%%%%%%%%%%%%%%%%%%%
where $\chi^2  =2.30,\,4.61,\,5.99$ corresponds to $68.3\%, \, 90\%, \, 95\%$ confidence level, respectively. In this work, we exclude points having a $\chi^2$ larger than $5.99$ to remain consistent with EWPOs constraints. With this, we conclude the discussion on constraints arising from the theoretical frontier and move to the limits arising from the experimental bounds in the next subsection.

%%%%%%%%%%%%%%%%%%%%%%%%%%%%%%%%%%%%%%%%%%%%%%%%%%%%%%%%%%%%%%%%%%%%%%%%%%%%%%%%%%
\subsection{Constraints from $W^\pm,\,Z^0$ bosons properties}
\label{subsec:constraints-from-SMbosons-properties}
%%%%%%%%%%%%%%%%%%%%%%%%%%%%%%%%%%%%%%%%%%%%%%%%%%%%%%%%%%%%%%%%%%%%%%%%%%%%%%%%%%
The decay widths (visible and invisible) of the EW gauge bosons are already measured with great precision for the SM \cite{ParticleDataGroup:2024cfk}. With several new BSM states, one would expect new decay processes like $W^\pm \to T^\pm T^0$, $Z^0\to T^\pm T^\mp,\, \ovr{\psi}\psi$, etc., for the present model. However, the discussion following Eq. (\ref{eq:mass-triplet}) effaces the possibility of decay into these triplet states $T^\pm, T^0$ as $m_{T^\pm}>m_{T^0}$ with $m_{T^0}\gg m_W, \, m_Z$. Further, $\psi$ being an SM gauge singlet, does not directly couple to $Z^0$. Nevertheless, mixing between $h,s$ states can yield new processes like $Z^0\to \ovr{f} f \ovr{\psi}\psi$ for suitable values of $m_\psi$ ($f$ being an SM fermion). Branching ratios $(Brs)$ for these four-body final states are typically mixing suppressed and thus, skipped in the current study. We note in passing that such channels may appear detectable in future colliders like GigaZ mode of the Linear Collider \cite{LinearColliderAmericanWorkingGroup:2001tzv} and TeraZ mode of the TLEP \cite{TLEPDesignStudyWorkingGroup:2013myl}, respectively.

%%%%%%%%%%%%%%%%%%%%%%%%%%%%%%%%%%%%%%%%%%%%%%%%%%%%%%%%%%%%%%%%%%%%%%%%%%%%%%%%%%
\subsection{Constraints from Higgs boson properties}
\label{subsec:constraints-from-Higgs-properties}
%%%%%%%%%%%%%%%%%%%%%%%%%%%%%%%%%%%%%%%%%%%%%%%%%%%%%%%%%%%%%%%%%%%%%%%%%%%%%%%%%%
  
The ATLAS and CMS collaborations have precisely measured certain properties (e.g., mass, reduced strengths, etc.) of a 125 GeV SM-like Higgs scalar boson \cite{ATLAS:2023oaq,CMS:2020xrn}. These measurements also constrained some of the model parameters. Here, we mention a few important constraints coming from the SM or BSM Higgs searches at colliders which need to be considered for a complete investigation.

(i) {\it Invisible Higgs decays:} 
The presence of new BSM states, $T^0,\,T^\pm,\, h_2$ and $\psi$, opens up the possibility of new visible (e.g., $T^\pm T^\mp,\, h_2 h_2$) and invisible ($T^0 T^0,\, \ovr{\psi} \psi$) decays for the 125 GeV SM-like Higgs state $h_1$, if allowed kinematically. The discussion mentioned following Eq. (\ref{eq:mass-triplet}), however, effaces the possibility of having $h_1\to T^0 T^0,\, T^\pm T^\mp$ kinematically viable. Besides, for this analysis, we consider $m_{h_2}$ to be the heavier state, as already stated after Eq. (\ref{eq:mass-eigen-HS}). Hence, the allowed BSM decay for $h_1$ will be $h_1\to \psi\ovr{\psi}$, i.e., an invisible one. So one needs to consider the constraints on the invisible Higgs decay $Br$ and the total decay width \cite{ParticleDataGroup:2024cfk,ATLAS:2023tkt,CMS:2023sdw}

%%%%%%%%%%%%%%%%%%%%%%%%%%%%%%%%%%%%%%%
\bea
\label{eq:constrains:Higgs-invisible-psi}
Br (h_1\to \ovr{\psi}\psi)= \frac{\Gamma(h_1 \rightarrow \psi \ovr{\psi})}{\Gamma^{\rm{tot}}_{h_1} + \Gamma(h_1 \rightarrow \psi \ovr{\psi})} < 0.107,
\eea
%%%%%%%%%%%%%%%%%%%%%%%%%%%%%%%%%%%%%%%
where $\Gamma^{\rm {tot}}_{h_1},\,\Gamma(h_1 \rightarrow \psi \ovr{\psi})$ represent the total decay width of $h_1$ into \textit{all the SM} modes and the decay width into $\psi \ovr{\psi}$ decay channel, respectively, of course assuming $2 m_{\psi} < m_{h_1}$. Apart from Eq. (\ref{eq:constrains:Higgs-invisible-psi}), one also needs to check whether 
$\Gamma^{\rm{tot}}_{h_1} + \Gamma(h_1 \rightarrow \psi \ovr{\psi})< 3.7^{+1.9}_{-1.4}$ MeV \cite{ParticleDataGroup:2024cfk,ATLAS:2023dnm,CMS:2022ley}. The concerned decay width is
given by
%%%%%%%%%%%%%%%%%%%%%%%%%%%%%%%%%%%%%%%
\bea
\label{eq:constraints:Higgs-invisible-psi-decay-width}
\Gamma(h_1 \rightarrow \psi \ovr{\psi}) ~~=~~ \frac{g_S^2 m_{h_1} \sin^2\theta }{8 \pi} \left({1 - \frac{4 m_{\psi}^2}{m_{h_1}^2}}\right)^{3/2}.
\eea
%%%%%%%%%%%%%%%%%%%%%%%%%%%%%%%%%%%%%%%
Clearly, for small $g_S$ and small $\sin\theta$, i.e., high mass separation between $h_1,\,h_2$ states, one can minimized $Br(h_1\to \psi \ovr{\psi})$, consistent with the experimental bounds \cite{ParticleDataGroup:2024cfk,ATLAS:2023tkt,CMS:2023sdw}. We want to point out that although we will consider the region of $m_{\psi} < m_{h_1}/2$ for a full numerical scan, we will be mostly focusing on the parameter space with $m_{\psi} > m_{h_1}/2$, thereby Higgs invisible decay constraint hardly affect our findings. For numerical analyses of the Higgs sector and to apply LHC constraints, we used publicly available packages {\tt HiggsBounds-5.10.2} \cite{Bahl:2022igd,Bahl:2021yhk,Bechtle:2020pkv,Bechtle:2015pma,Bechtle:2013wla,Bechtle:2012lvg,Bechtle:2011sb,Bechtle:2008jh} and {\tt HiggsSignals-2.6.2} \cite{Bahl:2022igd,Bechtle:2020uwn,Bechtle:2014ewa,Stal:2013hwa,Bechtle:2013xfa}.
 
(ii) {\it Higgs signal strength measurements:} Even if the new Higgs decays remain kinematically forbidden, the presence of new BSM states can potentially alter Higgs signal strengths, defined as $\mu^i_j=(\sigma_i \times Br_j)^{\rm{obs}}/(\sigma_i \times Br_j)^{\rm SM}$, where one estimates the observed production cross-section for the $i$-th channel, $\sigma_i$, (e.g., gluon fusion) times the decay $Br$ for some particular $j$-th process (e.g., $\ovr{b} b$) relative to the same in the SM. As we have identified $h_1$ as the 125 GeV SM-like Higgs (see subsection \ref{subsec:scalarmassmixing}), the estimated $\mu^i_j$, for different $h_1$ production channels and various $h_1$ decay processes like $\gamma\gamma,\, WW^*,\, ZZ^*$ must respect the existing experimental bounds \cite{ParticleDataGroup:2024cfk}. In other words, the existing bounds will check {\textit{purity}} of the $h_1$ state, i.e., put a bound on $\sin\th$, the singlet-doublet Higgs mixing parameter. Thus, a large mixing angle is experimentally disfavored from $\mu^i_j$ measurements and restricts us to $|\sin\th|<0.2$ for most of the parameter space. Let us illustrate a bit more for an elucidated understanding of how $\mu^i_j$ changes in the presence of new BSM states. The presence of new charged state, $T^\pm$, will affect $h_1\rightarrow {\gamma\gamma}$ \cite{YaserAyazi:2014jby, Gunion:1989we}, compared to the SM. Hence, to claim a phenomenologically viable parameter space or to present benchmark points, one must ensure whether the concerned Higgs signal strength remains in the measured range of $1.10 \pm 0,06$ \cite{ParticleDataGroup:2024cfk}, following the latest available LHC limits \cite{ATLAS:2022tnm,CMS:2022dwd}. For the implementation of $\mu^i_j$ constraints in our numerical analyses we use {\tt HiggsSignals-2.6.2} \cite{Bahl:2022igd,Bechtle:2020uwn,Bechtle:2014ewa,Stal:2013hwa,Bechtle:2013xfa} and {\tt Lilith} \cite{Bernon:2015hsa} packages.

(iii) {\it Heavy Higgs searches:} The singlet-doublet mixing allows a second CP-even Higgs state $h_2$ in this framework. Our demand about the lighter CP-even scalar state $h_1$ having a predominant doublet composition, i.e., $h_1$ to be the SM-like $125$ GeV Higgs (see subsection \ref{subsec:scalarmassmixing}), makes $h_2$ the {\textit{true}} BSM heavy Higgs. A small but non-zero value of $\sin\theta$ 
(see Eq. (\ref{eq:hsmixtheta})), nevertheless, allows SM decays for the 
$h_2$ state which is tightly constrained by the existing collider bounds. For example, a production of $h_2$ pairs at the LHC can be probed via 
$\ovr{b}b\tau^+\tau^-,\, \ovr{b}b\ovr{b}b, \ovr{b}b\gamma\gamma$ final states.
Interested readers are requested to see Refs~\cite{ATLAS:2021jki,AEDGE:2019nxb,ATLAS:2016paq,CMS:2016jvt,CMS:2016cma,
ATLAS:2015sxd,CMS:2015hra,ATLAS:2015oxt,ATLAS:2015pre,CMS:2013vyt} for further study of these channels. The decay width of $h_2$ into the SM states ($\Gamma(h_2 \rightarrow {\rm SM\, SM})$) will be analogous to the ones in the SM ($\Gamma^{\rm SM} (h_2 \rightarrow {\rm SM\, SM})$), assuming $h_2$ couples to the SM states just like the $h_1$, but scaled by a factor of $\sin^2\th$ (see Eq. (\ref{eq:hsmixtheta})). Mathematically,
%%%%%%%%%%%%%%%%%%%%%%%%%%%%%%%%%%%%%%%%%%
\bea
\label{eq:constraints:heavy-higgs:SM-partial}
\Gamma(h_2 \rightarrow {\rm SM\, SM}) = \sin^2\th\,\Gamma^{\rm SM} (h_2 \rightarrow {\rm SM\, SM}) = \sin^2\th Br(h_2 \rightarrow {\rm SM\, SM})_{\rm{SM}} \Gamma^{\rm SM}_{h_2},
\eea
%%%%%%%%%%%%%%%%%%%%%%%%%%%%%%%%%%%%%%%%%%
where $Br(h_2 \rightarrow {\rm SM\, SM})_{\rm{SM}}$ denotes branching ratio of $h_2$ into any possible two-body SM states,
assuming that it couples just like the $h_1$, and $\Gamma^{\rm SM}_{h_2}$ represents the total decay width of $h_2$ into the SM states. When allowed kinematically, the total decay width of $h_2$ $(\Gamma_{h_2}^{\rm tot})$, for all possible two-body decays, is given as
%%%%%%%%%%%%%%%%%%%%%%%%%%%%%%%%%%%%%%%%%
\beq
\label{eq:constraints:heavy-higgs:total-decay}
\Gamma_{h_2}^{\rm tot} = \sin^2\th \Gamma_{h_2}^{\rm SM} + \Gamma(h_2 \rightarrow h_1 h_1) + \Gamma(h_2 \rightarrow T^0 T^0) + \Gamma(h_2 \rightarrow T^+ T^-) + \Gamma(h_2 \rightarrow \psi \ovr{\psi}),
\eeq
%%%%%%%%%%%%%%%%%%%%%%%%%%%%%%%%%%%%%%%%%
with
%%%%%%%%%%%%%%%%%%%%%%%%%%%%%%%%%%%%%%%%%
\bea
\label{eq:constraints:heavy-higgs:partial-widths}
\Gamma(h_2 \rightarrow X\,X) &&= \frac{q_X \l^2_{h_2 X X}}{32 \pi m_{h_2}} \sqrt{1 - \frac{4 m^2_X}{m^2_{h_2}}}\,, \nn \\
\Gamma(h_2 \rightarrow \psi\,\ovr{\psi}) &&= \frac{g_S^2 m_{h_2} \cos^2\th}{8 \pi} \left({1 - \frac{4 m^2_{\psi} }{m^2_{h_2}}}\right)^{3/2}\,,
\eea
%%%%%%%%%%%%%%%%%%%%%%%%%%%%%%%%%%%%%%%%%
where $X=h_1, T^0, T^{\pm}$ and $q_{h_1} = q_{T^0} = 1, q_{T^{\pm}} =2$. The corresponding scalar triple couplings $\l_{h_2 X X}$, using Eqs. (\ref{eq:pot:different-part}),  (\ref{eq:pot:interaction-potential}), (\ref{eq:scalar-field-basis}), and Eq. (\ref{eq:hsmixtheta}), are given by,
%%%%%%%%%%%%%%%%%%%%%%%%%%%%%%%%%%%%%%%%
\bea
\label{eq:constraints:heavy-higgs:lh2XX}
\l_{h_2 h_1 h_1} &&= (\mu_{HS} + \l_{SH} v_s) \cos^3\th + 2 v (\l_{SH} - 3 \l_H) \cos^2\th \sin\th \nn \\
&&- 2 (\mu_{HS} + \mu_3 + \l_{SH} v_s - 3 \l_S v_s) \cos\th \sin^2\th - \l_{SH} v \sin^3\th, \nn\\
\l_{h_2 T^0 T^0} &&= \l_{h_2 T^+ T^-} = - \l_{HT} v \sin\th + \mu_{ST} \cos\th + \l_{ST} v_s \cos\th.
\eea
%%%%%%%%%%%%%%%%%%%%%%%%%%%%%%%%%%%%%%%%
In our study, we utilize the package {\tt HiggsBounds-5.10.2} \cite{Bahl:2022igd,Bahl:2021yhk,Bechtle:2020pkv,Bechtle:2015pma,Bechtle:2013wla,Bechtle:2012lvg,Bechtle:2011sb,Bechtle:2008jh} to implement the experimental limits from colliders on the BSM heavy neutral CP-even Higgs searches \cite{ParticleDataGroup:2024cfk} Finally, we need to consider constraints on $T^\pm$ decays which are somewhat special as $m_{T^\pm} - m_{T^0} \approx 166$ MeV \cite{Cirelli:2005uq}.

%%%%%%%%%%%%%%%%%%%%%%%%%%%%%%%%%%%%%%%%%%%%%%%%%%%%%%%%%%%%%%%%%%%%%%%%%%%%%%%%%%
\subsection{Disappearing charged track}
\label{subsec:constraints:disapperaing-charged-track}
%%%%%%%%%%%%%%%%%%%%%%%%%%%%%%%%%%%%%%%%%%%%%%%%%%%%%%%%%%%%%%%%%%%%%%%%%%%%%%%%%%
  
In the case of the $SU(2)_L$ inert triplet ${\bm{T}}$, a tiny mass splitting between 
the charged and the neutral components makes it hardly possible to follow the 
conventional search strategies used for the singly charged scalar. The unstable $T^\pm$ decay offers rather different collider signals, e.g., $T^\pm\to T^0 \pi^\pm$\footnote{Other possible, but sub-dominant decay modes are $T^\pm  \to T^0 e^- \ovr{\nu}_e,\,T^0 e^+ \nu_e$, $T^0 \mu^- \ovr{\nu}_\mu,\,T^0 \mu^+ \nu_\mu$ \cite{Cirelli:2005uq}.}. The resultant $\pi$ is too soft to be reconstructed, as $\Delta m=m_{T^\pm}-m_{T^0}\approx 166$ MeV. Thus, together with the other decay product $T^0$, one of the two DM candidates, $T^\pm$ decay can produce a disappearing charge track at the detector, a phenomenon that has attained interest in recent times. Using the disappearing track signature, recently in Ref. \cite{Chiang:2020rcv}, it was shown that the $13$ TeV LHC \cite{ATLAS:2017oal}\footnote{The CMS collaboration also did an analysis \cite{CMS:2016kce} in this direction.} excludes a real triplet lighter than $275$ $(248)$ GeV for integrated luminosity $\mathcal{L} = 36\,\rm fb^{-1}$ with $\Delta m = 160\,(172)$ MeV. This exclusion limit can reach up to $590$ $(535)$ GeV, and $745$ $(666)$ GeV for  $\mathcal{L} = 300\, \rm fb^{-1}, 3000\, fb^{-1}$, respectively with $\Delta m = 160 (172)$ MeV. However, given a systematic uncertainty of $30\%$ \cite{Chiang:2020rcv} at the LHC with high luminosity, the coverage shifts to $382\,(348)$ GeV and $520\, (496)$ GeV for $\mathcal{L} = 300\, \rm fb^{-1}$ and $3000\, \rm fb^{-1}$, respectively, considering the specified $\Delta m$. In our study, we adopt a conservative approach by setting a lower mass limit on the triplet based on the current LHC constraints, focusing on the mass region where $m_{T^0} > 300$ GeV\footnote{This partly excludes the desert region \cite{Cirelli:2005uq, Fischer:2011zz,Araki:2011hm, Khan:2016sxm} below $300$ GeV.}.
%%%%%%%%%%%%%%%%%%%%%%%%%%%%%%%%%%%%%%%%%%%%%%%%%%%%%%%%%%%%%%%%%%%%%%%%%%%%%%%%%%
\subsection{Constraints from the Dark Matter}
\label{subsec:constraints:DM-constraints}
%%%%%%%%%%%%%%%%%%%%%%%%%%%%%%%%%%%%%%%%%%%%%%%%%%%%%%%%%%%%%%%%%%%%%%%%%%%%%%%%%%
  
We have thus far addressed various possible collider bounds, besides plausible theoretical constraints, on the chosen framework.
Besides these limits, the concerned model also receives constraints from the different DM measurables. To start with we consider the DM relic abundance $\Omega_{\rm DM}^{\rm exp} h^2$ as measured by the PLANCK collaboration \cite{Planck:2018vyg},
%%%%%%%%%%%%%%%%%%%%%%%%%%%%%%%%%%%%
\bea
\label{eq:constraints:DM-relic-Planck}
\Omega_{\rm DM}^{\rm exp} h^2 = 0.1198 \pm 0.0012.
\eea
%%%%%%%%%%%%%%%%%%%%%%%%%%%%%%%%%%%%
In addition to this, the observed and projected sensitivity reaches of the existing and upcoming DM direct search experiments, e.g., XENON1T \cite{XENON:2019gfn}, XENOXnT \cite{XENON:2023cxc} LZ-2022 \cite{LZ:2022lsv}, recently updated to LZ-2024 \cite{LZCollaboration:2024lux}, and DARWIN \cite{DARWIN:2016hyl} further restrict the model parameter space. 
The DD limits are often complemented by bounds from the ID of the DM that are detectable either at the Earth-based set-ups like H.E.S.S. \cite{Abdalla:2016olq,Abdallah:2016ygi,HESS:2015cda,Abramowski:2014tra,Abramowski:2011hc}, CTA \cite{CTA:2015yxo,CTAConsortium:2010umy}
or at satellite-based experiments like FERMI-LAT \cite{Fermi-LAT:2016uux,Fermi-LAT:2015kyq,Fermi-LAT:2015att,Ackermann:2015tah,Ackermann:2013yva,Ackermann:2013uma,
Abramowski:2012au,Ackermann:2012nb,Fermi-LAT:2011vow,Abdo:2010ex}, AMS \cite{AMS:2021nhj,AMS:2002yni}, etc.. The DD and ID limits, in the context of multi-components DM, will be rescaled by the relative relic abundance parameters $f_{i}$ and $f_{i}^2$ \cite{Cao:2007fy,Aoki:2012ub,Bhattacharya:2013hva} respectively, as
%%%%%%%%%%%%%%%%%%%%%%%%%%%%%%%%%%%%
\bea
\label{eq:constraints:DM-relic-rescaled}
f_i = \frac{\Omega_i h^2}{\Omega^{\rm exp}_{\rm{DM}} h^2},
\eea
%%%%%%%%%%%%%%%%%%%%%%%%%%%%%%%%%%%%
with $\Omega_{i} h^2$ corresponding to individual relic density for $i=\{T^0, \psi\}$. We note that, the combined relic density, $\Omega_{\rm{tot}} h^2 = \Omega_{T^0} h^2 + \Omega_\psi h^2$ must obey Eq. (\ref{eq:constraints:DM-relic-Planck}) to remain experimentally viable while individual contributions may remain underabundant.
For the fermionic DM $\psi$, the annihilation rate into the SM final states remains $p$-wave suppressed \cite{Agrawal:2010fh}. Hence, we do not consider ID bounds for $\psi$.\footnote{The same can yield a sizable ID rate via a parity-violating interaction between $\psi$ and a pseudoscalar $a$, i.e., $i \ovr{\psi} \gamma_5 \psi a$ \cite{Esch:2013rta,Bagherian:2014iia,Franarin:2014yua,Kim:2016csm,Ghorbani:2014qpa,Balazs:2015boa,GAMBIT:2018eea,Kim:2018uov} which is not present in the current framework. }
%%%
The scalar triplet DM $T^0$ can, dominantly, annihilate to $W^+ W^-$ with an exchange of $h_1/h_2$ and $T^{\pm}$ via $s$- and $t$-channel, respectively. The calculated annihilation cross-section, $\langle \sigma v \rangle_{WW}^{T^0}$ should be compared with the upper limit derived from the analysis of data on Dwarf Spheroidal Galaxies by the Fermi-LAT satellite and the MAGIC collaborations \cite{Balazs:2015boa,MAGIC:2016xys,Reinert:2017aga}. As previously studied in Refs. \cite{Cirelli:2005uq,Cirelli:2007xd,FileviezPerez:2008bj,Cirelli:2009uv,Cirelli:2015bda, Chiang:2020rcv}, for the case of the minimal DM where the real triplet is the only DM and $m_{T^0}$ is a free parameter (with triplet-Higgs quartic coupling, $\l_{HT} \approx 0$), the DM relic density would require $m_{T^0} \simeq 2.5$ TeV and this parameter region has already been ruled out from DM indirect search limits. However, with $\l_{HT} \neq 0$ this limit may change, for e.g., see Refs.~\cite{Khan:2016sxm,Jangid:2020qgo}.  
Nevertheless, in the present analysis, the presence of the scaling factor $f_{T^0}^2$, as depicted in Eq.(\ref{eq:constraints:DM-relic-rescaled}), reduces the aforesaid upper bound on $m_{T^0}$ in the sub-TeV mass range ($<1$ TeV) where $\Omega_{T^0}h^2$ remains underabundant. Therefore, The ID constraints remain suppressed to rule out any parameter space for the triplet DM with $m_{T^0} < 1$ TeV. We continue discussing further details of the DM phenomenology in the next section.

%%%%%%%%%%%%%%%%%%%%%%%%%%%%%%%%%%%%%%%%%%%%%%%%%%%%%%%%%%%%%%%%%%%%%%%%%%%%%%%%%%
\section{Dark matter phenomenology}\label{sec:DM-pheno}
%%%%%%%%%%%%%%%%%%%%%%%%%%%%%%%%%%%%%%%%%%%%%%%%%%%%%%%%%%%%%%%%%%%%%%%%%%%%%%%%%%
The $\mathbb{Z}_2\times \mathbb{Z}'_2$ charge assignments, as already depicted in Table \ref{tab:discrete-charge}, admit two DM candidates in the present scenario. These DM candidates, $T^0,\,\psi$, have non-trivial $\mathbb{Z}_2,\,\mathbb{Z}'_2$ charges. The combined relic density of these two DM candidates must follow Eq.~(\ref{eq:constraints:DM-relic-Planck}), i.e., the value measured by the PLANCK collaboration \cite{Planck:2018vyg}, although an individual component may appear underabundant. The latter often appears useful to evade experimental bounds and hence, allures phenomenological analysis of the model with more than one DM candidate over the minimal version. At this point, we perform a brief survey of the DM sector of the ITM model and address the motivations for moving beyond. Afterwards, we start discussing certain general aspects of the DM productions in the early Universe for the chosen framework.

For the ITM model with a hypercharge zero triplet, the DM annihilation is mainly driven by the gauge couplings and the DM-Higgs coupling $\l_{HT}$ \cite{Araki:2011hm, Khan:2016sxm}. 
The quasi-degeneracy between $m_{T^\pm}, m_{T^0}$ at one-loop level, i.e.,  $m_{T^\pm} - m_{T^0} \approx 166$ MeV, besides cancellations in $T^0 T^0 \rightarrow W^+W^-, ZZ$, also inevitably leads to compulsory coannihilations \cite{Griest:1990kh}. As a result, $T^0$ remains underabundant up to $\sim 1.9$ TeV \cite{Khan:2016sxm}, larger compared to the existing LHC bounds \cite{MonoJetSearch2021}. Further, null results in the DM direct searches impose stringent constraints on the $\l_{HT}$ coupling. Similar to DD, ID limits also impose strong constraints on the minimal triplet DM model. As previously iterated in subsection~\ref{subsec:constraints:DM-constraints}, for a DM mass of $\simeq 2.5$ TeV with triplet-Higgs quartic coupling $\l_{HT} \approx 0$, where relic density is satisfied, the parameter space is excluded by ID searches \cite{Cirelli:2005uq,Cirelli:2007xd,Cirelli:2009uv,Cirelli:2015bda}. However, with non-zero $\l_{HT} ~(\sim \mathcal{O}(0.1))$, the ID excluded upper limit on the triplet DM mass is reduced to $\simeq 1.2$ TeV \cite{Khan:2016sxm,Jangid:2020qgo}. 

The DD and ID constraints can be relaxed when the DM segment of the ITM is minimally augmented with one fermionic DM $\psi$. This fermionic DM interacts with the SM sector only via an SM singlet scalar $S$ following an interaction $g_S \ovr{\psi} \psi S$ (see Eq. (\ref{eq:Lag:Fermion})). Thus, for $\psi$, the DM annihilation is governed by Higgs particle exchange channels. As a consequence, the coupling $g_S$ and the Higgs mixing angle $\sin\th$ play pivotal roles in the DM annihilation and DD, while other allowed couplings, such as $\l_{SH}$, $\l_S$ (or $\mu_{HS}$), and $\mu_3$, have lesser impacts in the DM phenomenology. Including the fermionic DM $\psi$ helps to elevate the total DM relic density to match Eq. (\ref{eq:constraints:DM-relic-Planck}) in the region of underabundant triplet DM, i.e., $m_{T^0} \lesssim 1.9$ TeV. A similar relaxation is also witnessed for the DD. Thus, by minimally augmenting the ITM DM sector, one can rescue the sub-TeV ``desert'' region, consistent with collider searches, i.e., $300\,{\rm GeV} \lesssim m_{T^0} \lesssim 1000\,{\rm GeV}$, while allowing for the exploration of parameters that support a successful FOPT. We continue these discussions in the following subsections and present the findings of our numerical analyses.

%%%%%%%%%%%%%%%%%%%%%%%%%%%%%%%%%%%%%%%%%%%%%%%%%%%%%%%%%%%%%%%%%%%%%%%%%%%%%%%%%%
\subsection{DM relic density and direct searches}
\label{subsec:DM-pheno-relic-DD-ID}
%%%%%%%%%%%%%%%%%%%%%%%%%%%%%%%%%%%%%%%%%%%%%%%%%%%%%%%%%%%%%%%%%%%%%%%%%%%%%%%%%%

We start our discussions by writing the possible annihilation and coannihilation channels \cite{Griest:1990kh} for the triplet DM $T^0$, which govern the DM relic abundance. Afterwards, we present the feasible annihilation channels for the fermionic DM $\psi$. Besides, one will also get DM conversion \cite{Liu:2011aa,Belanger:2011ww} between the two DM candidates, $T^0$ and $\psi$.
%%%%%%%%%%%%%%%%%%%%%%%%%%%%%%%%%%%%%%%%%
%%%%%%%%%%%%%%%%%%%%%%%%%%%%%%%%%%%%%%%%%
\begin{align*} 
	\begin{array}{l l}
		{\underline {T^0- \text{\it annihilation}}:} & {\underline{T^0- \text{\it co-annihilations}}}:  \\[0.5cm]
		T^0 T^0 ~\to SM ~SM \equiv f\bar{f},\, ZZ,\, W^\pm W^\mp, ~~~~~& T^0 T^\pm~\to SM ~SM' \equiv f f',\, W^\pm Z,\, W^\pm \gamma,  \\
		T^0 T^0 ~\to h_i h_j, ~~~~~& T^0 T^\pm~\to W^\pm h_i,  \\
        & T^\mp T^\pm \to SM ~SM \equiv f \bar{f},\, ZZ,\, W^\pm W^\mp, \\
         & T^\mp T^\pm ~\to Z h_i,\, h_i h_j, \\
		 \\[0.2cm]
        {\underline {\psi- \text{\it annihilation}}:} &   \\[0.3cm]
		\psi \bar{\psi} ~\to f\bar{f},\, ZZ,\, W^\pm W^\mp,\, h_i h_j,  \\
       \\ [0.2cm] 
		\multicolumn{2}{l}{\underline{\text{\it DM conversion}}:} \\[0.3cm]
		\multicolumn{2}{l}{\psi~\ovr{\psi} ~\to T^0 ~T^0, \hspace{0.3cm} \text{when} \hspace{0.2cm} m_{\psi} > m_{T^0},} \\
        \multicolumn{2}{l}{T^0 T^0 \to \psi~\ovr{\psi}, \hspace{0.3cm} ~~\text{when} \hspace{0.2cm} m_{T^0} > m_{\psi},} \\
	\end{array}
\end{align*}
%%%%%%%%%%%%%%%%%%%%%%%%%%%%%%%%%%%%%%%%%
where $h_{i(j)}=h_{1(2)}$ and $f,\,f'$ represents any SM fermion. For reference, we present a few Feynman diagrams representing various DM annihilation and coannihilation processes in Figs. \ref{fig:feynman-diagram:psi-anni}, \ref{fig:feynman-diagram:Trip-anni} and \ref{fig:feynman-diagram:Trip-coanni} of Appendix \ref{appx:feynman-diagrams}.

To proceed further in the DM analyses, one needs to solve the coupled Boltzmann equations (BEQs) within the chosen framework to determine the co-moving number densities $Y_{T^0} = n_{T^0}/s,\, Y_{\psi} = n_{\psi}/s$, i.e., the ratio of the DM number density $n_{T^0,\,\psi}$ to the entropy density. 
With two DMs, we define a dimensionless parameter $x$, expressed as $x= \frac{\mu_{\rm red}}{T}$, $T$ being the temperature. The latter denotes the reduced mass of the concerned two DM systems and is written as $\mu_{\rm red} = \frac{m_{T^0} m_{\psi}}{m_{T^0} + m_{\psi}}$. In terms of $x, Y_{i=T^0,\, \psi}$, Boltzmann equations are written as,
%%%%%%%%%%%%%%%%%%%%%%%%%%%%%%%%%%%%%%%%%%%%%%%%%%
\begingroup
\allowdisplaybreaks
\bea
\label{eq:DM-pheno-BEQs-1}
\frac{d Y_{T^0}}{d x} &=& -\frac{s(x)}{x ~\mathcal{H}(x)} \Big[ \langle \sigma v \rangle_{T^0 T^0 \rightarrow SM~SM} \left( Y^2_{T^0} - Y^2_{T^0, eq} \right) \nn\\
%&&+  \langle \sigma v \rangle_{T^0 T^0 \rightarrow h_i h_j} \left( Y^2_{T^0} - Y_{X, eq} Y_{Y, eq} \right) \nn \\ % \Theta (2 m_{T^0} - m_X - m_Y) \nn \\ 
&&+  \langle \sigma v \rangle_{T^0 T^0 \rightarrow h_i h_j} \left( Y^2_{T^0} - Y_{{h_i}, eq} Y_{{h_j}, eq}\right) \nn \\
%% co-annihilation
&&+ \langle \sigma v \rangle_{T^0 T^{\pm} \rightarrow SM ~SM'} \left( Y_{T^0} Y_{T^{\pm}} - Y_{T^0, eq} Y_{T^{\pm}, eq} \right) \nn \\
&&+ \langle \sigma v \rangle_{T^0 T^0 \rightarrow \psi \ovr{\psi}} \left( Y^2_{T^0} - \frac{
Y^2_{T^0, eq}}{Y^2_{\psi, eq}} Y^2_{\psi} \right) %\Theta (m_{T^0} - m_{\psi})
\Big], \\
%% Second BE
\label{eq:DM-pheno-BEQs-2}
\frac{d Y_{\psi}}{d x} &=& -\frac{s(x)}{x ~\mathcal{H}(x)} \Big[ \langle \sigma v \rangle_{\psi \ovr{\psi} \rightarrow SM ~SM} \left( Y^2_{\psi} - Y^2_{\psi, eq} \right) \nn\\
%&&+ \langle \sigma v \rangle_{\psi \ovr{\psi} \rightarrow h_i h_j} \left( Y^2_{\psi} - Y_{X, eq} Y_{Y, eq} \right) \nn \\ %\Theta (2 m_{\psi} - m_X - m_Y) \nn \\
&&+ \langle \sigma v \rangle_{\psi \ovr{\psi} \rightarrow h_i h_j} \left( Y^2_{\psi} - Y_{{h_i}, eq} Y_{{h_j}, eq}\right) \nn \\ %\Theta (2 m_{\psi} - m_X - m_Y) \nn \\
&&+ \langle \sigma v \rangle_{ \psi \ovr{\psi} \rightarrow T^0 T^0 } \left( Y^2_{\psi} - \frac{
	Y^2_{\psi, eq}}{Y^2_{T^0, eq}} Y^2_{T^0} \right) %\Theta (m_{\psi} - m_{T^0})
    \Big]. 
\eea
\endgroup
%%%%%%%%%%%%%%%%%%%%%%%%%%%%%%%%%%%%%%%%%%%%%%%%%%
where $Y_{T^0, eq},\, Y_{\psi, eq},$ $Y_{T^\pm, eq},\, Y_{{h_i}(h_j), eq},$ are the co-moving number densities for $T^0,\,\psi$, $T^\pm, h_i(h_j)$, respectively, at the equilibrium, $s(x)$ is the comoving entropy density, and $\mathcal{H}(x)$ is the Hubble parameter. The last two quantities are expressed as
%%%%%%%%%%%%%%%%%%%%%%%%%%%%%%%%%%%%%%%%%%%%%%%%%%
\bea
\label{eq:DM-pheno-BEQ-s-H}
s(x) = \frac{2 \pi^2}{45} g_s \frac{\mu_{\rm red}^3} {x^3}, ~~~~\mathcal{H}(x) = \sqrt{\frac{\pi^2 g_{\rho}}{90}} \frac{\mu_{\rm red}^2}{x^2 {M_{\rm Pl}}},
\eea
%%%%%%%%%%%%%%%%%%%%%%%%%%%%%%%%%%%%%%%%%%%%%%%%%%
where $g_s$ and $g_{\rho}$ are the relativistic {\it d.o.f.} associated with the entropy and matter, $M_{\rm Pl} = 2.4 \times 10^{18}$ GeV is the reduced Planck mass and $\langle \sigma v \rangle$s are the thermally averaged annihilation cross-section inclusive of both (co)annihilation and DM conversion processes. The co-moving equilibrium density of the $i^{th}$ DM species takes the form:
%%%%%%%%%%%%%%%%%%%%%%%%%%%%%%%%%%%%%%%%%%%%%%%%%%%
\bea
\label{eq:DM-pheno-Yeq}
Y_{i, eq} = 0.145 \frac{g_i}{g_s} x^{3/2} \left(\frac{m_i}{\mu_{\rm red}}\right)^{3/2} e^{-x \left(\frac{m_i}{\mu_{\rm red}}\right)},
\eea
%%%%%%%%%%%%%%%%%%%%%%%%%%%%%%%%%%%%%%%%%%%%%%%%%%%
where $g_i$ denotes {\it d.o.f.} for the $i^{th}$ DM species, having a mass $m_i$. It is worth mentioning that although $g_s$ and $g_{\rho}$ vary slightly during the evolution of the Universe, we can, nevertheless,  treat them as constants, and it equals to the effective {\it d.o.f.} defined as $g_*(T)^{1/2} = \frac{g_s}{\sqrt{g_{\rho}}} \left(1 +  \frac{1}{3}  \frac{T}{g_s}\frac{d g_s}{d T} \right)$ \cite{Gondolo:1990dk}.

The relic density of the $i^{th}$ DM species, using the aforesaid pieces of information, is obtained as,
%%%%%%%%%%%%%%%%%%%%%%%%%
\beq
    \Omega_i h^2=2.74\times 10^{8}\left(\frac{m_i}{\rm GeV}\right)Y_i({x=\infty}).
\eeq
%%%%%%%%%%%%%%%%%%%%%%%%%%%%%%
Various $\Omega_i h^2$ and $\langle \sigma v \rangle$s are numerically evaluated using {\tt micrOMEGAs-5.3.41} \cite{Belanger:2001fz}, after implementing the model (see Eqs. (\ref{eq:Lag:total-scalar})-(\ref{eq:Lag:Fermion})) in {\tt FeynRules} \cite{Christensen:2008py,Alloul:2013bka}.
%
%%Direct Detection
Besides the relic abundance, one should also consider constraints from the DD searches, e.g., XENON1T, LZ-2022, LZ-2024, DARWIN, etc. Null results in these searches, thus far, put bounds (i) the DM mass and/or (ii) the coupling between the DM and the SM. One, however, needs to rescale the experimental bound accordingly for a model with more than one DM. Following the idea behind Eq. (\ref{eq:constraints:DM-relic-rescaled}), the rescaled upper limit on the SI DM-nucleon (N) scattering cross-section is written as \cite{Cao:2007fy},
%%%%%%%%%%%%%%%%%%%%%%%%%%%%%%%%%%%
\bea
\label{eq:DM-pheno:rescaled-DD}
\sigma^{\rm SI}_{\rm tot} = \frac{f_{T^0}}{m_{T^0}} \sigma^{\rm SI}_{T^0} + \frac{f_{\psi}}{m_{\psi}} \sigma^{\rm SI}_{\psi} < \frac{\sigma^{\rm SI}_{\rm exp}}{m_{\rm DM}},
\eea
%%%%%%%%%%%%%%%%%%%%%%%%%%%%%%%%%%%
where $ \sigma^{\rm SI}_{i=T^0,\,\psi}$ represents the SI DM-N scattering cross-section for individual DM species. $f_{i=T^0,\,\psi}$ are defined in Eq. (\ref{eq:constraints:DM-relic-rescaled}), $\sigma^{\rm SI}_{\rm exp}$ is the experimental limit \cite{XENON:2019gfn,XENON:2023cxc,LZ:2022lsv,LZCollaboration:2024lux,DARWIN:2016hyl} and $m_{\rm DM}$ is the DM mass. 

\begin{figure*}[!h]
	\hspace*{0.25cm}
	\centering
	\subfigure[]{\includegraphics[height=3.5cm,width=3.6cm]{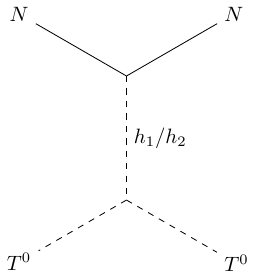}}
	\hspace*{0.95cm}
	\subfigure[]{\includegraphics[height=3.5cm,width=3.6cm]{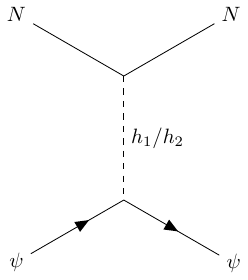}}
	\caption{Possible $t$-channel processes responsible for the DM-N scattering.}
	\label{fig:DM-pheno:DM-nucleon}
\end{figure*}

The DM-N scattering in this model can arise via a $t$-channel exchange of
$h_1$ or $h_2$, as depicted in Fig. \ref{fig:DM-pheno:DM-nucleon}. For the 
triplet scalar DM, the effective SI-DD cross-section is written as
%%%%%%%%%%%%%%%%%%%%%%%%%%%%%%%%%%%%%%%%%%%%%%%%%%%%%%%%
%%%%%%%%%%%%%%%%%%%%%%%%%%%%%%%%%%%%%%%%%%%%%%%%%%%%%%%%
%%%%%%%%%%%%%%%%%%%%%%%%%%%%%%%%%%%%%%%%%%%%%%%%%%%%%%%
\bea
\label{eq:DM-pheno:T0-DD-SI-appx}
\sigma^{\rm SI}_{T^0}~ \simeq~ \frac{f^2_N m^2_N \mu^2_{T^0\rm N}}{4 \pi m^2_{T^0} v^2} \left[ \frac{\l_{h_1 T^0 T^0} \cos\th}{m^2_{h_1} } - \frac{\l_{h_2 T^0 T^0} \sin\th}{m^2_{h_2}} \right]^2,
\eea
%%%%%%%%%%%%%%%%%%%%%%%%%%%%%%%%%%%%%%%%%%%%%%%%%%%%%%%
%\sout{$f_{T^0}$ is the fractional DM density,} 
where $\mu_{T^0\rm N}$ is the reduced mass of the $T^0$-$N$ system, defined as $\mu_{T^0\rm N} = \frac{m_{T^0} m_N}{m_{T^0} + m_N}$ with $m_N = 0.946$ GeV and the nucleon form factor, which depends on the hadronic matrix elements and, approximately, is given as $f_N =0.28$ \cite{Alarcon:2012nr}. The trilinear couplings $\l_{h_1 T^0 T^0}$ and $\l_{h_2 T^0 T^0}$ are expressed as
%%%%%%%%%%%%%%%%%%%%%%%%%%%%%%%%%%%%%%%%%%%%%%%%%%%%%%%
\bea
\label{eq:DM-pheno:higgs-T0-trlinear-couplings}
\l_{h_1 T^0 T^0} &&= \l_{HT} v \cos\th + \mu_{ST} \sin\th + \l_{ST} v_s \sin\th, \nn\\
\l_{h_2 T^0 T^0} &&= -\l_{HT} v \sin\th + \mu_{ST} \cos\th + \l_{ST} v_s \cos\th.
\eea
%%%%%%%%%%%%%%%%%%%%%%%%%%%%%%%%%%%%%%%%%%%%%%%%%%%%%%%
An analysis similar to Eq. (\ref{eq:DM-pheno:T0-DD-SI-appx}), but for $\psi$ yields \cite{Lopez-Honorez:2012tov}
%%%%%%%%%%%%%%%%%%%%%%%%%%%%%%%%%%%%%%%%%%%%%%%%%%%%%%%
\bea
\label{eq:DM-pheno:psi-DD-SI-appx}
\sigma^{\rm SI}_{\psi}~ \simeq~ \frac{f^2_N m^2_N \mu^2_{\psi \rm N}}{4 \pi v^2} g^2_S \sin^2 2 \th \left[ \frac{1}{m^2_{h_1} } - \frac{1}{m^2_{h_2}} \right]^2,
\eea
%%%%%%%%%%%%%%%%%%%%%%%%%%%%%%%%%%%%%%%%%%%%%%%%%%%%%%%
where the reduced mass of $\psi$-N system is defined as $\mu_{\psi\rm N} = \frac{m_{\psi} m_N}{m_{\psi} + m_N}$. Estimation of $\sigma^{\rm SI}_{T^0},\,\sigma^{\rm SI}_{\psi}$, once again, are done using {\tt micrOMEGAs-5.3.41}.

\begin{figure*}[!h]
	\hspace*{-0.20cm}
	\centering
	\subfigure{\includegraphics[height=6.3cm,width=8.3cm]{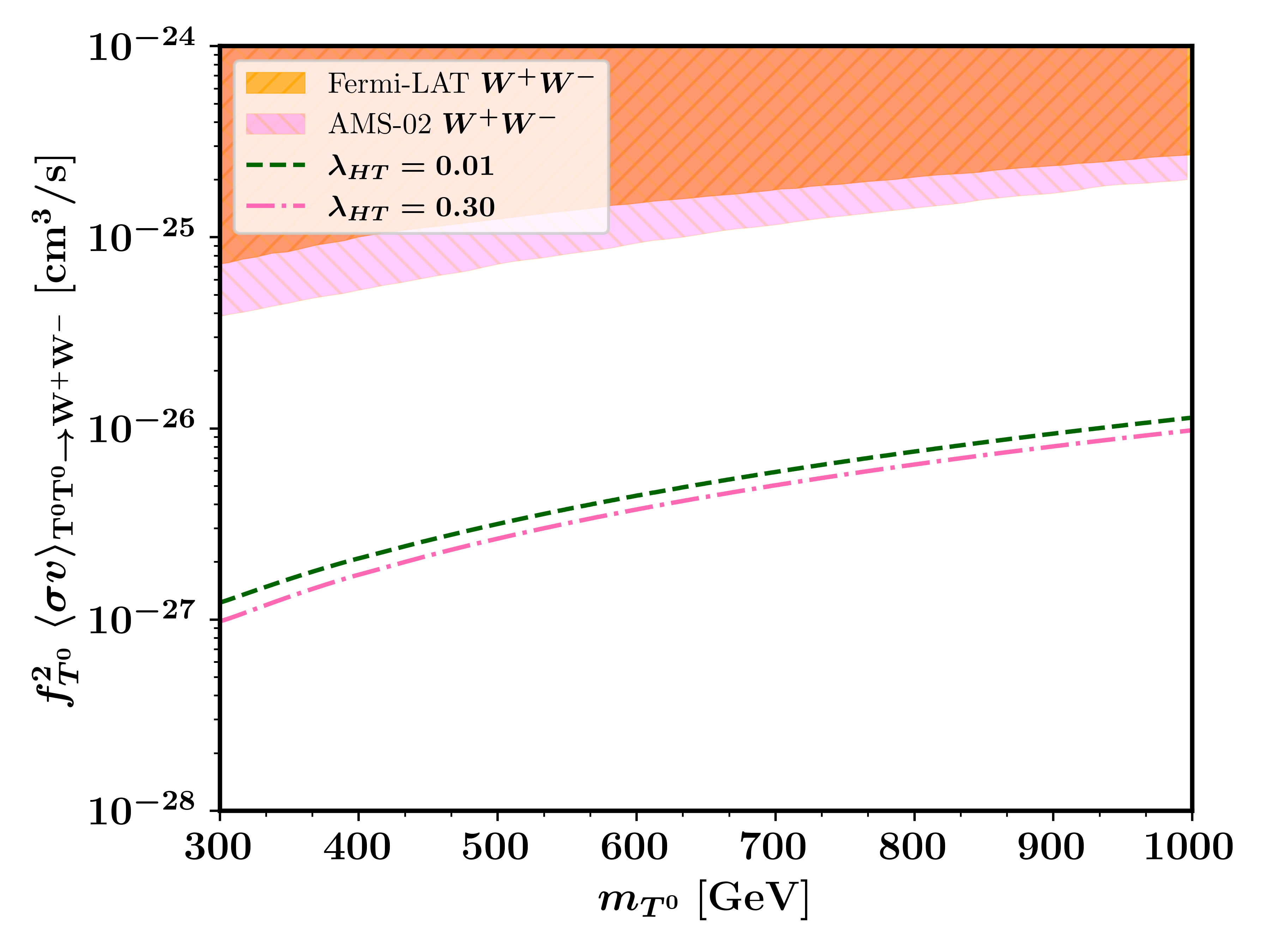}}
	\caption{Bounds on the scalar Triplet mass $m_{T^0}$ from indirect searches of $W^+W^-$ channel by Fermi-LAT \cite{Balazs:2015boa, MAGIC:2016xys} (depicted by golden-yellow coloured patterned region) and AMS-02 \cite{AMS:2016oqu} (depicted by pink coloured patterned region). The two differently coloured dashed lines correspond to two different choices of the triplet SM-like Higgs coupling parameter, $\l_{HT}$, as depicted on the plot.}
	\label{fig:DM-pheno:sigmavT0WW-ID}
\end{figure*}

%%%% ID fig-----------------------
Besides direct searches, the DM can also be detected indirectly by observing $\gamma$-rays produced from the DM annihilation in galaxies. 
Being $p$-wave suppressed, as already stated, ID constraints would not be considered for the fermionic DM. For the scalar triplet DM, on the other hand, annihilations are primarily to gauge boson final state. Thus, $T^0 T^0 \to W^+ W^-$ can dominantly constrain the parameter space for the triplet DM. However, for $m_{T^0} < 1 ~\rm TeV$, the relative relic abundance of the triplet DM remain as $f_{T^0} < 1$. As a result, after rescaling by $f^2_{T^0}$, the overall thermally averaged annihilation cross section, $\langle \sigma v \rangle_{W^+W^-}$, is suppressed and remains below the experimental sensitivity reaches, as illustrated in Fig. \ref{fig:DM-pheno:sigmavT0WW-ID}\footnote{In preparing Fig. \ref{fig:DM-pheno:sigmavT0WW-ID}, we assumed a pure $Y=0$ triplet DM scenario with $\lambda_{HT}$ and $m_{T^0}$ as the free inputs, and calculated $f_{T^0}$ following Eq.~(\ref{eq:constraints:DM-relic-rescaled}). In the upcoming analyses (for e.g., see subsection~\ref{subsec:scan}), we will observe that $f_{T^0}$ indeed remains below 1 in a phenomenologically viable scenario. Hence, the current ID bounds hardly affect the triplet DM physics.}.
%%%%%%%%%%%%%%%%%%%%%%%%%%%%%%%%%%%%%%%%%%%%%%%%%%%%%%%%
%%%%%%%%%%%%%%%%%%%%%%%%%%%%%%%%%%%%%%%%%%%%%%%%%%%%%%%% 
In this figure, we compare the predicted cross-section for the process $T^0 T^0 \to W^+ W^-$, calculated for two different values of $\l_{HT} = 0.01$ (green coloured line) and $0.30$ (hot-pink coloured line), with the upper bounds on $\langle \sigma v \rangle_{W^+ W^-}$ derived from satellite-based Fermi-LAT data \cite{Balazs:2015boa, MAGIC:2016xys} (represented by the
golden-yellow coloured patterned region)
and AMS-02 data \cite{AMS:2016oqu} (depicted by the pink-coloured patterned region). The upward trend of the cross-section lines in Fig. \ref{fig:DM-pheno:sigmavT0WW-ID} can be attributed to the fact that, as $m_{T^0}$ increases, the relic density also rises, leading to an increase in the ratio $f_{T^0}$ (see Eq. (\ref{eq:constraints:DM-relic-rescaled})). This compensates for the annihilation cross-section $\langle \sigma v \rangle_{T^0 T^0 \to W^+ W^-}$ in the increasing mass region, resulting in a slight increase in the annihilation cross-section. Furthermore, Fig. \ref{fig:DM-pheno:sigmavT0WW-ID} clearly shows that the predicted cross-section for the dominant annihilation channel remains unaffected by the current ID bounds on this channel. This indicates that in the sub-TeV regime 
of $m_{T^0}$, where $f_{T^0} < 1$, the ID limits on the triplet DM will be ineffectual for the concerned two-component DM model.
%%%--------------------------
%%%%%%%%%%%%%%%%%%%%%%%%%%%%%%%%%%%%%%%%%%%%%%%%%%%%%%%%
\begin{figure*}[!h]
	\hspace*{0.25cm}
	\centering
	\subfigure{\includegraphics[height=3.6cm,width=5.0cm]{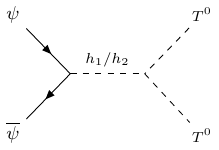}}
	\caption{$\psi \overline{\psi} \leftrightarrow T^0 T^0$ conversion in the chosen model.}
	\label{fig:DM-pheno:DM-DM-interaction}
\end{figure*} 
%%%%%%%%%%%%%%%%%%%%%%%%%%%%%%%%%%%%%%%%%%%%%%%%%%%%%%%%

Finally, before concluding this subsection, it is important to highlight that the key couplings relevant for the triplet DM phenomenology are $\l_{HT}, \l_{ST}$ and $\mu_{ST}$, while the ones for the fermion DM are, primarily, $g_S$, and $\sin\th$. The two DM components interacts through the Yukawa interaction: $g_S S \ovr{\psi} \psi$, and also through the Higgs portal interactions: $\sim \l_{HT} |H|^2 {\rm Tr}[{\bm{T}}^\dagger {\bm{T}}]$, $\l_{ST} S^2 {\rm Tr}[{\bm{T}}^\dagger {\bm{T}}]$ and $\mu_{ST} S {\rm Tr}[{\bm{T}}^\dagger {\bm{T}}]$.
The DM-DM interaction in this model via the exchange of $h_1/h_2$ is shown in Fig. \ref{fig:DM-pheno:DM-DM-interaction}. Note that, the Higgs portal couplings $\l_{HT}$ and $\l_{ST}$ are strongly constrained by the direct searches \cite{FileviezPerez:2008bj,Araki:2011hm,YaserAyazi:2014jby,Khan:2016sxm,Chiang:2020rcv,DuttaBanik:2020jrj}. For sufficiently small values of $\l_{HT}$ and $\l_{ST}$, $\sim \mathcal{O}(0.01)$, the DM-DM conversion would mostly depend on $\{g_S, \sin\th, \mu_{ST}\}$ in addition to DM and mediator masses. We continue this discussion on the parameter dependence of various DM observables in further detail in the next subsection, focusing on the relic density and SI DD cross-section.
%%%%%%%%%%%%%%%%%%%%%%%%%%%%%%%%%%%%%%%%%%%%%%%%%%%%%%%%%%%%%%%%%%%%%%%%%%%%%%%%%%
\subsection{Parameter dependence of the DM observables}
\label{subsec:DM-param-dependence}
%%%%%%%%%%%%%%%%%%%%%%%%%%%%%%%%%%%%%%%%%%%%%%%%%%%%%%%%%%%%%%%%%%%%%%%%%%%%%%%%%%

\begin{figure*}[!h]
	%\hspace*{-0.2cm} %% This will shift the figure
	\centering
	\subfigure[]{\includegraphics[height=5.84cm,width=7.5cm]{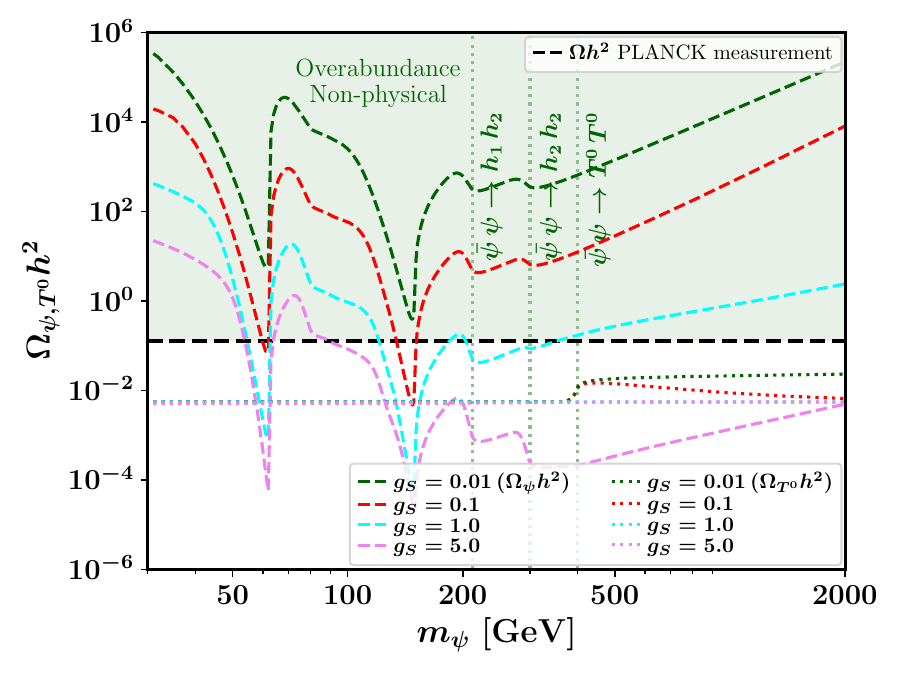}}
	\subfigure[]{\includegraphics[height=5.8cm,width=7.5cm]{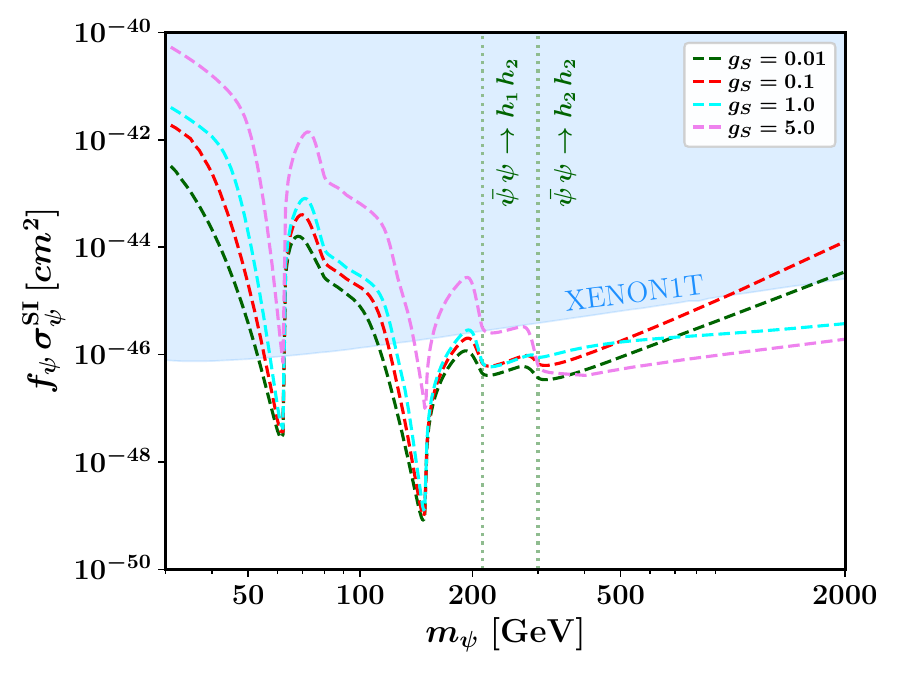}}
		\subfigure[]{\includegraphics[height=5.8cm,width=7.5cm]{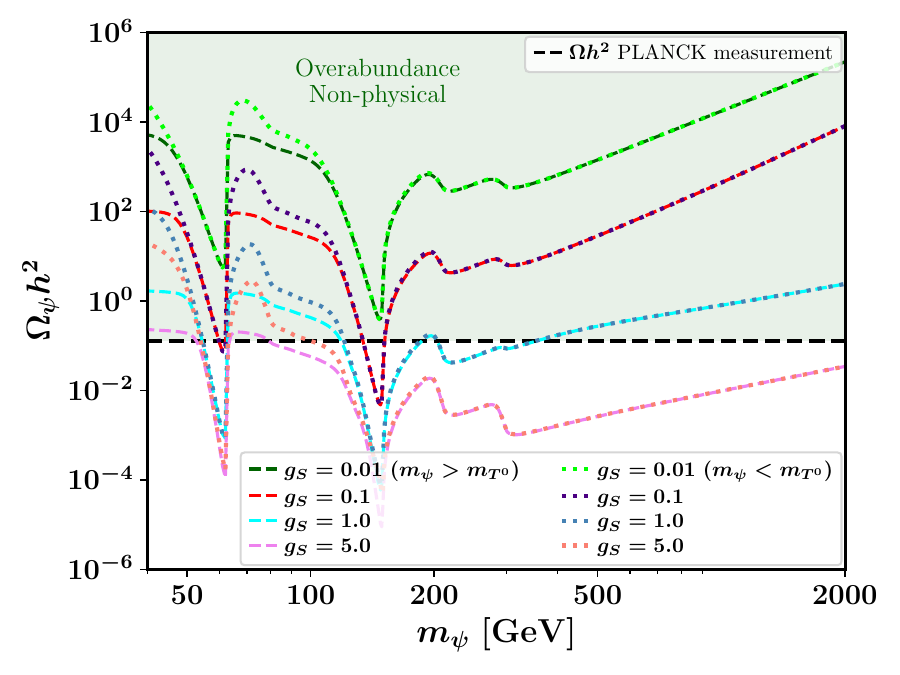}}
	\subfigure[]{\includegraphics[height=5.8cm,width=7.5cm]{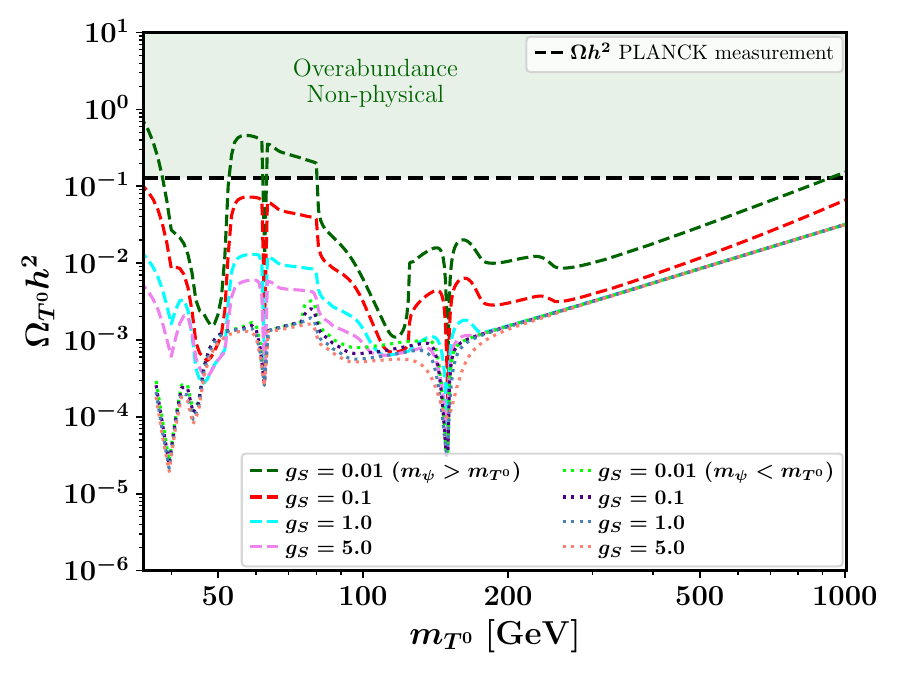}}
		\caption{The top row represents the effect of parameter $g_S$ on the $m_\psi-\Omega_{\psi, T^0} h^2$ plane (a) and $m_\psi - f_\psi \sigma^{\rm{SI}}_\psi$ plane (b) for parameter values detailed in Eq. (\ref{eq:DM-pheno:fixed-param-grid-scan}). The left plot of the bottom row (c) depicts the same for the $m_\psi-\Omega_{\psi}h^2$ plane but for two possible hierarchies between $m_\psi$ and $m_{T^0}$. Lastly, for similar mass hierarchies between $m_\psi,\,m_{T^0}$, the right plot of the bottom row (d) shows $g_S$ dependence on the $m_{T^0}-\Omega_{T^0} h^2$ plane. The region excluded by the XENON1T experiment is shaded in cyan colour. Other details are given on individual plots.}
	\label{fig:DM-pheno:grid-scan-gS}
\end{figure*}

In this subsection, we start our discussion by probing how relic densities (total and individual) and the DD limits vary with the DM mass parameters, $m_{T^0}$ and $m_\psi$. Such variations, needless to say, do depend on all other relevant or connected parameters. However, to understand how a particular parameter affects the DM observables, we vary only a few at one go, for fixed assignations of the remaining parameters. The effect of Yukawa coupling $g_S$ on the $m_\psi$ - $\Omega_{\psi,\, T^0} h^2$-plane is depicted in Fig. \ref{fig:DM-pheno:grid-scan-gS} (a) for four discrete values of $g_S$, namely, $\{0.01, 0.1, 1.0, 5.0\}$ (represented with different colour codes), keeping fixed values for other crucial parameters as follows:
%%%%%%%%%%%%%%%%%%%%%%%%%%%%%%%%%%%%%%%%%%%%%%%%%%%%%%%%
\bea
\label{eq:DM-pheno:fixed-param-grid-scan}
&&\l_S = 1.0, ~\sin\th = 0.1, \l_{HT} = \l_{ST} = 0.01,\nn \\
&&m_{h_2} = v_s = 300 ~{\rm GeV}, ~m_{T^0} = 400 ~{\rm GeV}, \mu_{ST} = \mu_3 = -50 ~{\rm GeV},
\eea
%%%%%%%%%%%%%%%%%%%%%%%%%%%%%%%%%%%%%%%%%%%%%%%%%%%%%%
where for $m_{T^0}$, we have considered a value higher than the existing
collider bound of $\mathcal{O} (300)$ GeV, as already detailed in subsection
\ref{subsec:constraints:disapperaing-charged-track}. The parameter
$m_\psi$ is varied between $10-2000$ GeV.
%%%%%%%%%%%%%%%%%%%%%%%%%%%%%%%%%%%%%%%%%%%%%%%%%%%%%%%%
%%%%%%%%%%%%%%%%%%%%%%%%%%%%%%%%%%%%%%%%%%%%%%%%%%%%%%%%
%%%%%%%%%%%%%%%%%%%%%%%%%%%%%%%%%%%%%%%%%%%%%%%%%%%%%%%%
The black-coloured dashed line in Fig. \ref{fig:DM-pheno:grid-scan-gS} corresponds to the experimental value of relic density (see Eq.~(\ref{eq:constraints:DM-relic-Planck})). The DM remains overabundance above this line. The key observations from  Fig. \ref{fig:DM-pheno:grid-scan-gS} (a) are detailed below:
%%%%%%%%%%%%%%%%%%%%%%%%%%%%%%%%%%%%%%%%%%%%%%%%%%%%%%%%
\begin{itemize}
	\item
	Except for Higgs resonances, the fermionic DM typically remains overabundance for smaller $g_S$ values, e.g., $g_S\sim \mathcal{O}$ $(10^{-2})$, given fixed assignations of other relevant parameters as of Eq.~(\ref{eq:DM-pheno:fixed-param-grid-scan}). Larger $g_S$ values, $g_S\gsim \mathcal{O}$ $(1)$, on the contrary, make the DM underabundant. This is connected to the fact that
	smaller $g_S$ values yield a lower $\langle \sigma v \rangle$, which decreases as
	the DM mass grows. Consequently, this leads to relatively high values of $\Omega_{\psi} h^2$, i.e., overabundance, and keeps increasing with $m_\psi$ as apparent from the figure. At Higgs resonances, i.e., $m_{\psi} \simeq {m_{h_{1,2}}/2}$, DM annihilations via
	$h_{1,2}$ results dips in $\Omega_{\psi}h^2$, making it accessible for or below the PLANCK threshold. In the same plot, we also plot the variations
	of $\Omega_{T^0}h^2$ with $m_{\psi}$ which, for a fixed $m_{T^0}$ of $400$ GeV, shows non-trivial variation when $m_{\psi}$ exceeds $m_{T^0}$ through DM-conversion process.
	\item  For $g_S = 1.0$, the relic curve $\Omega_{\psi} h^2$ (cyan coloured) intersects the observed relic density line at multiple points for parameters fixed as of Eq.~(\ref{eq:DM-pheno:fixed-param-grid-scan}). Thus, for $g_S$ values of $\mathcal{O}(1)$ or more, the correct relic density appears for several $m_\psi$ values. This is different from a case with a small $g_S$ value, where a similar observation is typically possible at Higgs resonances only.
\end{itemize}
%%%%%%%%%%%%%%%%%%%%%%%%%%%%%%%%%%%%%%%%%%%%%%%%%%%%%%%%

New number-changing processes become accessible beyond certain threshold values when the DM mass grows. This results in a reduction of the relic density. For $m_\psi < m_{h_1}$, the DM abundance is primarily governed by the number-changing processes $\psi \ovr{\psi} \to {SM} {SM}$, mediated by $h_1$ or $h_2$. A sharp drop in the DM relic density is observed near $h_1$ (SM-like Higgs) resonance at $m_{\psi} \sim m_{h_1}/2$. With increasing DM mass, we notice a wider and shallow dip at around $80$-$90$ GeV as $\psi \ovr{\psi} \to W^+ W^-$ and $\psi \ovr{\psi} \to Z Z$ processes become kinematically accessible, resulting in an enhancement in $\langle \sigma v \rangle$ and reducing the DM density slightly. For $m_{\psi} > m_{h_1}$, another sharp drop in $\Omega_{\psi} h^2$ occurs at the second Higgs resonance $h_2$, due to the enhancement in $\psi \ovr{\psi} \to h_1 h_1$ process via a $h_1/h_2$ mediated $s$-channel process. It can also be noticed that $m_{\psi} \in [m_{h_1}, m_{h_2}]$ is generally disfavoured from the observed relic limits, except for large $g_S$ ($ > 1.0$). Two new dips appear in $\Omega_\psi h^2$ for $m_{\psi} \sim (m_{h_1} + m_{h_2})/2$ and $m_{\psi} \sim m_{h_2}$ for  $2 m_{\psi} > m_{h_i} + m_{h_j}$ with $i,\,j = 1,\,2$. These happen with the opening of new processes that start contributing to the relic density.
For large $g_S$ values, processes $\psi \ovr{\psi} \to h_i h_j, i,j=1,\,2$ increase $\langle \sigma v \rangle$ and thereby, reduce $\Omega_{\psi} h^2$, as evident from the plot. Pushing
$m_\psi$ beyond $m_{T^0}$ (fixed at $400$ GeV as given by Eq. (\ref{eq:DM-pheno:fixed-param-grid-scan})), triggers the process $\psi \bar{\psi} \to T^0 T^0$. This process yields enhancement in $\Omega_{T^0} h^2$, especially for smaller $g_S$ values, $\sim \mathcal{O}(0.01)$ (see Fig. \ref{fig:DM-pheno:grid-scan-gS} (a)), since, for smaller $g_S$, the overabundant fermion DM component $\psi$ already saturates the Universe, allowing further conversion. For larger $g_S$, on the contrary, $\psi$ is less abundant, resulting in fewer conversions and a negligible impact on $\Omega_{T^0} h^2$.

Continuing with the fermionic DM $\psi$, in Fig. \ref{fig:DM-pheno:grid-scan-gS} (b) we show the $g_S$ dependence in the $m_\psi-f_\psi \sigma^{\rm{SI}}_\psi$ plane. We also compare
the $\sigma^{\rm{SI}}_\psi$ values with the XENON1T bounds, as a case study. Limits from the other DD searches, e.g., LZ-2022, LZ-2024, DARWIN, etc., will, nevertheless, be considered later for the full numerical scan. As anticipated from Eq.~(\ref{eq:DM-pheno:psi-DD-SI-appx}), smaller $g_S$ values allow the fermion DM to evade DD limits rather easily. The dips in this plot are due to the same reasons discussed above for Fig. \ref{fig:DM-pheno:grid-scan-gS} (a). For larger $g_S$ values, we notice a slight decline in the $\sigma^{\rm{SI}}_\psi$ curve for heavier mass, $m_{\psi} \gtrsim 1.0$ TeV. This appears as the decrease in $\sigma^{\rm SI}_{\psi}$ for heavier DM mass is compensated by the relic density factor $f_{\psi}$ (see Eq. (\ref{eq:DM-pheno:rescaled-DD})). For $g_S > \mathcal{O}(1.0)$, the fermion DM remains underabundant for heavier $m_\psi$ beyond the Higgs resonances by a factor of approximately $ \mathcal{O}(10^{-3})$, as can be seen in Fig. \ref{fig:DM-pheno:grid-scan-gS} (a). The $\sigma^{\rm SI}_{\psi}$, as a consequence is reduced by the factor $f_{\psi} = \Omega_{\psi} h^2/\Omega_{\rm DM}^{\rm exp} h^2$. Therefore, larger $g_S$ with a heavy DM mass can still bypass the DD limits, however, such scenarios are less favoured by the relic density constraint and require full numerical investigation. On the other hand, $g_S \sim \mathcal{O}(1.0)$  is generally preferred by both the SI DD and DM relic density bounds, as illustrated in Fig.  \ref{fig:DM-pheno:grid-scan-gS} (a) and (b).

It is important to address the validity of observations made in the context of Fig. \ref{fig:DM-pheno:grid-scan-gS} (a) and (b) as the concerned numerical analysis was performed following Eq.~(\ref{eq:DM-pheno:fixed-param-grid-scan}) for fixed assignations of parameters, relevant for the DM phenomenology. This is certainly true and, besides variations of the necessary parameters, also depend on $m_{T^0},\,m_\psi$, their hierarchy and mass scales of the possible mediators. Nevertheless, our observations about $g_S$ seem to be rather generic as can be verified through Fig. \ref{fig:DM-pheno:grid-scan-gS} (c)  and (d) where we also vary $m_{T^0}$ from $10-1000$ GeV\footnote{Note that, in preparing the figure, we chose the mass range below 300 GeV only for illustration. For our full numerical analysis, see subsection \ref{subsec:scan}, we have omitted $m_{T^0} \lesssim 300$ GeV, as this mass range is excluded from the current LHC searches of disappearing charged tracks (see subsection \ref{subsec:constraints:disapperaing-charged-track}).}. Besides, to understand the effect of the DM conversion between two species, we consider two sample mass hierarchies,
(i) $m_{\psi} > m_{T^0}$ with $m_{\psi}/m_{T^0} = 1.2$ and (ii) $m_{\psi} < m_{T^0}$ with $m_{T^0}/m_{\psi} = 1.2$. The parameter $g_S$ is varied as before with fixed assignations of the other relevant parameters as of Eq.~(\ref{eq:DM-pheno:fixed-param-grid-scan}).

In a scenario with $T^0$ as the lightest DM species, the $g_S$ interaction significantly impacts both $\Omega_{\psi}h^2$ and $\Omega_{T^0}h^2$. Due to the DM conversion, $\Omega_{\psi} h^2$ is reduced (by a factor of $\sim 10^2$ or smaller, for the chosen parameters) whereas $\Omega_{T^0} h^2$ enhances near Higgs resonances compared to Fig. \ref{fig:DM-pheno:grid-scan-gS} with $m_{T^0}$ fixed at $400$ GeV. Beyond the heavy Higgs resonances, i.e., $m_{h_2}/2$, the DM annihilation to all other channels are possible, if kinematically allowed. However, it also gets suppressed by the DM mass and hence, almost resembles the case with a fixed $m_{T^0}$. Overabundance of $\psi$ for smaller $g_S$ values ($\lsim 0.1$) leads to more $\psi \ovr{\psi} \to T^0 T^0$ conversion as can be seen from Fig. \ref{fig:DM-pheno:grid-scan-gS} (a). This, consequently, enhances $\Omega_{T^0}h^2$ as evident from comparing Fig. \ref{fig:DM-pheno:grid-scan-gS} (a) and (d), particularly around $h_1$ resonance. In fact, the relic curve for the triplet touches the observed relic line multiple times for this case, as shown in Fig. \ref{fig:DM-pheno:grid-scan-gS} (d). However, this situation is ruled out by both DM relic constraints
and by disappearing charged track constraints. The former requires $\Omega_{\rm tot} h^2 \lsim \Omega_{\rm DM}^{\rm exp} h^2$ while the latter demands $m_{T^0}\gsim 280$ GeV. Thus, within our region of interest for the triplet DM, i.e., $300 < m_{T^0} < 1000$ (in GeV), $\Omega_{T^0} h^2$ remains underabundant to comply with
the relic density constraint (see Eq. (\ref{eq:constraints:DM-relic-Planck})) which is evident from the Fig. \ref{fig:DM-pheno:grid-scan-gS} (d). The dips in $\Omega_{T^0}h^2$ around $m_{T^0}=40$ GeV and $45$ GeV arise
as a consequence of the $W^\pm$ (via co-annihilation) and $Z$ resonances, respectively.

In the second case, when $\psi$ appears to be the lightest DM candidate, the conversion channel $T^0 T^0 \to \psi \ovr{\psi}$ opens up. This conversion assists enhancement of $\Omega_{\psi}h^2$ while
lowers $\Omega_{T^0}h^2$ as depicted in Figs. \ref{fig:DM-pheno:grid-scan-gS} (c) and (d). In Fig. \ref{fig:DM-pheno:grid-scan-gS} (c), for $g_S = 5.0$, indicated by the dotted violet line, we observe a slight shift towards heavy DM mass that meets the $\Omega_{\rm DM}^{\exp} h^2$. Therefore, when the mediator mass changes, i.e., $m_{h_2}$, the process $T^0 T^0 \to \psi \ovr{\psi}$ can shift the relic-allowed fermion DM mass to a high mass region near at $m_{h_2}/2$ compared to the situation when this conversion is absent. Large $g_S$ value, e.g., $5.0$, however, would face constraints from the DD limits, especially for $m_\psi \lsim m_{h_2}$, barring pole regions as shown in Fig. \ref{fig:DM-pheno:grid-scan-gS} (b). Therefore, for both $m_\psi > m_{T^0}$ and $m_{T^0}> m_\psi$ configurations except at Higgs resonances, the preferred mass region would be beyond $m_{h_2}$, assuming $m_{h_2} > m_{h_1}$. Moreover, for a scenario with $m_{h_2} \sim m_{h_1}$, we expect that there would be larger parameter space for the fermion DM that could satisfy both the DM relic and DD bounds.

\begin{figure*}[!h]
	%\hspace*{-0.2cm} %% This will shift the figure
	\centering
	\subfigure[]{\includegraphics[height=5.84cm,width=7.5cm]{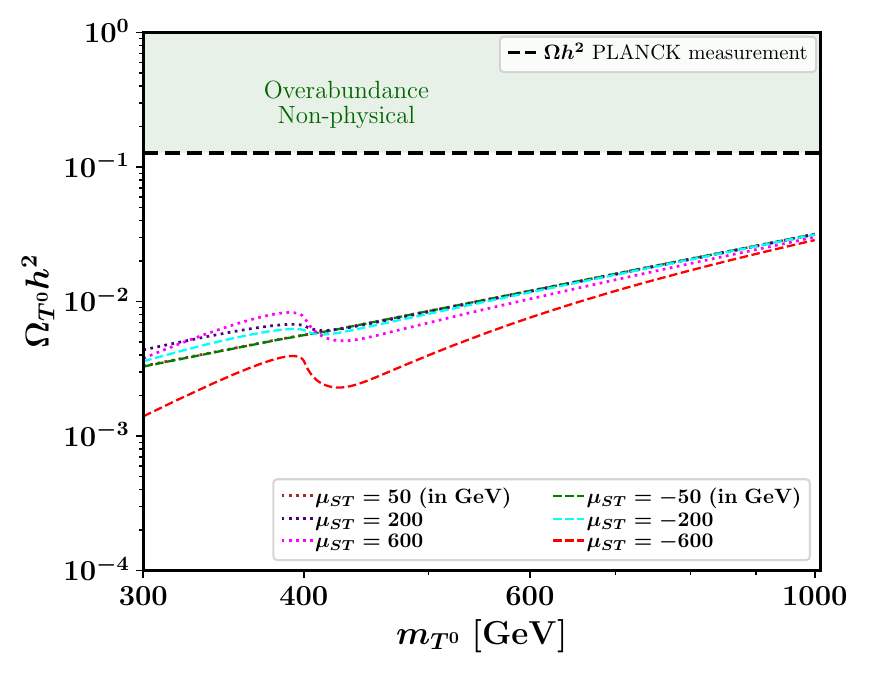}}	
	\subfigure[]{\includegraphics[height=5.8cm,width=7.5cm]{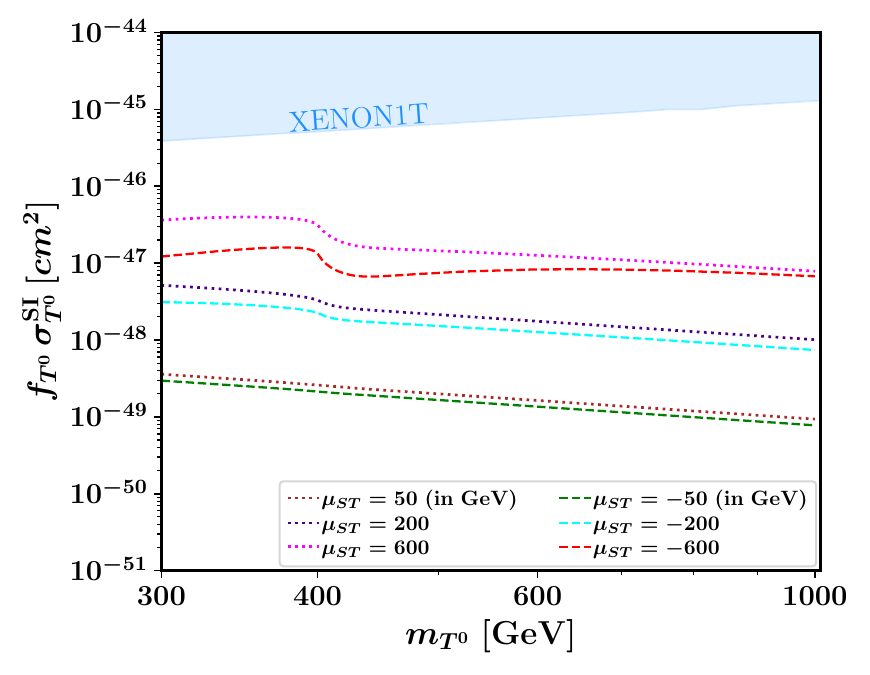}}
	\subfigure[]{\includegraphics[height=5.8cm,width=7.5cm]{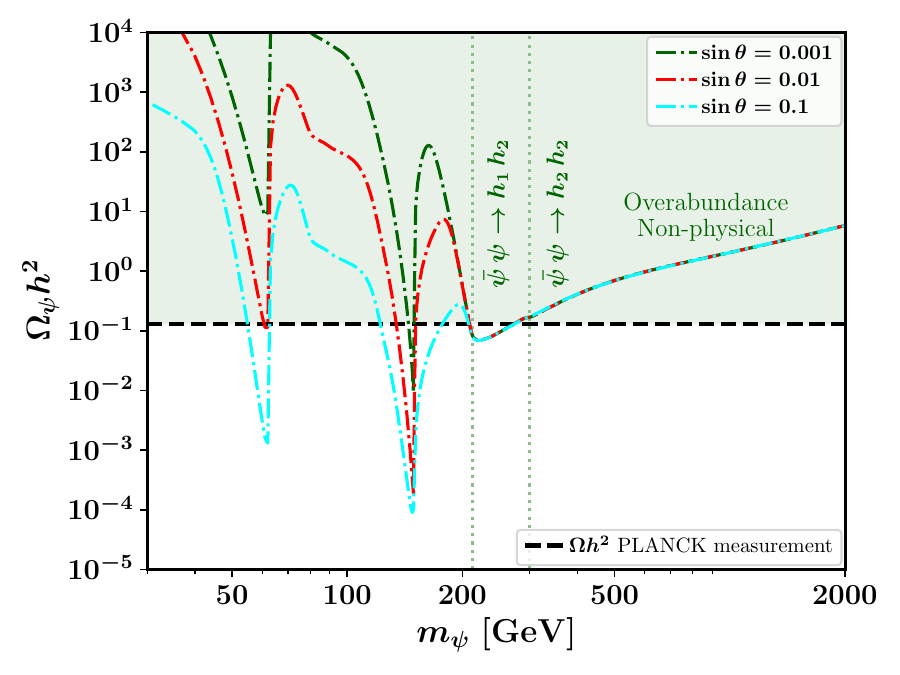}}
	\subfigure[]{\includegraphics[height=5.8cm,width=7.5cm]{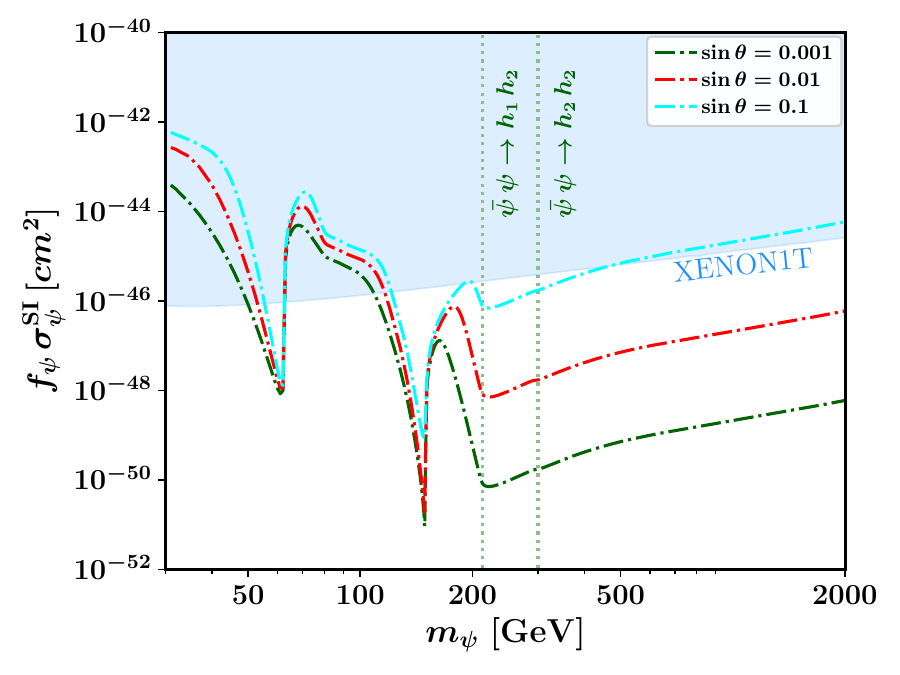}}
	\caption{The top row represents $\mu_{ST}$ dependence of the
	$m_{T^0}-\Omega_{T^0} h^2$ plane (a) and $m_{T^0}-f_{T^0}\sigma^{\rm{SI}}_{T^0}$ plane (b). The bottom row depicts the effect of $\sin\th$ for $\psi$-DM, in the $m_{\psi}-\Omega_{\psi} h^2$ plane (c) and $m_{\psi}-f_{\psi}\sigma^{\rm{SI}}_{\psi}$ plane (d). The region excluded by the XENON1T experiment is shaded in cyan colour. Other details are given on individual plots.}
	\label{fig:DM-pheno:grid-scan-muST-theta}
\end{figure*}

In Fig. \ref{fig:DM-pheno:grid-scan-muST-theta}, we present variations
of the relic densities $\Omega_{T^0,\,\psi}h^2$ and duly weighted SI DD cross-sections $f_{T^0,\,\psi} \sigma^{\rm{SI}}_{T^0,\,\psi}$ with two other
key parameters controlling the DM  phenomenology, namely, $\mu_{ST}$ and
$\sin\th$. The parameter $\mu_{ST}$ is pivotal for the triplet DM phenomenology and is essential for the DM conversion processes, as can be realized
from Eq. (\ref{eq:pot:interaction-potential}) and Eq. (\ref{eq:Lag:Fermion}).
For Figs. \ref{fig:DM-pheno:grid-scan-muST-theta} (a) and (b),
we consider $300 \leq m_{T^0} \leq 1000$ (in GeV), $g_S=1$, $m_\psi$ fixed at $400$ GeV and kept other parameters as of Eq. (\ref{eq:DM-pheno:fixed-param-grid-scan}), barring $\mu_{ST}$. For all choices of $\mu_{ST}$, $\Omega_{T^0}h^2$ increases with
$m_{T^0}$. At $m_{T^0}> m_\psi$, one observes a mild dip in $\Omega_{T^0} h^2$, owing to the onset of the DM conversion process $T^0 T^0 \to \psi \ovr{\psi}$. This is apparent
from Fig. \ref{fig:DM-pheno:grid-scan-muST-theta} (a). This dip is more prominent for larger $|\mu_{ST}|$ values. The latter phenomenon is anticipated from Eq.~(\ref{eq:DM-pheno:higgs-T0-trlinear-couplings}) and Fig. \ref{fig:DM-pheno:DM-DM-interaction} which hint a larger conversion rate for higher $|\mu_{ST}|$ values.
%% Plot::muST, sin theta scan
%%%%%%%%%%%%%%%%%%%%%%%%%%%%%%%%%%%%%%%%%%%%%%%%%%%%%%%%
%%%%%%%%%%%%%%%%%%%%%%%%%%%%%%%%%%%%%%%%%%%%%%%%%%%%%%%%
Larger $|\mu_{ST}|$ values, consequently, also enhances, the corresponding $\sigma^{\rm SI}_{T^0}$ as shown in Fig. \ref{fig:DM-pheno:grid-scan-muST-theta} (b). However, even with a large $|\mu_{ST}|$ value, the weighted $\sigma^{\rm{SI}}_{T^0}$ remains well below the XENON1T limit due to the presence
of $f_{T^0}$ factor. The near flat asymptotic behaviour of $\sigma^{\rm SI}_{T^0}$ at large $m_{T^0}$ suggests that the decline in $\sigma^{\rm SI}_{T^0}$ with increasing $m_{T^0}$ is compensated by the corresponding increase in $\Omega_{T^0} h^2$. This trend is clearly depicted in Fig. \ref{fig:DM-pheno:grid-scan-muST-theta} (a). The splitting of the SI cross-section curve for different signs of $\mu_{ST}$ is solely due to the influence of $f_{T^0}$. For large $|\mu_{ST}|$ values, the difference in $\Omega_{T^0} h^2$ due to its sign becomes more pronounced compared to the small $|\mu_{ST}|$ values, thereby resulting in a greater split in $f_{T^0} \sigma^{\rm SI}_{T^0}$ for large $|\mu_{ST}|$ values than for its small values.

The effects of $\sin\th$ on $\Omega_\psi h^2$ and $\sigma^{\rm{SI}}_{\psi}$ are displayed in Figs. \ref{fig:DM-pheno:grid-scan-muST-theta} (c) and (d), respectively. We consider the following values of $\sin\th = \{0.001,0.01,0.1\}$, keeping other parameters as of Eq.~(\ref{eq:DM-pheno:fixed-param-grid-scan}). The effect of $\sin\th$ on $\Omega_{\psi}h^2$ is apparent: lower values of $\sin\th$ generally lead to an overproduction of fermion DM, except around Higgs resonances.  This occurs because, the annihilation cross-section is suppressed for smaller values $\sin\th$, leading to an increase in $\Omega_{\psi} h^2$. In contrast, smaller $\sin\th$
values are instrumental in bypassing the DD constraints as it reduces $\sigma^{\rm{SI}}_\psi$ drastically, which can be observed from Eq.~(\ref{eq:DM-pheno:psi-DD-SI-appx}). As the DM mass increases beyond the $h_2$ resonance, processes like $\psi \ovr{\psi} \to h_1 h_2,\, h_2 h_2$ become kinematically feasible. These channels lead to a significant decrease in $\Omega_\psi h^2$ compared to the values at smaller $m_\psi$. For $m_\psi$ beyond the $h_2$ resonances, $\Omega_\psi h^2$ appears insensitive to changes in $\sin\th$, as depicted in Fig. \ref{fig:DM-pheno:grid-scan-muST-theta} (c). This happens as higher $m_\psi$ suppresses
$\langle \sigma v \rangle$, outweighing the effect of $\sin\th$ on $\Omega_\psi h^2$. Consequently, $\Omega_{\psi} h^2$ increases with the $m_\psi$ as expected and observed in Fig. \ref{fig:DM-pheno:grid-scan-muST-theta} (c). On the other hand, as observed in Fig. \ref{fig:DM-pheno:grid-scan-muST-theta} (d), for $m_\psi$ beyond $h_2$ resonance,
smaller $\sin\th$ reduces $\sigma^{\rm{SI}}_\psi$, as expected from Eq. (\ref{eq:DM-pheno:psi-DD-SI-appx}).

The remaining parameters of Eq. (\ref{eq:DM-pheno:fixed-param-grid-scan}),
namely, $\mu_3, v_s$, and $\l_S ~({\rm or}~\mu_{HS})$ do not significantly impact the DM phenomenology of the fermion DM, as the annihilation into final states $f \bar{f}, W^+ W_, ZZ$, $h_1 h_1, h_1 h_2, h_2 h_2$ are predominantly controlled by $\sin\th,g_S$, masses of the DM only. In addition, these parameters have negligible role in $\Omega_{T^0}h^2$ and $\sigma^{\rm{SI}}_{T^0}$ for $m_{T^0}\lsim 1$ TeV. The contributions from these parameters in the DM phenomenology will be detailed in the next subsection. These parameters, however, can play a non-trivial role in the PT dynamics, which we shall address in the upcoming section. These two observations allow us to address the experimentally viable DM phenomenology primarily through parameters $g_S,\, \sin\th, m_\psi$ and $m_{T^0}$ and, at the same time, accommodates the observed matter-antimatter asymmetry of the Universe through the remaining
parameters that have little effect on the physics of the dark sector.
%%%%%%%%%%%%%%%%%%%%%%%%%%%%%%%%%%%%%%%%%%%%%%%%%%%%%%%%%%%%%%%%%%%%%%%%%%%%%%
\subsection{Numerical scan}\label{subsec:scan}
%%%%%%%%%%%%%%%%%%%%%%%%%%%%%%%%%%%%%%%%%%%%%%%%%%%%%%%%%%%%%%%%%%%%%%%%%%%%%%
%%%%%%%%%%%%%%%%%%%%%%%%%%%%%%%%%%%%%%%%%%%%%%%%%%%%%%%%%%%%%%%%%%%%%
We continue our investigation of how different independent couplings, as
given by Eq. (\ref{eq:model-parameter}), affect the DM observables $\Omega_{T^0,\,\psi} h^2,\, \sigma^{\rm{SI}}_{T^0,\,\psi}$ in this subsection also, using the outcome of a random scan over $500,000$ points. We also explore plausible correlations among these parameters. The ranges of these parameters are listed in Table \ref{tab:par_ranges}. We consider $h_2$ to be the heavier Higgs always and
$m_{T^0}$ within $300-1000$ GeV. The lower bound on $m_{T^0}$ arises
from the disappearing charged track measurement \cite{ATLAS:2017oal} while
the upper threshold is driven by the urge to restore the ``desert'' region, keeping it below 1 TeV.
%%%%%%%%%%%%%%%%%%%%%%%%%%%%%%%%%%%%%%%%%%%%%%%%%%%
\begin{table}[t]
	\centering    
	\begin{tabular}{cccc}
		\toprule
		Parameter & Minimum & Maximum  \\ \midrule
		$m_{h_2}$ & $130\,\textnormal{GeV}$ & $2000\,\textnormal{GeV}$  \\[1.2mm]
		$m_{T^0}$ & $300\,\textnormal{GeV}$ & $1000\,\textnormal{GeV}$  \\[1.2mm]
		$m_\psi$ & $20\,\textnormal{GeV}$ & $2000\,\textnormal{GeV}$  \\[1.2mm]
		$v_S$ & $-1000\,\textnormal{GeV}$ & $1000\,\textnormal{GeV}$  \\[1.2mm]
		$\mu_3$ & $-1000\,\textnormal{GeV}$ & $1000\,\textnormal{GeV}$  \\[1.2mm]
		$\mu_{ST}$ & $-1000\,\textnormal{GeV}$ & $1000\,\textnormal{GeV}$  \\[1.2mm]
		$\sin\theta$ & $-0.2$ & $0.2$ \\
		$\l_S$ & $0.001$ & $3.5$ \\
		$g_S$ & $0.01$ & $3.5$ \\
		$\l_{HT}$ & $0.01$ & $0.3$ \\ 
		$\l_{ST}$ & $0.01$ & $0.3$ \\
		\bottomrule
	\end{tabular}    
	\caption{Scan ranges of the independent parameters (see Eq. (\ref{eq:model-parameter})) of our model. The parameter $\l_{\bm{T}}$ has negligible impact on the triplet DM and thus, kept fixed at a value of $0.05$ throughout the scan.}
	\label{tab:par_ranges}
\end{table}
%%%%%%%%%%%%%%%%%%%%%%%%%%%%%%%%%%%%%%%%%%%%%%%%%%%  
The range for $\sin\theta$ is chosen by keeping in mind the constraints from Higgs boson properties, as discussed in subsection \ref{subsec:constraints-from-Higgs-properties}.
%%%%%%%%%%%%%%
After implementing all theoretical and experimental constraints (see Sec. \ref{sec:constraints} for details), except the ones relevant to the DM phenomenology, only $\approx 31\%$ points survive. These points are subsequently scrutinized using bounds from the DM observables and are tabulated in Table \ref{tab:cutflow}. The bounds from different direct searches no doubt dominate over the ones listed in Sec. \ref{sec:constraints},
however, the most stringent constraint appears from the requirement of the correct relic density. As expected, the DD bound tightens up as one moves from XENON1T to the DARWIN.
%%%%%%%%%%%%%%%%%%%%%%%%%%%%%%%%%%%%%%%%%%%%%%%%%%%%%%%%%%%%%%%%%%%%%
\begin{table}[ht]
\centering
\begin{tabular}{ccc}
\toprule
\textbf{Constraints} & \textbf{Surviving points} & \textbf{$\%$ reduction}\\
\midrule
%Total Data scanned          & 500,000  &  -\\
Theory + Exp. (Sec. \ref{sec:constraints})
                            & 154,448  & $\approx 31\%$\\
XENON1T                     & 139,448  & $\approx 28\%$\\
LZ-2022                      & 128,805  & $\approx 26\%$\\
LZ-2024                      & 105,071  & $\approx 21\%$\\
DARWIN                      & 61,410    & $\approx 12\%$\\
$3\sigma$ relic density (see Eq. (\ref{eq:constraints:DM-relic-Planck}))                     & 4,813    & $\approx 0.96\%$\\
XENON1T + $3\sigma$ relic density (see Eq. (\ref{eq:constraints:DM-relic-Planck}))           & 4,398   &  $\approx 0.88\%$\\
LZ-2022  + $3\sigma$ relic density (see Eq. (\ref{eq:constraints:DM-relic-Planck}))          & 3,996  & $\approx 0.80\%$  \\
LZ-2024  + $3\sigma$ relic density (see Eq. (\ref{eq:constraints:DM-relic-Planck}))          & 3,166  & $\approx 0.63\%$  \\
DARWIN + $3\sigma$ relic density (see Eq. (\ref{eq:constraints:DM-relic-Planck}))           & 1,734   &  $\approx 0.35\%$\\
\bottomrule
\end{tabular}
\caption{Constraints cut flow and rate of survival for the given scan (see Table \ref{tab:par_ranges} for scan ranges) with
$500,000$ points.}
\label{tab:cutflow}
\end{table}
%%%%%%%%%%%%%%%%%%%%%%%%%%%%%%%%%%%%%%%%%%%%%%%%%%%%%%%%%%%%%%%%%%%%%

\begin{figure*}[!h]
	%\hspace*{-0.2cm} %% This will shift the figure
	\centering
	\subfigure[]{\includegraphics[height=5.7cm,width=7.2cm]{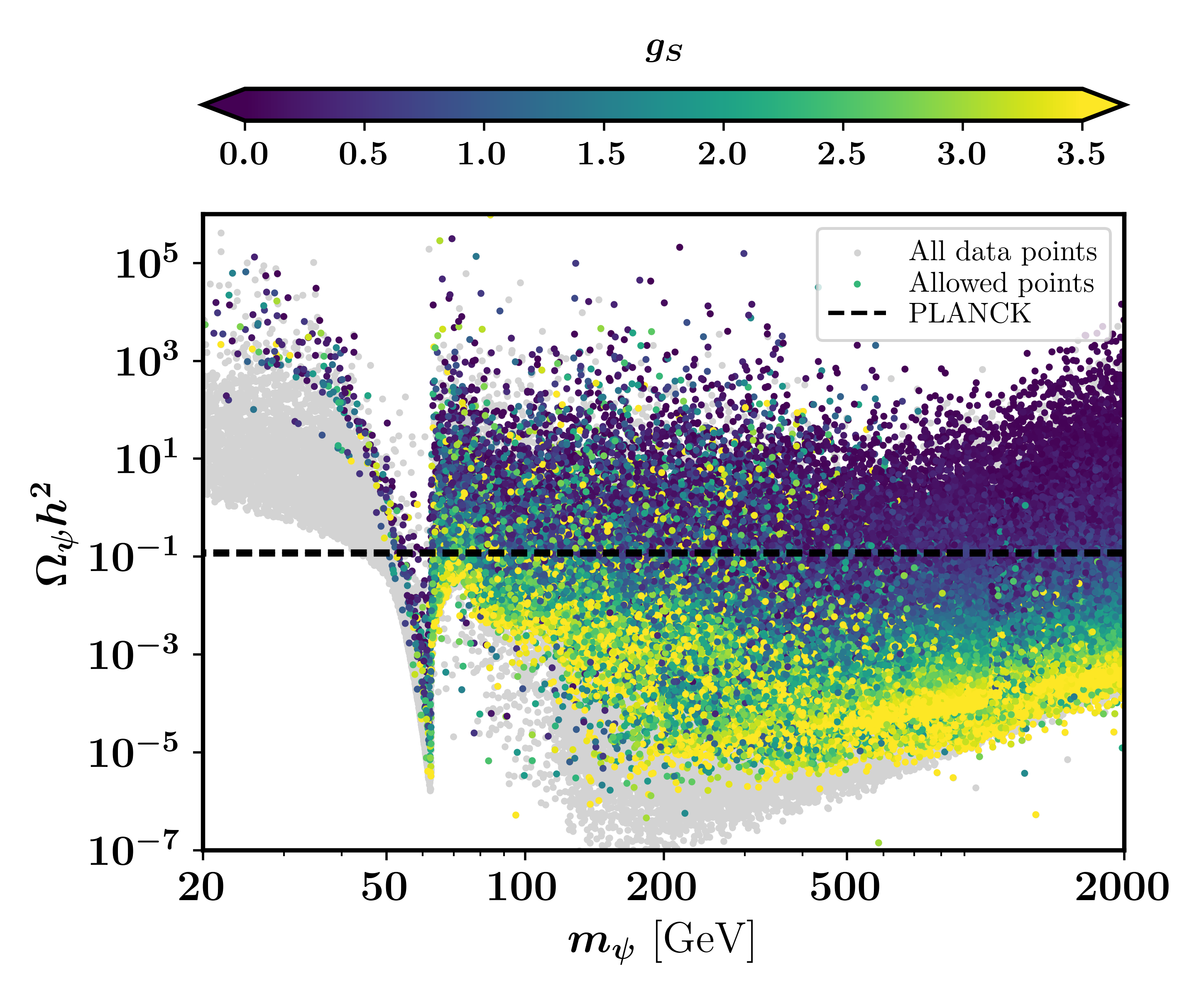}}
	\subfigure[]{\includegraphics[height=5.7cm,width=7.2cm]{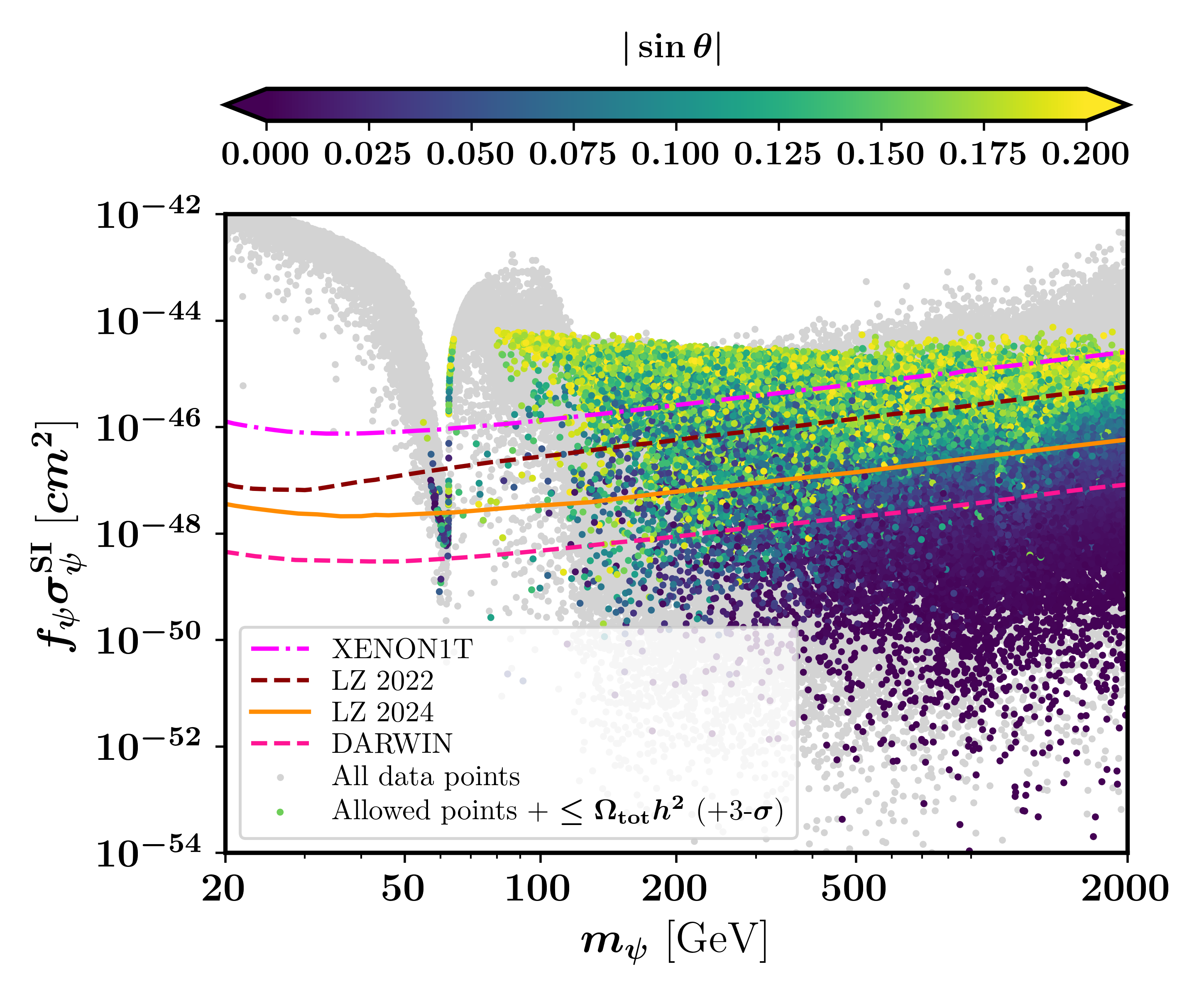}}
	\hspace*{0.15cm}
	\subfigure[]{\includegraphics[height=5.3cm,width=7.15cm]{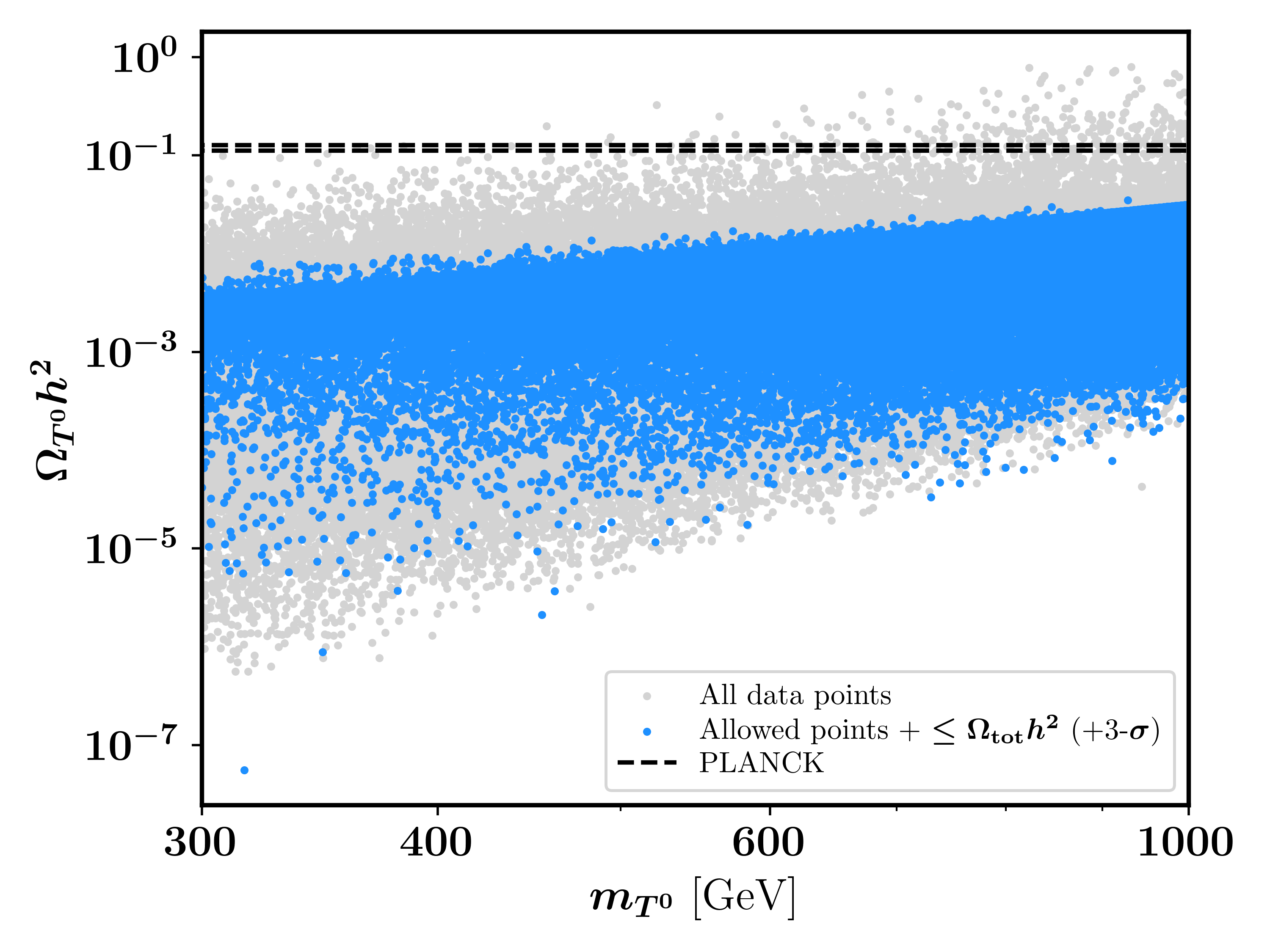}}
    \subfigure[]{\includegraphics[height=5.3cm,width=7.15cm]{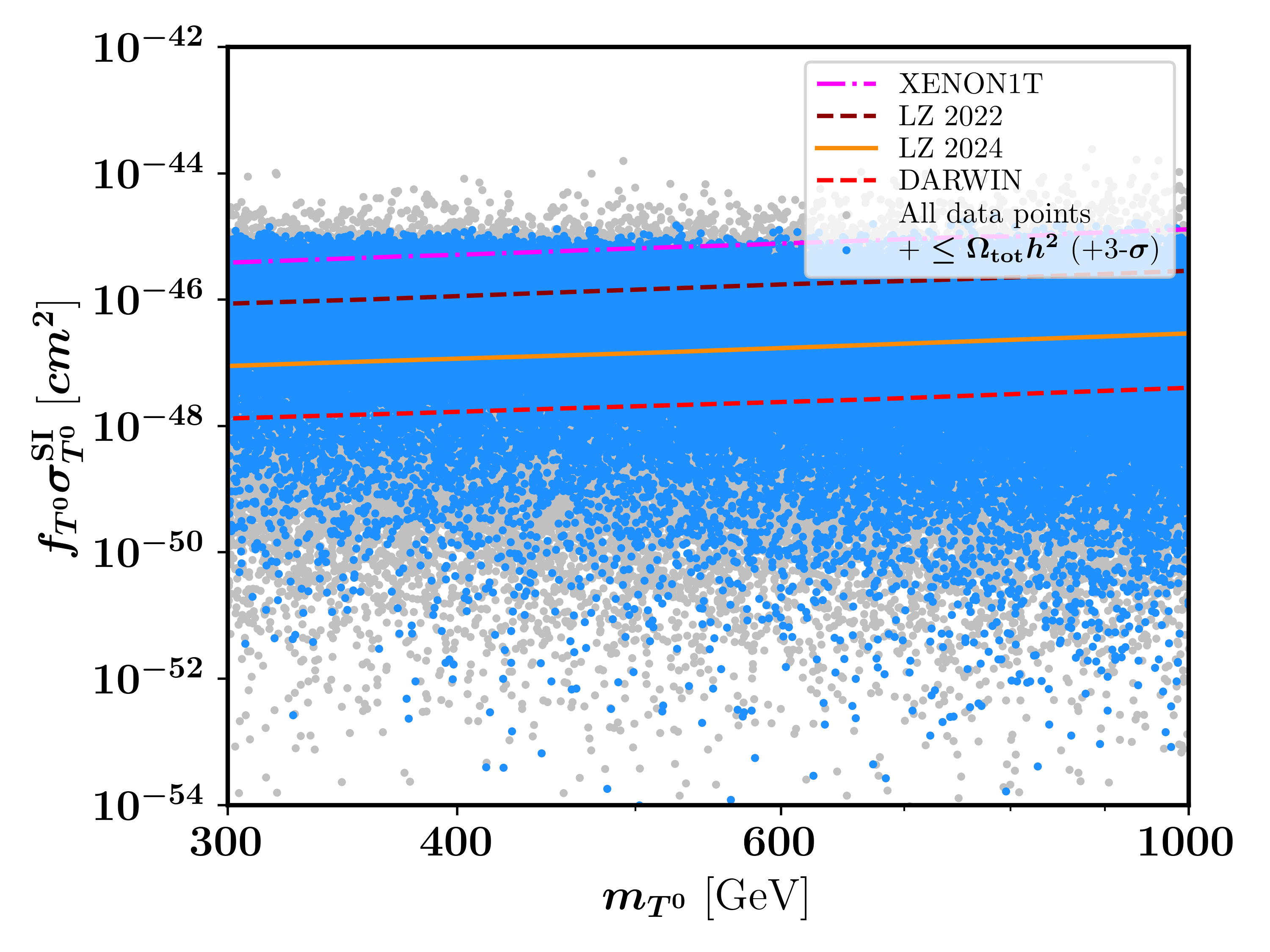}}
	\caption{Top row left plot shows $g_S$ dependence (depicted above in graded colour bar) in the $m_\psi - \Omega_\psi h^2$ plane whereas the top right plot shows the same for parameter $\sin\th$ in the $m_\psi - f_{\psi}\sigma^{\rm{SI}}_\psi$ plane. Detail specifications of these graded coloured points are given in the text body. The black-coloured dashed line corresponds to the $3\sigma$ band of $\Omega_{\rm{tot}}h^2$ as shown in Eq. (\ref{eq:constraints:DM-relic-Planck}). The bottom row plots depict variations in the $m_{T^0}-\Omega_{T^0}h^2$ plane (left) and
	in $m_{T^0} - f_{T^0}\sigma^{\rm{SI}}_{T^0}$ plane (right) with (blue coloured points) and without (light-grey coloured points) constraints which are detailed in the text body and also on the plots. The four differently styled differently coloured lines in the right row plots represent bounds from various DD experiments, as detailed on the plots. Other relevant details are given on the individual plots.}
	\label{fig:DM-pheno:relic-sigma-psi-T0}
\end{figure*}

In Fig. \ref{fig:DM-pheno:relic-sigma-psi-T0}, we
present scatter plots showing variations of $\Omega_{\psi}h^2,\,\Omega_{T^0} h^2$ and $f_\psi \sigma^{\rm{SI}}_{\psi},\, f_{T^0} \sigma^{\rm{SI}}_{T^0}$
with $m_{\psi}$ and $m_{T^0}$, respectively, subjected to various experimental
constraints. In these plots, the light-grey coloured points obey the list of constraints mentioned in Sec. \ref{sec:constraints}, barring the one on Higgs properties and of course, bounds from the DM sector. Fig. \ref{fig:DM-pheno:relic-sigma-psi-T0} (a) illustrates the impact of parameter $g_S$ on the $m_\psi - \Omega_\psi h^2$ plane while Fig. \ref{fig:DM-pheno:relic-sigma-psi-T0} (b) depicts the impact of parameter $\sin\th$ on the $m_{\psi} - f_\psi \sigma^{\rm{SI}}_{\psi}$ plane.
For $m_\psi$ below $h_1$ resonance, the process $h_1\to \psi \bar{\psi}$
appears feasible. This rules out a significant number of the light-grey coloured points,
thanks to bounds from the invisible Higgs decay as already stated in subsection \ref{subsec:constraints-from-Higgs-properties}. A small region below $h_1$ resonance can still escape the invisible $h_1$ decay bounds, owing to small
values of $g_S$ and $\sin\th$ (see Eq.~(\ref{eq:constraints:Higgs-invisible-psi-decay-width})). Nevertheless, barring a very few points that lie within the black-coloured strip, most of the points with smaller $g_S,\,\sin\th$ values yield overabundance for $\psi$ and hence, are ruled out from the relic density bound. It is evident from Fig. \ref{fig:DM-pheno:relic-sigma-psi-T0} (a) that $g_S \sim \mathcal{O}(0.1)$ (relatively dark green shade) generates overabundance for $\psi$, except the $h_1$ resonance while $g_S \gsim 2.0$ (relatively light green and yellow shade) normally yield underabundance for $\psi$. These features are reconfirmations of our findings of the last subsection.
%%%%%%%%%%%%%%%%%%%%%%%%%%%%%%%%%%%%%%%%%%%%%%%%%%%%%%%%
%%%%%%%%%%%%%%%%%%%%%%%%%%%%%%%%%%%%%%%%%%%%%%%%%%%%%%%%
Beyond the $h_1$-resonance, as shown in Fig. \ref{fig:DM-pheno:relic-sigma-psi-T0} (a),
the correct relic density (depicted by black coloured dashed line) can be accommodated for a large range of $m_\psi$. Here bounds on SM-like $h_1$
couplings and reduced signal strengths, as detailed in Sec. \ref{sec:constraints}, rule out a certain number of light-grey coloured points.
The graded coloured points for Fig. \ref{fig:DM-pheno:relic-sigma-psi-T0} (a)
correspond to different $g_S$ values which escape all constrains stated in Sec. \ref{sec:constraints}. Needless to say, the invisible Higgs decay, i.e., $h_1\to \bar{\psi} \psi$, remains kinematically forbidden in this region of $m_\psi$. From this plot, it appears that lower (higher) values of $g_S$ are preferred by higher (lower) values of $m_\psi$. In a similar way, Fig. \ref{fig:DM-pheno:relic-sigma-psi-T0} (c) shows the variation of $\Omega_{T^0} h^2$ with $m_{T^0}$ where overabundance is observed for $m_{T^0}\gsim 500$ GeV. This happens because of an additional source for $T^0$ relic through the DM conversion process $\psi \bar{\psi} \to T^0 T^0$. This conversion is more favourable at low $g_S$ values as depicted in Fig. \ref{fig:DM-pheno:relic-sigma-psi-T0} (a). The blue coloured points of Fig. \ref{fig:DM-pheno:relic-sigma-psi-T0} (c) obey all constraints mentioned in Sec. \ref{sec:constraints}
together with the $3\sigma$ upper bound on $\Omega_{\rm{tot}}h^2$ as given by Eq. (\ref{eq:constraints:DM-relic-Planck}). For this plot, unlike Fig. \ref{fig:DM-pheno:relic-sigma-psi-T0} (a), we do not show the surviving points with graded colours as the parameter $g_S$ has less significant effect on the $T^0$ phenomenology compared to $\psi$. The same observation holds true for  Fig. \ref{fig:DM-pheno:relic-sigma-psi-T0} (d) when compared with Fig. \ref{fig:DM-pheno:relic-sigma-psi-T0} (b), in the context of $\sin\th$ parameter. Combining the information gathered from Fig. \ref{fig:DM-pheno:relic-sigma-psi-T0} (a) and (c), one can see that $T^0$ remains mostly underabundant in the
window of $300-1000$ GeV while $\psi$ may appear overabundant in the same mass window, especially for $g_S \lsim 1.0$.

\begin{figure*}[!h]
	\hspace*{0.2cm} %% This will shift the figure
	\centering
	\subfigure{\includegraphics[height=6.3cm,width=8.3cm]{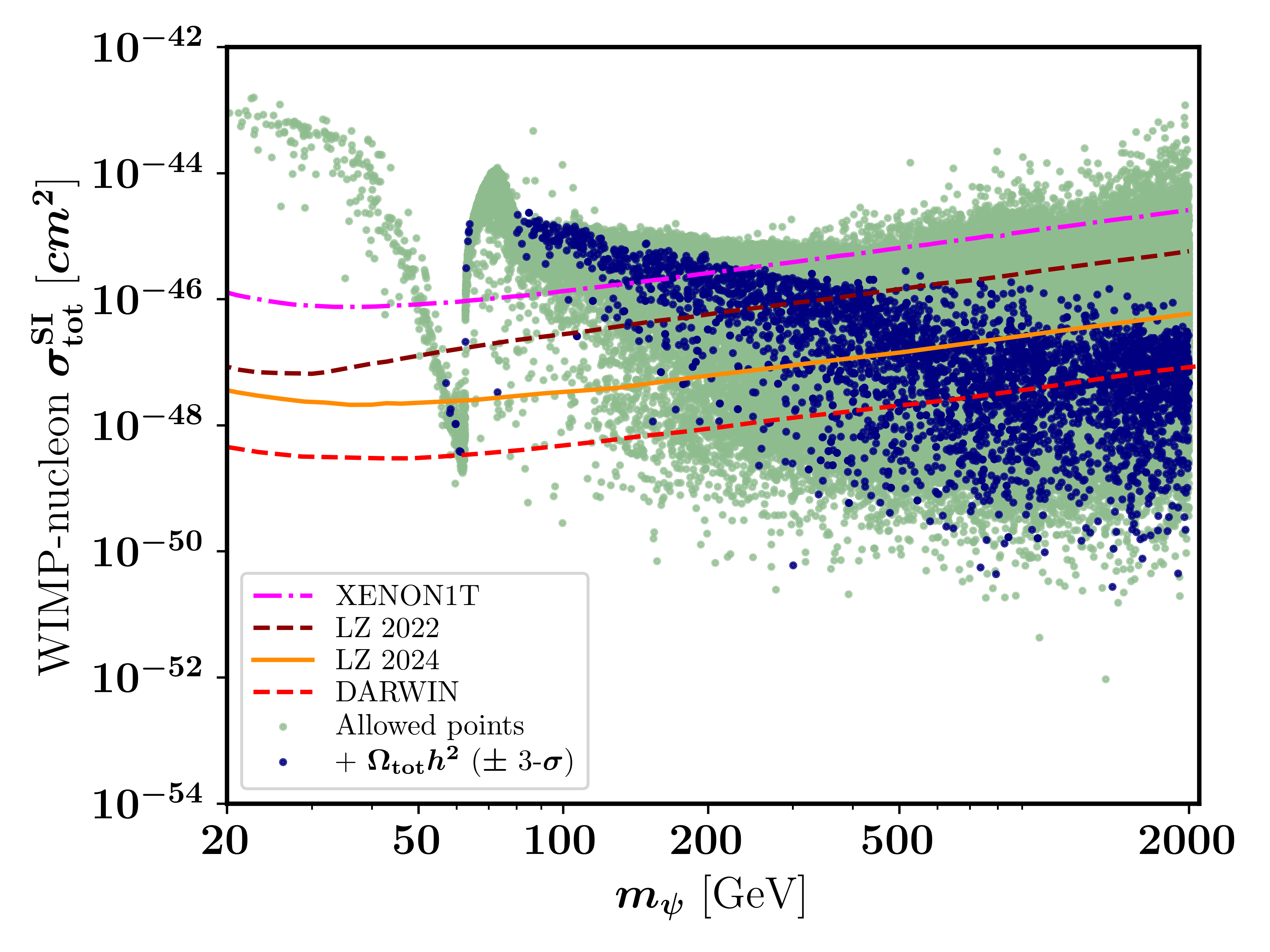}}
	\caption{$\sigma^{\rm{SI}}_{{\rm{tot}}}$ versus $m_\psi$ plot where four differently styled differently coloured lines represent bounds from four different DD experiments, as detailed on the plot. The light green coloured points survive all
constraints mentioned in Sec. \ref{sec:constraints}. The dark blue points,
besides the same, also obey the $3\sigma$ bounds on the relic density.}
	\label{fig:DM-pheno:combined-sigma}
\end{figure*}

In Fig. \ref{fig:DM-pheno:relic-sigma-psi-T0} (b) and (d), we show predictions for the weighted (see Eq.~(\ref{eq:DM-pheno:rescaled-DD})) SI DD cross-section for $\psi$ and $T^0$, respectively. For Fig. \ref{fig:DM-pheno:relic-sigma-psi-T0} (b(d)), unlike (a(c)), the graded (blue) coloured points satisfy not only all constraints presented in Sec. \ref{sec:constraints} but also remain under the $3\sigma$ upper bound on $\Omega_{\rm{tot}}h^2$ as given by Eq. (\ref{eq:constraints:DM-relic-Planck}). In these plots, bounds from various
direct searches are shown by differently styled coloured lines. Application of these experimental limits requires a combination of the duly weighted contributions of individual DM species as shown by Eq.~(\ref{eq:DM-pheno:rescaled-DD}). It is evident from Fig. \ref{fig:DM-pheno:relic-sigma-psi-T0} (b) that escaping the existing and anticipated bounds on the SI DD cross-section favours $|\sin\th|\lsim 0.15$ and $m_\psi \gsim 150$ GeV. On the other hand, $|\sin\th|\gsim 0.16$ and/or $m_\psi \lsim 100$ GeV values are still possible when $g_S$ is small ($\lesssim \mathcal{O}(0.1)$) and/or $m_{h_2}$ is not too far from $m_{h_1}$. This observation is already apparent from Eq.~(\ref{eq:DM-pheno:psi-DD-SI-appx}) and remains applicable even for DARWIN, although for a
smaller window in $m_\psi \sim 100-400$ GeV. A simultaneous look at Figs. \ref{fig:DM-pheno:relic-sigma-psi-T0} (a) and (b) suggests that the fermionic DM, which can be probed by different DD experiments, will have $\Omega_{\psi}h^2\lsim 0.1234$ (see Eq. (\ref{eq:constraints:DM-relic-Planck})). This, however, is not true for $T^0$. Here, a similar simulation analysis of Figs. \ref{fig:DM-pheno:relic-sigma-psi-T0} (c) and (d) show that the triplet DM will remain always underabundant, even though various
DD experiments can probe the corresponding SI cross-section. This is the key
advantage of the two-component DM scenario where individual species may appear underabundant, however, together they satisfy the relic density bound and appear detectable at various direct searches. This observation is evident from Fig. \ref{fig:DM-pheno:combined-sigma} where $\sigma^{\rm{SI}}_{{\rm{tot}}}$ is plotted against $m_\psi$. Here, light green coloured points survive all constraints stated in Sec. \ref{sec:constraints} while the dark blue coloured points also respect the $3\sigma$ bounds on relic density. These points, clearly remain probable by various direct searches. The observed pattern can be understood from Fig. \ref{fig:DM-pheno:relic-sigma-psi-T0} which shows that the correct $\Omega_{\rm{tot}}h^2$, for the chosen framework, primarily depends on $\Omega_\psi h^2$ as $\Omega_{T^0} h^2$ remains underabundant throughout.
%%%%%%%%%%%%%%%%%%%%%%%%%%%%%%%%%%%%%%%%%%%%%%%%%%%%%%%%
%%%%%%%%%%%%%%%%%%%%%%%%%%%%%%%%%%%%%%%%%%%%%%%%%%

\begin{figure*}[!h]
	\hspace*{0.20cm} %% This will shift the figure
	\centering
	\subfigure[]{\includegraphics[height=5.3cm,width=7.0cm]{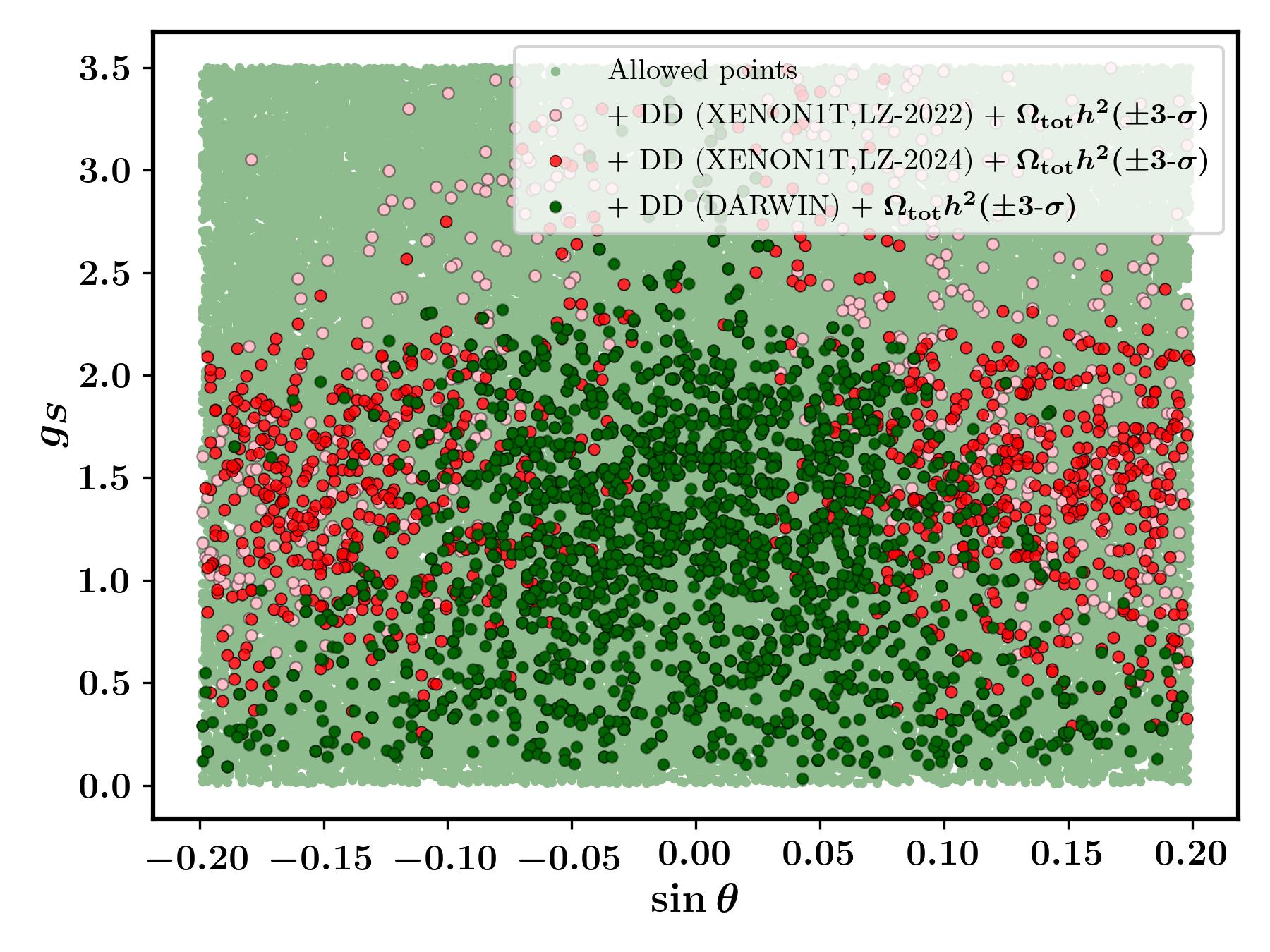}}
	\subfigure[]{\includegraphics[height=5.3cm,width=7.15cm]{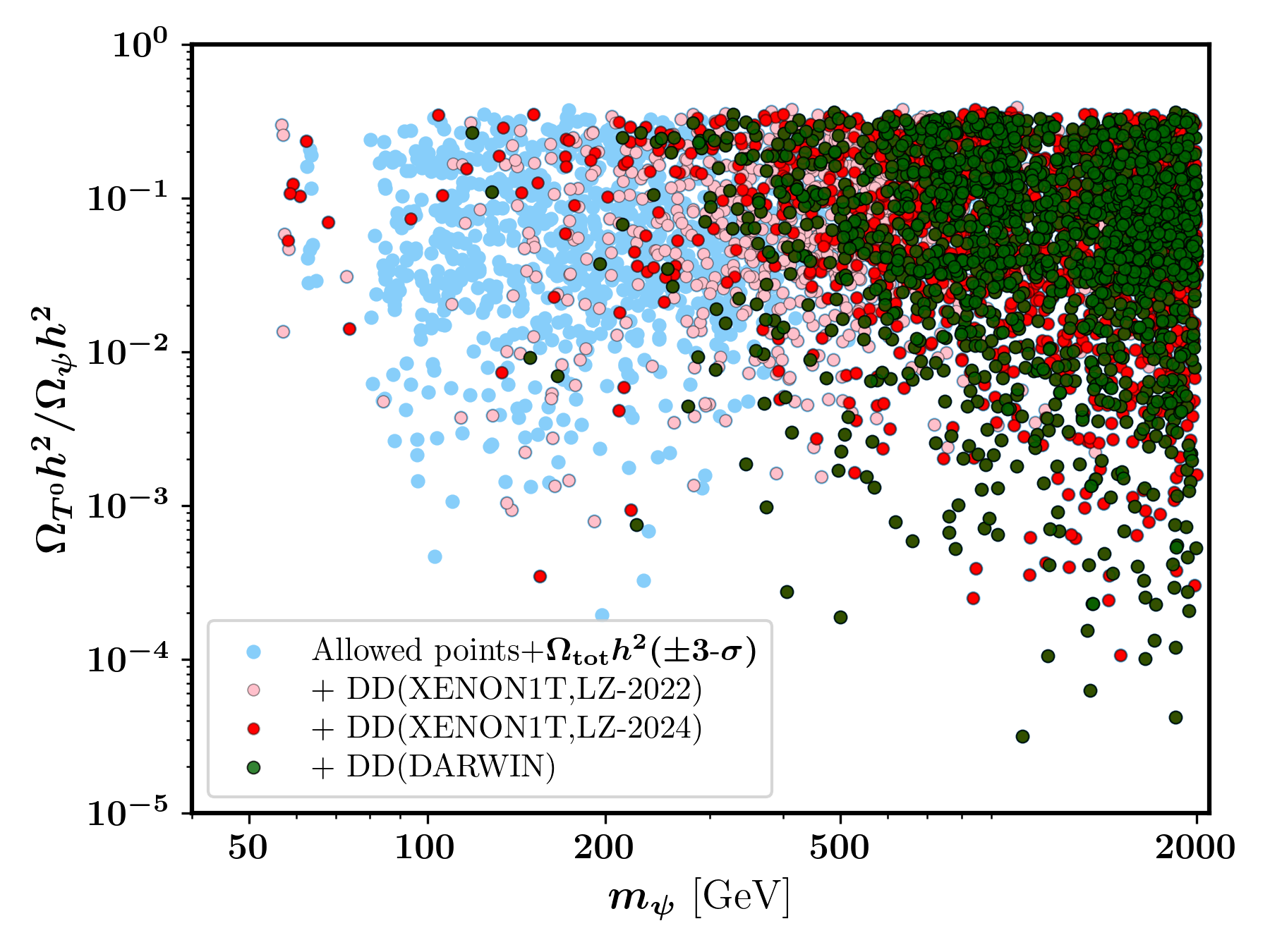}}
	\caption{The left plot depicts effects of various constraints in the
	$\sin\th-g_S$ plane whereas the right plot shows the same for
	$m_\psi -\Omega_{T^0}h^2/\Omega_\psi h^2$ plane. Here, the light green coloured points bypass all constraints mentioned in Sec. \ref{sec:constraints}. The light blue coloured points, for the right plot, depicts the same but together with the $3\sigma$ bound on the relic density (see Eq. (\ref{eq:constraints:DM-relic-Planck})). The pink-coloured points, besides constraints stated in Sec. \ref{sec:constraints}, also obey the relic density $3\sigma$ bound and are consistent with combined the DD limits reported by XENON1T and LZ-2022. The red coloured points represent the same but with updated bound from LZ-2024. Finally, dark green coloured points are used to depict the same using the projected sensitivity reach of the DARWIN experiment.}
	\label{fig:DM-pheno:parameter-correlation1}
\end{figure*}

In Figs. \ref{fig:DM-pheno:grid-scan-gS} and \ref{fig:DM-pheno:grid-scan-muST-theta} we have explored sensitivities of the DM observables $\Omega_{\psi}h^2,\, \Omega_{T^0}h^2$, $f_\psi \sigma^{\rm{SI}}_\psi,\,f_{T^0} \sigma^{\rm{SI}}_{T^0}$ on parameters $g_S$ and $\sin\th$. Now we
present the possible correlation between these two parameters, i.e., $\sin\th$ and $g_S$, in Fig. \ref{fig:DM-pheno:parameter-correlation1} (a). In this plot,
the light green coloured points bypass all constraints mentioned in Sec. \ref{sec:constraints}. The pink coloured points, besides constraints stated in Sec. \ref{sec:constraints}, also obey the relic density $3\sigma$ bound (see Eq. (\ref{eq:constraints:DM-relic-Planck})) and are consistent with the DD limits reported by XENON1T and LZ-2022. The red coloured points represent the same but with updated bound from LZ-2024. Finally, dark green coloured points are used to depict the same using the projected sensitivity reach of the DARWIN experiment. Fig. \ref{fig:DM-pheno:parameter-correlation1} (a) shows that the DD cross-section bounds from the XENON1T and LZ hardly put any restrictions on the $\sin\th$ values. These limits, however, disfavours $g_S\gsim 2.5$,
barring the $-0.10 \lsim \sin\th \lsim 0.10$ window, where $g_S\gsim 2.5$ appears admissible due to $h_1$ resonance. Larger $\sin\th$ values. i.e., $0.10 \lsim |\sin\th|\lsim 0.20$, nevertheless, can co-exist with $g_S<2.5$ region, owing to $h_2$ resonance. Including the sensitivity reach of the DARWIN experiment offers somewhat stronger bounds for $\sin\th$ values. Here, to comply with the bounds of relic density and DD-cross-section, one ends up getting  a visible correlation in the $\sin\th-g_S$ plane, which clearly favours the
$-0.10 \lsim \sin\th \lsim 0.10$ window with $g_S$\lsim 2.5. These observations are consistent with the findings of schematic analyses performed in the last subsection.
%%%%%%%%%%%%%%%%%%%%%%%%%%%%%%%%%%%%%%%%%%%%%%%%%%%%%%%%
%%%%%%%%%%%%%%%%%%%%%%%%%%%%%%%%%%%%%%%%%%%%%%%%%%%%%%%%

The last subsection clearly demonstrated the relative contributions of $\Omega_\psi h^2$ and $\Omega_{T^0}h^2$ to assure the correct relic density.
With the full parameter scan, as detailed in Table \ref{tab:par_ranges},
one can re-ensure the same by looking at Fig. \ref{fig:DM-pheno:parameter-correlation1} (b). Here, we plot the variations of $\Omega_{T^0} h^2/\Omega_\psi h^2$ with $m_\psi$. The colour
specifications remain almost the same as Fig. \ref{fig:DM-pheno:parameter-correlation1} (a), except the light blue coloured points where besides Sec. \ref{sec:constraints} constraints, $3\sigma$ bound on the relic density is also imposed. As evident from the last subsection,
unlike $\Omega_\psi h^2$, $\Omega_{T^0}h^2$ remains underabundant throughout the chosen scan range. Hence, the ratio $\Omega_{T^0} h^2/\Omega_\psi h^2$ always remains below $1$, as shown in Fig. \ref{fig:DM-pheno:parameter-correlation1} (b). The maximum
share of $\Omega_{T^0}h^2$ in $\Omega_{\rm tot} h^2$ is observed to be
$26\%$ which is evident from Fig. \ref{fig:DM-pheno:parameter-correlation1} (b). The underabundance of $T^0$, often by five orders of magnitude compared to that of $\psi$, is essential to evade bounds on SI DD cross-section. This
observation is consistent with Fig. \ref{fig:DM-pheno:relic-sigma-psi-T0} (d)
which shows that the majority of the allowed parameter space for the triplet DM remain available and can be effectively explored in various DD experiments. We also want to point out that, in a pure $Y=0$ triplet DM scenario \cite{Cirelli:2005uq}, $\Omega_{T^0}h^2$ is primarily determined by $T^0 T^0 \to W^+ W^-, Z Z$ modes where the DM mass $m_{T^0}$ plays the pivotal role and other relevant parameters, such as $\lambda_{HT}$, has a minuscule contribution. As a consequence, $\Omega_{T^0}h^2$ remains severely constrained for $m_{T^0}\lsim 1$ TeV and can share only $10\%$ of
$\Omega_{\rm{tot}}h^2$ \cite{DuttaBanik:2020jrj}. For the chosen two-component DM framework, within a minimal augmentation of the pure $Y=0$ triplet DM model, however, the presence of $\psi \bar{\psi} \to T^0 T^0$ with a proper choice of $g_S$, allows $\Omega_{T^0}h^2$ to reach $26\%$ of $\Omega_{\rm{tot}}h^2$ which was previously only $10\%$.
%%%%%%%%%%%%%%%%%%%%%%%%%%%%%%%%%%%%%%%%%%%%%%%%%%%%%%%%
\begin{figure*}[!ht]
	\hspace*{0.2cm} %% This will shift the figure
	\centering
	\subfigure{\includegraphics[height=16.3cm,width=14.2cm]{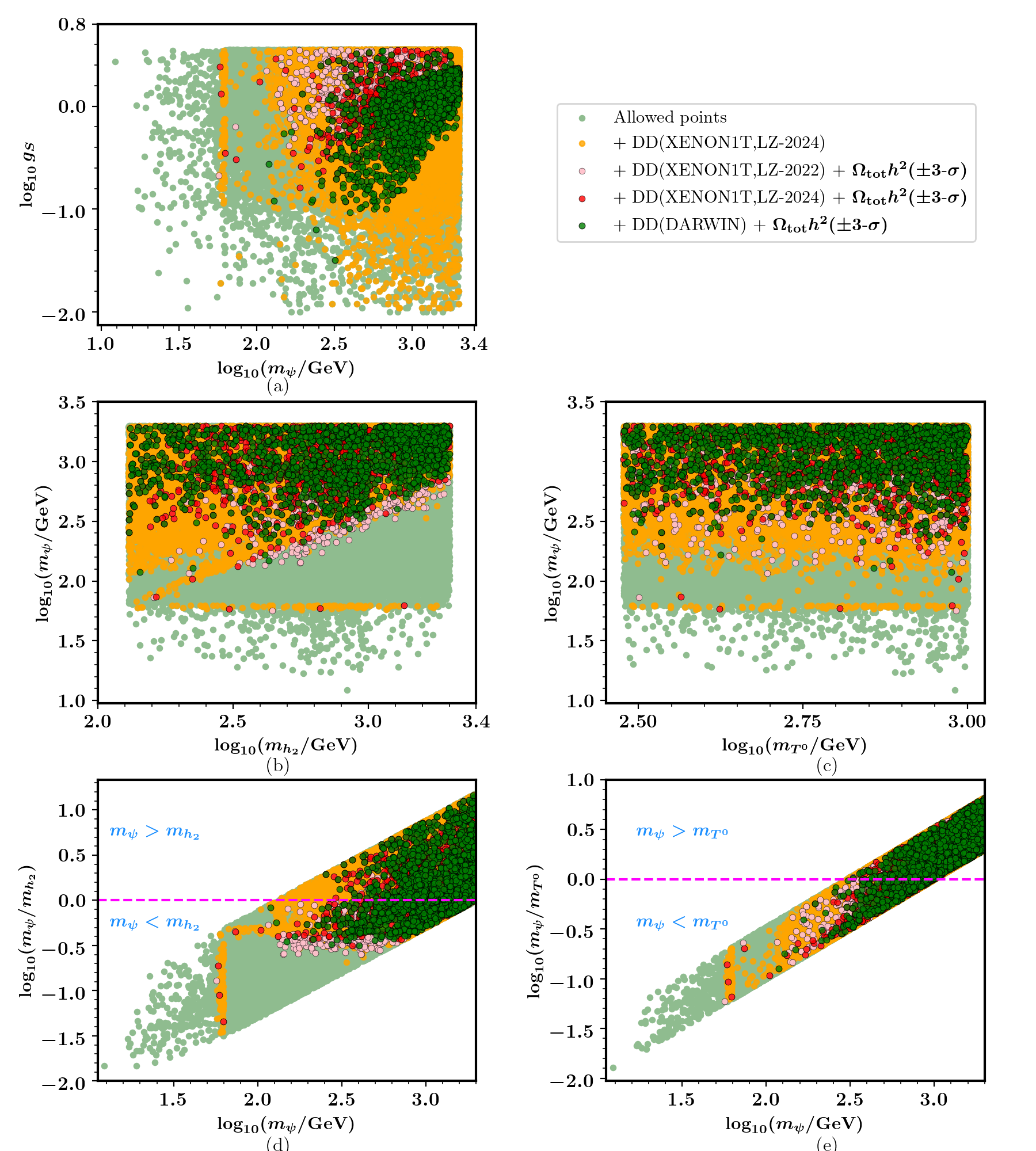}}
	\caption{Plots showing certain salient features of the chosen two-component DM model which are detailed in the text body. The colour codes are the same as of Fig. \ref{fig:DM-pheno:parameter-correlation1} (a), except the orange ones which are used to reflect the importance of the combined direct search bounds in the context of XENON1T and LZ-2024 results. The magenta-coloured dashed line in the bottom row plots
is used to represent a pair of mass degenerate states, i.e., $m_\psi = m_{h_2}$ (d) and $m_\psi = m_{T^0}$ (e).}
	\label{fig:DM-pheno:parameter-correlation2}
\end{figure*}
%%%%%%%%%%%%%%%%%%%%%%%%%%%%%%%%%%%%%%%%%%%%%%%%%%%%%%%%

Finally, in Fig. \ref{fig:DM-pheno:parameter-correlation2}, we present certain interesting features of the chosen two-component DM model. Here, for an elucidated understanding of certain features, plots are made with $\log_{10}$ scale. The colour codes are the same as of Fig. \ref{fig:DM-pheno:parameter-correlation1} (a), except the orange ones which are used to reflect
the importance of direct search bounds in the context of the combined XENON1T and LZ-2024 results without the relic density constraints. The magenta coloured dashed line in the bottom row plots
is used to represent a pair of mass degenerate states, i.e., $m_\psi = m_{h_2}$ (for Fig. \ref{fig:DM-pheno:parameter-correlation1} (d)) and $m_\psi = m_{T^0}$ (for Fig. \ref{fig:DM-pheno:parameter-correlation1} (e)). The correlation between $m_\psi$ and $g_S$ is plotted in Fig. \ref{fig:DM-pheno:parameter-correlation2} (a). Clearly, here the $3\sigma$ bound on the relic density leads over the limits from direct searches. These two constraints, together, set a lower bound on $m_\psi$ as $150$ GeV, except
for $h_1,\,h_2$ resonances. This lower bound on $m_\psi$ moves to $250$ GeV, once we consider the projected sensitivity reach from the DARWIN which also ruled out the $h_1$ resonance region. All these observations agree with our previous findings concerning Figs. \ref{fig:DM-pheno:grid-scan-gS}, \ref{fig:DM-pheno:relic-sigma-psi-T0} and \ref{fig:DM-pheno:combined-sigma}.

In Fig. \ref{fig:DM-pheno:parameter-correlation2} (b), we illustrate the correlation between $m_{\psi}$ and the heavy Higgs mass $m_{h_2}$. Here,
even the non-DM observables, as detailed in Sec. \ref{sec:constraints},
favour $m_\psi$ above the $h_1$ resonance. Application of the combined direct search
bounds from XENON1T and LZ-2024, except for the $h_1$ resonance, largely
rules out points below the $2m_\psi = m_{h_2}$ line. The broadening of the
said line is arising from a variation of $\sin\th$ near the $h_2$ resonance
as already shown in Fig. \ref{fig:DM-pheno:grid-scan-muST-theta} (c). The relic density constraint again appears to be the leading one and rules out a significant number of points around the $h_1,\,h_2$ resonances as well as in between which escapes the DD bounds. The ruled-out region between two Higgs resonances is connected with small $g_S$ values ($\lesssim \mathcal{O}(0.1)$) which easily escape the DD bounds but yield overabundant $\psi$-DM. The large $g_S$ values, on the contrary, yield underabundant $\psi$-DM which fails
to bypass DD bounds. as already depicted in Figs. \ref{fig:DM-pheno:grid-scan-gS}, \ref{fig:DM-pheno:relic-sigma-psi-T0} and \ref{fig:DM-pheno:combined-sigma}.

The triplet DM, as already depicted in Fig. \ref{fig:DM-pheno:relic-sigma-psi-T0} (c), always remains underabundant for the
entire scan range of $300\lsim m_{T^0}~({\rm GeV})\lsim 1000$. Besides,  Fig. \ref{fig:DM-pheno:relic-sigma-psi-T0} (d) shows that $m_{T^0}$ has no distinct correlation with the DD bounds.
This is exactly what one observes in Fig. \ref{fig:DM-pheno:parameter-correlation2} (c) which aims to study correlation in the $m_{T^0}-m_\psi$-plane.
The relic density bound, as usual, appears to be the dominant one. This, however, hardly affects $T^0$-DM which remains underabundant throughout in the window of $300-1000$ GeV. This $m_{T^0}$ mass window is ruled out for a pure $Y=0$ ITM scenario \cite{Cirelli:2005uq}. For the chosen framework, this mass window, however, survives the DM constraints as here pivotal contributions in the DM phenomenology are coming through $\sin\th,\, g_S$ which hardly affect the $T^0$-DM. Additionally, we have chosen $\l_{HT}$ in a range such that the triplet DM is barely affected by the DD constraints, thereby leaving the entire triplet DM mass range, considered in this analysis, unconstrained by the DM observables.
The bottom panel of Fig. \ref{fig:DM-pheno:parameter-correlation2} shows the mass hierarchy correlations between the fermionic DM $\psi$ and the heavy Higgs $h_2$ in (d), as well as the triplet DM $T^0$ in (e). Fig. \ref{fig:DM-pheno:parameter-correlation2} (d) shows that parameter space, consistent with all kinds of bounds (the ones stated in Sec. \ref{sec:constraints} + DM ones), favours $m_\psi$ in the window of $0.5 m_{h_2}$ to $10 m_{h_2}$. Once again, a few points near the $h_1$ resonance, survive all kinds of constraints which are contained in the line, parallel to the y-axis. The annihilation channel $\psi \bar{\psi} \to h_2 h_2$ plays a crucial role in acquiring the correct relic density in the $m_\psi > m_{h_2}$ region. The other hierarchy, i.e., $m_\psi < m_{h_2}$, also remains
consistent with the bounds on the DM observables, owing to the $h_2$ resonance, below the magenta-coloured line.

Fig. \ref{fig:DM-pheno:parameter-correlation2} (e) demonstrates the possible mass hierarchy between the two DM species. One can see that the combined DD limits from XENON1T and LZ-2024 rule out a reasonable number of parameter points that survive Sec. \ref{sec:constraints} constraints. Besides, these limits appear rather insensitive to the DM mass hierarchy. Application of the DM relic density bound, however, clearly favours $m_\psi > m_{T^0}$, barring Higgs resonances. Anticipated sensitivity reach from the DARWIN also trails the same pattern, although it completely rules out the $h_1$ resonance. For the $m_{\psi} < m_{T^0}$ regime, the DM conversion process $T^0 T^0 \to \psi \ovr{\psi}$ enhances $\psi$-DM concentration at the cost of lowering the same for the $T^0$ species. Thus, in this region, it is hardly possible that the contribution from the $T^0$-DM will reach the upper threshold as depicted in Fig. \ref{fig:DM-pheno:parameter-correlation1} (b). Besides, to avoid overabundant $\psi$-DM, one needs to opt for $g_S \gsim 1$ in this region as already shown in Fig. \ref{fig:DM-pheno:relic-sigma-psi-T0}. With $m_{T^0}$
in the window of $300-1000$ GeV, higher contribution from the triplet DM to the relic density is feasible only when $m_\psi > m_{T^0}$.

To summarize, in this subsection, we show that one can partially revive the ``desert'' region for the triplet DM by extending the $Y=0$ inert triplet DM model with another real scalar singlet $S$ and a Dirac fermion DM $\psi$. This enriched framework allows a large region of the parameter space that avoids various theoretical and experimental bounds including the ones from the relic density and direct searches. Our investigation reveals that the combined constraints, relic density, direct searches and others, as detailed in Sec. \ref{sec:constraints}, restrict the choice of $m_\psi$ and $g_S$, except for the $h_1$ resonance, as follows:
%%%%%%%%%%%%%%%%%%%%%%%%%%%%%%%%%%%%%
\beq\label{eq:gs-mpasi-ranges-restricted}
m_{h_2}/2 \lesssim m_{\psi}, ~~ 0.1 \lesssim g_S \lesssim 2.5.
\eeq
%%%%%%%%%%%%%%%%%%%%%%%%%%%%%%%%%%%%% 
Besides, our scan-based investigation also prefers $m_{\psi} > m_{T^0}$, to have more triplet DM contribution to the total relic density, and $m_{\psi} > m_{h_2}$ beyond the $h_2$-resonance. These limits are, however, not stringent
as these are roughly estimated by analysing the impact of different parameters on the DM phenomenology within the considered scan range as shown in Table \ref{tab:par_ranges}. Other parameters, e.g., $v_S, \lambda_S, \mu_3$ have insignificant effects on the DM phenomenology but appear crucial for the PT dynamics as will be addressed subsequently. In the coming sections, we will investigate finite temperature PT in the early Universe and their phenomenological implications, focusing particularly on the GW signals detectable by space-based observatories such as LISA, BBO, DECIGO, U-DECIGO, and others.

%%%%%%%%%%%%%%%%%%%%%%%%%%%%%%%%%%%%%%%%%%%%%%%%%%%%%%%%%%%%%%%%%%%%%%%%%%%%%%%%%%

%%%%%%%%%%%%%%%%%%%%%%%%%%%%%%%%%%%%%%%%%%%%%%%%%%%%%%%%%%%%%%%%%%%%%%%%%%%%%%%%%%
\section{Electroweak phase transition and gravitational waves} \label{sec:EWPT-GW}
%%%%%%%%%%%%%%%%%%%%%%%%%%%%%%%%%%%%%%%%%%%%%%%%%%%%%%%%%%%%%%%%%%%%%%%%%%%%%%%%%%
In this section, we will explore the origin and dynamics of the EWPT in the chosen two-component DM framework. Subsequently, we will also address the generation of GWs in this framework along with special emphasis on the detection prospects.
\subsection{One-loop effective potential}
\label{subsec:one-loop-Veff}
%%%%%%%%%%%%%%%%%%%%%%%%%%%%%%%%%%%%%%%%%%%%%%%%%%%%%%%%%%%%%%%%%%%%%%%%%%%%%%%%%%
Investigation of the EWPT dynamics requires the construction of the effective potential for the chosen model which depends on three CP-even scalar fields $\varphi \equiv h,s, T^0$, as shown in Eq. (\ref{eq:scalar-field-basis}). The full one-loop effective potential $V_{\rm eff}(\varphi,T) $ at finite temperature $T$ is written as,
%%%%%%%%%%%%%%%%%%%%%%%%%%%%%%%%%%%%%
\bea
\label{eq:pot:Eff-pot}
V_{\rm eff}(\varphi,T) = V_0(\varphi) + V_{\rm CW} (\varphi, T=0) + V_{\rm T}(\varphi,T),
\eea
%%%%%%%%%%%%%%%%%%%%%%%%%%%%%%%%%%%%%
where $V_0(\varphi)$ represents the tree-level potential, $V_{\rm CW} (\varphi, T=0)$ denotes the one-loop corrections at $T=0$, known as the Coleman-Weinberg (CW) potential \cite{Weinberg:1973am,Coleman:1973jx}, and finally, $V_{\rm T}(\varphi,T)$ is the finite temperature, i.e., $T\neq 0$ contribution. The tree-level potential, $V_0(\varphi)$, in the $\varphi_i=h,s,T^0$ basis, using Eqs. (\ref{eq:pot:total-scalar})-(\ref{eq:pot:interaction-potential}) and Eq. (\ref{eq:scalar-field-basis})\footnote{One needs to remain cautious while using Eq. (\ref{eq:scalar-field-basis}) as here one needs to use only the fields without VEVs.}, is given as,
%%%%%%%%%%%%%%%%%%%%%%%%%%%%%%%%%%%%%
\bea
\label{eq:pot:tree-level}
V_0(h,s,T^0) &=& -\frac{1}{2} \mu_{H}^2 h^2 + \frac{1}{4} \l_H h^4 - \frac{1}{2} \mu_S^2 s^2 + \frac{1}{4} \l_S s^4 - \frac{1}{2} \mu_{\bm{T}}^2 {T^0}^2 + \frac{1}{4} \l_{\bm{T}} {T^0}^{4} + \frac{1}{2} \mu_{ST}~s {T^0}^2\nn\\
&& + \frac{1}{2} \mu_{HS}\, h^2 s + \frac{1}{4} \l_{SH}\, h^2 s^2 - \frac{1}{3} \mu_3 s^3 + \frac{1}{4} \l_{HT}\, h^2 {T^0}^2 + \frac{1}{4} \l_{ST}\, s^2 {T^0}^2.
\eea
%%%%%%%%%%%%%%%%%%%%%%%%%%%%%%%%%%%%%

The inclusion of Coleman-Weinberg contributions may result in shifting the original physical minima of the effective potential from $T=0$. Therefore, the inclusion of counter-terms are necessary in Eq.~(\ref{eq:pot:Eff-pot}) to restore the physical minima and masses \cite{Cline:2011mm}. However, in this work, we choose to follow the on-shell renormalization scheme \cite{Delaunay:2007wb,Curtin:2014jma} in which the one-loop contributions do not disturb the tree-level minimization conditions. Therefore, it is not necessary to include any counter term potential ($V_{\rm CT}$) in the Eq.~(\ref{eq:pot:Eff-pot}) as conventionally done in the $\ovr{\rm MS}$ renormalization scheme \cite{Coleman:1973jx}. It should be noted that the effective potential explicitly depends on the gauge parameter $\xi$. However, the gauge-independent physical content of the effective potential can be identified using Nielsen identities \cite{Nielsen:1975fs,Fukuda:1975di}. It states that at the extrema of a field $\varphi_i$, i.e., $\widetilde{\varphi_i}$, the gauge dependence of the effective potential disappears, since
%%%%%%%%%%%%%%%%%%%%%%%%%%%%%%%%%%%%%%%%%%%%%%%%%%%%%%
\bea
\label{eq:pot:Nielsen-identity}
\frac{\partial V_{\rm eff} (\varphi, \xi)}{\partial \xi} \propto \frac{\partial V_{\rm eff} (\varphi, \xi)}{\partial \varphi_i},
\eea
%%%%%%%%%%%%%%%%%%%%%%%%%%%%%%%%%%%%%%%%%%%%%%%%%%%%%%%
and therefore,
%%%%%%%%%%%%%%%%%%%%%%%%%%%%%%%%%%%%%%%%%%%%%%%%%%%%%%%
\bea
\frac{d V_{\rm eff} (\widetilde{\varphi}, \xi)}{d \xi} = \frac{\partial V_{\rm eff} (\widetilde{\varphi}, \xi)}{\partial \xi} + \frac{d \varphi_i}{d \xi} \frac{\partial V_{\rm eff} (\widetilde{\varphi}, \xi)}{\partial \varphi_i} = 0.
\eea
%%%%%%%%%%%%%%%%%%%%%%%%%%%%%%%%%%%%%%%%%%%%%%%%%%%%%%%
However, the location of the extrema is gauge dependent, i.e., $\partial \varphi_i/\partial \xi \neq 0$. For more details on this issue and a general discussion on the gauge independent treatment, interested readers can follow, for e.g., Refs.~\cite{Patel:2011th,DiLuzio:2014bua,Fukuda:1975di,Laine:1994zq,Kripfganz:1995jx}.  In this work, we calculate the effective potential in the Landau gauge ($\xi = 0$), where the Goldstone boson contributions are decoupled from those of the massive gauge bosons, and ghosts have no impact. Moreover, it has been shown in the literature (see, for e.g., some recent works \cite{Kozaczuk:2019pet,Chiang:2019oms,Croon:2020cgk,Arunasalam:2021zrs,Borah:2023zsb} and references therein) that when barrier formation is achievable at the tree-level itself, the PT dynamics in the Landau gauge are similar to those in a gauge-dependent treatment of the effective potential. In the chosen model, the potential barrier can be obtained at the tree level itself. Therefore, we do not extend our analysis to include a gauge-independent treatment. However, as already mentioned, gauge independent treatment is available in the literature, and one should incorporate them for robust calculations. We leave this implementation in this present work due to the presence of tree-level barrier upliftment and consider the full analysis of $V_{\rm eff} (\varphi, T)$ only. The presence of the tree-level barrier appears
adequate to shed light on the parameter space of the chosen model and its interplay with the PT dynamics, even without performing a more rigorous and complete gauge-invariant analysis. The zero-temperature CW potential, in the on-shell renormalization scheme, can be written as \cite{Iliopoulos:1974ur,Sher:1988mj,Anderson:1991zb,Espinosa:1992kf},
%%%%%%%%%%%%%%%%%%%%%%%%%%%%%%%%%%%%
\bea
\label{eq:pot:V_CW}
V_{\rm CW} (\varphi, T=0) = \sum_{i} (-1)^{F_i} \frac{d_i}{64 \pi^2} \left[ m_i^4 (\varphi) \left(\log\frac{m_i^2(\varphi)}{m^2_{0i}} - \frac{3}{2}\right) + 2m_i^2(\varphi) m^2_{0i} \right],
\eea
%%%%%%%%%%%%%%%%%%%%%%%%%%%%%%%%%%%%
where $\varphi \equiv \{h,s,T^0\}$, the index $i$ runs over all particles contributing to the potential with $F_i = 0\,(1)$ for bosons (fermions), $d_i$ is the number of {\it d.o.f.} of the particle species, $m_i(\varphi)$ is the field dependent mass (see Appendix~\ref{appen:C1} for details) and $m_{0i}$ denotes the value of $m_i(\varphi)$ at the EW vacuum. The form of CW potential, as shown in Eq. (\ref{eq:pot:V_CW}), ensures that the zero temperature conditions are completely determined by the tree-level contribution \cite{Kirzhnits:1976ts}. In other words,
%%%%%%%%%%%%%%%%%%%%%%%%%%%%%%%%%%%%%%
\bea
\frac{\partial V_{\rm CW} (\varphi, T=0)}{\partial \varphi_i}\Big|_{\varphi_0} = 0 = \frac{\partial^2 V_{\rm CW} (\varphi, T=0)}{\partial\varphi_i \partial \varphi_j}\Big|_{\varphi_0},
\eea
%%%%%%%%%%%%%%%%%%%%%%%%%%%%%%%%%%%%%%
where $\varphi_0$ corresponds to the field values at the $T=0$ EW vacuum. For the chosen model, $\varphi_0 = \{v,v_s,0\}$, and $\varphi_{i,j} = \{h, s, T^0\}$.

It is important to emphasize that the contribution to the CW potential from the Goldstone bosons needs special care as the masses of such modes vanish at the physical minimum leading to infrared divergences of the effective potential. In general, this divergence is an artefact of the perturbative calculation \cite{Martin:2014bca,Elias-Miro:2014pca}, however, proper resummation of the Goldstone modes must be performed to avoid such divergences. One can deal with it by shifting the masses of the Goldstone modes by an infrared regulator, $\mu_{\rm IR}^2$ and choosing a small value for it, say $\simeq 1\, {\rm GeV^2}$. However, it has been shown in Ref. \cite{Martin:2014bca} that the numerical impact of the resummation procedure as a function of the renormalization scale is rather minuscule. Therefore, we do not include the contributions from the Goldstone boson in our numerical study.

At finite temperature, the one-loop contribution to the effective potential is encapsulated in $V_{\rm T} (\varphi, T)$ and is given by \cite{Dolan:1973qd},
%%%%%%%%%%%%%%%%%%%%%%%%%%%%%%%%%%%%%%
\bea
\label{eq:pot:thermal}
V_{\rm T} (\varphi, T) = \frac{T^4}{2 \pi} \left[ \sum_{i} n_i^B J_B\left(\frac{m_i^2(\varphi, T)}{T^2} \right) + \sum_{i} n_i^F J_F\left(\frac{m_i^2(\varphi, T)}{T^2} \right)\right], 
\eea
%%%%%%%%%%%%%%%%%%%%%%%%%%%%%%%%%%%%%%
where the two sums are over the bosonic and fermionic {\it d.o.f}. respectively, and the corresponding thermal functions \cite{Anderson:1991zb} are,
%%%%%%%%%%%%%%%%%%%%%%%%%%%%%%%%%%%%%%
\bea
\label{eq:pot:thermal-functions}
J_{B}(x) =  \int_{0}^{\infty} dy ~y^2 \ln \left(1 - e^{-\sqrt{y^2+x^2}}\right)\&~~
J_{F}(x) = - \int_{0}^{\infty} dy ~y^2 \ln \left(1 + e^{-\sqrt{y^2+x^2}}\right).
\eea
%%%%%%%%%%%%%%%%%%%%%%%%%%%%%%%%%%%%%%
Here, $B\,(F)$ stands for bosons (fermions) and $x^2 \equiv \frac{m_i^2(\varphi, T)}{T^2}$. A consistent treatment of the thermal corrections also requires resummation of the leading self-energy daisy diagrams, which shifts the field-dependent masses \cite{Curtin:2016urg}. Details of field-dependent masses, thermal masses and other relevant information are mentioned in the Appendix \ref{appx:appendixA}. Thermodynamic predictions based on the effective potential at the one-loop level have significantly been improved by the two-loop calculations \cite{Niemi:2021qvp,Schicho:2021gca,Schicho:2022wty,Ekstedt:2022bff} and non-perturbative approaches (for a recent update, see Ref. \cite{Gould:2022ran}). Moreover, a dimensionally reduced 3-dimensional effective field theory (3D EFT) approach is also being developed, which systematically includes thermal resummation, promising a gauge-independent evaluation of the effective potential while addressing several challenges of the traditional approaches \cite{Niemi:2021qvp,Schicho:2021gca,Schicho:2022wty,Gould:2022ran,Ekstedt:2022bff}. We
postpone a detailed investigation beyond the one-loop level or use a dimensionally reduced 3D EFT for future work.
%%%%%%%%%%%%%%%%%%%%%%%%%%%%%%%%%%%%%%%%%%%%%%%%%%%%%%%%%%%%%%%%%%%%%%%%%%%%%%%%%%
\subsection{PT Dynamics: nucleation and percolation}
\label{sec:Dynamics-of-PT}
%%%%%%%%%%%%%%%%%%%%%%%%%%%%%%%%%%%%%%%%%%%%%%%%%%%%%%%%%%%%%%%%%%%%%%%%%%%%%%%%%%
 Studying the characteristics of the EWPT in the early Universe using a particle physics model has two significant benefits. First, it can confirm whether the model has the potential to explain the origin of EWBG within a certain model parameter space. Second, it provides a complementary avenue to test a BSM model at GW detectors beyond the conventional collider searches. The methods for determining the thermodynamic properties of a thermal
FOPT in perturbative models are well established. At very high temperatures, thermal contributions to the effective potential dominate over other contributions, causing the fields to have vanishing VEVs at the origin of the field space. As the Universe cools and expands, additional minima appear in the scalar potential at finite field values. At a temperature $T_c$, known as the critical temperature, multiple minima of the thermal effective potential become degenerate. As temperature decreases ($< T_c$), PT can occur through tunnelling across the potential barrier separating the two minima, driven by thermal effects. A FOPT proceeds via nucleation of a broken phase in the space filled with an unstable phase. One needs to find the time of nucleation at which the probability of a true vacuum bubble forming within a horizon radius becomes significant \cite{Enqvist:1991xw}, i.e.,
%%%%%%%%%%%%%%%%%%%%%%%%%%%%%%%
\bea
\label{eq:nucleation:volume}
N(T_n) = \int_{T_n}^{T_c} \frac{dT}{T} \frac{\Gamma(T)}{H(T)^4} = 1,
\eea
%%%%%%%%%%%%%%%%%%%%%%%%%%%%%%%
where $T_n$ denotes the nucleation temperature, and
%%%%%%%%%%%%%%%%%%%%%%%%%%%%%%%
\bea
\label{eq:nucleation:probability}
\Gamma(T) = \left(\frac{S_E}{2\pi T}\right)^{3/2} T^4 e^{-S_E/T},
\eea
%%%%%%%%%%%%%%%%%%%%%%%%%%%%%%% 
is the nucleation probability per unit time and volume \cite{Linde:1977mm,Coleman:1977py,Callan:1977pt,Affleck:1980ac,Linde:1980tt,Linde:1981zj}. The quantity $S_E$ represents the three-dimensional Euclidean action corresponding to the bounce solution and can be written as \cite{Linde:1981zj}
%%%%%%%%%%%%%%%%%%%%%%%%%%%%%%
\bea
\label{eq:nucleation:euclidean-action}
S_E = \int_{0}^{\infty} 4 \pi r^2 dr \left(V_{\rm eff} (\varphi, T) + \frac{1}{2} \left(\frac{d\varphi(r)}{dr}\right)^2\right),
\eea
%%%%%%%%%%%%%%%%%%%%%%%%%%%%%%
with $r$ being the radial coordinate, and $\varphi$ corresponds to the dynamic fields of the model, i.e., $h,\,s,\,\,T^0$. The critical bubble profile $\varphi(r)$ can be derived by solving the classical field equation \cite{Affleck:1980ac,Linde:1977mm,Linde:1981zj}
%%%%%%%%%%%%%%%%%%%%%%%%%
\beq\label{eq:nucleation:phivariationr}
\frac{d^2 \varphi(r)}{dr^2} + \frac{2}{r} \frac{d\varphi(r)}{dr} = \frac{dV_{\rm eff} (\varphi, T)}{dr},
\eeq
%%%%%%%%%%%%%%%%%%%%%%%%%
and subsequently applying proper boundary conditions: $\frac{d\varphi(r)}{dr} = 0$ when $r \rightarrow 0$ and $\varphi(r) \rightarrow \varphi_{\rm false}$ when $r \rightarrow \infty$ \cite{Linde:1981zj}. Here $\varphi_{\rm false}$ represents the three-dimensional field values at the false vacua. Finally, $H(T)$ is the Hubble expansion rate of the Universe in a radiation-dominated epoch and can be expressed as,
%%%%%%%%%%%%%%%%%%%%%%%%%%%%%%%
\bea
\label{eq:nucleation:Hubble-rate}
H(T)^2 = \frac{\rho_{\rm rad}}{3 M_{\rm Pl}^2}, \quad \rho_{\rm rad} \equiv \frac{\pi^2}{30} g_{*}(T) T^4.
\eea
%%%%%%%%%%%%%%%%%%%%%%%%%%%%%%%
Here $M_{\rm Pl} = 2.4\times10^{18} {\rm ~GeV}$ is the reduced Planck mass and $\rho_{\rm rad}$ is the radiation energy density of the relativistic particle species. We have used tabulated data from the estimates of Ref.~\cite{Saikawa:2018rcs} to consider the temperature dependence of the number of {\it d.o.f.} $g_{*}(T)$. In order to compute the bounce for a model with a single field, one can use the shooting algorithm \cite{Matausek1973DirectSM} to obtain the solution. However, for a multi-scalar scenario, as in our case, i.e., $\varphi \equiv \{h,s, T^0\}$, the task becomes complicated, and if one further wishes to investigate the full parameter space of the model, it becomes more onerous. Therefore, in this work, we use the publicly available code
{\texttt{cosmoTransitions}} \cite{Wainwright:2011kj}, which uses {\texttt{PathDeformation}}  algorithm and, in principle, can deal with an arbitrary number of scalars \cite{Wainwright:2011kj,Wainwright:2013maa}.

The nucleation condition Eq.~(\ref{eq:nucleation:volume}), up to the leading order accuracy, for temperatures close to the EW scale, approximately translates to
%%%%%%%%%%%%%%%%%%%%%%%%%%%%%%%%%%%
\bea
\label{eq:nucleation:nucleation-criterion}
\frac{S_E}{T_n} \simeq 140,
\eea
%%%%%%%%%%%%%%%%%%%%%%%%%%%%%%%%%%%
which provides a good approximation for sufficiently weak transitions which can be used to estimate $T_n$. The {\tt cosmoTransitions} code already implements this criterion, and we have used this for a preliminary study of the model parameter space where an FOPT can occur. However, this nucleation criteria needs to be modified when a transition becomes strong. Besides, for a more complete treatment of the same one should include the vacuum contribution in the Hubble parameter and assess if nucleation is possible and/or whether it completes.

To examine the possibility of nucleation, we look for a solution that satisfies
%%%%%%%%%%%%%%%%%%%%%%%%%%%%%%%%%%
\bea
\label{eq:nucleation:possible-nucleation}
\Gamma(T_n) = H_{\rm total} (T_n)^4,
\eea
%%%%%%%%%%%%%%%%%%%%%%%%%%%%%%%%%%
and we define the total Hubble rate including the vacuum contribution to Eq. (\ref{eq:nucleation:Hubble-rate}) as
%%%%%%%%%%%%%%%%%%%%%%%%%%%%%%%%%%
\bea
\label{eq:nucleation:tota-Hubble-rate}
H(T)^2_{\rm total} = \frac{g_{*}(T) \pi^2 T^4}{90 M^2_{\rm Pl}} + \frac{\Delta V(T)}{3 M^2_{\rm Pl}},
\eea
%%%%%%%%%%%%%%%%%%%%%%%%%%%%%%%%%
where $\Delta V(T)$ corresponds to the potential difference between the false and true vacua, evaluated at $T$.

After assessing the possibility of nucleation, we need to check whether the transition completes \cite{Ellis:2018mja}. This can be determined by calculating the temperature at which the probability of a point remaining in the false vacuum falls below $0.71$ \cite{Lorenz:2001lor}, and then confirming that the volume of the false vacuum is indeed decreasing at that temperature. In other words, we need to check \cite{Lewicki:2021pgr, Ellis:2022lft}
%%%%%%%%%%%%%%%%%%%%%%%%%%%%%%%%%
\bea
\label{eq:percolation:percolation-condition}
I(T) = \frac{4 \pi}{3} \int_{T}^{T_c} \frac{dT^{\prime}}{H(T^{\prime})} \Gamma(T^{\prime}) \frac{r(T,T^{\prime})^3}{{T^{\prime}}^4} = 0.34, \quad T\frac{dI(T)}{dT} < -3,
\eea
%%%%%%%%%%%%%%%%%%%%%%%%%%%%%%%%%
where $I(t)$ represents the fragment of the space that has already been converted to the broken phase, and
%%%%%%%%%%%%%%%%%%%%%%%%%%%%%%%%%
\bea
\label{eq:percolation:comoving-radius}
r(t,t^{\prime}) = \int_{t}^{t^{\prime}} \frac{v_w(\tilde{t}) d\tilde{t}}{a(\tilde{t})}
\eea
%%%%%%%%%%%%%%%%%%%%%%%%%%%%%%%%%
corresponds to the comoving radius of the bubble, $a(t)$ is the scale factor and $v_w(t)$ is the bubble wall velocity. The temperature at which both of the aforementioned conditions are met is known as the percolation temperature, $T_p$.

Usually, it is assumed that the FOPTs are instant and complete at a temperature $T \simeq T_n$. Therefore, all the parameters determining GW signals are typically evaluated at this temperature. However, the percolation temperature $T_p$ provides a more accurate picture of when the transition completes. Furthermore, recent studies, e.g., Ref. \cite{Athron:2022mmm}, proposed that
$T_n$ may not be suitable for the cases when the transition is strong, and $T_p$ can reflect the PT process more accurately. Therefore, one should, in principle, evaluate all physical observables at $T=T_p$. However, calculating this is quite challenging because the comoving radius in Eq.~(\ref{eq:percolation:comoving-radius}) depends on $v_w$. Although the calculation of $v_w$ is involved, it is safe to approximate $v_w \approx 1$ instead of solving the Boltzmann equations leading to $v_w$ as a solution. One can take further inspiration from this approximation as it is the value used in the recent NANOGrav report \cite{NANOGrav:2023hvm}. However, there exists a good numerical approximation for $v_w$ using the thermal equilibrium as a starting assumption as mentioned in Ref. \cite{Lewicki:2021pgr}, and one can use the following appropriate analytical formula:
%%%%%%%%%%%%%%%%%%%%%%%%%%%%%%%
\begin{align}
	\label{eq:percolation:bubble-velo-appx}
	v_w^{\rm appx} &= 
	\begin{dcases}
		\sqrt{\frac{\Delta V(T)}{\alpha \rho_{\rm rad}}} \quad {\rm for} \quad  \sqrt{\frac{\Delta V(T)}{\alpha \rho_{\rm rad}}} < v_J(\alpha),\\
		~~1 \quad ~~~~~~~{\rm for} \quad \sqrt{\frac{\Delta V(T)}{\alpha \rho_{\rm rad}}} \geq v_J(\alpha),
	\end{dcases} 
\end{align}
%%%%%%%%%%%%%%%%%%%%%%%%%%%%%%%
where $\alpha \rho_{\rm rad}$ denotes the latent heat released during the transition, $\Delta V(T)$ is already defined following Eq. (\ref{eq:nucleation:tota-Hubble-rate}), and $v_J(\alpha)$ is the Chapman-Jouguet velocity \cite{Steinhardt:1981ct, Kamionkowski:1993fg, Espinosa:2010hh} expressed as,
%%%%%%%%%%%%%%%%%%%%%%%%%%%%%%%
\bea
\label{eq:percolation:Chapman-velo}
v_J(\alpha) = \frac{1}{\sqrt{3}} \frac{1 + \sqrt{3 \alpha^2 + 2 \alpha}}{1 + \alpha}.
\eea
%%%%%%%%%%%%%%%%%%%%%%%%%%%%%%%
In this work, for the GWs analyses, we consider $v_w\approx 1$. Nevertheless, for completeness, we will also evaluate $v_w$ using Eqs.~(\ref{eq:percolation:bubble-velo-appx}) and (\ref{eq:percolation:Chapman-velo}) to study $v_w$ sensitivity of the different PT parameters. The parameter $\alpha$ evaluates the strength of the PT and will be described later.

At this point, it is also important to define, quantitatively, the strength of a PT, i.e., criteria to consider a PT to be a strong one or weak one which can subsequently be used to examine the viability of EWBG. In general, it is known as the baryon washout condition \cite{Farrar:1993hn} which, at the critical temperature evaluation, translates into
%%%%%%%%%%%%%%%%%%%%%%%%%%%%%%%
\bea
\label{eq:PT-strength:Tc}
\xi_c = \frac{v_c}{T_c} \gtrsim 1.0,
\eea
%%%%%%%%%%%%%%%%%%%%%%%%%%%%%%%
with $v_c$ being the VEV of the SM-like Higgs $h_1$ at $T=T_c$. 
For the SFOPT, one needs $\xi_c > 1$ \cite{Quiros:1994dr,Moore:1998swa}.   Similarly, at the nucleation temperature calculation, the same criteria reads as
%%%%%%%%%%%%%%%%%%%%%%%%%%%%%%%
\bea
\label{eq:PT-strength:Tn}
\xi_n = \frac{v_n}{T_n} = \frac{\sqrt{\left(\langle \varphi^{lT}_i\rangle - \langle \varphi^{hT}_i\rangle \right)^2} }{T_n} \gtrsim 1.0,
\eea
%%%%%%%%%%%%%%%%%%%%%%%%%%%%%%%
with $\varphi_i \equiv \{h,s,T^0\}$ and the notation, $\langle\varphi^{lT} \rangle$  denotes the low temperature minimum at true vacuum while $\langle\varphi^{hT} \rangle$ is the high temperature minimum at false vacuum. For the EWBG to be favoured, one must have $\langle h^{hT} \rangle = 0$ along with Eqs.~(\ref{eq:PT-strength:Tc}) and (\ref{eq:PT-strength:Tn}). Note that, the sphaleron washout conditions, as mentioned above, are not absolute and in fact are subject to some debate which could lead to minor modification in the range of $\xi_c$ and $\xi_n$ as $0.5 - 1.5$ \cite{Quiros:1999jp,Funakubo:2009eg,Fuyuto:2014yia}. However, in our analysis, we shall stick to the criteria mentioned in Eqs.~(\ref{eq:PT-strength:Tc}) and (\ref{eq:PT-strength:Tn}) in order to investigate the model parameter space.

While concluding this subsection, let us define two more important parameters derived from the PT analysis and are crucial for the GW spectrum calculation. The first parameter is $\alpha$ which has already been mentioned in Eqs. (\ref{eq:percolation:bubble-velo-appx}) and (\ref{eq:percolation:Chapman-velo}). The formal definition of $\alpha$,
following Refs.~\cite{Caprini:2019egz,Lewicki:2021pgr}, is given as
%%%%%%%%%%%%%%%%%%%%%%%%%%%%%%%%
\bea
\label{eq:PT-GW:alpha}
\alpha = \frac{\Delta V_{\rm eff} (\varphi, T) - \frac{T}{4} \Delta \frac{\partial V_{\rm eff} (\varphi, T)}{\partial T}}{\rho_{\rm rad}}\Bigg|_{T = T_*}.
\eea
%%%%%%%%%%%%%%%%%%%%%%%%%%%%%%%%
Here $V_{\rm eff}(\varphi, T)$ is defined in Eq. (\ref{eq:pot:Eff-pot}) and $\Delta V_{\rm eff} (\varphi, T)$ is the difference of $V_{\rm eff} (\varphi, T)$ values between the false, i.e., $\varphi=\varphi_{\rm{false}}$, and the true, i.e., $\varphi=\varphi_{\rm{true}}$, vacua\footnote{Following Eq. (\ref{eq:nucleation:Hubble-rate}), one should remain careful while estimating $g_*(T)$ which, in general, differs for the true and the false vacua. We, however, use the
refined estimates of $g_{\ast}(T)$ from Ref. \cite{Saikawa:2018rcs} that accurately accounts for the relevant thermodynamic effects without explicitly distinguishing between the two vacua.} (and similarly for $\Delta \frac{\partial V_{\rm eff}(\varphi, T)}{\partial T}$). Here, $T_*$ is the reference transition temperature.
%Note that, following \cite{Lewicki:2021pgr}, we have also kept $1/4$ on the second term of the numerator in $\alpha$ since it specifies the difference in the normalized trace of the energy-momentum tensor rather than merely the difference in the normalized energy density.

Finally, the other important parameter is the PT inverse time duration and is quantified in the parameter $\beta$ \cite{Grojean:2006bp},
%%%%%%%%%%%%%%%%%%%%%%%%%%%%%%%
\bea
\label{eq:PT-GW:beta}
\frac{\beta}{H_*} = T_{*} \frac{d}{dT} \left(\frac{S_E}{T}\right)\Bigg|_{T = T_*}.
\eea
%%%%%%%%%%%%%%%%%%%%%%%%%%%%%%%
We have calculated both $\alpha$ and $\beta$ at $T_* = T_n$ and also at the temperature when, after the transition, the vacuum energy is converted into radiation \cite{Leitao:2015fmj,Cai:2017tmh,Ellis:2018mja},
%%%%%%%%%%%%%%%%%%%%%%%%%%%%%%%
\bea
\label{eq:PT-GW-transition-temp}
T_* = T_p \left(1 + \alpha(T_p)\right)^{1/4}.
\eea
%%%%%%%%%%%%%%%%%%%%%%%%%%%%%%%
However, for the case of weak ($\alpha \lesssim \mathcal{O}(0.01)$) and/or intermediate transitions ($\alpha \sim \mathcal{O}(0.1)$) \cite{Athron:2024xrh}, one can consider $T_{\star} \simeq T_p$. The pivotal role played by the parameters $\alpha$ and $\beta$ in
determining the GW spectrum will be discussed in detail in the subsequent
subsections. For a more detailed discussion on nucleation and percolation, interested readers can refer to the recent review \cite{Athron:2023xlk}, and references therein.
%%%%%%%%%%%%%%%%%%%%%%%%%%%%%%%%%%%%%%%%%%%%%%%%%%%%%%%%%%%%%%%%%%%%%%%%%%%%%%%%%%%
\subsubsection{PT patterns}
\label{subsubsection-PT-patterns}
%%%%%%%%%%%%%%%%%%%%%%%%%%%%%%%%%%%%%%%%%%%%%%%%%%%%%%%%%%%%%%%%%%%%%%%%%%%%%%%%%%%
 The presence of three dynamic fields in the chosen framework opens up the avenue of complex PT patterns. In principle, the PT can occur from the symmetric phase, $\mathscr{O} \equiv \{0,0,0\}$, at very high temperature ($T \gg T_c$), to any individual field direction, i.e., $\{h,s,T^0\}$, or along some mixed field directions. However, due to the presence of linear and cubic terms in $S$ in the potential (see Eqs. (\ref{eq:pot:different-part}), (\ref{eq:pot:interaction-potential}) ), the symmetry in the $s$-direction is not necessarily restored and a transition can occur from high-temperature vacuum $\mathscr{O}^{\prime} \equiv \{0,s_0,0\}$ to a low-temperature vacuum configuration.
Moreover, we consider the triplet mass parameter $\mu^2_{\bm{T}} < 0$ (see subsection \ref{subsec:scalarmassmixing}) to ensure that the potential $V_0(h,s, T^0)$ (see Eq. (5.2)) does not develop a minimum along the $T^0$-direction \cite{Niemi:2020hto}. While thermal corrections may induce a non-zero VEV in this direction, it must remain small, particularly at the end of the transition or at the EW minimum ($T = 0$), to satisfy constraints from the $\rho$-parameter \cite{Khan:2016sxm,ParticleDataGroup:2020ssz}. However, our chosen parameter space, with $m_{T^0} > 300$ GeV and $\lambda_{HT} < 0.3$, inherently excludes regions where a PT along the $T^0$-direction could occur \cite{Niemi:2020hto}. As outlined in
Sec. \ref{sec:constraints}, the lower bound on $m_{T^0}$ aligns with constraints from disappearing charged tracks, while the small Higgs-triplet coupling, i.e., $\l_{HT}<0.3,$ ensures compatibility with the stringent direct DM detection limits for $T^0$ as a DM candidate \cite{Bandyopadhyay:2021ipw}. Consequently, we find no transitions along the $T^0$-direction within the viable model parameter space. Therefore, it is safe to discard any transitions along the $T^0$-direction, and we ensure that all the transitions end with $\varphi_0 = \{v,v_s,0\}$ at $T=0$.
%\sout{Moreover, in our study, $T^0$ constitutes one of the DM species, hence we are not interested in any PT dynamics along $T^0$-direction at any temperature. Also, any development of VEV along the $T^0$-direction needs to be small, at least when the transition completes or at $T=0$ since its VEV is constrained from the $\rho$-parameter {Khan:2016sxm,ParticleDataGroup:2020ssz}. In fact, in our model, the majority of the parameter space does not show any PT pattern along the $T^0$-direction. Therefore, it is safe to discard all those transitions along $T^0$ directions (if there are any) irrespective of their values and we ensure that all the transitions end with $\varphi_0 = \{v,v_s,0\}$ at $T=0$.}

In this work, therefore, we shall consider transitions with the following PT patterns,
%%%%%%%%%%%%%%%%%%%%%%%%%%%%%%%%%%%%%%%%%
\bea
\label{eq:PT-pattern}
\mathscr{O} \xlongrightarrow{\text{I}} h', \quad \mathscr{O} \xlongrightarrow{\text{II}} h's', \quad \mathscr{O}^{\prime} \xlongrightarrow{\text{III}} s', \quad \mathscr{O}^{\prime} \xlongrightarrow{\text{IV}} h's',
\eea
%%%%%%%%%%%%%%%%%%%%%%%%%%%%%%%%%%%%%%%%%
where symbols $h',\,s'$ are used to depict certain low-temperature VEVs in the true vacua, $v',\,s'$, respectively.
So in Eq. (\ref{eq:PT-pattern}), $\mathscr{O} \xlongrightarrow{\text{II}} h's'$ implies a transition $(0,0) \rightarrow (v',s')$ while
$\mathscr{O}^{\prime} \xlongrightarrow{\text{IV}} h's'$ depicts
a transition $(0,s_0) \rightarrow (v',s')$, when temperature is lowered.
The other transition patterns bear similar meanings. Since we are only interested in the parameter regions that can accommodate EWBG, we consider all those parameter points that show at least one transition along the $h$-direction for subsequent analyses. Therefore, we do not plan to divide our parameter regions depending on different PT patterns mentioned in Eq.~(\ref{eq:PT-pattern}).
%%%%%%%%%%%%%%%%%%%%%%%%%%%%%%%%%%%%%%%%%%%%%%%%%%%%%%%%%%%%%%%%%%%%%%%%%%%%%%%%%%
\subsection{GWs from the SFOPT: contributors} \label{subsec:GW-from-SFOPT}
%%%%%%%%%%%%%%%%%%%%%%%%%%%%%%%%%%%%%%%%%%%%%%%%%%%%%%%%%%%%%%%%%%%%%%%%%%%%%%%%%%
In this section, we present detailed discussions about three possible mechanisms that can generate GWs. An SFOPT can give rise to stochastic GW background mainly through three different mechanisms: (i) bubble wall collisions \cite{Turner:1990rc,Kosowsky:1991ua,Kosowsky:1992rz}, (ii) sound waves \cite{Hindmarsh:2013xza,Hindmarsh:2015qta,Hindmarsh:2016lnk}, and (iii) magneto-hydrodynamic (MHD) turbulence in the plasma \cite{Kamionkowski:1993fg,Kosowsky:2001xp,Dolgov:2002ra,Gogoberidze:2007an,Caprini:2009yp}. These three processes, in general, coexist and the full GW signal is the sum of these three contributions (approximately),
%%%%%%%%%%%%%%%%%%%%%%%%%%%%%%%
\bea
\label{eq:GW:total-contribution}
\Omega_{\rm GW} h^2 \simeq \Omega_b h^2 + \Omega_{\rm sw} h^2 + \Omega_{\rm t} h^2, ~~{\rm respectively,}
\eea
%%%%%%%%%%%%%%%%%%%%%%%%%%%%%%%
where, $h=H_0/(100~{\rm km} \cdot {\rm sec}^{-1} \cdot {\rm Mpc}^{-1})$ with $H_0$ corresponding to Hubble’s constant at the present epoch. Below we provide a brief detail of these three different mechanisms.
%%%%%%%%%%%%%%%%%%%%%%%%%%%%%%%%%%%%%%%%%%%%%%%%%%%%%%%%%%%%%%%%%%%%%%%%%%%%%%%%%%
\subsubsection*{(i) Bubble collision}
Generally, the contribution from bubble collision to the total GW signal due to PT can be treated using the ``envelope approximation'' \cite{Kosowsky:1991ua,Kosowsky:1992rz,Kosowsky:1992vn}. Under this approximation, utilizing numerical simulation, the contribution to the total GW amplitude as a function of frequency ($f$) is given by \cite{Huber:2008hg},
%%%%%%%%%%%%%%%%%%%%%%%%%%%%%%%%%%%%%
\bea
\label{eq:GW:bubble-collision-amp}
\Omega_{\rm b} h^2 (f) = 1.67 \times 10^{-5} \left(\frac{H_*}{\beta}\right)^2 \left(\frac{\kappa_b \alpha}{1 + \alpha}\right)^2  \left(\frac{100}{g_*(T)}\right)^{\frac{1}{3}} \left(  \frac{0.11 v_w^3}{0.42 + v_w^2}\right) S_{\rm env}(f).
\eea
%%%%%%%%%%%%%%%%%%%%%%%%%%%%%%%%%%%%%
Here quantifies $\beta/H_*, \alpha$, $g_*(T)$ and $v_w$ are defined
following Eqs. (\ref{eq:PT-GW:beta}), (\ref{eq:PT-GW:alpha}), (\ref{eq:nucleation:Hubble-rate}) and (\ref{eq:percolation:comoving-radius}), respectively. The entity $\kappa_b$ is the fraction of latent heat transformed into the kinetic energy of the scalar field; $S_{\rm env} (f)$ encapsulates the spectral shape of the GW radiation and a fit to the simulation data yields,
%%%%%%%%%%%%%%%%%%%%%%%%%%%%%%%%%%%%%
\bea
\label{eq:GW:bubble-shape}
S_{\rm env} (f) = \frac{3.8(f/f_{\rm env})^{2.8}}{1 + 2.8 (f/f_{\rm env})^{3.8}},
\eea
%%%%%%%%%%%%%%%%%%%%%%%%%%%%%%%%%%%%%
with $f_{\rm env}$ denoting the peak frequency and it is defined as \cite{Caprini:2015zlo}
%%%%%%%%%%%%%%%%%%%%%%%%%%%%%%%%%%%%%
\bea
\label{eq:GW:bubble-peak-freq}
f_{\rm env} = 16.5 \times 10^{-6} {\rm Hz} \left(\frac{0.62}{1.8 - 0.1 v_w + v_w^2}\right) \left(\frac{\beta}{H_*}\right) \left(\frac{T_*}{100\,{\rm GeV}}\right) \left(\frac{g_*(T)}{100}\right)^{1/6},
\eea
%%%%%%%%%%%%%%%%%%%%%%%%%%%%%%%%%%%%%
where $T_*$ is introduced in Eq. (\ref{eq:PT-GW-transition-temp}). Note that the peak frequency in Eq.~(\ref{eq:GW:bubble-peak-freq}) and also the peak amplitude in Eq.~(\ref{eq:GW:bubble-collision-amp}) is redshifted to today. In deriving this, it is assumed that the Universe transitioned directly to a radiation-dominated phase following the PT and has since expanded adiabatically.
%%%%%%%%%%%%%%%%%%%%%%%%%%%%%%%%%%%%%%%%%%%%%%%%%%%%%%%%%%%%%%%%%%%%%%%%%%%%%%%%%%
\subsubsection*{(ii) Sound waves}
%%%%%%%%%%%%%%%%%%%%%%%%%%%%%%%%%%%%%%%%%%%%%%%%%
 Percolation generates bulk motion in the fluid, manifesting as sound waves. Using the results of lattice simulations to compute the GW signal, the contribution to the total GW density from sound waves can be parameterized as \cite{Hindmarsh:2013xza,Hindmarsh:2016lnk,Hindmarsh:2017gnf,Caprini:2019egz,Guo:2020grp},
%%%%%%%%%%%%%%%%%%%%%%%%%%%%%%%%
\bea
\label{eq:GW:soundwave-amp}
\Omega_{\rm sw} h^2 &=& 4.13 \times 10^{-7} \left(1 - \frac{1}{\sqrt{1+ 2 \tau_{\rm sw} H_*}}\right) \left(\frac{100}{g_*(T)}\right)^{\frac{1}{3}} S_{\rm sw}(f)\nn\\
	&& \times 
\begin{dcases}
	\left(\frac{\kappa_{\rm sw} \alpha}{1 + \alpha}\right)^2 (R_* H_*), ~~~~{\rm for} ~~~\left(R_* H_* \lesssim \sqrt{\frac{3}{4} \kappa_{\rm sw} \alpha/(1+\alpha)}\right)\\
	\left(\frac{\kappa_{\rm sw} \alpha}{1 + \alpha}\right)^{\frac{3}{2}} (R_* H_*)^2, ~~~{\rm for} ~~~\left(R_* H_* > \sqrt{\frac{3}{4} \kappa_{\rm sw} \alpha/(1+\alpha)}\right).
\end{dcases} 
\eea
%%%%%%%%%%%%%%%%%%%%%%%%%%%%%%%%
Here quantifies $\beta/H_*, \alpha$, $g_*(T)$ and $v_w$ are defined
following Eqs. (\ref{eq:PT-GW:beta}), (\ref{eq:PT-GW:alpha}), (\ref{eq:nucleation:Hubble-rate}) and (\ref{eq:percolation:comoving-radius}), respectively, $R_* = (8\pi)^{1/3} v_w/\beta$ and the spectral shape is given by,
%%%%%%%%%%%%%%%%%%%%%%%%%%%%%%%%
\bea
\label{eq:GW:soundwave-spectral-shape}
S_{\rm sw} (f) = \left(\frac{f}{f_{\rm sw}}\right)^3 \left[\frac{4}{7} + \frac{3}{7} \left(\frac{f}{f_{\rm sw}}\right)^2\right]^{-\frac{7}{2}}.
\eea
%%%%%%%%%%%%%%%%%%%%%%%%%%%%%%%%
The peak frequency, using the results of Ref. \cite{Saikawa:2018rcs}, is expressed as
%%%%%%%%%%%%%%%%%%%%%%%%%%%%%%%%
\bea
\label{eq:GW:soundwave-peak-freq}
f_{\rm sw} = 2.6 \times 10^{-5} {\rm Hz}\, (R_* H_*)^{-1} \left(\frac{T_*}{100\,{\rm GeV}}\right) \left(\frac{g_*(T)}{100}\right)^{\frac{1}{6}}.
\eea
%%%%%%%%%%%%%%%%%%%%%%%%%%%%%%%%
For the lifetime of the sound waves, $\tau_{\rm sw}$, we use the approximation, normalized to Hubble rate, following Refs. \cite{Hindmarsh:2017gnf,Ellis:2018mja,Ellis:2019oqb}
%%%%%%%%%%%%%%%%%%%%%%%%%%%%%%%%%
\bea
\label{eq:GW:soundwave-lifetime}
\tau_{\rm sw} H_* = \frac{H_* R_*}{U_f}, \quad U_f \simeq \sqrt{\frac{3}{4} \frac{\alpha}{1 + \alpha} \kappa_{\rm sw}},
\eea
%%%%%%%%%%%%%%%%%%%%%%%%%%%%%%%%%
where $R_*$ and $U_f$ are the mean bubble separation and
the root-mean-squared fluid velocity which can be obtained from a hydrodynamic analysis, respectively. Finally, using the fluid velocity and temperature profiles \cite{Lewicki:2021pgr}, the sound wave efficiency factor, $\kappa_{\rm sw}$, can be obtained from Ref. \cite{Espinosa:2010hh} as
%%%%%%%%%%%%%%%%%%%%%%%%%%%%%%%%%%
\bea
\label{eq:GW:soundwave-kappa_sw}
\kappa_{\rm sw} &=& \frac{3}{\alpha \rho_{\rm rad} v_w^3} \int w ~\widetilde{\xi}^2 \frac{v^2}{1-v^2} d\widetilde{\xi} \nn\\
&=& \frac{4}{\alpha v_w^3} \int \left(\frac{T(\widetilde{\xi})}{T_p}\right)^4 ~\widetilde{\xi}^2 \frac{v^2}{1-v^2} d\widetilde{\xi},
\eea
%%%%%%%%%%%%%%%%%%%%%%%%%%%%%%%%%%
where $w$ is the enthalpy of the plasma and the variable $\widetilde{\xi}$ has units of velocity \cite{Lewicki:2021pgr}. In principle, one can use numerical fits to obtain Eq.~(\ref{eq:GW:soundwave-kappa_sw}) \cite{Espinosa:2010hh}. For the details of the numerical fits of the efficiency coefficients that we have utilized in our work, see Appendix. \ref{appx:efficiency-coeff}.
%%%%%%%%%%%%%%%%%%%%%%%%%%%%%%%%%%%%%%%%%%%%%%%%%%%%%%%%%%%%%%%%%%%%%%%%%%%%%%%%%%
\subsubsection*{(iii) Turbulence}
The last piece that contributes to the GW source is turbulence. During the time of FOPT, the plasma becomes fully ionized resulting in MHD turbulence in the plasma leading to a stochastic GW background. For the turbulence contribution to $\Omega_{\rm GW} h^2$, we consider Ref.~\cite{Caprini:2009yp}
%%%%%%%%%%%%%%%%%%%%%%%%%%%%%
\beq
\label{eq:GW:turb-amp}
\Omega_{\rm tur} h^2 = 3.35 \times 10^{-4} \left( \frac{\beta}{H_{\ast}} \right)^{-1} v_w \left( \frac{\kappa_{\rm tur} \alpha}{1 + \alpha} \right)^{3/2} \left( \frac{100}{g_{\ast}(T)}\right)^{1/3} S_{\rm t}(f),
\eeq
%%%%%%%%%%%%%%%%%%%%%%%%%%%%%
Here, as before, the quantifies $\beta/H_*, \alpha$, $g_*(T)$ and $v_w$ are defined following Eqs. (\ref{eq:PT-GW:beta}), (\ref{eq:PT-GW:alpha}), (\ref{eq:nucleation:Hubble-rate}) and (\ref{eq:percolation:comoving-radius}), respectively, and the spectral shape is approximated by
%%%%%%%%%%%%%%%%%%%%%%%%%%%%%
\bea
\label{eq:GW:turb-shape}
S_{\rm t} (f) = \left[ \frac{\left( f/f_{\rm tur} \right)^3}{\left[ 1 + \left( f/f_{\rm tur} \right) \right]^{11/3} \left( 1 + \frac{8 \pi f}{h_{\ast}} \right)} \right],
\eea
%%%%%%%%%%%%%%%%%%%%%%%%%%%%%
with $\kappa_{\rm tur}$ denoting the fraction of latent heat that is transformed into MHD turbulence, and $h_*$ is the frequency corresponding to the wave number $k_*$, which equals the Hubble rate at the time of GW production redshifted by the expansion of the Universe up to the present time and is given by,
%%%%%%%%%%%%%%%%%%%%%%%%%%%%%
\bea
\label{eq:GW:turb-hstar}
h_{\ast} = 16.5 \times 10^{-6} \left( \frac{T_*}{100 {\rm ~GeV}} \right) \left( \frac{g_{\ast}(T)}{100} \right)^{1/6} {\rm Hz}.
\eea
%%%%%%%%%%%%%%%%%%%%%%%%%%%%%

Similar to the case of sound waves, the peak frequency is associated with the inverse of the characteristic length scale of the source, which is the bubble size $R_*$
near the end of the PT. Analytical arguments indicate that this relationship is due to the specific time de-correlation properties of the turbulent source \cite{Caprini:2009yp}. After red-shifting, one has $f_{\rm tur}$ as,
%%%%%%%%%%%%%%%%%%%%%%%%%%%%
\bea
\label{eq:GW:turb-peak-freq}
f_{\rm tur} = 2.7 \times 10^{-5} \frac{1}{v_w} \left( \frac{\beta}{H_{\ast}} \right) \left( \frac{T_*}{100 {\rm~GeV}} \right) \left( \frac{g_{\ast}(T)}{100} \right)^{1/6} {\rm Hz.}
\eea
%%%%%%%%%%%%%%%%%%%%%%%%%%%%
Note that, in our analysis, we set $\kappa_{\rm tur}=\epsilon \kappa_{\rm sw}$ with $\epsilon = 0.1$, which is motivated from simulations, and $\epsilon$ stands for the fraction of the bulk motion which is turbulent. 
%%%%%%%%%%%%%%%%%%%%%%%%%%%%%%%%%%%%%%%%%%%%%%%%%%%%%%%%%%%%%%%%%%%%%%%%%%%%%%%%%%

To conclude this subsection, we mention here that for careful analysis, one has to distinguish between three different types of PTs: (i) \textit{non-runaway PT in plasma}, (ii) \textit{runaway PT in plasma}, and (iii) \textit{runaway PT in vacuum}, to consider different contributions coming from three different sources to the total GW amplitude. However, in the chosen model framework, it is important to note that, the potential is polynomial (see, Eq.~(\ref{eq:pot:tree-level})), therefore, there is very less chance that one can expect a large $\alpha$, i.e., significant supercooling \cite{Ellis:2018mja}, implying that the bubbles will not be very energetic to contribute dominantly to $\Omega_{\rm GW} h^2$ \cite{Ellis:2019oqb,Lewicki:2019gmv,Lewicki:2020jiv,Lewicki:2020azd}. Therefore, one expects $\alpha < \alpha_{\infty}$, where $\alpha_{\infty}$ corresponds to the threshold value at which the walls of the scalar-field bubbles begin to {\it runaway}  (for details, see Appendix \ref{appx:efficiency-coeff}), in most of the parameter space. As a consequence, we can concentrate solely on the GWs sourced by the plasma motion \cite{Caprini:2015zlo, Caprini:2019egz}. However, as our PT dynamics is singlet-field driven, keeping $\Omega_{\rm b} h^2$ would not significantly impact our analysis and we will use the conventional $\kappa_b$ (Eq.~(\ref{eq:eff-fact:kappa_b})) for further calculation. In later discussions, we will show that, indeed, the contribution of $\Omega_{\rm b} h^2$ to the total $\Omega_{\rm GW} h^2$ is very small compared to that of $\Omega_{\rm sw} h^2$ and $\Omega_{\rm t} h^2$. Furthermore, although we have calculated $v_w$ analytically, following the expressions and assumptions mentioned in Ref.~\cite{Ellis:2022lft}, yet we will consider $v_w$ as an input parameter for the GW signal calculation and take $v_w \approx 1$ which maximizes the GW signal strength. For a conservative comparison, we will also present the GW analysis with analytical $v_w$ (see Eq.~(\ref{eq:percolation:bubble-velo-appx})) calculation.
%%%%%%%%%%%%%%%%%%%%%%%%%%%%%%%%%%%%%%%%%%%%%%%%%%%%%%%%%%%%%%%%%%%%%%%%%%%%%%%%%%
\subsection{Results}
\label{subsec:EWPT-GW-results}
%%%%%%%%%%%%%%%%%%%%%%%%%%%%%%%%%%%%%%%%%%%%%%%%%%%%%%%%%%%%%%%%%%%%%%%%%%%%%%%%%%
%%%%%%%%%%%%%%%%%%%%%%%%%%%%%%%%%%%%%%%%%%%%%%%%%%%%%%%%%%%%%%%%%%%%%%%%%%%%%%%%%%
As already discussed in Sec. \ref{sec:Model}, in this work, we have extended the scalar sector of the SM with a real scalar singlet $S$  and a $Y=0$ scalar ${\bm{T}}$, triplet under $SU(2)_L$. The singlet scalar $S$ is non-trivially charged under the $\mathbb{Z}_2\times \mathbb{Z}'_2$,  as mentioned in Table~\ref{tab:discrete-charge}. The given $\mathbb{Z}_2\times \mathbb{Z}'_2$ charge assignments of $S$, after the EWSB, yields linear ($h^2 s$) and  cubic ($s^3$) interactions for the singlet, as depicted in Eq.~(\ref{eq:pot:tree-level}).
These terms play important roles in barrier formation and generate PT along $h$- and/or $hs$- field direction. Moreover, to have SFOPT in a pure triplet extended scenario, one must have very large ($\gtrsim \mathcal{O}(1.0)$) triplet-Higgs quartic coupling, i.e., $\l_{HT}$ \cite{Niemi:2018asa,Niemi:2020hto}. However, in this work, as we consider $T^0$ to be a DM candidate and, hence, we kept the triplet-Higgs and triplet-singlet quartic couplings, i.e., $\l_{HT},\, \l_{ST}$ below $\mathcal{O} (0.1)$ such that (i) $T^0$ remains underabundant and easily escape the DD bounds, as detailed in Sec.~\ref{sec:DM-pheno}, and (ii) PT along the $T^0$ direction remains disfavored, as mentioned in \ref{subsubsection-PT-patterns}. Therefore, in our study, the PT dynamics are mainly driven by the scalar singlet. Although $T^0$ does not take an active part in deciding the fate of the PT dynamics, it can still modify the effective thermal potential $V_{\rm eff} (\varphi, T)$ through loop effects.

In this section we will investigate the PT dynamics of the chosen model, examine the correlation among the important model parameters with the SFOPT, and their plausible connections with the DM observables detailed in Sec. \ref{sec:DM-pheno}. Subsequently, we will assess the detectability of GWs arising from such an SFOPT. Before we proceed, it is worth mentioning that solving the bounce solution with multiple scalar fields is numerically expensive. Therefore, we avoid running {\tt cosmoTransitions} code over the total sample data we obtained after the general scan (see Table \ref{tab:cutflow}), which we used to study the DM phenomenology. Rather, we took a conservative approach and sampled out $50,000$ data points out of $139,863$ that evade all constraints stated in Sec. \ref{sec:constraints} and either escape the DD bound from XENON1T without caring about the relic density constraint ($139,448$ points) or respect the bound on the relic density but do not care about the XENON1T limit. These numbers are taken from Table \ref{tab:cutflow} and $\approx 28\%$ of $500,000$, the original number of scanned points. To get an adequate sample size for the subsequent analysis, we consider DD bounds only from the XENON1T, the weakest one, as apparent from Table \ref{tab:cutflow}. Also, since low scalar masses (typically $\sim \mathcal{O}(1~ \rm TeV)$) favour the SFOPT (see, for e.g., Ref. \cite{Ramsey-Musolf:2019lsf}), we focus on the regime of parameter space where $m_{h_2} \lsim 1$ TeV.\footnote{We have also explored the $m_{h_2}> 1$ TeV case separately where, compared to the $m_{h_2} \lsim 1$ TeV case, a lesser number of parameter points can accommodate the SFOPT.}
 Finally, besides the criteria of an SFOPT, as mentioned in Eqs.~(\ref{eq:PT-strength:Tc}) and ({\ref{eq:PT-strength:Tn}}), to identify an SFOPT we also exclude parameter points if,
%%%%%%%%%%%%%%%%%%%%%%%%%%%%%%%%%
\begin{itemize}
	\item the EW vacuum $\left(\langle H \rangle, \langle S \rangle, \langle T \rangle\right) = (v, v_s, 0)$ is not the true minimum of the potential at $T=0$.
	\item the $h$- and $s$-field values in the broken or symmetric phase during the transition is much larger than their EW vacuum values and the potential of Eq.~(\ref{eq:pot:tree-level}) remains unbounded from below.
\end{itemize}
%%%%%%%%%%%%%%%%%%%%%%%%%%%%%%%%%
\subsubsection{Formation of barrier and vacuum upliftment}\label{sec:PTresults-vacuum-upliftment}
%%%%%%%%%%%%%%%%%%%%%%%%%%%%%%%%%
A strong first-order EWPT typically happens when a sufficiently high and wide potential barrier separates the two degenerate vacua of the thermal effective potential (see Eq. (\ref{eq:pot:Eff-pot}) for this work) at the critical temperature $T_c$. The addition of new scalars could enhance the barrier between the vacua and thus, make the EWPT a strong first-order one which is otherwise not possible within the SM \cite{DOnofrio:2014rug}. In the chosen model, the barrier in the effective potential
of Eq. (\ref{eq:pot:Eff-pot}) can arise from the tree-level (see Eq. (\ref{eq:pot:tree-level})) as well as from the bosonic thermal corrections through Eq.~(\ref{eq:pot:thermal}). The tree-level barrier arises due to the presence of linear and cubic terms in $s$
as shown in Eq. (\ref{eq:pot:tree-level}), the tree-level potential. While the barrier due to the thermal corrections appears from the terms cubic in $x=m_i(\varphi)/T$ once we plug the expansion of Eq.~(\ref{eq:pot:thermal-functions}) in the high-temperature limit ($T \gg m_i$) in Eq. (\ref{eq:pot:thermal}) which contains terms cubic in both $h$ and $s$. To have a quantitative understanding of the barrier formation and vacuum upliftment of the true vacuum which can guarantee an SFOPT, one can evaluate the ratio $\Delta \mathcal{F}_0/|\mathcal{F}^{\rm SM}_0|$
\cite{Dorsch:2017nza, Goncalves:2021egx, Goncalves:2023svb}. It is a gauge independent quantity calculated at zero temperature and one can use this to examine its correlation with different contributions, i.e., tree-level ($\delta V^b_0$), one-loop at $T=0$ ($\delta V^b_1$) and thermal one-loop level ($\delta V^b_T$) to the barrier formation of the effective potential $V_{\rm eff} (\varphi, T)$, shown in Eq.~(\ref{eq:pot:Eff-pot}). The different contributions to the barrier formation can be estimated by calculating the energy-density difference between the potential at the barrier ($\varphi_{\rm barrier}$) and false vaccum phase ($\varphi_{\rm false}$) \cite{Dorsch:2017nza,Goncalves:2021egx,Goncalves:2023svb} as,
%%%%%%%%%%%%%%%%%%%%%%%%%%%%%%%%%%%%%%%%%%%%%%%%%%%%%%%%%%%%%%%
\bea
\label{eq:barrier-contributions}
\delta V^b_0 &=& V_0(\varphi_{\rm barrier}) - V_0(\varphi_{\rm false}),\nn\\
\delta V^b_1 &=& V_{\rm CW}(\varphi_{\rm barrier}, T=0) - V_{\rm CW}(\varphi_{\rm false}, T=0),\nn\\
\delta V^b_T &=& V_{\rm T}(\varphi_{\rm barrier}, T=T_c) - V_{\rm T}(\varphi_{\rm false}, T=T_c).
\eea
%%%%%%%%%%%%%%%%%%%%%%%%%%%%%%%%%%%%%%%%%%%%%%%%%%%%%%%%%%%%%%%
%
%
Here, $\varphi_{\rm barrier}$ and $\varphi_{\rm false}$ are estimated at $T=T_c$. Note that the position of the potential barrier ($\varphi_{\rm barrier}$) is the point where the effective potential reaches its maximum value along the tunneling path \cite{Goncalves:2021egx,Goncalves:2023svb}, as determined by solving Eq.~(\ref{eq:nucleation:phivariationr}) numerically. Whereas, the height of the barrier is defined as the difference between the values of $V_{\rm eff}(\varphi,\,T)$, Eq.~(\ref{eq:pot:Eff-pot}), evaluated at the barrier ($\varphi_{\rm barrier}$) and at the false vacuum ($\varphi_{\rm false}$) \cite{Goncalves:2021egx,Goncalves:2023svb}.

\begin{figure*}[!h]
	%\centering
	\hspace*{-0.25cm}
	\subfigure{\includegraphics[height=5.5cm,width=15.5cm]{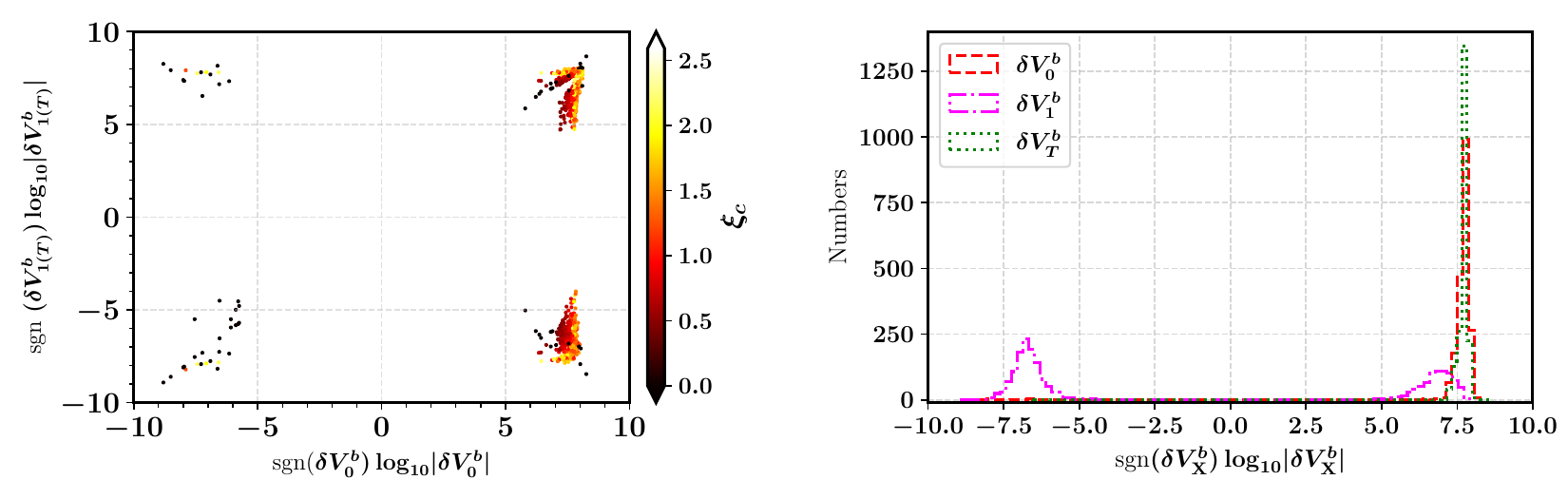}}
	\caption{The left plot depicts signed logarithmic correlations of different contributors (plotted as signed $\log_{10}$ of absolute values) to the potential barrier and how these affect the PT strength defined by Eq. (\ref{eq:PT-strength:Tc}). The right plot shows the relative sizes of different contributors (plotted as signed $\log_{10}$ of absolute values) to the potential barrier.}
	\label{fig:result:barrier-contri}
\end{figure*}

The quantity $\Delta \mathcal{F}_0/|\mathcal{F}^{\rm SM}_0|$ is defined as \cite{Dorsch:2017nza, Goncalves:2021egx, Goncalves:2023svb},
%%%%%%%%%%%%%%%%%%%%%%%%%%%%%%%%%%%%
\bea
\label{eq:result:PT-barrier-parameter}
\frac{\Delta \mathcal{F}_0}{|\mathcal{F}^{\rm SM}_0|} \equiv \frac{\mathcal{F}_0 - \mathcal{F}^{\rm SM}_0}{|\mathcal{F}^{\rm SM}_0|},
\eea
%%%%%%%%%%%%%%%%%%%%%%%%%%%%%%%%%%%%
where $\mathcal{F}_0$ is the vacuum energy density and, using \ref{subsubsection-PT-patterns} and Eq. (\ref{eq:pot:Eff-pot}), it can be written as,
%%%%%%%%%%%%%%%%%%%%%%%%%%%%%%%%%%%%
\bea
\label{eq:result:PT-barrier-F0}
\mathcal{F}_0 \equiv V_{\rm eff} (v, v_s, 0, T=0) - V_{\rm eff}(0,0,0,T=0),
\eea
%%%%%%%%%%%%%%%%%%%%%%%%%%%%%%%%%%%%
with $\mathcal{F}^{\rm SM}_0 = -1.25 \times 10^8\, {\rm GeV^4}$ \cite{Goncalves:2021egx, Goncalves:2023svb}. First, we present the correlations among different contributions to the potential barrier in Fig.~\ref{fig:result:barrier-contri}. 
%%%%%%%%%%%%%%%%%%%%%%%%%%%%%%%%%%%%%%%%%%%%%%%%%%%%%%%%
%%%%%%%%%%%%%%%%%%%%%%%%%%%%%%%%%%%%%%%%%%%%%%%%%%%%%%%%
It can be seen from the left panel of Fig.~\ref{fig:result:barrier-contri} that the potential barrier in the chosen framework is dominantly generated by the coalition between the tree-level and one loop-level components as $\delta V^b_0$ and $\delta V^b_{1(T)}$ positively contributing to the transition, i.e., $\delta V^b_0, \delta V^b_{1(T)} > 0$ for almost $90\%$ cases. The right panel of Fig.~\ref{fig:result:barrier-contri} represents the relative contribution of the same three sources to the barrier formation. As we have already anticipated due to the particular structure of our $V_{\rm eff}(\varphi, T)$, we observe that the dominant contribution is through the tree-level and thermal one-loop level. There is also some contribution from the $\delta V^b_1$ to enhance the barrier, however, mostly it plays a role in reducing the barrier, as evident from the size and the sign of $\delta V^b_1$ contributions. It is also evident that the leading
contributors are $\delta V^b_0$ and $\delta V^b_T$.

\begin{figure*}[!h]
	\hspace*{-0.25cm}
	%\centering
	\subfigure{\includegraphics[height=5.5cm,width=8.0cm]{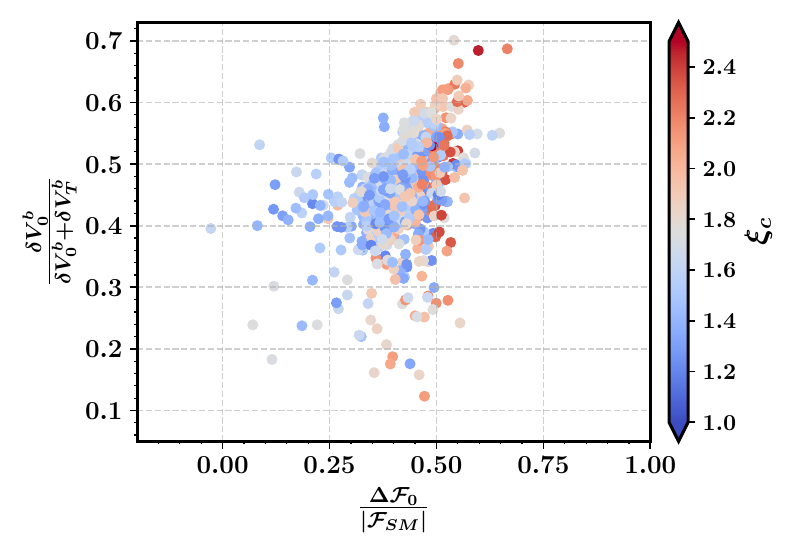}}
    \subfigure{\includegraphics[height=5.5cm,width=8.0cm]{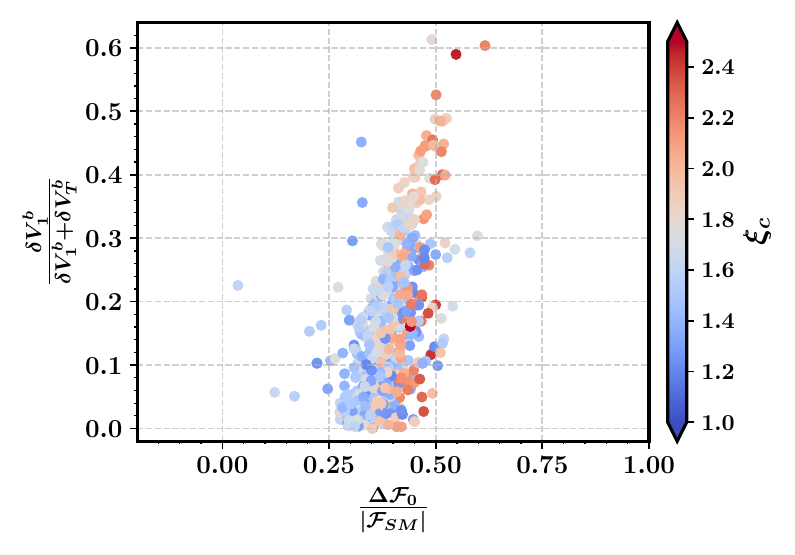}}
	\caption{Plots showing variations of the PT strength $\xi_c$ of
	Eq. (\ref{eq:PT-strength:Tc}) in the $\frac{\delta V^b_0}{\delta V^b_0 +\delta V^b_T}$ versus $\Delta \mathcal{F}_0/|\mathcal{F}^{\rm SM}_0|$ plane (left) and $\frac{\delta V^b_1}{\delta V^b_1 +\delta V^b_T}$ versus $\Delta \mathcal{F}_0/|\mathcal{F}^{\rm SM}_0|$ plane (right). Details are given in the text body.}
	\label{fig:result:barrier-F0}
\end{figure*}

In the left panel of Fig.~\ref{fig:result:barrier-F0}, we demonstrate how PT strength $\xi_c$ of Eq. (\ref{eq:PT-strength:Tc}) varies with the correlation between the zero temperature vacuum upliftment measure, as depicted by Eq. (\ref{eq:result:PT-barrier-parameter}), and the relative weightage of one of the leading contributors to the potential barrier, $\delta V^b_0$.
%%%%%%%%%%%%%%%%%%%%%%%%%%%%%%%%%%%%%%%%%%%%%%%%%%%%%%%%
%%%%%%%%%%%%%%%%%%%%%%%%%%%%%%%%%%%%%%%%%%%%%%%%%%%%%%%%
The large concentration of points in the window of $0.35 \lesssim \frac{\delta V^b_0}{\delta V^b_0 + \delta V^b_T} \lesssim 0.60$, suggests that both the tree-level and the thermal one-loop level contributes almost equally, verifying our earlier finding shown in the right plot of Fig.~\ref{fig:result:barrier-contri}. In the right panel of Fig.~\ref{fig:result:barrier-F0}, we show the relative size of $\delta V^b_1$ among the two possible one-loop contributors of the barrier formation with $\Delta \mathcal{F}_0/|\mathcal{F}^{\rm SM}_0|$ in the light of possible PT strength $\xi_c$. The large concentration of points for $\frac{\delta V^b_1}{\delta V^b_1+\delta V^b_T} \lesssim 0.30$ suggests that the one-loop thermal contribution dominates over the one-loop $T=0$ contribution. In Fig.~\ref{fig:result:barrier-F0}, we further notice a slight inclination of the data points towards the positive $\Delta \mathcal{F}_0/|\mathcal{F}^{\rm SM}_0|$ direction as the barrier contribution fraction increases.
This indicates that larger barrier contributions require higher vacuum upliftment measures. Similarly, in a general case, the higher the vacuum upliftment measure, the larger the PT strength. From Fig. \ref{fig:result:barrier-F0}, we can identify $\Delta \mathcal{F}_0/|\mathcal{F}^{\rm SM}_0| \gtrsim 0.25$ as the minimal lower bound for a strong FOPT, for the chosen framework. In our analysis, we observe, approximately, $0.25 \lesssim \Delta \mathcal{F}_0/|\mathcal{F}^{\rm SM}_0| \lesssim 0.60$ for the cases with an SFOPT. It should be noted that, when the vacuum upliftment measure is extremely large ($\gtrsim 0.90$), it can lead to a scenario where the tunneling from $\varphi_{\rm false}$ to $\varphi_{\rm true}$ becomes challenging, translating into no solution for Eq.~(\ref{eq:nucleation:nucleation-criterion}) \cite{Biekotter:2022kgf}. Therefore, the system would be trapped in the highly energetic EW symmetric vacuum with no physical vacuum. In our study, we did not find any such trapped vacuum scenario.
%%%%%%%%%%%%%%%%%%%%%%%%%%%%%%%%%%%%%%%%%%%%%%%%%%%%%%%%%%%%%%%%%%%%%%%%%%%%%%%%%%
\subsubsection{Parameter dependence of the PT} \label{sec:PTresults-PT-parameter-dependence}
%%%%%%%%%%%%%%%%%%%%%%%%%%%%%%%%%%%%%%%%%%%%%%%%%%%%%%%%%%%%%%%%%%%%%%%%%%%%%%%%%%

\begin{figure*}[!h]
	\hspace*{-0.5cm} %% This will shift the figure
	\centering
	\subfigure[]{\includegraphics[height=5.3cm,width=7.2cm]{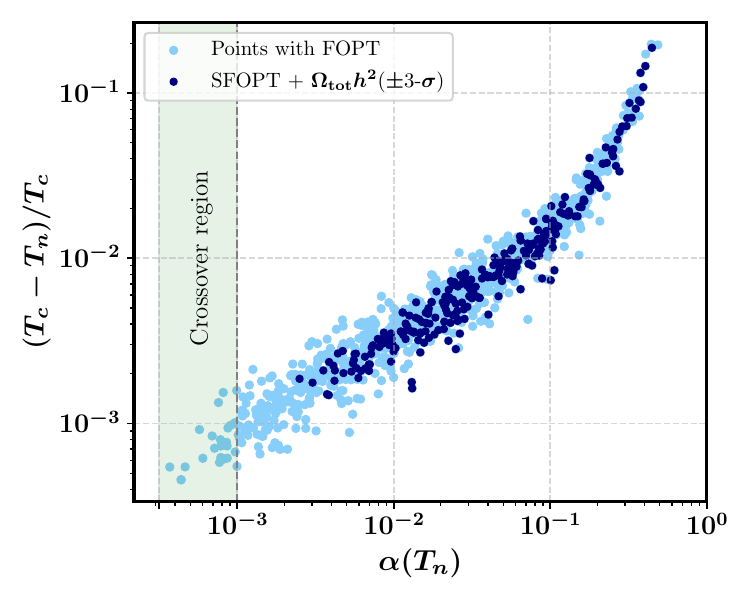}}
	\subfigure[]{\includegraphics[height=5.3cm,width=7.4cm]{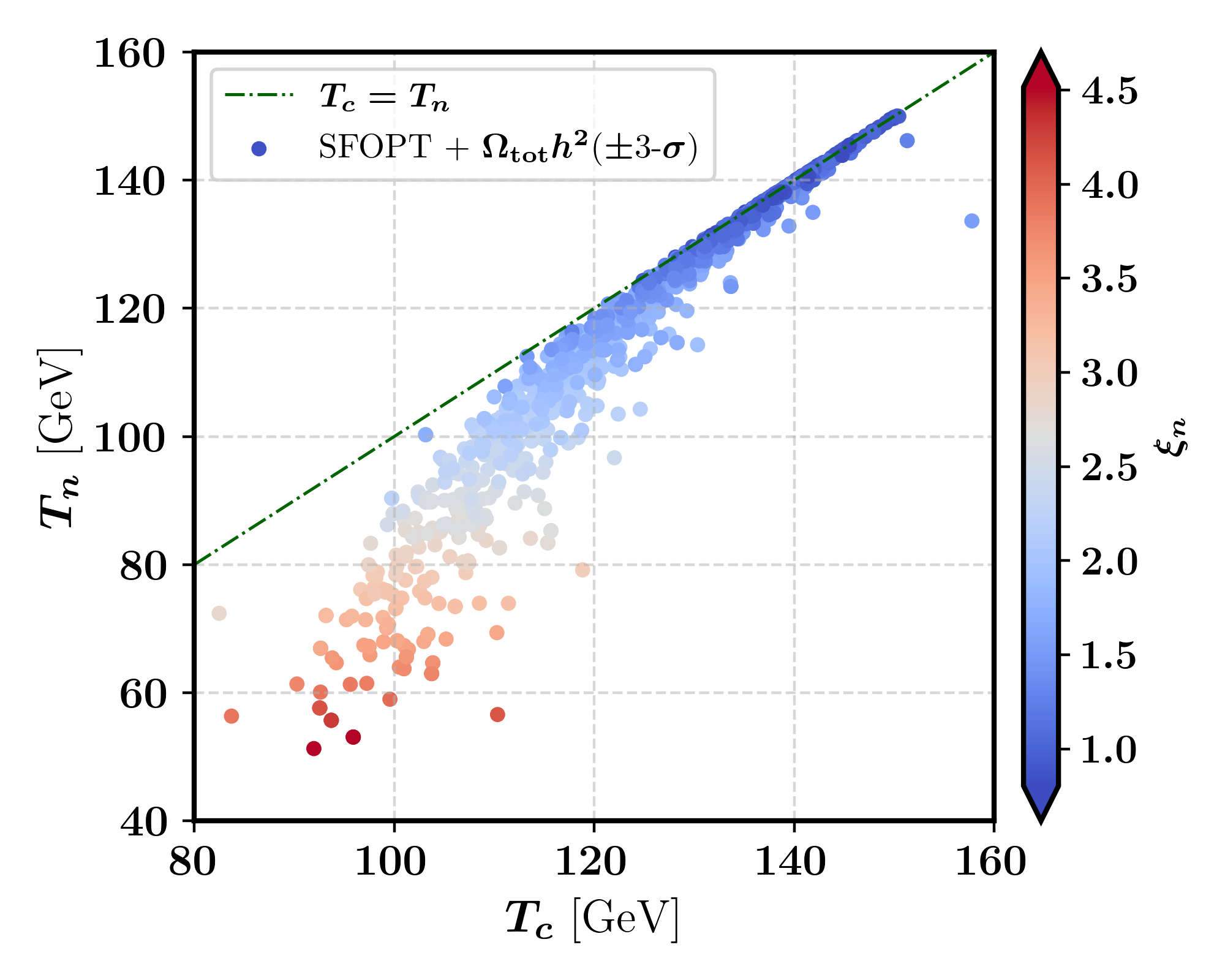}}\\
	\subfigure[]{\includegraphics[height=5.3cm,width=7.2cm]{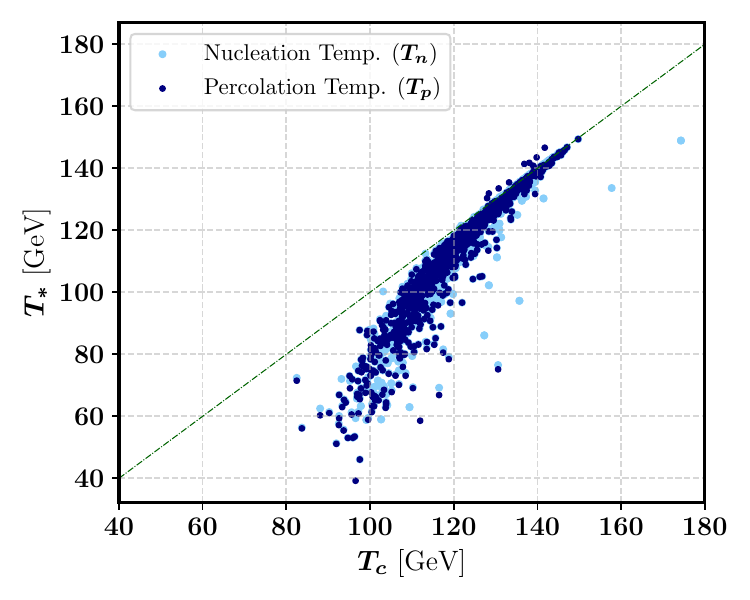}}    
	\caption{The top row left plot shows variations of $\frac{T_c-T_n}{T_c}$ with $\alpha(T_n)$. Here, the light-blue coloured points
	show an FOPT while the deep-blue coloured points, besides SFOPT,
	also obey the $3\sigma$ relic density limit. The top row right plot depicts the sensitivity of $\xi_n$ in the $T_c-T_n$ plane. All points
	of this plot show an SFOPT and obey the relic density bound.
	The bottom row plot shows how the transition temperature $T_*$ changes with $T_c$ when (i) $T_*=T_n$ (represented by the light-blue coloured points), and (ii) $T_*=T_p$ (represented by the dark-blue coloured points). The green-coloured dash-dotted line represents $T_c-T_n$ (top right) and  $T_*=T_c$ (bottom). The remaining details are depicted on individual plots.}
	\label{fig:result:param-depend-TcTnTp}
\end{figure*}

In this part, we study correlations between the parameters that characterize the PTs and also examine the dependence of model parameters on the PT dynamics. Fig.~\ref{fig:result:param-depend-TcTnTp} (a) shows a correlation of $1-T_n/T_c$ with the parameter $\alpha(T_n)$ for model points which satisfy the criteria of SFOPT. Most of the points yield a larger $\alpha$ when there is a sufficient gap between $T_c$ and $T_n$, meaning lower the $T_n$ higher is the $\alpha$ value. The light green coloured shaded region corresponds to very weak transitions, $\alpha \thickapprox 10^{-3}$, denoting as the crossover regime\footnote{The crossover phenomena is observed in the SM framework, as already addressed in the literature \cite{Kajantie:1996mn,Karsch:1996yh,Aoki:1996cu,Gurtler:1997hr,Laine:1998jb,Laine:2015kra}.}. One can see that inclusion of the relic density bound, as shown in Eq. (\ref{eq:constraints:DM-relic-Planck}), shifts $\alpha$ values almost
by a factor of $10$, which are depicted by deep-blue coloured points. The relic density bound also completely rules out
the crossover region. Next, we look for correlation in the $T_c$-$T_n$ plane as shown in Fig.~\ref{fig:result:param-depend-TcTnTp} (b) along with the strength of transition $\xi_n$ (see Eq. (\ref{eq:PT-strength:Tn})), depicted by colour bar. One can see that larger values of $\xi_n$ correspond to low $T_c$ and $T_n$ values. Intuitively, this behaviour can be understood as follows \cite{Benincasa:2022elt}: at large $T_n$, the value of $v_n$ can initially be small and then gradually evolve to its final value as the temperature decreases, leading to a small $\xi_n$. Conversely, a smaller $T_n$ necessitates a larger $v_n$, since the system has less time to gradually approach to its tree-level value. The strength of a point that undergoes a FOPT and is consistent with the $3\sigma$ relic density bound is not very apparent from  Fig. \ref{fig:result:param-depend-TcTnTp} (a), just by looking at the values of $\alpha(T_n)$. However, it is elucidated from Fig. \ref{fig:result:param-depend-TcTnTp} (b) that all the colour gradient points satisfy $\xi_n\gsim 1$, as mentioned in Eq. (\ref{eq:PT-strength:Tn}).

The percolation temperature ($T_p$), in the chosen framework, for the majority of the parameter region that incorporates an SFOPT, stays close to the nucleation temperature except for its low values, as apparent from Fig.~\ref{fig:result:param-depend-TcTnTp} (c).
%%%%%%%%%%%%%%%%%%%%%%%%%%%%%%%%%%%%%%%%%%%%%%%%%%%%%%%%
%%%%%%%%%%%%%%%%%%%%%%%%%%%%%%%%%%%%%%%%%%%%%%%%%%%%%%%%
For example, for a nucleation around $T_n \simeq 60\,{\rm GeV}$, we find $T_p$ to be close to $40\,{\rm GeV}$. It is to be noted here that, subsequently we will use $T_p$ as the reference temperature for the GW spectrum calculation which will be discussed in the upcoming subsection.
However, concerning the PT dynamics, we continue only with $T_c$ and $T_n$.

\begin{figure*}[!h]
	%\hspace*{-0.2cm} %% This will shift the figure
	\centering
	\subfigure{\includegraphics[height=5.3cm,width=7.2cm]{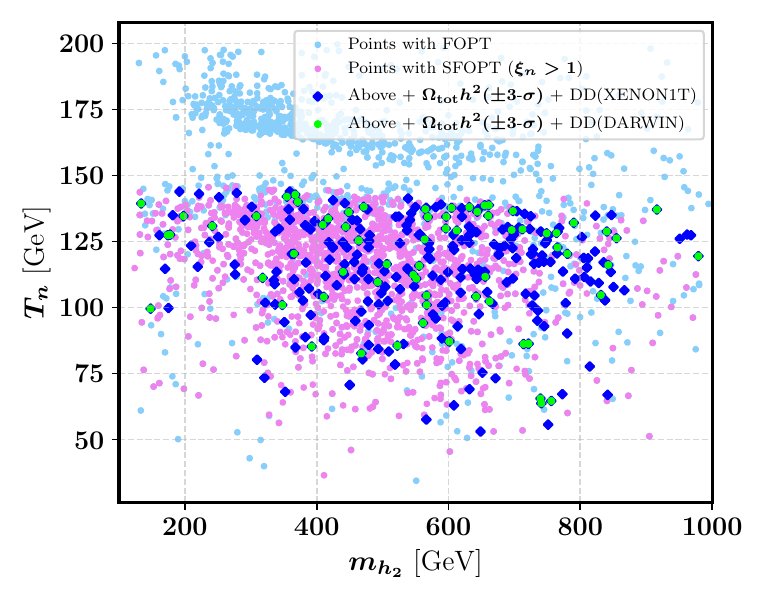}}
	\subfigure{\includegraphics[height=5.3cm,width=7.2cm]{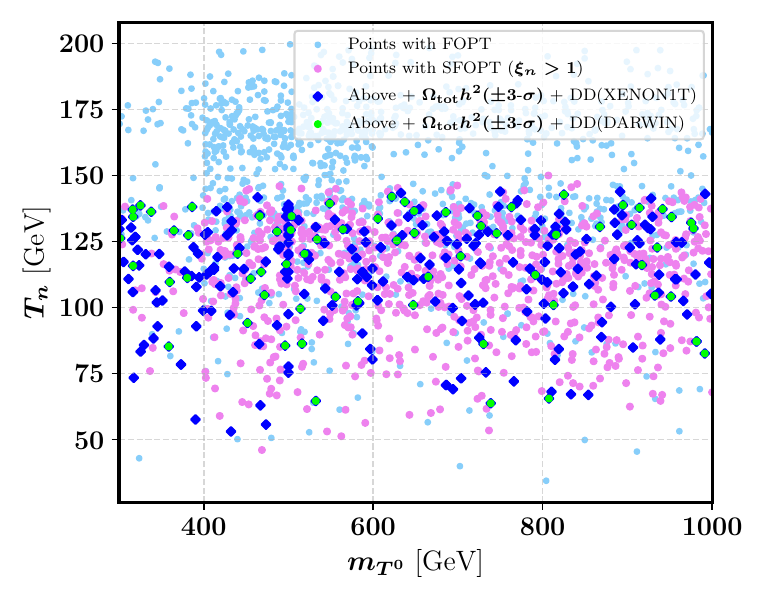}}
	\caption{The left (right) plot shows variations of model points in the $m_{h_2}-T_n$ ($m_{T^0}-T_n$) plane. Here, light-blue coloured points show an FOPT whereas the purple-coloured points undergo an SFOPT with $\xi_n>1$. The dark-blue (light-green) coloured squares (circles) depict model points that show an SFOPT, obey the $3\sigma$ relic density bound, and respect DD bounds from XENON1T (DARWIN).}
	\label{fig:result:param-depend-mSmT0vsDM}
\end{figure*}

\begin{figure*}[!h]
	\hspace*{0.2cm} %% This will shift the figure
	\centering
	\subfigure{\includegraphics[height=5.3cm,width=7.6cm]{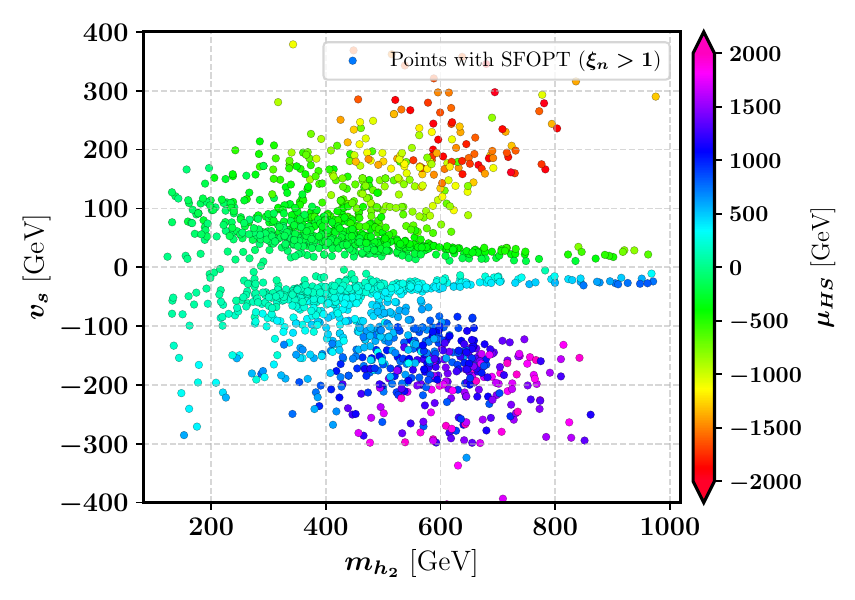}}
	\subfigure{\includegraphics[height=5.2cm,width=7.2cm]{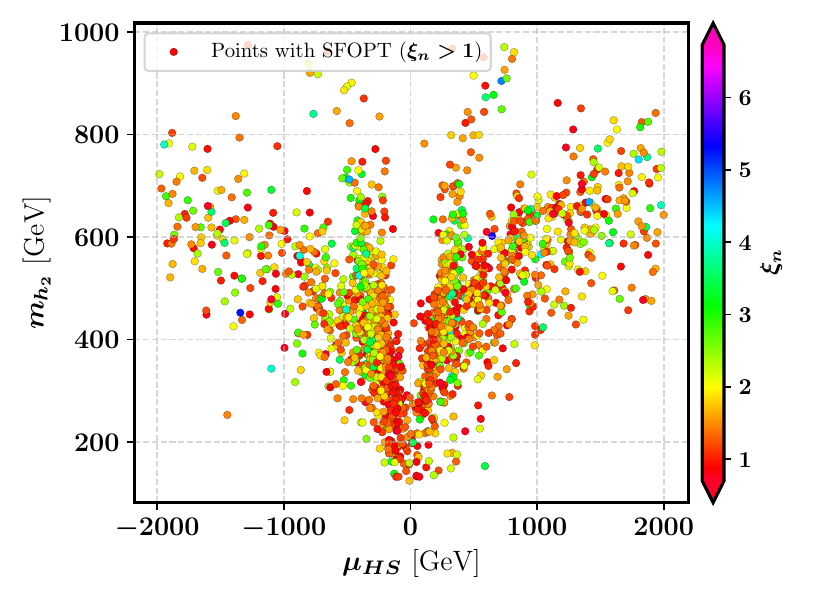}}
	\caption{The left (right) panel shows variations in the
	$m_{h_2}-v_s$ ($\mu_{HS}-m_{h_2}$) plane for different
	values of $\mu_{HS}~(\xi_n)$ for models points which can accommodate
	an SFOPT. }
	\label{fig:result:param-depend-vsmh2muhs}
\end{figure*}

In Figs.~\ref{fig:result:param-depend-mSmT0vsDM} and \ref{fig:result:param-depend-vsmh2muhs}, we present the distributions of a few model inputs and their correlation with the PT parameters.
It is important to emphasize here that not all independent inputs (see Table \ref{tab:par_ranges}) play a pivotal role in the PT dynamics.
Besides, fixing certain other parameters, e.g.,
$\l_{HT}, \l_{ST}$, at low values, i.e., $\lesssim \mathcal{O}(0.1)$,
automatically weakens their role in the PT dynamics.
The inputs $g_S,\,m_\psi$ are crucial for the DM phenomenology, especially to reach the threshold of the correct relic density, as detailed in Sec. \ref{sec:DM-pheno}. The parameter $m_\psi$ also contributes to the PT dynamics, but only through the loop-corrections. The anticipated effects of $m_\psi$, thus, will be loop-suppressed and, hence not considered in this analysis. Therefore, the set of inputs crucial for the PT dynamics is $m_{h_2}, v_s, \mu_3, \sin\theta$ and $\l_S$ which is anticipated as PT in this model, as previously stated in subsection \ref{subsec:GW-from-SFOPT}, is singlet driven.
%%%%%%%%%%%%%%%%%%%%%%%%%%%%%%%%%%%%%%%%%%%%%%%%%%%%%%%%
%%%%%%%%%%%%%%%%%%%%%%%%%%%%%%%%%%%%%%%%%%%%%%%%%%%%%%%%

Now let us summarize our key findings of Fig.~\ref{fig:result:param-depend-mSmT0vsDM}. From the left plot of Fig.~\ref{fig:result:param-depend-mSmT0vsDM} we observe that a FOPT is possible throughout the interested low mass region of $m_{h_2}$, i.e., below $1$ TeV, as stated in subsection \ref{subsec:EWPT-GW-results}. In this case, the nucleation temperature can go up to $\thickapprox 200\, {\rm GeV}$. However, an SFOPT is seen primarily in the range $50\,{\rm GeV} \lesssim T_n \lesssim 150\,{\rm GeV}$ and $m_{h_2} \lesssim 850\,{\rm GeV}$. Moreover, some model points, compatible with the SFOPT, also accommodate the correct relic density, although the latter appears to be the leading constraint as can be realized by looking at the relative concentrations of the purple and the dark-blue coloured points. Besides, as already stated in subsection \ref{subsec:EWPT-GW-results}, these points are also consistent with the DD bounds.
One, nevertheless, observes a significant shrinking of the parameter space that can simultaneously accommodate an SFOPT, the $3\sigma$ relic density and DD bounds when one moves from XENON1T (presented with dark-blue coloured squares) to DARWIN (depicted with light-green coloured circles)\footnote{XENON1T is chosen just for an optimal illustration. Other DD bounds, e.g., DARWIN, are more stringent and will yield a much lesser number of allowed points, as evident from Table \ref{tab:cutflow} and Fig.~\ref{fig:result:param-depend-mSmT0vsDM}.}.
Unlike the left plot of Fig.~\ref{fig:result:param-depend-mSmT0vsDM}, hardly any correlation is observed for the right plot, however, one can see that the whole mass region of $m_{T^0}$, i.e., $300\,{\rm GeV} \lesssim m_{T^0} \lesssim 1000,{\rm GeV}$, is now available for an SFOPT and remain consistent with the DM bounds. This feature was not possible in a pure $Y=0$ triplet extended model with its neutral part as the sole DM candidate \cite{Cirelli:2005uq}. Having a second DM species makes it possible and elevates the earlier caveats as seen in the literature \cite{Niemi:2018asa,Niemi:2020hto,Bandyopadhyay:2021ipw}.

The left plot of Fig.~\ref{fig:result:param-depend-vsmh2muhs} shows correlation between $m_{h_2}$ and $v_s$ for
model points showing an SFOPT, i.e., $\xi_n > 1.0$. The state
$h_2$, as already discussed in subsection \ref{subsec:scalarmassmixing},
contains leading singlet contribution while $h_1$ is chosen to be
the SM-like Higgs. Hence, $m_{h_2}$ can be dubbed as the ``singlet mass'', without any loss of generality. The colour palette indicates the values of the derived parameter $\mu_{HS}$ (see Eq.~(\ref{eq:muHS})). From Fig. \ref{fig:result:param-depend-mSmT0vsDM}, one observes that an SFOPT favours $m_{h_2}\lesssim 850$ GeV, which is also reflected in Fig.~\ref{fig:result:param-depend-vsmh2muhs}. This $m_{h_2}$ bound, in turn, shows that normally $|v_s|\lesssim 300$ GeV is favoured for an SFOPT
along with opposite signs compared to $\mu_{HS}$ parameter. Further, for a positive (negative) $v_s$ and a negative (positive) $\mu_{HS}$, allowed points show large
concentration for $0 \lesssim |v_s|~(\rm{GeV}) \lesssim 200$ and
$0 \lesssim |\mu_{HS}|~(\rm{GeV}) \lesssim 1000$. The same appears rather
spread out between $200 \lesssim |v_s|~(\rm{GeV}) \lesssim 300$ and $1000 \lesssim |\mu_{HS}|~(\rm{GeV}) \lesssim 2000$ for the reverse sign choice. One can also see that when $v_s$ values are very small, $\sim \mathcal{O}(10\rm GeV)$, $m_{h_2}$ can take any values over the entire allowed range.
%%%%%%%%%%%%%%%%%%%%%%%%%%%%%%%%%%%%%%%%%%%%%%%%%%%%%%%%
%%%%%%%%%%%%%%%%%%%%%%%%%%%%%%%%%%%%%%%%%%%%%%%%%%%%%%%%
On the contrary, when $|v_s| \gsim 200$ GeV, values of $m_{h_2}$ are restricted $\lesssim 850$ GeV, and prefers larger $|\mu_{HS}|$, i.e., $\gsim1.5$ TeV, to give rise an SFOPT. This behaviour can also be inferred from the right panel of  Fig.~\ref{fig:result:param-depend-vsmh2muhs} where one explores the $\mu_{HS}-m_{h_2}$ plane for model points showing an SFOPT. We do not find any particular correlation between $m_{h_2} - \mu_{HS}$ plane along with the PT strength $\xi_n$. We note that, for all ranges of $\mu_{HS}$, a strong PT can be obtained provided its nature with $v_s$ and $m_{h_2}$ is maintained as shown in the left plot of Fig.~\ref{fig:result:param-depend-vsmh2muhs}. Furthermore, the dependence of $\l_S$ in the PT properties is inherited in $\mu_{HS}$, through Eq.~(\ref{eq:muHS}). Therefore, the observations drawn for $\mu_{HS}$ can also be translated into $\l_S$ in accordance with Eq.~(\ref{eq:muHS}) and hence, not shown explicitly in this work. Similarly, the $\mu_3$ parameter plays an important role mostly in generating strong transitions along the $s$-direction. In fact, the presence of $\mu_3$ ensures that the singlet scalar field does not initially have to be zero, as mentioned in \ref{subsubsection-PT-patterns} for $\mathscr{O}^{\prime}$ transitions. Since, we are primarily interested in a strong transition along the $h$-direction, even if it falls under PT type-IV, $\mu_3$ does not put tighter constraints on such transitions. As a matter of fact, in our numerical scans, we find the whole parameter range $-1000 \lesssim \mu_3~(\rm{GeV}) \lesssim 1000$ remains favourable for an SFOPT. Therefore, similar to $\l_S$, we avoid providing any pictorial portrait of $\mu_3$.

\begin{figure*}[!h]
	\hspace*{0.2cm} %% This will shift the figure
	\centering
		\subfigure{\includegraphics[height=7.3cm,width=8.3cm]{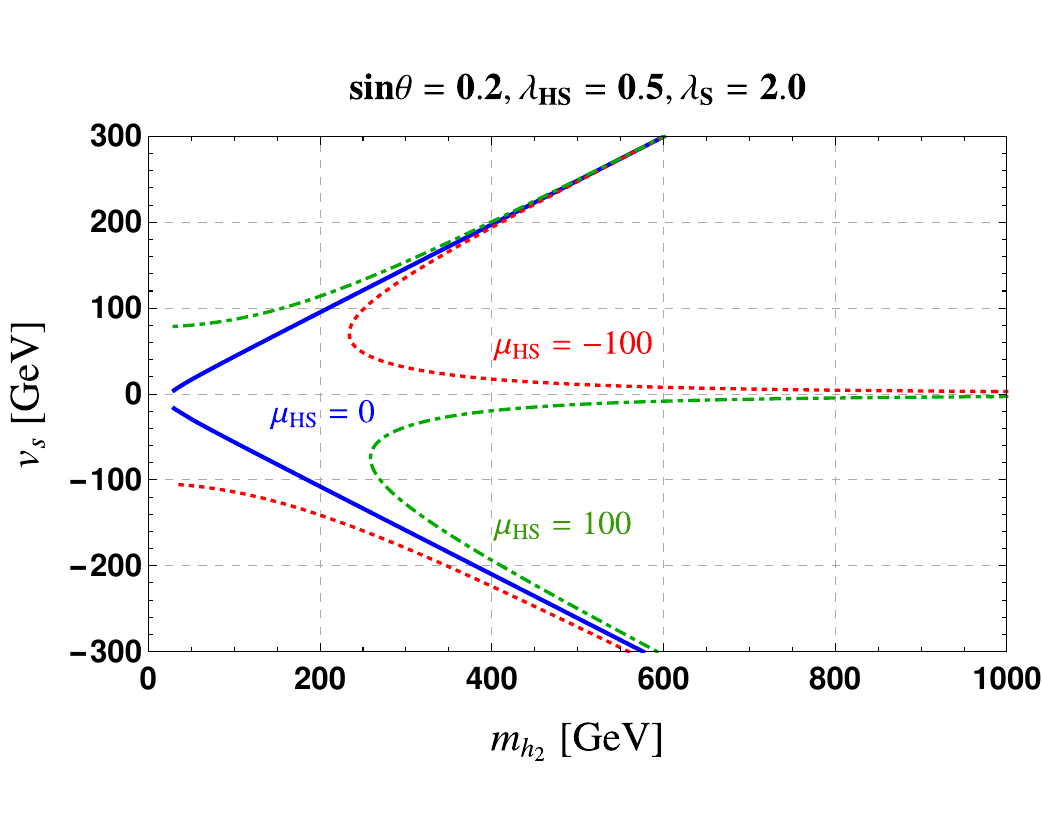}}
		%\subfigure{\includegraphics[height=10.3cm,width=12.6cm]{DM-Fig/correlation_masses_gs.png}}
		%\subfigure{\includegraphics[height=16.3cm,width=14.2cm]{DM-Fig/parameter-correlation.png}}
	\caption{Plot showing $\mu_{HS}$ sensitivity of the $m_{h_2}-v_s$ plane for a few discrete $\mu_{HS}$ values, keeping other relevant parameters fixed as written on the top of the plot. Different $\mu_{HS}$ values are depicted with differently styled differently coloured lines as written on the plot.}
	\label{fig:result:param-depend-analytic-msvs}
\end{figure*}

Fig.~\ref{fig:result:param-depend-analytic-msvs} demonstrates the relation between $m_{h_2}$ and $v_s$ for discrete $\mu_{HS}$ values, e.g., $\mu_{HS}=\{0, -100, 100\}$ in GeV, where we have fixed $\sin\theta=0.2, \l_{SH}=0.5,\l_S=2.0$ and $\mu_3 = 0$\footnote{\small We kept $\mu_3=0$ for this analysis as it is primarily relevant for a transition along the $s$-directions and hardly affects the behaviour of $\mu_{HS}$.}.
%%%%%%%%%%%%%%%%%%%%%%%%%%%%%%%%%%%%%%%%%%%%%%%%%%%%%%%%
%%%%%%%%%%%%%%%%%%%%%%%%%%%%%%%%%%%%%%%%%%%%%%%%%%%%%%%%
It is evident from Fig.~\ref{fig:result:param-depend-analytic-msvs} that, for $\mu_{HS}=0$ (represented by blue coloured solid line) there is no tail-like feature as one can observe in Fig.~\ref{fig:result:param-depend-vsmh2muhs}, while it appears for non-zero $\mu_{HS}$ values. Its appearance can be inferred from Eq.~(\ref{eq:mass-elements}), where the term, $-\frac{\mu_{HS} v^2}{2v_s}$ guarantees large singlet mass, i.e., large $m_{h_2}$ values (see Eq.~(\ref{eq:mass-eigen-HS})) in the limit $v_s \to 0$ but $\mu_{HS} \neq 0$. For a BSM scenario where the SM is extended
with an SM singlet real scalar $S$, $\mu_{HS}=0=\mu_3$ introduces
a new $\mathbb{Z}_2$ symmetry for the singlet sector. For such a framework, low $m_{h_2}$ and large $\sin\theta$ values are preferred for an SFOPT \cite{Profumo:2007wc,Carena:2019une}. Besides,
such BSM theories cannot account for both the observed DM abundance and the matter-antimatter asymmetry simultaneously \cite{Beniwal:2017eik} using the scalar Higgs portal coupling. This is because, an SFOPT
that leads to a successful EWBG, needs large values of the Higgs portal coupling whereas small values are crucial to evade DD bounds. For the chosen framework, non-zero $\mu_{HS}$, however, ameliorates such requirements on the ranges of $m_{h_2}$ and $\sin\th$, favoured for an SFOPT and the correct DM abundance. Moreover, the relative sign difference between $v_S$ and $\mu_{HS}$ is also apparent as it is necessary to assure a non-tachyonic $h_2$.
%%%%%%%%%%%%%%%%%%%%%%%%%%%%%%%%%%%%%%%%%%%%%%%%%%%%%%%%
\begin{figure*}[!h]
	\hspace*{-0.7cm} %% This will shift the figure
	\centering
    \subfigure[]{\includegraphics[height=5.4cm,width=6.7cm]{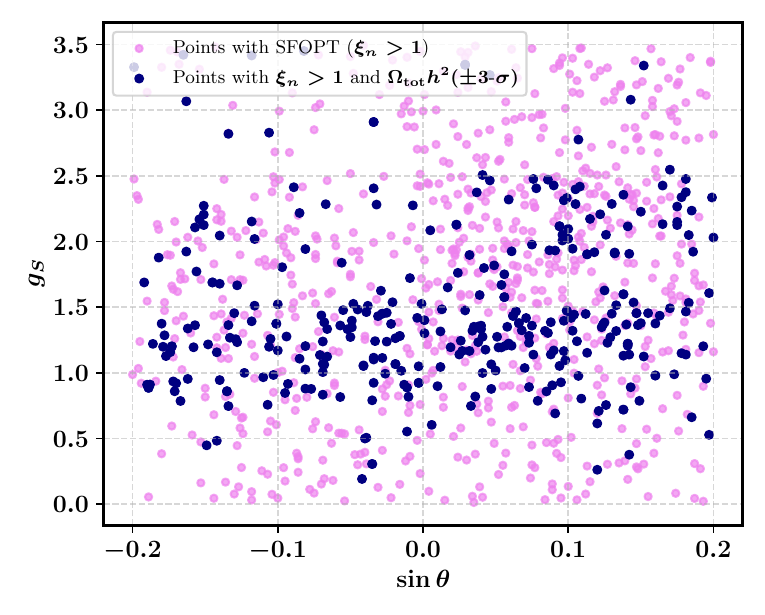}}
    \subfigure[]{\includegraphics[height=5.4cm,width=6.7cm]{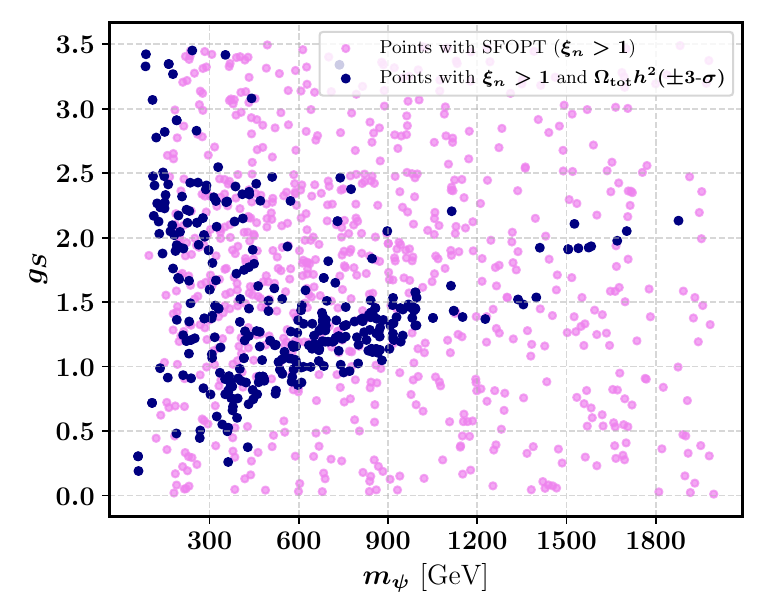}}
	\subfigure[]{\includegraphics[height=5.9cm,width=7.8cm]{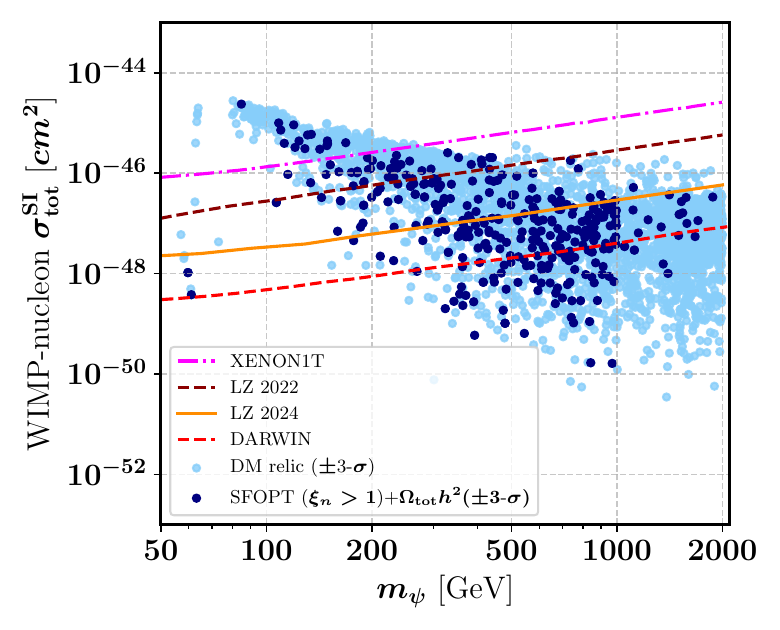}}   
	\caption{The left (right) plot of the top row shows variations in the $\sin\th-g_S$ ($m_\psi-g_S$) plane. The bottom row plot depicts
	a variation in the $m_\psi$ versus $\sigma^{\rm SI}_{\rm tot}$ plane (see Eq. (\ref{eq:DM-pheno:rescaled-DD})). Here, the purple-coloured points represent model points that can accommodate an SFOPT while the dark-blue coloured points, besides an SFOPT, can also accommodate the correct relic density. For the bottom row, the light-blue coloured points obey the $3\sigma$ bound of the correct relic density but do not assure an SFOPT.
    The four differently styled differently coloured lines in the bottom row plot represent bounds from various DD experiments, as detailed on the plot.}
	\label{fig:result:param-depend-DM-PT}
\end{figure*}
%%%%%%%%%%%%%%%%%%%%%%%%%%%%%%%%%%%%%%%%%%%%%%%%%%%%%%%%
The problem of Higgs portal coupling models is effaced in the chosen framework at the cost of an enhanced independent set of parameters.
For example, $\mu_{HS}$ and $\mu_3$ play a pivotal role in the SFOPT, however, have a minuscule impact on the DM phenomenology, as already addressed in Sec. \ref{sec:DM-pheno}. Thus, the chosen framework
can simultaneously account for the observed DM abundance, bypassing the DD and ID bounds, and the matter-antimatter asymmetry of the Universe.

In Fig.~\ref{fig:result:param-depend-DM-PT} (a) and (b),
we plot correlations in the $\sin\theta-g_S$ plane and in the $m_\psi-g_S$ plane, respectively. Here the purple-coloured points, as in Fig.~\ref{fig:result:param-depend-mSmT0vsDM}, represent
model points that can accommodate an SFOPT while the dark-blue coloured
points, besides an SFOPT, can also accommodate the correct relic density. These two plots, again establish the dominance of the relic density bound over the requirement of an SFOPT which is evident from the lesser concentration of the dark-blue coloured points over the purple-coloured ones. The density of dark-blue coloured points suggests $g_S\lsim 2.5$\footnote{This upper limit of $g_S$ is consistent with Eq. (\ref{eq:gs-mpasi-ranges-restricted}).} and $m_\psi \lsim 1.1$ TeV as the favoured ranges for simultaneous occurrence of an SFOPT with the correct relic abundance. We, however, do not claim these limits as the absolute ones as the sample data used for the investigation of the PT dynamics is only $\approx 28\%$ of the data (see subsection \ref{subsec:EWPT-GW-results}) that evades the theoretical and phenomenological bounds stated in Sec. \ref{sec:constraints}, including the ones from direct searches or relic abundances.

Fig.~\ref{fig:result:param-depend-DM-PT} (c) is very similar to
Fig. \ref{fig:DM-pheno:combined-sigma}. The dark-blue coloured points depict the same as of Fig.~\ref{fig:result:param-depend-DM-PT} (a), whereas the light-blue coloured points accommodate a correct value of the relic density without assuring an SFOPT.  We observe that there exist sufficient parameter points, having the correct relic abundance, that can evade various DD bounds including the projected sensitivity reach of the DARWIN and also accommodate an SFOPT. This observation suggests that the chosen model can safely accommodate an SFOPT and experimentally viable DM sector in a significant area of the parameter space, avoiding all the other relevant constraints mentioned in Sec.~\ref{sec:constraints}.

In summary, the presence of additional couplings, like $\mu_{HS},\,\mu_3$ and the absence of a third $\mathbb{Z}_2$ symmetry in the singlet scalar sector, boosts the barrier formation in the effective potential, leading to an SFOPT. Besides, the presence of two DM candidates lessens the constraints on accommodating an SFOPT, consistent with the DM observables, in the same corner of the parameter space. In particular, we have successfully shown that the triplet dark matter, $T^0$, in the ``desert'' mass region, $300\,{\rm GeV} \lesssim m_{T^0} \lesssim 1000\,{\rm GeV}$, safely accommodates SFOPT even with smaller quartic couplings, i.e., $\l_{HT} = \l_{ST} \lesssim \mathcal{O}(0.1)$. This feature was absent in literature for a pure or extended $Y=0$ triplet scalar models with its neutral part, $T^0$, being the only DM candidate \cite{Niemi:2018asa,Niemi:2020hto,Bandyopadhyay:2021ipw}. Our work elevates this shortcoming with a second DM species, which is selected to be a fermion.

%%%%%%%%%%%%%%%%%%%%%%%%%%%%%%%%%%%%%%%%%%%%%%%%%%%%%%%%%%%%%%%%%%%%%%%%%%%%%%%
\subsubsection{The GW signal}\label{sec:GW-results}
%%%%%%%%%%%%%%%%%%%%%%%%%%%%%%%%%%%%%%%%%%%%%%%%%%%%%%%%%%%%%%%%%%%%%%%%%%%%%%%
We have already shown that the SFOPT can co-exist with the experimentally viable DM sector for the chosen model, for a significant region of the model parameter space. In this and the subsequent text body, we will explore the correlations between GW and the thermodynamic parameters of the PTs; projections of the parameter points against different space-based current and future GW detectors; and also address the detectability prospects of the GWs.
%%%%%%%%%%%%%%%%%%%%%%%%%%%%%%%%%%%%%%%%%%%%%%%%%%%%%%%%
%%% Relation of alpha, beta, Tp
\begin{figure*}[!h]
	%\centering
	\hspace*{-0.43cm}
	\subfigure{\includegraphics[height=5.1cm,width=16.0cm]{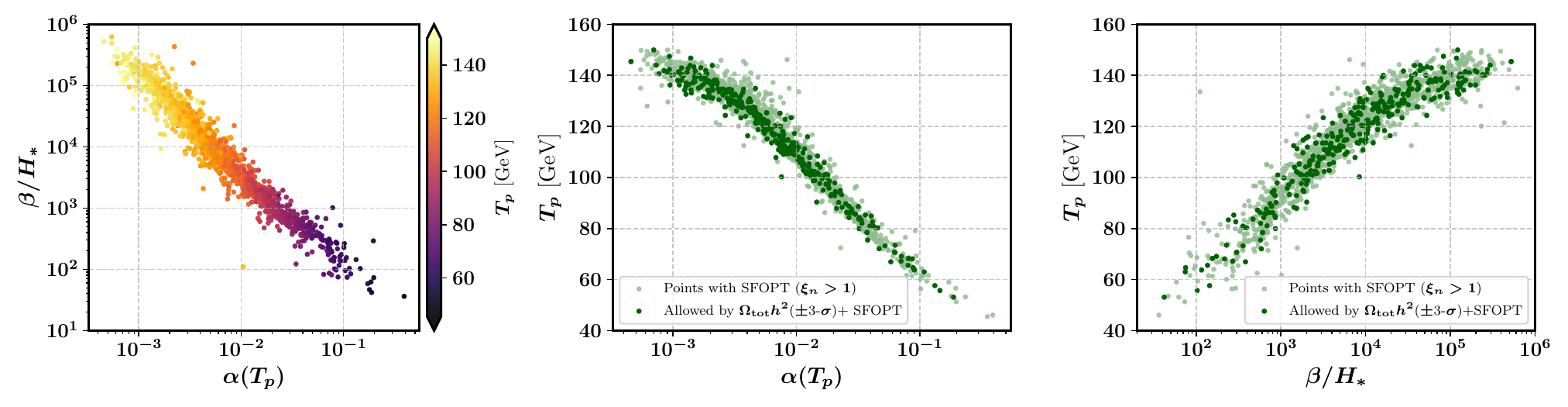}}
	\caption{The left plot shows percolation temperature $(T_p)$ sensitivity in the
	$\alpha(T_p)-{\beta/H_*}$ plane. The middle (right) plot depicts variations of $T_p$ with $\alpha(T_p)~(\beta/H_*)$. For the last two plots, the dark-green coloured points can accommodate an SFOPT and also obey the bounds on the DM observables whereas the light-green coloured points only accommodate an SFOPT. Both these points are, however, consistent with all constraints listed in Sec. \ref{sec:constraints}. Also, as already stated in subsection \ref{subsec:EWPT-GW-results}, all the model points presented in these three points evade at least XENON1T DD bounds. }
	\label{fig:result:GW-PT-param-dependance}
\end{figure*} 
%%%%%%%%%%%%%%%%%%%%%%%%%%%%%%%%%%%%%%%%%%%%%%%%%%%%%%%%
Before we proceed, we note here that, since an SFOPT can give rise to a sufficiently strong GW signal, therefore, in this part, we will provide GW analysis only for those parameter points that exhibit an SFOPT along the $h$-direction and/or $hs$-direction. Additionally, as noted, our data sample for PT analysis already meets various constraints discussed in Sec.~\ref{sec:constraints} \footnote{Except the points that satisfy correct relic, we didn't put the DD constraints on them, which can also be seen in Fig.~\ref{fig:result:param-depend-DM-PT} (c)}, therefore, the remaining points with SFOPT also satisfy these constraints.

The left panel of Fig.~\ref{fig:result:GW-PT-param-dependance} shows scatter plot in the
$\alpha (T_p)-\beta/H_*$ plane where the colour palette implies the variation of percolation temperature $T_p$. We see that $\alpha(T_p)$ and $\beta/H_*$\footnote{Note that, PT parameters relevant for the GW spectrum estimation in our analysis are calculated at the transition temperature $T_{\ast} \approx T_p$, including $\beta/H_{\ast}.$} are inversely related, which is expected, as larger values of $\alpha$ indicate a greater separation between the false and the true vacua (see Eq. (\ref{eq:PT-GW:alpha})), leading to a longer transition time between them, and vice versa.
Besides, higher values of $\alpha$, following Eq. (\ref{eq:percolation:bubble-velo-appx}), correspond to greater emissions of latent heat during the transition, which in turn increases the Hubble parameter. Similarly, the lower the transition temperature compared to $T_c$, the higher the latent heat released. These behaviours are also visible in the middle and the right panel plots of Fig.~\ref{fig:result:GW-PT-param-dependance}. These two plots also highlight the dark-green coloured points that survive the DM relic constraints, besides accommodating an SFOPT. We note that, although in Fig.~\ref{fig:result:GW-PT-param-dependance} we present $\alpha$ and $\beta/H_*$ calculated at $T_p$, however, the strength of the PT is still evaluated with Eq.~(\ref{eq:PT-strength:Tn}). Since, in the chosen model, as previously demonstrated in Fig.~\ref{fig:result:param-depend-TcTnTp} (b), the percolation temperature and nucleation temperature stay close to each other in the majority of the parameter space, therefore, our findings inferred from Fig.~\ref{fig:result:GW-PT-param-dependance} would not alter much for the evaluation of $\alpha,\, \beta/H_*$ at the nucleation temperature $T_n$.

From the phenomenological perspective, one of the most relevant quantities to investigate the GW power spectrum is its peak amplitude, denoted as $\Omega_{\rm GW}^{\rm peak} h^2$, as well the corresponding peak frequency $f_{\rm peak}$. These peak amplitude and the peak frequency can be easily obtained from  Eqs.~(\ref{eq:GW:bubble-collision-amp}), (\ref{eq:GW:soundwave-amp}) (\ref{eq:GW:turb-amp}), and Eqs.~(\ref{eq:GW:bubble-peak-freq}), (\ref{eq:GW:soundwave-peak-freq}), (\ref{eq:GW:turb-peak-freq}), respectively, for the three different sources. In what follows, below we show scatter plots of the same SFOPT points shown in Fig.~\ref{fig:result:GW-PT-param-dependance},  focusing on their different characteristics.
%%%%%%%%%%%%%%%%%%%%%%%%%%%%%%%%%%%%%%%%%%%%%%%%%%%%%%%%
%% GW spectrum
\begin{figure*}[!ht]
	\centering
	\hspace*{-0.50cm}
	\subfigure[]{\includegraphics[height=8.1cm,width=10.5cm]{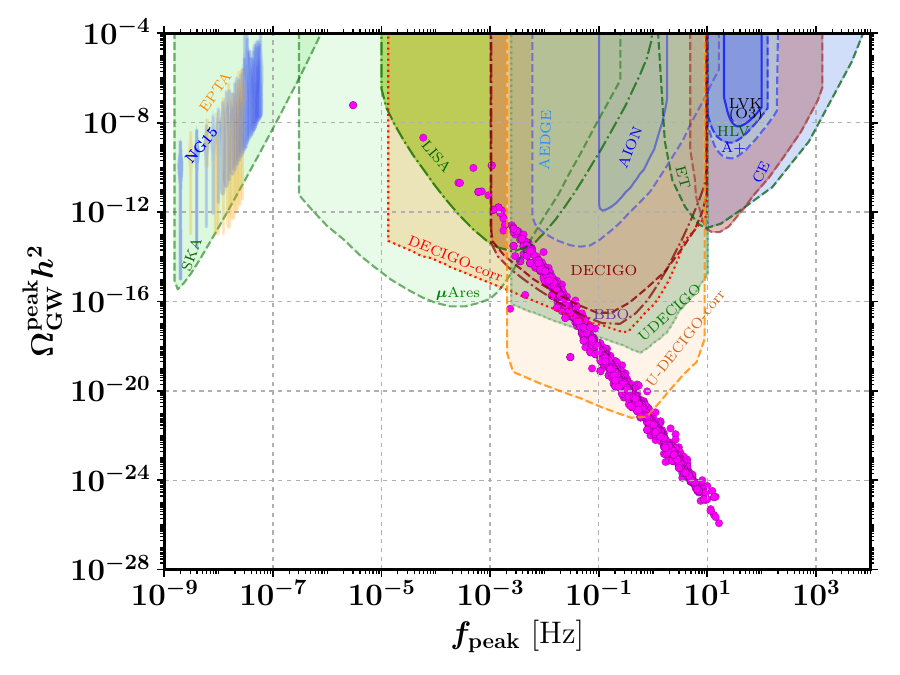}}\\
	\subfigure[]{\includegraphics[height=6.3cm,width=7.6cm]{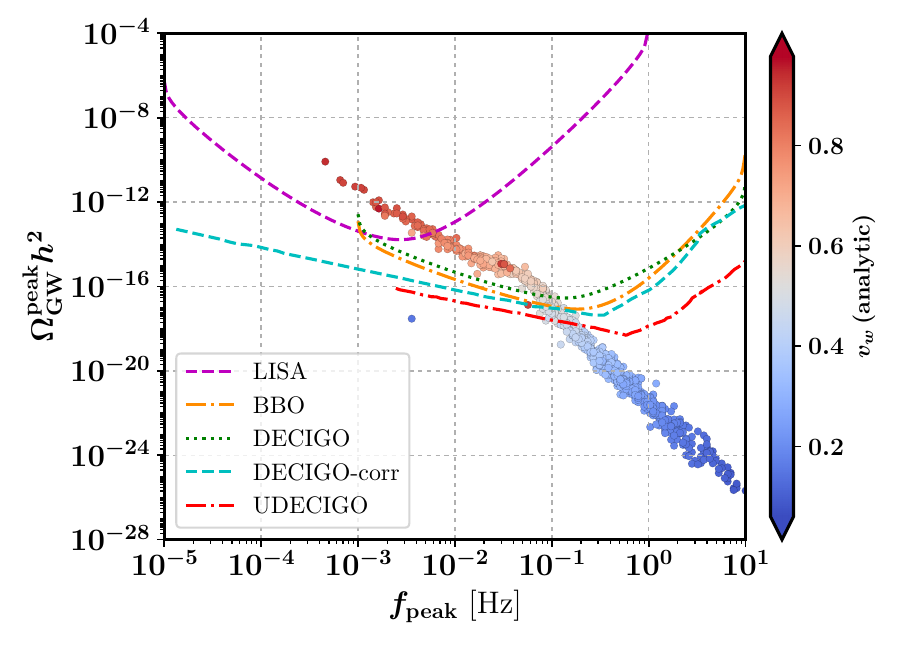}}
	\subfigure[]{\includegraphics[height=6.3cm,width=7.4cm]{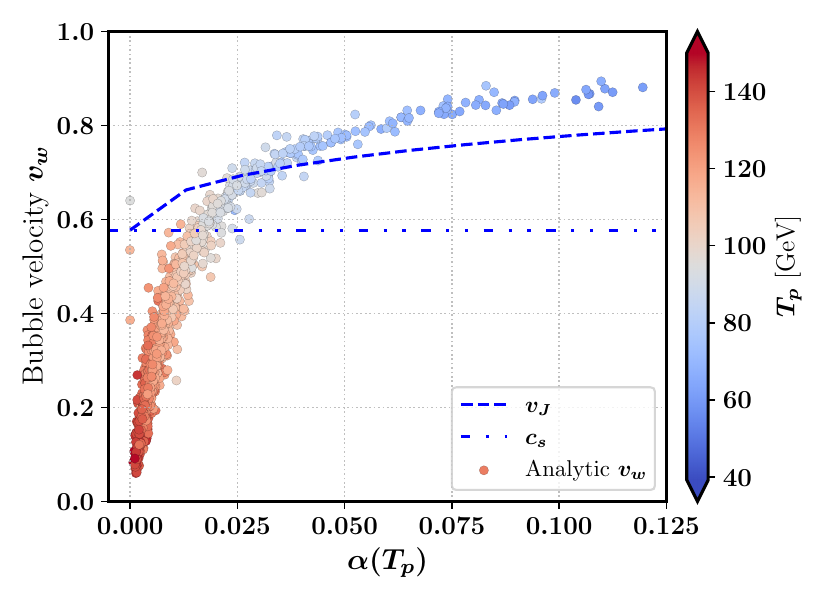}}
	\caption{The top row plot shows variations of the model points in the $\Omega_{\rm GW}^{\rm peak} h^2-f_{\rm peak}$ plane, that can accommodate an SFOPT,  along with the sensitivity reach of different GW detection experiments, shown with differently styled and differently coloured lines. In the bottom row, the left (right) plot depicts the sensitivity of the analytic estimate of $v_w$ ($T_p$) in the $\Omega_{\rm GW}^{\rm peak} h^2-f_{\rm peak}$ $(\alpha(T_p)-v_w)$ plane. Other relevant details are mentioned on the plots.}
	\label{fig:result:GW-spectrm-total}
\end{figure*} 
%%%%%%%%%%%%%%%%%%%%%%%%%%%%%%%%%%%%%%%%%%%%%%%%%%%%%%%%

In Fig.~\ref{fig:result:GW-spectrm-total} (a)\footnote{For this plot, we used Refs.~\cite{Schmitz:2020syl,Fu:2024rsm} to obtain the power-law sensitivity curves data for different GW experiments, except for DECIGO with correlation, U-DECIGO and U-DECIGO with correlation which are taken from Ref.~\cite{Nakayama:2009ce}.}, we display GW signals supported by our model in the $\Omega_{\rm GW}^{\rm peak} h^2$ versus $f_{\rm peak}$ plane for points that exhibits an SFOPT. We also show the power-law integrated sensitivity curves of the upcoming and proposed GW detectors, such as SKA \cite{Janssen:2014dka}, $\mu$-Ares \cite{Sesana:2019vho}, LISA \cite{eLISA:2013xep,LISA:2017pwj}, BBO \cite{Crowder:2005nr,Corbin:2005ny,Harry:2006fi}, DECIGO, U-DECIGO, and U-DECIGO-corr \cite{Kudoh:2005as, Yagi:2011wg,Kawamura:2020pcg}, AEDGE \cite{AEDGE:2019nxb}, AION \cite{Badurina:2019hst}, CE \cite{LIGOScientific:2016wof}, ET \cite{Hild:2008ng}, future upgrades of LVK \cite{KAGRA:2021kbb, Jiang:2022uxp, LIGOScientific:2022sts}, along with the recent results from pulsar timing arrays NANOGrav \cite{NANOGrav:2023gor, NANOGrav:2023hvm} and EPTA \cite{EPTA:2023sfo, EPTA:2023fyk} for comparison. We observe that most of the GW signals can be probed by LISA, BBO, DECIGO and its variants. For the points that fall within these detectors' sensitivity domain, we also need to compute the signal-to-noise ratio (SNR) which we will discuss subsequently. Note that, while preparing this plot, we assumed $v_w \approx 1$. However, as mentioned in subsection ~\ref{sec:Dynamics-of-PT}, we have also calculated the bubble wall velocity using the approximation of Eq.~(\ref{eq:percolation:bubble-velo-appx}), and we present the correlation of $v_w$ (analytic) and the GW signals in Fig.~\ref{fig:result:GW-spectrm-total} (b). We see that the bubble velocity that maximizes the GW peak amplitude, takes values between $0.60$ and $0.95$. These signals with maximum peak amplitude can be probed by LISA, DECIGO and BBO. However, points with low bubble velocities ($v_w \lesssim 0.60$) are mostly in the sensitivity domain of the proposed U-DECIGO-corr detector. The relationship between the peak value of the GW spectrum and $v_w$ is proportional, i.e., $\Omega_{\rm GW}^{\rm peak} h^2$ gradually rises as the wall velocity increases. Conversely, the peak frequency tends to decrease with increasing $v_w$. We note that the aforementioned range of the bubble velocity i.e., $0.60 \lsim v_w \lsim 0.95$, also satisfies the requirement that it is above the Chapman-Jouguet velocity, given by Eq.~(\ref{eq:percolation:Chapman-velo}), such that the formalism presented in Ref.~\cite{Caprini:2019egz} also applies to our case for a single-step transition, either $h$- or $hs$-type. This behaviour can also be seen in Fig.~\ref{fig:result:GW-spectrm-total} (c), where we present the correlation between the analytically calculated bubble velocity and the $\alpha$ parameter with the percolation temperature in the colour bar. The trend of the bubble velocity with the percolation temperature can be understood from Eq.~(\ref{eq:percolation:bubble-velo-appx}). Since, $v_w \sim T_*^{-2}$, therefore, the smaller the transition temperature, the larger the wall velocity, and vice versa. Depending on the relation among the wall velocity, speed of sound in plasma ($c_s \approx 1/\sqrt{3}$) and Chapman-Jouguet velocity $v_J(\alpha)$ (see Eq. (\ref{eq:percolation:Chapman-velo})), one can identify three different regions of bubbles' motion in the plasma.
If $v_w < c_s$, then the explosive growth of the bubble in this region is called a ``deflagration''. In between the Chapman-Jouguet velocity (dashed blue coloured line) and the speed of sound (dash-dotted blue coloured line), the motion of the bubble is composed of a shock discontinuity in front of the wall and a rarefaction wave behind it, and this regime is known as ``hybrid''. Whereas, if $v_w > v_J(\alpha)$, the bubble attains its maximum speed and falls into the category of ``detonation''. It is the detonation region that gives the strongest GW amplitude depending on the $\alpha$ and $\beta/H_*$ parameters. In our findings, SFOPTs that lead to the strong GW signals which can be probed by LISA, fall under the category of the detonation region based on our analytic estimation of $v_w$. However,
this estimate of $v_w$ is an approximation and not the general case. One has to consider a full hydrodynamic treatment in order to evaluate $v_w$ from the first principle. Nevertheless, in both cases, i.e., with $v_w \approx 1$ and analytic $v_w$, the evaluated GW signals in our model can be probed by LISA, BBO and DECIGO.
%%%%%%%%%%%%%%%%%%%%%%%%%%%%%%%%%%%%%%%%%%%%%%%%%%%%%%%% DADA
%% Relation of alpha, beta, Tp with spectrum
\begin{figure*}[!ht]
	%\centering
	\hspace*{-0.98cm}
	\subfigure{\includegraphics[height=6.5cm,width=16.9cm]{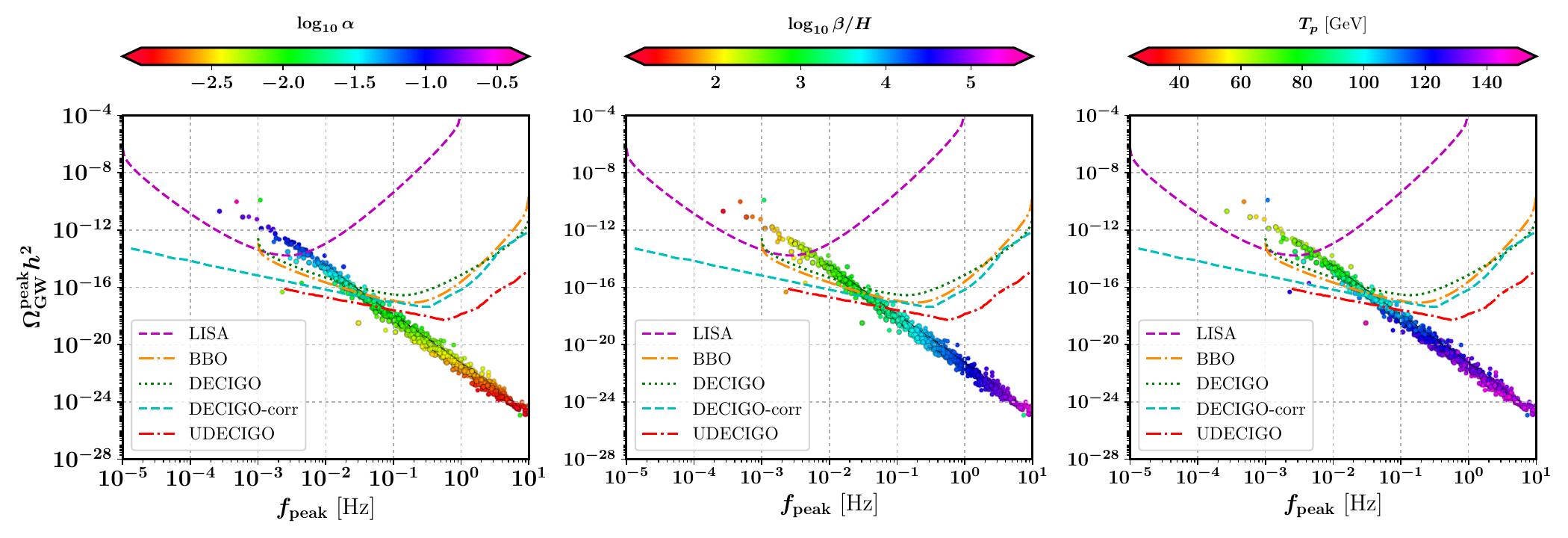}}
	\caption{Plots showing sensitivity of the correlation in the $\Omega_{\rm GW}^{\rm peak} h^2-f_{\rm peak}$ plane on the parameter $\alpha$ (left), $\beta/H_*$ (middle) and $T_p$ (right). All these model points can accommodate an SFOPT. The sensitivity reaches of different GW detectors are shown with differently styled and differently coloured lines.}
	\label{fig:result:GW-spectrum-PT-param}
\end{figure*} 
%%%%%%%%%%%%%%%%%%%%%%%%%%%%%%%%%%%%%%%%%%%%%%%%%%%%%%%%

Finally, in Fig.~\ref{fig:result:GW-spectrum-PT-param}, we present the correlation between the spectral GW peak signal, $\Omega_{\rm GW}^{\rm peak} h^2$, and the peak frequency, $f_{\rm peak}$, with the colour axis representing relevant GW observables for the signal's dynamics, specifically: the PT strength $\alpha (T_p)$ (left panel), the inverse time-scale of the PT (middle panel), and the percolation temperature (right panel). As anticipated, there is a strong correlation between the GW observables and the signal strength. Higher values of $\alpha(T_p)$ indicate a stronger FOPT and assist for a large GW amplitude, while lower $\beta/H_*$ and $T_p$ are favoured for a strong GW signal. The reciprocal behaviour of $\alpha$ and $\beta/H_{\ast}$ is evident from Fig.~\ref{fig:result:GW-PT-param-dependance} (left) plot. In general, in the chosen framework, a parameter point falls within the sensitivity range of LISA if it satisfies, approximately, $\alpha \gtrsim 0.1$, $\beta/H_* \lesssim 600$, and $T_p \lesssim 70$ GeV. However, these do not guarantee the detectability of those GW signals and we must compare their SNR value with the SNR-threshold values of the respective detectors. In the following text body, we present the estimation of SNR and address the detection prospects of the GW signals that fall within the sensitivity reach of LISA, BBO and DECIGO.
%%%%%%%%%%%%%%%%%%%%%%%%%%%%%%%%%%%%%%%%%%%%%%%%%%%%%%%%%%%%%%%%%%%%%%%%%%%%%%%
\subsubsection{Detectability of GW signal}\label{sec:GW-results-detectability}
%%%%%%%%%%%%%%%%%%%%%%%%%%%%%%%%%%%%%%%%%%%%%%%%%%%%%%%%%%%%%%%%%%%%%%%%%%%%%%%
The projections of the GW signals, obtained from the SFOPT in the chosen model, are displayed against the power-law integrated curves (PLIs) of different detectors in
\ref{sec:GW-results}. However, those projections do not inherently tell us about the true detectability prospects in the respective detectors. These GW signals need to be compared to the noise spectrum of the relevant experiment to determine the SNR \cite{Maggiore:1999vm, Allen:1997ad},
%%%%%%%%%%%%%%%%%%%%%%%%%%%%%%%%%%%%%%%
\bea
\label{eq:GW-result:SNR}
{\rm SNR} \equiv \rho = \left[\delta\,\times t_{\rm obs} \int\limits_{f_{\rm min}}^{f_{\rm max}} \frac{df}{\rm Hz} \left(\frac{\Omega_{\rm GW} h^2 (f)}{\Omega_{\rm noise} h^2 (f)}\right)^2 \right]^{1/2},
\eea
%%%%%%%%%%%%%%%%%%%%%%%%%%%%%%%%%%%%%%%
where $\delta$ corresponds to the number of independent channels to distinguish between the different detectors by means of auto-correlation ($\delta = 1$) or cross-correlation ($\delta = 2$) measurements in order to determine the stochastic origin of GW. For LISA, $\delta$ is $1$, whereas for BBO, DECIGO and its variants, it is set as $2$. The parameter $t_{\rm obs}$ defines the duration of the experimental mission in years. In this work, we calculate Eq.~(\ref{eq:GW-result:SNR}) assuming the duration to be $4$ years. The denominator in the integrand, $\Omega_{\rm noise} h^2$ denotes the effective power spectral density of the experiment's strain noise data and, the numerator $\Omega_{\rm GW} h^2$ corresponds to the GW signal calculated in the chosen model.
%%%%%%%%%%%%%%%%%%%%%%%%%%%%%%%%%%%%%%%%%%%%%%%%%%%%%%%%
%% Relation of alpha, beta, Tp with spectrum
\begin{figure*}[!ht]
	%\centering
	\hspace*{-0.99cm}
	\subfigure{\includegraphics[height=5.8cm,width=16.9cm]{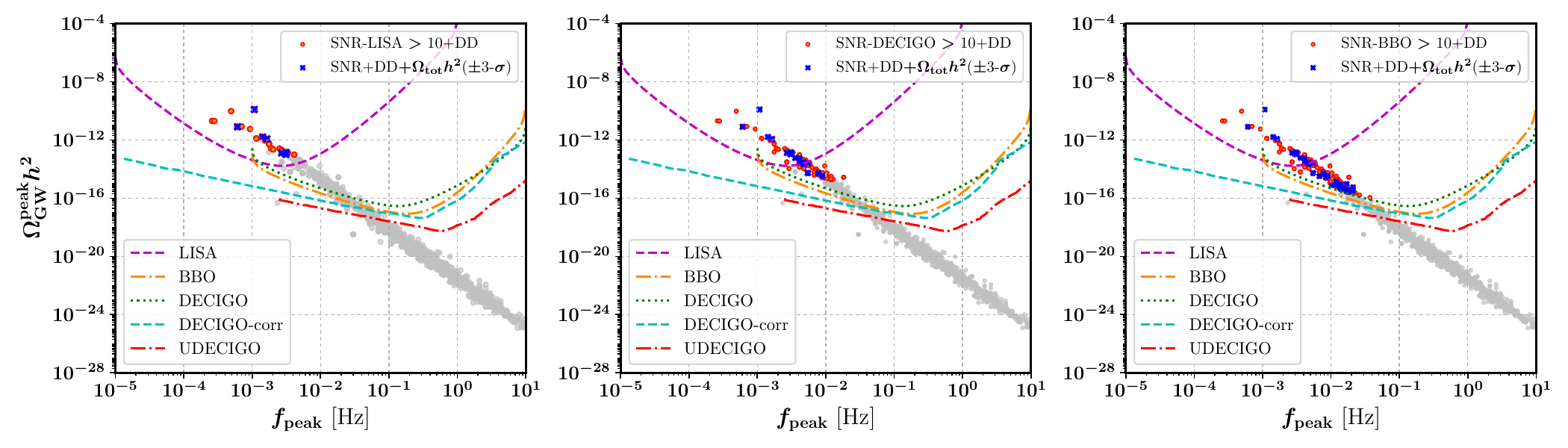}}
\caption{Plots showing model points in the $\Omega_{\rm GW}^{\rm peak} h^2-f_{\rm{peak}}$ plane for LISA (left), DECIGO (middle) and BBO (right), respectively. Here, the light-grey coloured points survive all constraints of Sec. \ref{sec:constraints}. The orange-coloured	points yield detectable GW signals, following an SFOPT, at the various detectors. The blue-coloured crosses, along with a detectable GW signal from an SFOPT, also respect the relic density. Note that, all these model points, i.e., orange, blue and including the light-grey coloured points, also satisfy DD bounds, at least from XENON1T.  The sensitivity reaches of different GW detectors are shown with differently styled and differently coloured lines.}
	\label{fig:result:GW-spectrum-PLI-SNR}
\end{figure*} 
%%%%%%%%%%%%%%%%%%%%%%%%%%%%%%%%%%%%%%%%%%%%%%%%%%%%%%%%
For these experiments to have successful detection prospects, the SNR values must exceed the threshold ($\rho_{\rm thr}$) specific to the experiment's configuration. For instance, the recommended threshold is approximately $50$ for a four-link LISA setup, whereas a six-link design permits a significantly lower threshold of around $10$ \cite{LISA:2017pwj, Robson:2018ifk}. In the present analysis, we will consider a signal to be detectable at LISA, if SNR is $> 10$. For DECIGO and BBO, we also keep the same threshold, i.e., $\rho_{\rm thr} > 10$ \cite{Sato:2017dkf,Crowder:2005nr,Yagi:2011yu, Yagi:2013du}. Additionally, we also exclude model points in the cross-over regime with $\alpha < 10^{-3}$ from our SNR calculations, as they predominantly correspond to $\beta/H_{\ast} \gtrsim 10^5$ (see Fig.~\ref{fig:result:GW-PT-param-dependance}) and typically produce weak GW signals \cite{Caprini:2015zlo} that fall far below the sensitivity reach of the upcoming detectors.

In Fig.~\ref{fig:result:GW-spectrum-PLI-SNR}, we display the earlier obtained GW signals in the $f_{\rm peak}-\Omega_{\rm GW}^{\rm peak}$ plane, and represent the points that pass the threshold value of SNR, calculated using Eq.~(\ref{eq:GW-result:SNR}), for the three different detectors, LISA, DECIGO and BBO, respectively. Note that, in this exercise, we have to calculate the SNR for each data point individually and compare it to the SNR threshold. The GW signals that pass the respective thresholds are shown in orange-coloured dots, whereas the points that further satisfy DM relic constraints are shown by blue-coloured crosses. Note that, all these model points, including the grey points, satisfy DD bounds, at least from XENON1T. A quick comparison of Fig. \ref{fig:result:GW-spectrum-PT-param} and Fig. \ref{fig:result:GW-spectrum-PLI-SNR} clearly reflect the importance of SNR to judge the true detectability prospects of the GWs in a specific detector. These figures indicate that model points above or near the PLI curves are generally more likely to be detected at the corresponding detectors.

%\sout{It is important to mention here that the PLIs we have used here for comparing our GW signal to different experiments, have limitations {Alanne:2019bsm, Schmitz:2020syl}.}
However, it is important to mention that while the PLIs used in our comparison, in general, serve as useful indicators for identifying potential model points that may lead to detectable signals, they have certain limitations \cite{Alanne:2019bsm, Schmitz:2020syl}.
%%%%%%%%%%%%%%%%%%%%%%%%%%%%%%%%%%%%%%%%%%%%%%%%%%%%%%%%
%% Final PICs curves
\begin{figure*}[!ht]
	\hspace*{-0.15cm} %% This will shift the figure
	\centering
	\subfigure{\includegraphics[height=5.8cm,width=7.5cm]{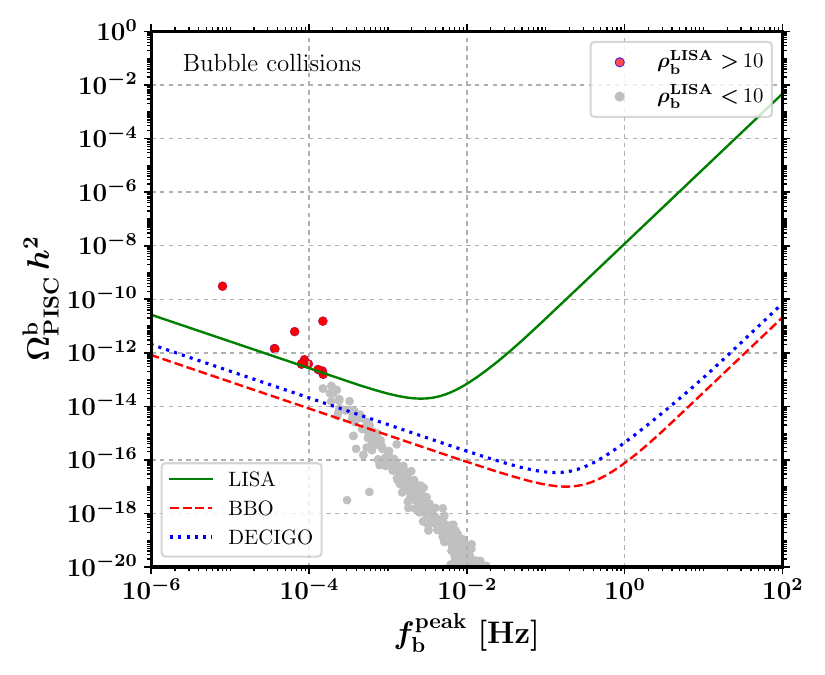}}
	\subfigure{\includegraphics[height=5.8cm,width=7.5cm]{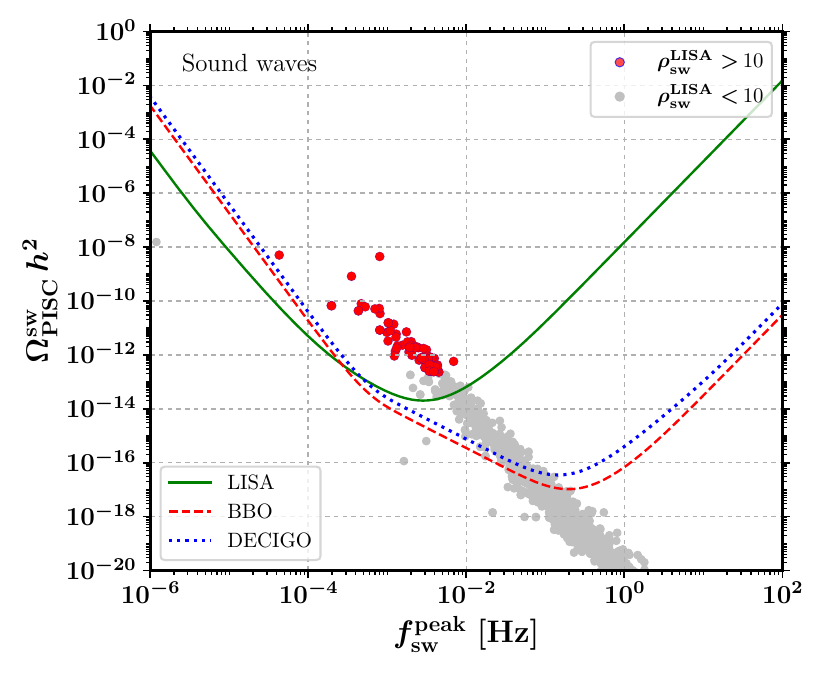}}\\
	\hspace*{-0.15cm}
	\subfigure{\includegraphics[height=5.8cm,width=7.5cm]{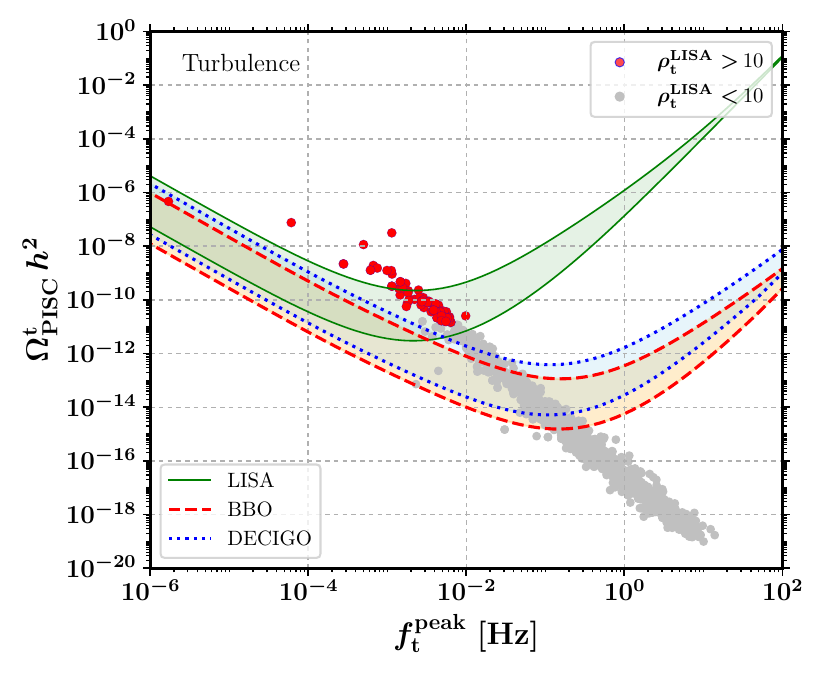}}
	\subfigure{\includegraphics[height=5.8cm,width=7.5cm]{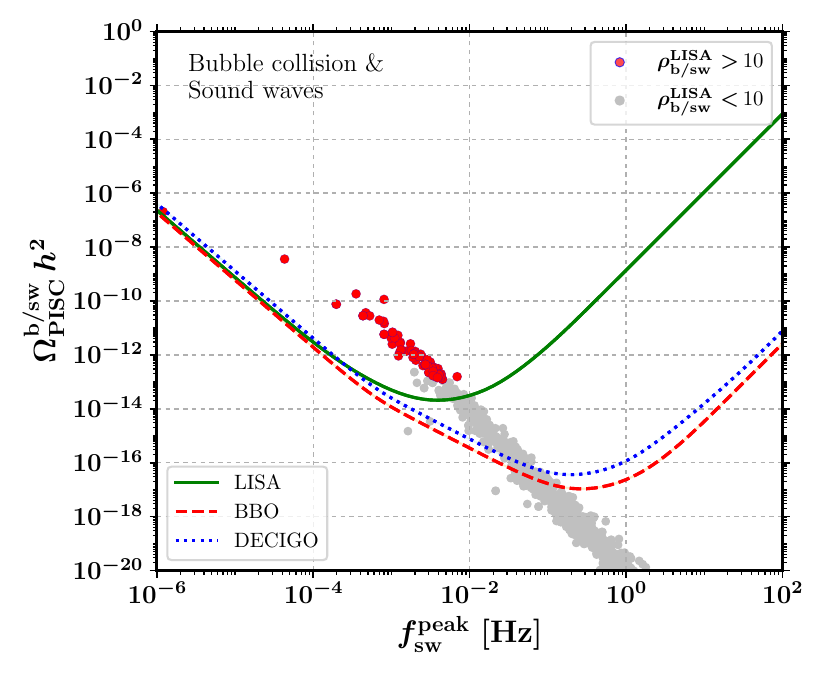}}\\
	\hspace*{-0.15cm}
	\subfigure{\includegraphics[height=5.8cm,width=7.5cm]{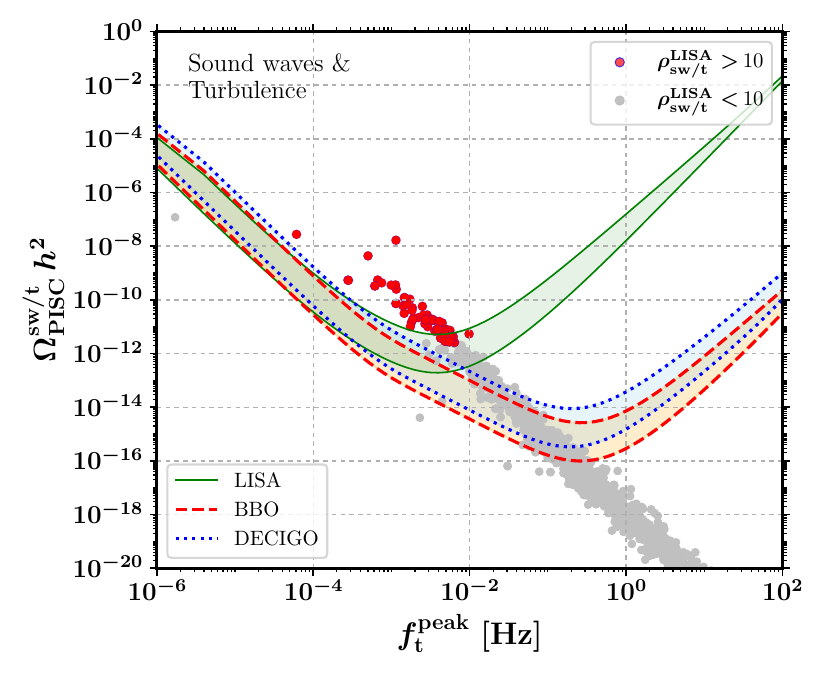}}
	\subfigure{\includegraphics[height=5.8cm,width=7.5cm]{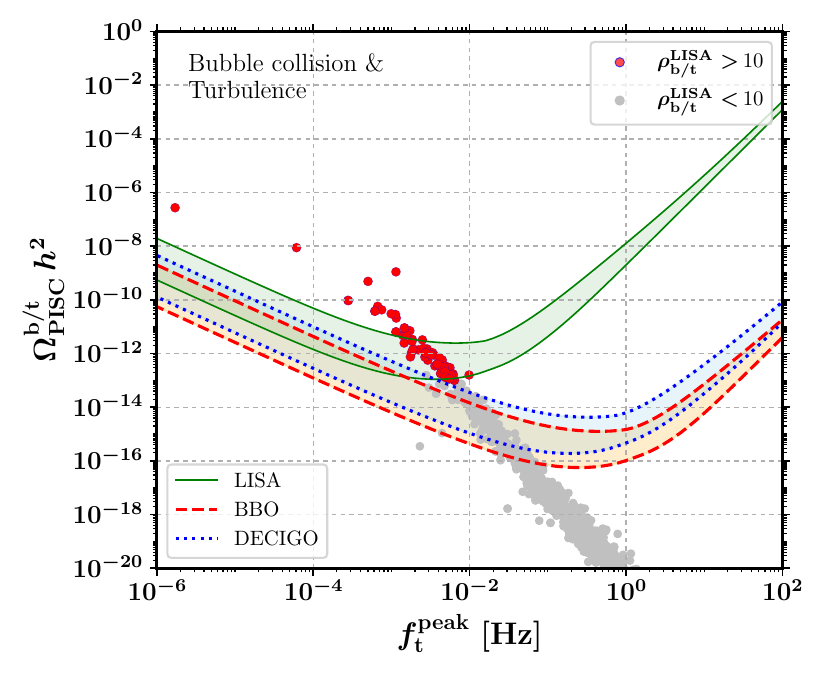}}
	\caption{Plots showing variations of various PISCs driven peak amplitudes, as mentioned in Eq. (\ref{eq:GW-result:PICs-eqn}), with the corresponding peak frequencies for the LISA detector. The light-grey (red) coloured points correspond to the value of the threshold smaller (larger) than $10$.  The sensitivity reach of different GW detectors is shown with differently styled and differently coloured lines or bands, as appropriate. All these model points show SFOPT and are free from constraints discussed in Sec. \ref{sec:constraints}, including DD bounds (at least from XENON1T, see subsection~\ref{subsec:EWPT-GW-results}). Some of the red and light-grey coloured points may also accommodate the correct relic density.}
	\label{fig:result:GW-PISCs}
\end{figure*}
%%%%%%%%%%%%%%%%%%%%%%%%%%%%%%%%%%%%%%%%%%%%%%%%%%%%%%%%
Particularly, the calculation of the peak frequency and peak amplitude of individual points carries no inherent information about the SNR. Moreover, the PLIs only have a well-defined statistical meaning for GW signals that are described by a power law, which is maximally violated close to the ``peak''  in the GW spectrum arising from an SFOPT.

In a recent development, Refs. \cite{Alanne:2019bsm, Schmitz:2020syl} proposed a new approach, named ``peak-integrated sensitivity curves'' (PISCs) to investigate the GW signal region and the experimental reach of different GW detectors by elevating the shortcomings of the conventional PLIs, as outlined before. The advantage of PISCs is that when the expected signal shape is known, as in the case of the FOPT, the integration over the spectral shape can be carried out. This results in the SNR being uniquely determined by the peak energy densities and peak frequencies, which are governed by the model-specific PT parameters only. Therefore, the PISCs are constructed in a way such that they retain full information on the SNR in contrast to PLIs. For more details on PISCs, interested readers can look at Refs. \cite{Alanne:2019bsm, Schmitz:2020syl}. In this formalism, the SNR in Eq.~(\ref{eq:GW-result:SNR}) can be re-written as,
%%%%%%%%%%%%%%%%%%%%%%%%%%%%%%%%%%%%%%%%%%
\bea
\label{eq:GW-result:PICs-eqn}
\frac{\rho^2}{t_{\rm obs}} &=& \left(\frac{\Omega_{\rm b}^{\rm peak} h^2}{\Omega^{\rm b}_{\rm PISCs} h^2}\right)^2 + \left(\frac{\Omega_{\rm sw}^{\rm peak} h^2}{\Omega^{\rm sw}_{\rm PISCs} h^2}\right)^2 + \left(\frac{\Omega_{\rm t}^{\rm peak} h^2}{\Omega^{\rm t}_{\rm PISCs} h^2}\right)^2 \nn \\
&& + \left(\frac{\Omega_{\rm b/sw}^{\rm peak} h^2}{\Omega^{\rm b/sw}_{\rm PISCs} h^2}\right)^2 + \left(\frac{\Omega_{\rm sw/t}^{\rm peak} h^2}{\Omega^{\rm sw/t}_{\rm PISCs} h^2}\right)^2 + \left(\frac{\Omega_{\rm b/t}^{\rm peak} h^2}{\Omega^{\rm b/t}_{\rm PISCs} h^2}\right)^2.
\eea
%%%%%%%%%%%%%%%%%%%%%%%%%%%%%%%%%%%%%%%%%%
Here, it is understood that the integration over the frequency range has already been carried out implicitly,
%%%%%%%%%%%%%%%%%%%%%%%%%%%%%%%%%%%%%%%%%%
\bea
\label{eq:GW-result:PICs-integration}
\Omega_{\rm PISCs}^{i/j} h^2 = \left[(2-\delta_{ij})\, \delta \times 4\,{\rm yr} \int_{f_{\rm min}}^{f_{\rm max}} df \frac{S_i(f)\,S_j(f)}{(\Omega_{\rm noise} h^2(f))^2} \right]^{-1/2},
\eea
%%%%%%%%%%%%%%%%%%%%%%%%%%%%%%%%%%%%%%%%%%
where the factor of $2$ for $i \neq j$ results from Eq.~(\ref{eq:GW-result:SNR}) and $i,j$ corresponds to $\{\rm b, sw, t\}$, $S_{i/j}(f)$ are the individual shape profiles mentioned in subsection~\ref{subsec:GW-from-SFOPT}. Finally, the mixed peak amplitudes are defined as the geometric means,
%%%%%%%%%%%%%%%%%%%%%%%%%%%%%%%%%%%%%%%%%%
\bea
\label{eq:GW-result:PISC-geometric-mean}
\Omega_{i/j}^{\rm peak} = \left(\Omega_i^{\rm peak} h^2 \Omega_j^{\rm peak} h^2\right)^{1/2}.
\eea
%%%%%%%%%%%%%%%%%%%%%%%%%%%%%%%%%%%%%%%%%%
As mentioned earlier, once the integral of Eq.~(\ref{eq:GW-result:PICs-integration}) is carried out, the SNR is uniquely determined by the peak energy densities and the corresponding peak frequencies. We further note that, unlike the analysis presented in Refs.~\cite{Alanne:2019bsm,Schmitz:2020syl} which considered $t_{\rm obs} = 1~{\rm yr},\, \rho_{\rm thr} = 1$ for demonstration, we have extended our analysis with $t_{\rm obs} = 4~{\rm yr}, \,\rho_{\rm thr} = 10$, as already stated in \ref{sec:GW-results}.

In Fig.~\ref{fig:result:GW-PISCs}, we show each PISC curve for a different combination of $i/j$ as a function of the corresponding peak frequency, and for each PISC curve, $\Omega^{i/j}_{\rm PISCs} h^2$, we also present the peak amplitudes $\Omega_{i/j}^{\rm peak} h^2$ and peak frequencies of the data points in the chosen model that exhibits an SFOPT. In this way, we obtain six PISC scatter plots as shown in Fig.~\ref{fig:result:GW-PISCs}. Each panel of this figure illustrates one of the contributions to the total SNR as shown in Eq.~(\ref{eq:GW-result:PICs-eqn}), with each point representing an entire GW spectrum of a specific physical origin. One can understand the six panels of Fig.~\ref{fig:result:GW-PISCs} as follows: in general convention, one compares the GW signal calculated for a particular choice of the parameter point of a given particle physics model to the PLIs. Subsequently, this exercise has to be repeated for each data point.  However, in the case of PISCs, we just need to check if a given point lies above any of the six PISCs of a particular experiment. In that case, the SNR will naturally exceed the predefined threshold of the respective experiments. On the other hand, points below the PISCs may still exceed the SNR threshold if the sum of all plausible contributions is greater than the threshold, however, we need to check them individually. Therefore, in Fig.~\ref{fig:result:GW-PISCs}, in all panels, points with red colour satisfy the SNR threshold. In our case, $\rho_{\rm thr}^{\rm LISA} > 10$, and points with light-grey colour needs separate inspection for the SNR threshold. It is also important to note that for any contribution dependent on more than one peak frequency, e.g., turbulence (t)\footnote{The spectrum shape profile for turbulence involves two characteristic frequencies, i.e., $h_{\ast}, f_{\rm tur}$, see Eqs.~(\ref{eq:GW:turb-shape}), (\ref{eq:GW:turb-hstar}) and (\ref{eq:GW:turb-peak-freq}).}, sound wave/turbulence (sw/t), and bubble collisions/turbulence (b/t) channels, a single PISC cannot be represented in a two-dimensional plot. For these cases, similar to Refs.~\cite{Alanne:2019bsm,Schmitz:2020syl}, we draw peak-integrated sensitivity ``bands'' and we choose $h_*/f_{\rm t} \in [2.4\times 10^{-3}, 2.0]$ for t, sw/t and b/t PISCs, while we take $f_{\rm sw}/f_{\rm t} \in [6.0 \times 10^{-1}, 6.8\times10^{-1}]$ and $f_{\rm b}/f_{\rm t} \in [1.4 \times 10^{-1}, 1.6 \times 10^{-1}]$ for sw/t and b/t, respectively. We denote $f_{\rm b}$ and $f_{\rm t}$ as short notations for the peak frequencies of bubble collision (Eq.~(\ref{eq:GW:bubble-peak-freq})) and turbulence (Eq.~(\ref{eq:GW:turb-peak-freq})), respectively\footnote{The chosen ranges for $h_{\ast}/f_t$, $f_{\rm sw}/f_t$, and $f_b/f_t$ are guided by Ref. \cite{Schmitz:2020syl}. The range for $h_{\ast}/f_t$ is chosen to cover model points within the PISCs, ensuring coverage near the LISA sensitivity region, with the upper limit approximately reflecting the maximum span in the (peak)amplitude-frequency plane. We adopt $f_{\rm sw}/f_t \approx 0.7$ as suggested in Ref. \cite{Schmitz:2020syl} (see Eq.~(2.16) therein) and restrict it to $[0.60, 0.68]$ to mitigate numerical instabilities. Lastly, the range for $f_b/f_t$ is chosen based on the relation $f_b/f_t \approx 0.14$, assuming $v_w = 1$ \cite{Schmitz:2020syl}.}. Furthermore, we kept $v_w \approx 1$ while preparing these plots. In our study, the most constraining channel is the bubble collision, displayed in the top left panel of Fig.~\ref{fig:result:GW-PISCs}, which is not very surprising as we have already mentioned in subsection~\ref{subsec:GW-from-SFOPT} that for a scenario with a dominant GW production due to the plasma motion, the contribution from the $\Omega_{\rm b} h^2$ is suppressed. Although we see a shift in $f_{\rm b}^{\rm peak}$ toward the lower value, however, there are fewer data points that exceed the $\rho_{\rm thr}^{\rm LISA} > 10$ limit. However, for the case with sound waves and turbulence, the number of data points above the PISCs is rather large, suggesting their dominant contributions to the source of GW spectrum. Therefore, the key idea behind the PISCs approach to decompose the total SNR into six partial SNRs gives a better picture of detectability compared to PLIs. In fact, PISCs provide a clear distinction between the separate contributions to the total GW amplitude that are dominant or negligible.

To summarize, in this part, we provide details of the detection prospects of the GW signals arising from an SFOPT in the chosen model following two different strategies: the conventional PLIs and the newly proposed PISCs. In both of the cases, we find a sufficient number of parameter points that can be probed in LISA, BBO, and DECIGO. We find it is the bubble collision that contributes less to the GW amplitude while the contributions from sound waves and turbulence are dominant. The chosen framework successfully accommodates SFOPT favouring EWBG, generates GW spectra that can be probed at upcoming and proposed space-based GW detectors, and also satisfies DM constraints to accommodate two DM particles in the same parameter region. Our investigation serves as a complementary probe of the BSM physics at the GW and DM frontiers apart from the collider experiments. 

%%%%%%%%%%%%%%%%%%%%%%%%%%%%%%%%%%%%%%%%%%%%%%%%%%%%%%%%%%%%%%%%%%%%%%%%%%%%%%%%%%
%%%%%%%%%%%%%%%%%%%%
\section{Summary and Conclusion}\label{sec:conclu}
%%%%%%%%%%%%%%%%%%%%

In this work, we have explored a bipartite DM framework with a scalar and a fermion DM, $T^0$ and $\psi$, respectively. The scalar DM belongs to a hyperchrageless $SU(2)_L$ triplet while the fermion DM is an SM gauge singlet Dirac fermion. Another SM singlet scalar, having a non-zero VEV, is introduced to this framework which mixes with the SM $SU(2)_L$ doublet Higgs. Such a mixing yields two mass eigenstates $h_1,\,h_2$ which offer a Higgs portal for the fermion DM. The scalar DM $T^0$, owing to its $SU(2)_L$ charge, can undergo both the gauge and Higgs portal interactions. The presence of more than one DM species modifies the relic density and direct search analyses, making it easier to evade the concerned experimental bounds. Individual components, however, may remain experimentally disfavoured. Collectively, the concerned two-component DM framework remains experimentally feasible for a large region of the model parameter space, besides, reviving the ``desert'' region encountered in BSM scenarios with $T^0$ as the sole DM contender. Needless to mention, the aforementioned region also obeys various other plausible theoretical constraints and experimental limits, e.g., from colliders, precision Higgs physics, etc.

The presence of cubic interactions in the tree-level scalar potential offers prospects of having an SFOEWPT in the various plausible field directions. With $T^0$ being a DM candidate, the intricacy of understanding the PT dynamics is not very complex in the chosen bipartite DM framework as the concerned filed space is only two-dimensional.
An SFOEWPT can lead to EWBG and also generate detectable GW signatures. A significant region of the parameter space is observed to comply with: (1). The DM relic density bound and limits from various direct searches, including the projected sensitivity reach of the DARWIN; (2). SFOEWPT in the $SU(2)_L$ singlet or doublet or mixed directions, and (3). Detection potentials at the various GW interferometers. These three different tasks are accomplished in the chosen model at the cost of introducing an optimal set of BSM inputs. Some of these entities play pivotal roles in depicting the experimentally admissible DM phenomenology while having an inconsequential impact on the PT dynamics. In a similar way the inputs that guide the PT dynamics and consequently, the generation of the GWs, have a marginal influence on the DM phenomenology. A good number of model points, which show an SFOEWPT and have the correct relic density, can be probed either at the ongoing or upcoming direct searches, e.g., LZ-2024 or DARWIN, as well as in the GW interferometers like LISA.

The detection aspect of the GW signatures can be improved with the recently proposed PISCs which uniquely assess the signal-to-noise ratio, in terms of the peak amplitudes and peak frequencies, through a known signal shape. Nevertheless, for completeness and comparison, we have also presented detectability prospects of the GWs using the traditional PLIs. Unlike PLIs, the PISCs provide an excellent portrayal of the GW detection prospects at any specific interferometer, e.g., LISA or BBO or DECIGO, with a simultaneous understanding of the relative dominance of various possible GW generation mechanisms.

In a nutshell, our analysis of the chosen bipartite DM framework shows how GWs, following an SFOEWPT, can provide a complementary detection prospect of the BSM physics beyond the relic abundance, direct searches and collider probes. Our numerical exercise, in this work, shows revival, although partial, of the ``desert region'' in the considered sub-TeV range, i.e., $300$ GeV $\lsim m_{T^0} \lsim 1000$ GeV. Besides, the conducted scan also seems to favour $g_S \lsim 2.5$, $m_\psi \lsim 1.1$ TeV, and $m_{h_2}\lsim 850$ GeV, for model points where an SFOEWPT can co-exist with the correct DM phenomenology. Some of these points, eventually, will also yield detectable GW signals. It is compelling to see that the model points permissible by various possible theoretical bounds and different relevant experimental limits, including colliders, precision Higgs physics, relic density, direct searches and GWs following an SFOEWPT, collectively favour a corner of the parameter space where all possible BSM states remain within the reach of the LHC, justifying the true aspect of the complementarity. A definite claim of these upper bounds, however, should be tested with a more extensive numerical analysis.

%%%%%%%%%%%%%%%%%%%%%%%%%%%%%%%%%%%%%%%%%%%%%%%%%%%%%%%%%%%%%%%%%%%%%%%%%%%%%%%%%%
%%%%%%%%%%%%%%%%%%%%%%%%%%%%%%%%%%%%%%%%%%%%%%%%%%
%%%%%%%%%%%%%%%%%%%%%%%%%%%%%%%%%%%%%%%%%%%%%%%%%%%%%%%%%%%%
\section*{Acknowledgements}
%%%%%%%%%%%%%%%%%%%%%%%%%%%%%%%%%%%%%%%%%%%%%%%%%%%%%%%%%%%%
P. B. acknowledges the financial support from the Indian Institute of Technology, Delhi (IITD) and as a Senior Research Fellow and {\it{Hydra}} high-performance computing facility at the MS-516 HEP-PH Laboratory, Department of Physics, IITD.
P. G. acknowledges the IITD SEED grant support IITD/Plg/budget/2018-2019/21924, continued as IITD/Plg/budget/2019-2020/173965, IITD Equipment Matching Grant IITD/ IRD/MI02120/ 208794, Start-up Research Grant (SRG) support SRG/2019/000064, and Core Research Grant (CRG) support CRG/2022/002507 from the Science and Engineering Research Board (SERB), Department of Science and Technology, Government of India. A.K.S acknowledges the postdoctoral fellowship from the Institute of Physics, Bhubaneswar.

%%%%%%%%%%%%%%%%%%%%%%%%%%%%%%%%%%%%%%%%%%%%%%%%%%%%%%%%%%%%%%%%%%%%%%%%%%%%%%%%%%
\appendix
%%%%%%%%%%%%%%%%%%%%%%%%%%%%%%%%%%%%%%%%%%%%%%%%%%%%%%%%%%%%%%%%%%%%%%%%%%%%%%%%%%
\section{Tree level unitarity} \label{appx:unitarity}
%%%%%%%%%%%%%%%%%%%%%%%%%%%%%%%%%%%%%%%%%%%%%%%%%%%%%%%%%%%%%%%%%%%%%%%%%%%%%%%%%%
 In this section of the appendix, we discuss the perturbative unitarity limits on the quartic couplings present in the scalar sector of our model as depicted by Eqs. (\ref{eq:pot:different-part}), (\ref{eq:pot:interaction-potential}). As discussed in subsection~\ref{subsec:unitarity}, the unitarity bounds on the extended scalar sector can be determined from the scattering matrix (S-matrix) of various processes. For any $2 \rightarrow 2$ process, the scattering amplitude $\mathcal{M}_{2\rightarrow2}$ and the scattering cross-section $\sigma$, using Legendre polynomials $P_l$, can be written as \cite{Lee:1977eg},
%%%%%%%%%%%%%%%%%%%%%%%%%%%%%%%%%%%%%%%
\bea
\label{eq:appx:unitarity:S-amp-Legendre}
\mathcal{M}_{2\rightarrow2} = 16\pi \sum_{l=0}^{\infty} (2 l + 1) P_l (\cos\alpha) a_l, \quad \sigma = \frac{16 \pi}{s} \sum_{l=0}^{\infty} (2l+1) |a_l(s)|^2,\, {\rm respectively}.
\eea
%%%%%%%%%%%%%%%%%%%%%%%%%%%%%%%%%%%%%%%
Here, $\alpha$ is the scattering angle, $P_l(\cos\alpha)$ is the Legendre polynomial of order $l$ associated with the $l^{\rm {th}}$ partial wave whose amplitude is given by $a_l$, and $s=4E_{\rm {CM}}^2$ is the Mandelstam variable with $E_{\rm CM}$ as the centre-of-mass energy of the incoming particles. At high energies, the dominant contributions to the scattering amplitude come from diagrams that involve the scalar quartic couplings. Additionally, due to the equivalence theorem \cite{Lee:1977eg,Yao:1988aj,Veltman:1989ud,He:1992nga}, as mentioned earlier in subsection~\ref{subsec:unitarity}, gauge bosons can be substituted with their corresponding Goldstone bosons, simplifying the estimation of the scattering amplitude matrix for scalars alone. Therefore, to constrain our model from unitarity, it is sufficient to construct the S-matrix with only the scalar quartic couplings.

The post EWSB mixing between the SM singlet scalar $s$ and the SM doublet Higgs $h$
makes the scalar quartic couplings complicated functions of various $\l$'s in the mass/physical basis. One can, nevertheless, use a unitary transformation \cite{Kanemura:1993hm} to express the S-matrix in the pre-mixing basis, i.e., $h,s, T^{0}, T^{\pm}, G^{0}, G^{\pm}$.
Thus, the different scalar quartic couplings can be obtained just by expanding the scalar potential of Eq.~(\ref{eq:pot:total-scalar}), using Eq. (\ref{eq:scalar-field-basis}). With this exercise, one gets {\it thirteen} neutral scalar (NS) combinations of two-particle states in the bases,
%%%%%%%%%%%%%%%%%%%%%%%%%%%%%%%%%%%%%%%%%%%%%
\bea
\label{eq:appx:unitarity:neutral-basis}
{\rm NS-I} &&\equiv \{\frac{h h }{\sqrt{2}}, \frac{s s}{\sqrt{2}}, \frac{T^0 T^0 }{\sqrt{2}}, \frac{G^0 G^0 }{\sqrt{2}}, T^+ T^-, G^+G^-\}, \nn \\
{\rm NS-II} &&\equiv \{hs, h T^0, s T^0, h G^0, s G^0, T^0 G^0, T^+ G^- ({\rm or~} T^- G^+)\},
\eea
%%%%%%%%%%%%%%%%%%%%%%%%%%%%%%%%%%%%%%%%%%%%%
and {\it eight} combinations of singly charged scalar (CS) two-particle states in the bases,
%%%%%%%%%%%%%%%%%%%%%%%%%%%%%%%%%%%%%%%%%%%%%
\bea
\label{eq:appx:unitarity:charged-basis}
{\rm CS} \equiv \{hG^+, h T^+, s G^+, s T^+, T^0 G^+, T^0 T^+, G^0 G^+, G^0 T^+\}.
\eea
%%%%%%%%%%%%%%%%%%%%%%%%%%%%%%%%%%%%%%%%%%%%%
The $1/\sqrt{2}$ factor appearing in $\rm NS-I$ basis is due to the statistics of identical particles. The scattering amplitude matrix $\mathcal{M}$ can now be written in a block diagonal form by decomposing it into the neutral and charged sectors, as shown below:
%%%%%%%%%%%%%%%%%%%%%%%%%%%%%%%%%%%%%%%%%%%%%%
\begin{equation}
\label{eq:appx:unitarity:full-scattering-matrix}
	\mathcal{M}_{21 \times 21} =
	\begin{pmatrix}
		\mathcal{M}^{\rm NS}_{13 \times 13} & 0 \\
		0 & \mathcal{M}^{\rm CS}_{8 \times 8}
	\end{pmatrix},
\end{equation}
%%%%%%%%%%%%%%%%%%%%%%%%%%%%%%%%%%%%%%%%%%%%%%
where $\mathcal{M}^{\rm NS}_{13 \times 13}$ can be decomposed further as,
%%%%%%%%%%%%%%%%%%%%%%%%%%%%%%%%%%%%%%%%%%%%%%
\begin{equation}
	\label{eq:appx:unitarity:scattering-matrix-NS}
	\mathcal{M}^{\rm NS}_{13 \times 13} =
	\begin{pmatrix}
		\mathcal{M}^{\rm NS-I}_{6 \times 6} & 0 \\
		0 & \mathcal{M}^{\rm NS-II}_{7 \times 7}
	\end{pmatrix}.
\end{equation}
%%%%%%%%%%%%%%%%%%%%%%%%%%%%%%%%%%%%%%%%%%%%%%
The sub-matrix $\mathcal{M}^{\rm NS-I}_{6 \times 6}$ in basis $\rm NS-I$ is written as,
%%%%%%%%%%%%%%%%%%%%%%%%%%%%%%%%%%%%%%%%%%%%%%
\begin{equation}
\label{eq:appx:unitarity:scattering-matrix-NS-I}
\mathcal{M}^{\rm NS-I}_{6 \times 6} =
\begin{pmatrix}
	3 \lambda_H & \frac{\lambda_{{SH}}}{2} & \frac{\lambda_{{HT}}}{2} & \lambda_H & \frac{\lambda_{{HT}}}{\sqrt{2}} & \sqrt{2} \lambda_H \\
	\frac{\lambda_{{SH}}}{2} & 3 \lambda_S & \frac{\lambda_{{ST}}}{2} & \frac{\lambda_{{SH}}}{2} & \frac{\lambda_{{ST}}}{\sqrt{2}} & \frac{\lambda_{{SH}}}{\sqrt{2}} \\
	\frac{\lambda_{{HT}}}{2} & \frac{\lambda_{{ST}}}{2} & 3 \l_{\bm{T}} & \frac{\lambda_{{HT}}}{2} & \sqrt{2} \l_{\bm{T}} & \frac{\lambda_{{HT}}}{\sqrt{2}} \\
	\lambda_H & \frac{\lambda_{{SH}}}{2} & \frac{\lambda_{{HT}}}{2} & 3 \lambda_H & \frac{\lambda_{{HT}}}{\sqrt{2}} & \sqrt{2} \lambda_H \\
	\frac{\lambda_{{HT}}}{\sqrt{2}} & \frac{\lambda_{{ST}}}{\sqrt{2}} & \sqrt{2} \l_{\bm{T}} & \frac{\lambda_{{HT}}}{\sqrt{2}} & 4 \l_{\bm{T}} & \lambda_{{HT}} \\
	\sqrt{2} \lambda_H & \frac{\lambda_{{SH}}}{\sqrt{2}} & \frac{\lambda_{{HT}}}{\sqrt{2}} & \sqrt{2} \lambda_H & \lambda_{{HT}} & 4 \lambda_H \\
\end{pmatrix}.
\end{equation}
%%%%%%%%%%%%%%%%%%%%%%%%%%%%%%%%%%%%%%%%%%%%%%
The eigenvalues of $\mathcal{M}^{\rm NS-I}_{6 \times 6}$ are $2 \l_H, 2 \l_H, 2 \l_{\bm{T}},\,x_1,x_2,x_3$ with $x_{1,2,3}$ being the solutions of Eq. (\ref{eq:unitary:x-values}).
%%%%%%%%%%%%%%%%%%%%%%%%%%%%%%%%%%%%%%%%%%%%%
%\bea
%x^3 &&+ x^2 (-12 \l_{H} -6 \l_S -10 \l_T) \nn \\
%&&+ x (-12 \l_{HT}^2 +72 \l_H \l_S -4 \l_{SH}^2 -3 \l_{ST}^2 + 120 \l_H \l_T + 60 \l_S \l_T) \nn \\
%&&+ (72 \l_{HT}^2 \l_S - 24 \l_{HT} \l_{SH} \l_{ST} + 36 \l_H \l_{ST}^2 - 720 \l_H \l_S \l_T + 40 \l_{SH}^2 \l_T) = 0
%\eea
%%%%%%%%%%%%%%%%%%%%%%%%%%%%%%%%%%%%%%%%%%%%%
Finally, the matrices $\mathcal{M}^{\rm NS-II}_{7 \times 7}$ and $\mathcal{M}^{\rm CS}_{8 \times 8}$ are diagonal in the bases $\rm NS-II$ and $\rm CS$, respectively, and can be expressed as,
%%%%%%%%%%%%%%%%%%%%%%%%%%%%%%%%%%%%%%%%%%%%%%
\bea
\label{eq:appx:unitarity:scattering-matrix-NS-II-CS}
\mathcal{M}^{\rm NS-II}_{7 \times 7} &&= {\rm diag}\{\l_{SH}, \l_{HT}, \l_{ST}, 2\l_H, \l_{SH}, \l_{HT}, \l_{HT}\},\nn \\
\mathcal{M}^{\rm CS}_{8 \times 8} &&= {\rm diag}\{2 \l_H, \l_{HT}, \l_{SH}, \l_{ST}, \l_{HT}, 2 \l_{\bm{T}}, 2 \l_H, \l_{HT} \}.
\eea
%%%%%%%%%%%%%%%%%%%%%%%%%%%%%%%%%%%%%%%%%%%%%%
Hence, using Eqs.~(\ref{eq:appx:unitarity:scattering-matrix-NS-I}), (\ref{eq:appx:unitarity:scattering-matrix-NS-II-CS}), and demanding that the eigenvalues of the scattering amplitude matrices are less than $< 8\pi$, one can reproduce the limits mentioned in Eq.~(\ref{eq:constraints:unitarity}).

%%%%%%%%%%%%%%%%%%%%%%%%%%%%%%%%%%%%%%%%%%%%%%%%%%%%%%%%%%%%%%%%%%%%%%%%%%%%%%%%%%
\section{Electroweak precision test constraints} \label{appx:EWPO-constraints}
%%%%%%%%%%%%%%%%%%%%%%%%%%%%%%%%%%%%%%%%%%%%%%%%%%%%%%%%%%%%%%%%%%%%%%%%%%%%%%%%%%
 Following Refs. \cite{Forshaw:2001xq,Khan:2016sxm,Chakrabarty:2021kmr}, the contributions to the oblique parameters for a $SU(2)_L$ scalar triplet, with vanishing hypercharge, is given by
%%%%%%%%%%%%%%%%%%%%%%%%
\bea\label{eq:STU-inertTriplet}
\Delta S_{\textrm{IT}} &&= 0,\nn\\
\Delta T_{\textrm{IT}} &&= \frac{1}{16 \pi s^2_W m^2_W}
\Big[\frac{m^2_{T^\pm}+m^2_{T^0}}{2}+\frac{m^2_{T^\pm}\,m^2_{T^0}}{m^2_{T^\pm}-m^2_{T^0}}\mathrm{ln} \Big(\frac{m^2_{T^0}}{m^2_{T^{\pm}}}\Big) \Big], \nn\\
%F(m^2_{T^+},m^2_{T^0})\nn\\
&& \simeq \frac{(\Delta m)^2}{24 \pi s^2_W m^2_W} \quad \mathrm{for}\, \Delta m 
= m_{T^\pm}-m_{T^0} \ll m_{T^0},\nn\\
\Delta U_{\textrm{IT}} &&= -\frac{1}{3 \pi} \Big[m^4_{T^0} \mathrm{ln}\Big(\frac{m^2_{T^0}}{m^2_{T^{\pm}}}\Big) \frac{3 m^2_{T^{\pm}} - m^2_{T^0}}{(m^2_{T^0}-m^2_{T^{\pm}})^3} + \frac{5(m^4_{T^0} + m^4_{T^{\pm}}) - 22 m^2_{T^0} m^2_{T^{\pm}}}{6(m^2_{T^0} - m^2_{T^{\pm}})^2}\Big]\nn\\
&& \simeq \frac{\Delta m}{3 \pi m_{T^{\pm}}},
\eea
%%%%%%%%%%%%%%%%%%%%%%%%%%%%%%%%%%
where, $\Delta \mathcal{X}_a \equiv \mathcal{X}_a - \mathcal{X}_a^{\rm{SM}}$ for $\mathcal{X} \in (S,T,U)$, and $a={\textrm{IT}, {\textrm{rS}}}$, as already introduced in subsection \ref{subsec:EWPO}. 
In the expression of $\Delta U_{\textrm{IT}}$, a term $\mathcal{O}\sim (m_Z/m_{T^\pm})$ is dropped \cite{Forshaw:2001xq}. The angle $\theta_W$ is Weinberg angle \cite{ParticleDataGroup:2024cfk}. The quantity $\Delta S_{{\textrm{IT}}}$ is zero. 
With a heavy $m_{T^0}$, i.e., $\gsim 1$ TeV, $\Delta U_{{\textrm{IT}}}$ also appears tiny, since $\Delta m \approx 166$ MeV \cite{Cirelli:2005uq}.  Finally, with $m_W=80.3692$ GeV \cite{ParticleDataGroup:2024cfk} and $\Delta m \approx 166$ MeV \cite{Cirelli:2005uq}, $\Delta T_{{\textrm{IT}}}$ also appears insignificant. A small but non-zero value for the $T$-parameter is related to the one-loop corrections to the $\rho$ parameter which otherwise agrees with the SM value as the triplet procures no VEV at the zero temperature.

Similar corrections for the oblique parameters also appear with the inclusion of the real singlet scalar $S$. Following Refs. \cite{Baek:2011aa,Beniwal:2018hyi}, those are written as 
%%%%%%%%%%%%%%%%%%%%%%%%%%%%%%%%%%%%%%%%%
\bea\label{eq:STU-realScalar}
\Delta S_{\textrm{rS}} &&= \frac{\sin^2\theta}{2 \pi} \Big[f_S\Big(\frac{m^2_{h_2}}{m^2_Z}\Big) - f_S\Big(\frac{m^2_{h_1}}{m^2_Z}\Big)\Big],\nn\\
\Delta T_{\textrm{rS}} &&= \frac{3\sin^2\theta}{16 \pi s^2_W} \Big[ f_T\Big(\frac{m^2_{h_2}}{m^2_W}\Big) - f_T\Big(\frac{m^2_{h_1}}{m^2_W}\Big) -\frac{1}{c^2_W}\Big\{ f_T\Big(\frac{m^2_{h_2}}{m^2_Z}\Big) - f_T\Big(\frac{m^2_{h_1}}{m^2_Z}\Big)\Big\}\Big],\nn\\
%\Delta U_{\textrm{rS}} &&= \frac{\sin^2\theta}{2 \pi} \Big[f_S\Big(\frac{m^2_{h_2}}{m^2_W}\Big) - f_S\Big(\frac{m^2_{h_2}}{m^2_Z}\Big) - \Big\{f_S\Big(\frac{m^2_{h_1}}{m^2_W}\Big) - f_S\Big(\frac{m^2_{h_1}}{m^2_Z}\Big) \Big\}\Big],
\Delta U_{\textrm{rS}} &&= \frac{\sin^2\theta}{2 \pi} \Big[f_S\Big(\frac{m^2_{h_2}}{m^2_W}\Big) - f_S\Big(\frac{m^2_{h_1}}{m^2_W}\Big)\Big] - \Delta S_{\textrm{rS}},
\eea
%%%%%%%%%%%%%%%%%%%%%%%%%%%%%%%%%%%%%%%%%
where,
%%%%%%%%%%%%%%%%%%%%%%%%%%%%%%%%%%%%%%%%%
\begingroup
\allowdisplaybreaks
\begin{align}
	f_T(x) &= \frac{x\log x}{x-1}, \\[1.5mm]
	f_S(x) &= 
	\begin{dcases}
		\frac{1}{12} \left[ -2 x^2 + 9 x + \left( x^2 - 6x - \frac{18}{x-1} + 18 \right) x \log x \right. \\ 
		\left. + 2 \sqrt{(4-x) x} \left(x^2-4 x+12\right)    
		\left( \tanh^{-1} \frac{\sqrt{x}}{\sqrt{x-4}} - \tanh^{-1} \frac{\sqrt{x-2}}{\sqrt{(x-4)x}}   \right) \right], \quad 0 < x < 4, \\
		\frac{1}{12} \left[-2 x^2 + 9 x + \left(x^2 - 6x - \frac{18}{x-1} + 18 \right) x \log x \right. \\
		\left. + \sqrt{(x-4) x} \left(x^2 - 4x + 12\right)  \log \frac{1}{2} \left( {x - \sqrt{(x-4)x} - 2} \right)\right], \quad x \geq 4. 
	\end{dcases} 
\end{align}
\endgroup
%%%%%%%%%%%%%%%%%%%%%%%%%%%%%%%%%%%%%%%%%
Unlike Eq. (\ref{eq:STU-inertTriplet}), for the real scalar singlet it is not apparent
that corrections to the oblique parameters are insignificant. However, one can
recast all three contributions of Eq. (\ref{eq:STU-realScalar}) in a compact form as
%%%%%%%%%%%%%%%%%%%%%%%%%%%%%%%%%%%%%%%%
\bea
\Delta \mathcal{X}_{\textrm{rS}} = \sin^2\theta \Big[f_2(m_{h_2}) - f_1(m_{h_1})\Big],
%\Delta \mathcal{X} = (1 - \cos^2\th) \Big[\mathcal{O}_{\rm SM}(m_{h_2}) - \mathcal{O}_{\rm SM}(m_{h_1})\Big].
\eea
%%%%%%%%%%%%%%%%%%%%%%%%%%%%%%%%%%%%%%%%
where, $f_1,\, f_2$ are nothing but the generalized compact form of various $f_T,\, f_S$
functions. So vanishingly small values for all three $\Delta \mathcal{X}_{\textrm{rS}}$
needs either $m_{h_2}\approx m_{h_1}$, i.e., maximal mixing scenario with $\theta \rightarrow\pi$ or $m_{h_2} \gg m_{h_1}$, i.e., minimal mixing scenario with $\theta\rightarrow 0$. For more details, see the Refs. \cite{Baek:2011aa,Beniwal:2018hyi}, and for a recent discussion and constraints on $(\th, m_{h_2})$ in the context of EWPOs, see Ref. \cite{Ellis:2022lft}.
%%%%%%%%%%%%%%%%%%%%%%%%%%%%%%%%%%%%%%%%%%%%%
%%%%%%%%%%%%%%%%%%%%%%%%%%%%%%%%%%%%%%%%%%%%%%%%%%%%%%%%%%%%%%%%%%%%%%%%%%%%%%%%%%
\section{Field dependent and thermal masses}\label{appx:appendixA}
%%%%%%%%%%%%%%%%%%%%%%%%%%%%%%%%%%%%%%%%%%%%%%%%%%%%%%%%%%%%%%%%%%%%%%%%%%%%%%%%%%
\subsection{Field dependent masses}\label{appen:C1}
%%%%%%%%%%%%%%%%%%%%%%%%%%%%%%%%%%%%%%%%%%%%%%%%%%%%%%%%%%%%%%%%%%%%%%%%%%%%%%%%%%
 The field-dependent mass matrices are obtained from the tree-level effective potential $V_{\rm{eff}(\varphi, T=0)}\equiv V_0(\varphi)$ (see Eqs. (\ref{eq:pot:Eff-pot}) and (\ref{eq:pot:tree-level})) at $T=0$,
%%%%%%%%%%%%%%%%%%%%%%%%%%%%%%%%%%%%%%%%%%%%
\bea
\widetilde{\mathcal{M}}^2_{ij}(\varphi) = \frac{\partial^2 V_0(\varphi)}{\partial \varphi_i \partial \varphi_j}.
\eea
%%%%%%%%%%%%%%%%%%%%%%%%%%%%%%%%%%%%%%%%%%%%
In the basis $\varphi = (\varphi_1,\,\varphi_2,\,\varphi_3) \equiv \{h, s, T^0\}$, the elements of the $3 \times 3$ scalar squared mass matrix are,
%%%%%%%%%%%%%%%%%%%%%%%%%%%%%%%%%%%%%%%%%%%%
\begingroup
\allowdisplaybreaks
\bea
\label{eq:appx:field-dependent-scalar-neutral}
\widetilde{\mathcal{M}}^2_{11} &=& -\mu_{H}^2 + \mu_{HS} s + 3 \l_{H} h^2 + \frac{1}{2} \l_{SH} s^2 + \frac{1}{2} \l_{HT} {T^0}^2, \nn \\
\widetilde{\mathcal{M}}^2_{12} &=& \l_{SH} s h + \mu_{HS} h, \nn \\
\widetilde{\mathcal{M}}^2_{13} &=& \l_{HT} h T^0, \nn \\
\widetilde{\mathcal{M}}^2_{21} &=& \widetilde{\mathcal{M}}^2_{12}, \nn \\
\widetilde{\mathcal{M}}^2_{22} &=& -\mu_S^2 - 2 \mu_3 s + 3 \l_S s^2 + \frac{1}{2} \l_{SH} h^2 + \frac{1}{2} \l_{ST} {T^0}^2, \nn \\
\widetilde{\mathcal{M}}^2_{23} &=& \l_{ST} s T^0 + \mu_{ST} T^0 ,\nn \\
\widetilde{\mathcal{M}}^2_{31} &=& \widetilde{\mathcal{M}}^2_{13}, \nn \\
\widetilde{\mathcal{M}}^2_{32} &=& \widetilde{\mathcal{M}}^2_{23}, \nn \\
\widetilde{\mathcal{M}}^2_{33} &=& -\mu_{\bm{T}}^2 + \mu_{ST} s + 3 \l_{\bm{T}} {T^0}^2 + \frac{1}{2} \l_{ST} s^2 + \frac{1}{2} \l_{HT} h^2.
\eea
\endgroup
%%%%%%%%%%%%%%%%%%%%%%%%%%%%%%%%%%%%%%%%%%%%
At the zero temperature physical vacuum, i.e., $\{v, v_s, 0\}$, the above $3 \times 3$ mass matrix reduces to a $2 \times 2$ matrix in the $\{h,s\}$ basis and reproduces the mass matrix $\mathcal{M}^2$ as mentioned in Eq. (\ref{eq:scalar-mass-HS-new}). Besides, using the corresponding tadpole equations, one can show $\widetilde{\mathcal{M}}^2_{11} = \mathcal{M}^2_{hh}$, $\widetilde{\mathcal{M}}^2_{12} = \mathcal{M}^2_{hs}$ and $\widetilde{\mathcal{M}}^2_{22} = \mathcal{M}^2_{ss}$, as already
shown in Eq.~(\ref{eq:mass-elements}). With a vanishing triplet VEV, the element $\widetilde{\mathcal{M}}^2_{33}$ decouples from the said $3\times 3$ matrix and corresponds to $m^2_{T^0}$, as indicated in Eq.~(\ref{eq:mass-triplet}). The field dependent masses for the scalars, i.e., $m_{h_1}(\varphi), m_{h_2}(\varphi),~{\rm and}~ m_{T^0}(\varphi)$ are obtained from the eigenvalues of the above $3 \times 3$ matrix. However, the closed-form expressions are cumbersome and lengthy, so we have opted not to present them here. Instead, we calculate the corresponding eigenvalues of this scalar squared matrix numerically using {\tt Python} routines. The field-dependent masses of the charged triplet scalar (${T^{\pm}}$) and the SM Goldstone bosons ($G^{0,\pm}$) are given as
%%%%%%%%%%%%%%%%%%%%%%%%%%%%%%%%%%%%%%%%%%%
\bea
\label{eq:appx:field-dependent-scalar-charged}
m^2_{T^{\pm}} (\varphi) &=& -\mu_{\bm{T}}^2 + \mu_{ST} s + \l_{\bm{T}} {T^0}^2 + \frac{1}{2} \l_{ST} s^2 + \frac{1}{2} \l_{HT} h^2, \nn \\
m^2_{G^{0,\pm}} (\varphi) &=& -\mu_{H}^2 + \mu_{HS} s + \l_{H} h^2 + \frac{1}{2} \l_{SH} s^2 + \frac{1}{2} \l_{HT} {T^0}^2,
\eea
%%%%%%%%%%%%%%%%%%%%%%%%%%%%%%%%%%%%%%%%%%%
Note that, with a vanishing triplet VEV, we have $\widetilde{\mathcal{M}}^2_{33} \equiv m^2_{T^0} = m^2_{T^\pm}$, at the tree-level, as in Eq.~(\ref{eq:mass-triplet}).
Finally, the field-dependent fermion DM ($\psi$) mass is,
%%%%%%%%%%%%%%%%%%%%%%%%%%%%%%%%%%%%%%%%%%%
\bea
m_{\psi} (\varphi) = \mu_{\psi} + g_S s.
\eea
%%%%%%%%%%%%%%%%%%%%%%%%%%%%%%%%%%%%%%%%%%%
Among the SM fermions, the top quark contribution dominates over others, owing to the largest Yukawa coupling ($y_t$). Therefore, in this work, we only consider the contribution of the top quark and neglect others, and the field-dependent mass of the same is given as,
%%%%%%%%%%%%%%%%%%%%%%%%%%%%%%%%%%%%%%%%%%%%
\bea
m^2_t(\varphi) = \frac{y_t^2}{2} h^2.
\eea
%%%%%%%%%%%%%%%%%%%%%%%%%%%%%%%%%%%%%%%%%%%%
On the other hand, the field-dependent masses of the EW gauge bosons are \cite{Niemi:2020hto}
%%%%%%%%%%%%%%%%%%%%%%%%%%%%%%%%%%%%%%%%%%%%
\bea
m^2_W(\varphi) = \frac{g_2^2}{4} (h^2+4 {T^0}^2), ~~m^2_Z(\varphi) = \frac{g_1^2 + g_2^2}{4} h^2,
\eea
%%%%%%%%%%%%%%%%%%%%%%%%%%%%%%%%%%%%%%%%%%%%
with $g_1,g_2$ being the $U(1)_Y,~SU(2)_L$ gauge couplings, respectively.
%%%%%%%%%%%%%%%%%%%%%%%%%%%%%%%%%%%%%%%%%%%%%%%%%%%%%%%%%%%%%%%%%%%%%%%%%%%%%%%%%%%
\subsection{Thermal masses}\label{appen:C2}
%%%%%%%%%%%%%%%%%%%%%%%%%%%%%%%%%%%%%%%%%%%%%%%%%%%%%%%%%%%%%%%%%%%%%%%%%%%%%%%%%%%
As previously mentioned in subsection \ref{subsec:one-loop-Veff}, the resummation of the leading self-energy Daisy diagrams shifts the field-dependent masses,
%%%%%%%%%%%%%%%%%%%%%%%%%%%%%%%%%%%%%%%%%%%%%%%
\bea\label{eq:field_mass:daisy-coeff}
m_i^2 (\varphi, T) = m^2_i(\varphi) + \Pi_i (T), ~~ \Pi_i (T)=c_i T^2,
\eea
%%%%%%%%%%%%%%%%%%%%%%%%%%%%%%%%%%%%%%%%%%%%%%%
where, $\Pi_i(T)$ are the thermal masses for the $i^{th}$ bosons with $c_i$ representing the Daisy coefficients \cite{Dolan:1973qd, Kirzhnits:1976ts, Parwani:1991gq, Espinosa:1992gq, Arnold:1992rz, Croon:2020cgk, Schicho:2021gca}. The non-zero thermal masses in this model are given by
%%%%%%%%%%%%%%%%%%%%%%%%%%%%%%%%%%%%%%%%%%%%%%%
\bea
\label{eq:appx:thermal-masses}
\Pi_{h_1}(T) &=& \left(\frac{g_1^2}{16} + \frac{3g_2^2}{16} + \frac{y_t^2}{4} + \frac{\l_{H}}{2} + \frac{\l_{SH}}{24} + \frac{\l_{HT}}{8}\right)~T^2, \nn \\
\Pi_{h_2}(T) &=& \left(\frac{g_S^2}{6} + \frac{\l_{S}}{4} + \frac{\l_{SH}}{6} + \frac{\l_{ST}}{4}\right)~T^2, \nn \\
\Pi_{T^0}(T) &=& \left(\frac{\l_{HT}}{6} + \frac{\l_{ST}}{12} + \frac{5 \l_{\bm{T}}}{12}\right)~T^2, ~~{\rm and}, \nn \\
\Pi_{G^0}(T) &=& \Pi_{G^{\pm}}(T) = \Pi_{h_1}(T), ~\Pi_{T^{\pm}}(T) = \Pi_{T^0}(T).
\eea
%%%%%%%%%%%%%%%%%%%%%%%%%%%%%%%%%%%%%%%%%%%%%%%
The terms inside the parentheses in Eq.~(\ref{eq:appx:thermal-masses}), are the corresponding Daisy coefficients, as depicted in Eq. (\ref{eq:field_mass:daisy-coeff}). Finally, at a non-zero temperature, $T \neq 0$, the longitudinal modes of the $W$-boson, $Z$-boson and photon, $\gamma$, also receive thermal mass corrections. For $W$-boson, the thermal correction reads \cite{Niemi:2018asa},
%%%%%%%%%%%%%%%%%%%%%%%%%%%%%%%%%%%%%%%%%%%%%%%%%
\bea
m_W^2 (\varphi, T) = m_W^2 (\varphi) + \Pi_W (T), ~~~\Pi_W (T) = \frac{13}{6} g_2^2 T^2,
\eea
%%%%%%%%%%%%%%%%%%%%%%%%%%%%%%%%%%%%%%%%%%%%%%%%%
whereas, for the $Z$-boson and photon, $\gamma$, these corrections are obtained by the eigenvalues of the squared mass matrix provided below \cite{Niemi:2018asa,Beniwal:2018hyi,Niemi:2020hto}, 
%%%%%%%%%%%%%%%%%%%%%%%%%%%%%%%%%%%%%%%%%%%%%%
\bea
m^2_{Z/\gamma}(\varphi, T) =
\begin{pmatrix}
	~\frac{1}{4}g_2^2 h^2 + \frac{11}{6} g_2^2 T^2~ & ~ -\frac{1}{4} g_1 g_2 h^2 ~ \\
	~-\frac{1}{4} g_1 g_2 h^2 ~ & ~ \frac{1}{4}g_1^2 h^2 + \frac{11}{6} g_1^2 T^2~
\end{pmatrix}.
\eea
%%%%%%%%%%%%%%%%%%%%%%%%%%%%%%%%%%%%%%%%%%%%%%
%%%%%%%%%%%%%%%%%%%%%%%%%%%%%%%%%%%%%%%%%%%%%%%%%%%%%%%%%%%%%%%%%%%%%%%%%%%%%%%%%%
\section{Efficiency coefficients}\label{appx:efficiency-coeff}
%%%%%%%%%%%%%%%%%%%%%%%%%%%%%%%%%%%%%%%%%%%%%%%%%%%%%%%%%%%%%%%%%%%%%%%%%%%%%%%%%%
In this section, we comment on the efficiency coefficients appearing in the GW amplitudes of the three different GW production mechanisms, as mentioned in subsection \ref{subsec:GW-from-SFOPT}. First, we consider $\kappa_b$. In most general cases one usually approximates it as \cite{Kamionkowski:1993fg},
%%%%%%%%%%%%%%%%%%%%%%%%%%%%%%%%%%%%%%%%%%
\bea
\label{eq:eff-fact:kappa_b}
\kappa_b = \frac{1}{1+A\alpha} \left(A \alpha + \frac{4}{27} \sqrt{\frac{3 \alpha}{2}}\right),
\eea
%%%%%%%%%%%%%%%%%%%%%%%%%%%%%%%%%%%%%%%%%%
with $A = 0.715$ and $\alpha$ is already given by Eq. (\ref{eq:PT-GW:alpha}). However, one must take care while considering $\kappa_b$ in the fitted result of $\Omega_b h^2(f)$, as shown in Eq.~(\ref{eq:GW:bubble-collision-amp}). For example, the contribution of bubble collision to the total GW signal becomes important when energy deposited into the scalar field becomes significant and this occurs when there is a ``runaway'' PT in the plasma. This situation arises for $\alpha$ values larger than $\alpha_\infty$, where $\alpha_{\infty}$ marks the threshold value at which the walls of the scalar-field bubbles begin to ``runaway''. However, $\alpha_{\infty}$ is model dependent and can be derived using \cite{Espinosa:2010hh},
%%%%%%%%%%%%%%%%%%%%%%%%%%%%%%%%%%%%
\bea
\label{eq:eff-fact:alpha_inf}
\alpha_{\infty} \simeq \frac{30}{24 \pi^2 g_*(T) T_*^2} \left[\sum_{b} g_b \Delta m_b^2 - \frac{1}{2} \sum_{f} g_f \Delta m_f^2  \right],
\eea
%%%%%%%%%%%%%%%%%%%%%%%%%%%%%%%%%%%%
where $b,f$ represents bosons and fermions, respectively, with $g_b$ and $g_f$ corresponding to the numbers of {\it d.o.f.} for the species $b$ and $f$, respectively, and $\Delta m_i = m_i\big|_{\varphi_{\rm true}} - m_i\big|_{\varphi_{\rm false}}$ with $i = \{b,f\}$. For the case of an SFOPT with SM-like particle content, one finds, approximately \cite{Caprini:2015zlo},
%%%%%%%%%%%%%%%%%%%%%%%%%%%%%%%%%%%%
\bea
\label{eq:eff_fact:alpha_inf_appx}
\alpha_{\infty} \simeq 4.9 \times 10^{-3} \left(\frac{\varphi_*}{T_*}\right)^2,
\eea
%%%%%%%%%%%%%%%%%%%%%%%%%%%%%%%%%%%%
where, $\frac{\varphi_*}{T_*}$ denotes the PT strength at a reference temperature. One can safely use this approximation if the SM Higgs field is weakly coupled to the new physics sector. Otherwise, for the case of a runaway PT, the ratio of $\alpha_{\infty}$ and the actual $\alpha$ determines the efficiency factor, and is given by \cite{Espinosa:2010hh}
%%%%%%%%%%%%%%%%%%%%%%%%%%%%%%%%%%%%
\bea
\label{eq:eff-fact:kappa_b-final}
\kappa_b = 1 - \frac{\alpha_{\infty}}{\alpha}.
\eea
%%%%%%%%%%%%%%%%%%%%%%%%%%%%%%%%%%%%
It is commonly believed that the runaway PTs occur for $\alpha > \alpha_{\infty}$. However, recent findings \cite{Ellis:2020nnr,Hoche:2020ysm} indicate that the efficiency of converting vacuum energy into the scalar field is reduced by additional Lorentz factor ratios, making the efficiency lower than the previously estimated one. This results in most PTs, previously classified as runaway types, now being categorized as non-runaway types unless they are strongly supercooled, as confirmed in
Refs.~\cite{Alanne:2019bsm,Schmitz:2020syl}. Since, in the interested parameter region of our work, we did not find any supercooled transition, we can safely use the approximation of Eq.~(\ref{eq:eff-fact:kappa_b}) for $\kappa_b$, for further calculations.

Next, we briefly review the fits, as mentioned in Ref.~\cite{Espinosa:2010hh}, to the function $\kappa(\xi_w, \alpha_N)$ which resembles $\kappa_{\rm sw}$ as in Eq.~(\ref{eq:GW:soundwave-kappa_sw}) with $\xi_w, \alpha_N$ bearing similar meaning as the wall velocity and the PT strength, respectively. These fits provide easier use of the efficiency factors without solving the hydrodynamic flow equations with a precision $\sim\mathcal{O}(15\%)$ \cite{Espinosa:2010hh}, and one gets
%%%%%%%%%%%%%%%%%%%%%%%%%%%%%%%%%%%%%
\bea
\label{eq:eff-fact:kappa-sw-fit}
\kappa_{\rm sw}(v_w,\alpha) &&=
\begin{dcases}
	\frac{c_s^{11/5} \kappa_A \kappa_B}{(c_s^{11/5} - v_w^{11/5}) \kappa_B + v_w c_s^{6/5} \kappa_A}, ~~~{\rm if} \quad (v_w \lesssim c_s)\nn\\
	\kappa_B + (v_w - c_s) \delta k + \left(\frac{v_w - c_s}{v_J - c_s}\right)^3 \left[\kappa_C - \kappa_B - (v_J - c_s) \delta k\right], ~~{\rm if} \quad (c_s < v_b < v_J) \nn \\
	\frac{(v_J - 1)^3 v_J^{5/2} v_w^{-5/2} \kappa_C \kappa_D }{[(v_J -1)^3 - (v_w - 1)^3] v_J^{5/2} \kappa_C + (v_w - 1)^3 \kappa_D}, ~~{\rm if} \quad (v_J \lesssim v_w),
\end{dcases} 
\eea
%%%%%%%%%%%%%%%%%%%%%%%%%%%%%%%%%%%%%
where $c_s = 1/\sqrt{3}$, is the speed of sound in plasma, $v_J\equiv v_J(\alpha)$ is Chapman-Jouguet velocity defined in Eq.~(\ref{eq:percolation:Chapman-velo}), and $\delta k \simeq -0.9 \ln \frac{\sqrt{\alpha}}{1+\sqrt{\alpha}}$. For small wall velocities ($v_w \ll c_s$), one gets
%%%%%%%%%%%%%%%%%%%%%%%%%%%%%%%%%%%%%
\bea
\label{eq:eff-fact:kappa_A}
\kappa_A = v_w^{6/5} \frac{6.9 \alpha}{1.36 - 0.037 \sqrt{\alpha} + \alpha},
\eea
%%%%%%%%%%%%%%%%%%%%%%%%%%%%%%%%%%%%%
while, for a transition from the subsonic to supersonic deflagration, when $v_w = c_s$, gives
%%%%%%%%%%%%%%%%%%%%%%%%%%%%%%%%%%%%%
\bea
\label{eq:eff-fact:kappa_B}
\kappa_B = \frac{\alpha^{2/5}}{0.017 + (0.997 + \alpha)^{2/5}}.
\eea
%%%%%%%%%%%%%%%%%%%%%%%%%%%%%%%%%%%%%
In the case when $v_w = v_J\equiv v_J(\alpha)$, known as Jouguet detonations, and for very large bubble velocity ($v_w \rightarrow 1$), one gets
%%%%%%%%%%%%%%%%%%%%%%%%%%%%%%%%%%%%
\bea
\label{eq:eff-fact:kappab_CD}
\kappa_C = \frac{\sqrt{\alpha}}{0.135 + \sqrt{0.98 + \alpha}} \quad \rm{and}\quad \kappa_D = \frac{\alpha}{0.73 + 0.083 \alpha + \alpha}.
\eea
%%%%%%%%%%%%%%%%%%%%%%%%%%%%%%%%%%%%
%%%%%%%%%%%%%%%%%%%%%%%%%%%%%%%%%%%%%%%%%%%%%%%%%%%%%%%%%%%%%%%%%%%%%%%%%%%%%%%%%%
\section{A few Feynman diagrams for the DM annihilation processes}\label{appx:feynman-diagrams}
%%%%%%%%%%%%%%%%%%%%%%%%%%%%%%%%%%%%%%%%%%%%%%%%%%%%%%%%%%%%%%%%%%%%%%%%%%%%%%%%%%
%%%%%%%%%%%%%%%%%%%%%%%%%%%%%%%%%%%%%%%%%%%%%%%%%%%%%%%%
\begin{figure}[H]
	% \hspace*{0.2cm} %% This will shift the figure
	%\centering
	{\includegraphics[height=5.8cm,width=16.0cm]{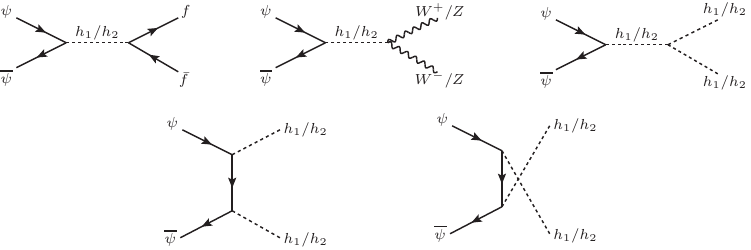}}
	\caption{Feynman diagrams of different annihilation channels for fermionic DM $\psi$, see subsection \ref{subsec:DM-pheno-relic-DD-ID} for more details.}
	\label{fig:feynman-diagram:psi-anni}
\end{figure}
%%%%%%%%%%%%%%%%%%%%%%%%%%%%%%%%%%%%%%%%%%%%%%%%%%
%%%%%%%%%%%%%%%%%%%%%%%%%%%%%%%%%%%%%%%%%%%%%%%%%%%%%%%%
\begin{figure}[!h]
	% \hspace*{0.2cm} %% This will shift the figure
	%\centering
	{\includegraphics[height=5.2cm,width=16.0cm]{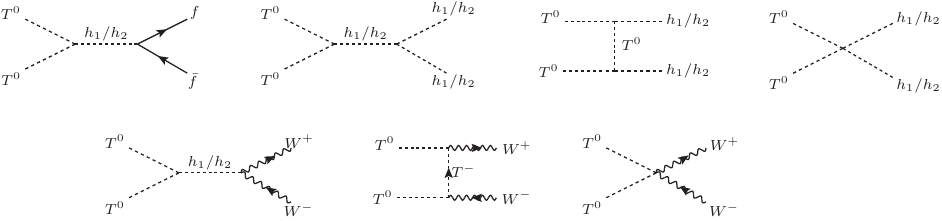}}
	\caption{Annihilation channels for triplet scalar DM $T^0$, see subsection \ref{subsec:DM-pheno-relic-DD-ID}. For the first and the last figures of the bottom row, replacing $W^+,\, W^-$ with $Z$s yields two more possible configurations. For the middle figure of the bottom row, a charge conjugate diagram exists with $T^+$. }
	\label{fig:feynman-diagram:Trip-anni}
\end{figure}
%%%%%%%%%%%%%%%%%%%%%%%%%%%%%%%%%%%%%%%%%%%%%%%%%%
%%%%%%%%%%%%%%%%%%%%%%%%%%%%%%%%%%%%%%%%%%%%%%%%%%%%%%%%
\begin{figure}[!h]
	\hspace*{0.2cm} %% This will shift the figure
	%\centering
	{\includegraphics[height=6.8cm,width=16.0cm]{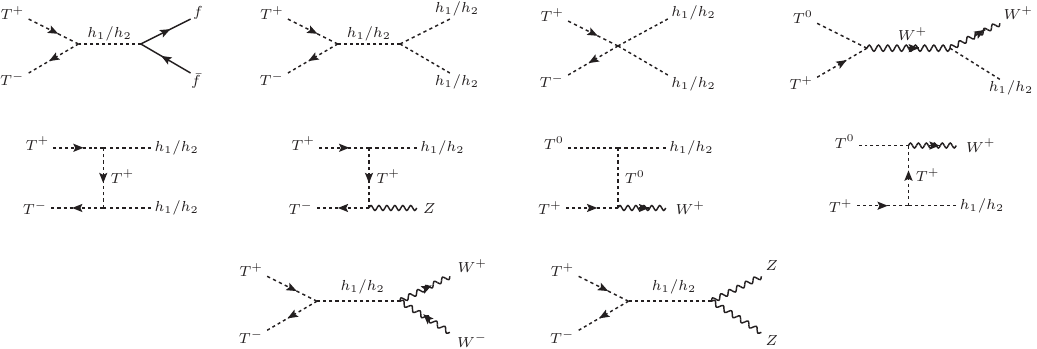}}
	\caption{A few coannihilation (annihilation) channels for the triplet scalar DM $T^0 (T^\pm)$, see subsection \ref{subsec:DM-pheno-relic-DD-ID} for details. Note that, charge conjugate diagrams exist with $T^\pm \to T^\mp$ and $W^+ \to W^-$. For the remaining processes that do not have any $h_i$ final states or are mediated by the same, see Ref.~\cite{DuttaBanik:2020jrj}.}
	\label{fig:feynman-diagram:Trip-coanni}
\end{figure}
%%%%%%%%%%%%%%%%%%%%%%%%%%%%%%%%%%%%%%%%%%%%%%%%%%
%%%%%%%%%%%%%%%%%%%%%%%%%%%%%%%%%%%%%%%%%%%%%%%%%%%%%%%%%%%%%%%%%%%%%%%%%%%%%%%	
%%%%%%%%%%%%%%%%%%%%%%%%%%%%%%%%%%%%%%%%%%
	\bibliographystyle{JHEP}
	\bibliography{refv1}  
%%%%%%%%%%%%%%%%%%%%%%%%%%%%%%%%%%%%%%%%%%
	
\end{document}